\documentclass[11pt]{elsarticle}

\usepackage{lipsum}
\usepackage{graphicx}
\usepackage{epstopdf}
\usepackage{amssymb}
\usepackage{amsmath}
\usepackage{amsfonts}
\usepackage{mathrsfs}
\usepackage{natbib}
\usepackage{xcolor}
\usepackage[draft]{changes}
\usepackage{algorithm}
\usepackage[noend]{algpseudocode}
\usepackage{float}
\usepackage{amssymb}
\usepackage{bigints}
\usepackage{mathtools}
\usepackage{bbm}
\usepackage{dsfont}
\usepackage{amsmath}

\usepackage{lineno}

\makeatletter
\def\ps@pprintTitle{%
 \let\@oddhead\@empty
 \let\@evenhead\@empty
 \def\@oddfoot{}%
 \let\@evenfoot\@oddfoot}
\makeatother

\begin{document}
\begin{frontmatter}

\title{Probabilistic Forecast of Multiphase Transport under Viscous and Buoyancy Forces in Heterogeneous Porous Media}

% \author[]{Farzaneh Rajabi}
% \author[]{Hamdi A. Tchelepi}
% \address{Energy Resources Engineering Department, Stanford University, Stanford, CA 94305, USA}

%\author[1]{Farzaneh \snm{Rajabi}\corref{cor1}}
\author[1]{Farzaneh {Rajabi}\corref{cor1}}
\cortext[cor1]{Corresponding author: 
frajabi@stanford.edu}
%% Third author's email
% \ead{frajabi@stanford.edu}
\author[1]{Hamdi A. {Tchelepi}}
% \fntext[fn1]{This is author footnote for second author.}  
%\author[2]{Given-name3 \snm{Surname3}}
%% Third author's email
% \ead{tchelepi@stanford.edu}
%\author[2]{Given-name4 \snm{Surname4}}

\address[1]{Department of Energy Resources Engineering, Stanford University, Stanford, CA 94305, USA}
%\address[2]{Affiliation 2, Address, City and Postal Code, Country}

\begin{abstract}
In this study, we develop a probabilistic approach to map the parametric uncertainty to the output state uncertainty in first-order hyperbolic conservation laws. We analyze this problem for nonlinear immiscible two-phase transport in heterogeneous porous media in the presence of a stochastic velocity field. The uncertainty in the velocity field can arise from the incomplete description of either porosity field, injection flux, or both. The uncertainty in the total-velocity field leads to the spatiotemporal uncertainty in the saturation field. Given information about the spatial/temporal statistics of the correlated heterogeneity, we leverage method of distributions to derive deterministic equations that govern the evolution of single-point CDF of saturation. Unlike Buckley Leverett equation, the equation for the raw CDF function is linear in space and time. Hereby, we give routes to circumventing the computational cost of Monte Carlo scheme while obtaining the full statistical description of saturation. We conduct a set of numerical experiments and compare statistics of saturation computed with the method of distributions, against those obtained using the statistical moment equations approach and kernel density estimation post-processing of high-resolution Monte Carlo simulations. This comparison demonstrates that the CDF equations remain accurate over a wide range of statistical properties, i.e. standard deviation and correlation length of the underlying random fields, while the corresponding low-order statistical moment equations significantly deviate from Monte Carlo results, unless for very small values of standard deviation and correlation length.
\end{abstract}


\end{frontmatter}


\section{Introduction}
%------------------------------- motivation-applications- why stochastic?
Long-term resource allocation, perspective planning, and risk mitigation of available energy resources rely on the accurate modeling of physical phenomena, including fluid flow and transport systems in the complex large-scale geological formations, e.g. oil reservoirs as well as carbon capture and sequestration equipment. Deterministic modeling of flow and transport in such systems has been widely studied for both single-phase and multiphase flow by \cite{fan2012fully},  \cite{yousefzadeh2020numerical}, \cite{yousefzadeh2017physics}. Rigorous modeling of these systems primarily depends on sophisticated strategies for uncertainty quantification and stochastic treatment of the system under study. Such an uncertainty is inherent to, and critical for any physical modeling, essentially due to the incomplete knowledge of state of the world, noisy observations, and limitations in systematically recasting physical processes in a suitable mathematical framework. These sources of uncertainty, render stochastic modeling of partial differential equations particularly crucial. To this end, accurate predictions of outputs (e.g. saturation fields) from reservoir simulations guarantee precise oil recovery forecasts. These quantitative predictions rely on the quality of the input measurements/data, such as the reservoir permeability and porosity fields as well as forcings, such as initial and boundary conditions. However, the available information about a particular geologic formation (e.g. from well logs and seismic data of an outcrop) is usually sparse and inaccurate compared to the size of the natural system and the complexity arising from multi-scale heterogeneity of the underlying system. Eventually, the uncertainty in the flow prediction can have a huge impact on the oil recovery. Therefore, we can no longer rely on a naive deterministic model, and we need to develop probabilistic flow predictions. 
Thereupon, because of inherent uncertainties in the input data, originating from sparsity or inaccuracy in the measurements, a stochastic framework is usually employed to describe their variability. Such a probabilistic framework is the key building block for making quantitative analysis of the uncertainty propagation from system inputs to the output state variable of interest. Full probability distribution of the random variables are then the essential metric to quantify the uncertainty in the outputs. Within this stochastic framework, predicting the saturation fields consists of determining the saturation probability distributions, by leveraging the knowledge of probabilistic distribution of the spatiotemporal uncertain random inputs.

% --------------------     single phase flow stochastic analysis using SME method
Probabilistic analysis of single-phase flow and transport has been explored and well-addressed by many researchers. Different approaches have been adopted for this purpose, among which earlier works of \cite{gelhar1993stochastic}, and \cite{rubin2003applied} have presented a comprehensive study for the stochastic hydrology. Such stochastic frameworks quantify the associated uncertainty of the previously studied deterministic models. 
Later works have focused on deriving deterministic equations for leading statistical moments of the system states, called Statistical Moments Equations (SME) method. This approach employed for single-phase contaminant transport has been investigated by \cite{kitanidis1988prediction}, \cite{graham1989stochastic}, \cite{winter2003moment}, \cite{li2005conditional}, \cite{dongxiao1999moment}. 
Even though this scheme provides the first two statistical moments (ensemble mean and variance) to forecast a system's average response, they fail to predict the probability of rare events, which is critical for risk assessment. 

%-----------------------   single phase flow stochastic analysis using PDF method
This deficiency in predictive capabilities of SME methods can be tackled by using the probabilistic method of distributions, i.e. probability density function (PDF) or cumulative density function (CDF) methods. The scheme was first adopted in the turbulence flow literature by \cite{pope1985pdf}. The PDF/CDF methods provide a full probabilistic distribution by deriving deterministic equations for PDF or CDF of the system states. PDF methods, sometimes require dealing with closures that are problem dependent and sometimes not conveniently attainable. This method has been utilized for the particle-base single-phase transport of stochastic total-velocity field using a Markovian approximation by \cite{meyer2010particle}, \cite{meyer2010joint}, and \cite{meyer2013fast}. A More recent study by \cite{yang2019probabilistic} has presented a comparison between CDF method, moment equations (SME) approach, and Monte Carlo scheme. There is a key advantage in using CDF method over their PDF counterparts, as formulation of boundary conditions for PDF method is not unique, and the closure problems might require burdensome manipulation. In contrast, the boundary conditions for CDF scheme tend to be uniquely formulated.

%---multiphase flow stochastic analysis (streamline, SME, stochastic Galerkin, Monte Carlo, PDf method)
Stochastic treatment of immiscible multiphase transport i.e. Buckley Leverett model in highly heterogeneous media, poses additional challenges compared to their single-phase analogue. This originates from nonlinearity of the transport equation and coupling of the total-velocity, pressure and saturation fields. The nonlinearity stems from the fact that the uncertain velocity field depends on the system state itself. Previous studies focused on using streamline-base strategies for solving flow (pressure and velocity), and transport (saturation) problems, i.e. Buckley Leverett equation, and provided the statistics for the travel-time metric. Several studies in this context were conducted by \cite{ibrahima2015distribution} for single-point distribution of one-dimensional highly-heterogeneous porous media, in which the cumulative distribution function obtained from the streamline-based simulation approach was presented to be more efficient than Monte Carlo modeling of the nonlinear transport equation. Moreover, \cite{ibrahima2018efficient} investigated single-point multi-dimensional heterogeneous formations by using the distribution scheme called frozen (time independent) streamline method (FROST). Another study by \cite{ibrahima2017multipoint} extended the single-point CDF approach to multi-point distribution of saturation in two-phase transport, where they have computed the covariance function of saturation in  order to build more robust confidence intervals. Another study by \cite{fuks2019analysis} extended the argument of \cite{ibrahima2015distribution} to estimate the statistics of fluid saturations in the channelized porous systems by using the frozen streamline assumption of FROST scheme.
The SME methods were extended to be used in the multiphase flow problems as well. Numerous studies were performed by \cite{zhang2000stochastic} for a Lagrangian treatment of moment equations subject to a stochastic permeability field, and also for the Eulerian moment equations by \cite{jarman2003eulerian}, for forward and inverse modeling by \cite{likanapaisal2010statistical}.
% , macro-dispersion for two-phase flow by \cite{langlo1994macrodispersion}, \cite{teodorovich2011stochastic}.
Method of distributions, discussed previously for the uncertainty quantification of single-phase flow, remains as one of the most robust and accurate approaches when extended to multiphase flow models. Distribution methods provide full probabilistic description of the saturation field. \cite{wang2013cdf} proposed a framework to estimate one-point CDF of saturation for one-dimensional Buckley Leverett model, in which the total velocity field is random with a known given distribution, due to the uncertain total flux as a function of time. They derive a general deterministic equation for the single-point CDF of saturation for a horizontal reservoir, in which gravity segregation plays no role. In this framework, a deterministic equation is derived for a quantity called raw CDF, which conceptualizes the Heaviside function over the saturation domain per se. 
% Because of the spatiotemporal correlations of saturation, solving the Buckley Leverett equation requires estimating the probability distribution of saturation at multiple locations and times simultaneously. 

%------------------------------          polynomial chaos expansion
More theoretical approaches have used polynomial chaos expansion method for uncertainty  assessment of both single-phase and multiphase flow physical frameworks. These methods have attracted interest because of their faster predictive power compared to the more commonly used Mote Carlo scheme. 
% \cite{cinnella2010robust}, Ghanem and Dham \cite{ghanem1998stochastic} have proposed using this approach as an alternative to computationally expensive methods.
Stochastic spectral methods have been also utilized to estimate the statistics of the state variable with a direct probabilistic collocation method by \cite{li2009efficient}.
Besides the collocation methods, Pettersson and Tchelepi have recently applied stochastic Galerkin method to derive statistics of the first moment for saturation. The method is based on truncating the Karhunen-Loeve (KL) expansion of random processes. A major advantage of these methods is the fact that convergence analysis can be conducted. However, in practice, they suffer from two disadvantages; one is the curse of dimensionality, the other is the  fact that they require further sophistication for highly nonlinear problems as proposed by \cite{pettersson2014stochastic}.

%--------------------------------       Monte Carlo
Monte Carlo method has been widely used to solve stochastic differential equations describing two-phase immiscible flow in heterogeneous media. A large number of equiprobable realizations of the reservoir description serve as input for the corresponding deterministic governing equation. MC methods, even though suffering from slow convergence rate (inversely proportional to the square root of the number of realizations) of statistics that define dynamic quantities of interest, are one of the simplest methods to implement. PDF and CDF of the output variable from MC simulations is then computed by some statistical post-processing. The robustness and easy implementation makes them one of the first candidates to be chosen. Also their convergence is guaranteed by the law of large numbers.
Monte Carlo simulations can be accelerated using Multilevel Monte Carlo (MLMC) approach. \cite{muller2013multilevel} used MLMC to accelerate the uncertainty quantification of streamline simulation-based two-phase flow model with the stochastic underlying permeability field.

%------------------------------------- deterministic gravity segregation    
Numerous studies including \cite{tchelepi1995viscous}, \cite{li2013influence}, \cite{li2014unconditionally}, \cite{li2015nonlinear}, \cite{kwok2008convergence}, \cite{wang2013trust}, \cite{zaleski2017model}, \cite{lie2019introduction}, \cite{brenier1991upstream} have investigated deterministic behavior of buoyancy forces interacting with viscous and capillary forces in multiphase hyperbolic transport models.

%----------------------------------   what we are doing in this work:
Moreover, while uncertainty assessment of the hyperbolic conservation laws (kinematic wave models) has been explored in several studies by \cite{cheng2019efficient}, \cite{zaleski2017model} and \cite{leveque2002finite}, to the best of our knowledge, no study has been conducted on evaluating the single-point distribution of saturation for stochastic nonlinear first-order hyperbolic conservation laws in two-phase flow models, in which buoyancy forces have been taken into account. Our study extends the idea of \cite{wang2013cdf} to propose an efficient distribution method for estimating the single-point cumulative distribution function (CDF) of saturation for the Buckley Leverett model subject to the gravitational forces. More broadly, we have investigated the horizontal, updip and downdip flooding as well as pure gravity segregation scenarios.
This framework results in a full probabilistic description of the saturation field for different physical set ups. While high heterogeneity is traditionally regarded as an obstacle to designing stochastic methods, this distribution method exploits the physics of highly heterogeneous porous formation. The method identifies the underlying random fields that explain the saturation distribution, together with the nonlinear mapping from these random fields to the saturation domain. We show that the statistics and distributions of these random fields can be estimated very efficiently and simply, leading to a method that is computationally inexpensive compared to the exhausting Monte Carlo (MC) simulations.

%--------------------------------------- structure of the paper
In section \ref{section: problemformulation}, we review the governing equations for the Buckley Leverett model, and explain the differences in flux functions of horizontal vs. gravitational multiphase case studies. We then explain the CDF method in section \ref{section: Single-point CDF equation for Saturation}, and derive a general deterministic equation for the (single-point) CDF of saturation for the case studies with gravitational forces, subject to a random porosity field and/or random injection flux.
In section \ref{section : Numerical experiments}, we illustrate the numerical aspects of the nonlinear hyperbolic laws with three different approaches, Monte Carlo (MC), method of distributions (MD) and statistical moment equation (SME) approach. We perform the Monte Carlo simulation for the Riemann problem of nonlinear BL equation using the mass conservative finite volume Godunov scheme, and then compare its solution for several test cases against those of our MD method and the low-order SME approximation, in order to validate our MD framework. A more comprehensive analysis of low-order approximation for different random quantities is offered in section \ref{section: LOA}. We conclude this work by analyzing the convergence and accuracy of our method in sections \ref{section: convergence},\ref{section:accuracy}, followed by the conclusion section in \ref{section:conclusion}.
%%%%%%%%%%%%%%%%%%%%%%%%%%%%%%%%%%%%%%%%%%%%%%%%%%%%%%%%%%%%%%%%%%%%%%%%%%%%%%%%%%%%%%%
%                            PROBLEM FORMULATION
%%%%%%%%%%%%%%%%%%%%%%%%%%%%%%%%%%%%%%%%%%%%%%%%%%%%%%%%%%%%%%%%%%%%%%%%%%%%%%%%%%%%%%%

\section{Problem formulation} \label{section: problemformulation}
We consider the nonlinear, incompressible, two-phase (Darcy) flow in a heterogeneous reservoir with intrinsic heterogeneous permeability $K$ and spatially-varying porosity $\phi(x)$ fields, where there are two immiscible phases; a wetting phase (e.g. water) is injected to displace a non-wetting phase (e.g. oil) through an imbibition process such as  waterflooding. The key underlying physical mechanisms are viscous and gravitational forces, whereas capillarity, chemical reactions, diffusion, and changes of state are neglected. We are primarily interested in estimating saturation fields while quantifying the uncertainty associated with their description for horizontal domains as well as vertical reservoirs in which gravity segregation plays a major role throughout the displacement process. 

%---------------------- input parameters
The inevitable uncertainty in the response parameters stems from several sources of uncertainty in the given input information of the reservoir. These input specifications include initial saturations of the wetting and non-wetting phases ($S_{wi}$ and $S_{oi}$), boundary conditions for the wetting phase ($S_w(x=0,t)$ at the inlet of domain), parameters representing the fluid properties, such as wetting and non-wetting phase viscosities ($\mu_w$ and $\mu_o$), parameters describing rock properties, such as porosity and permeability ($\phi(x)$ and $K(x)$), parameters outlining the fluid-rock interactions, such as relative permeabilities ($K_{rw}(S_w)$, $K_{ro}(S_o)$), as well as the injection and production pressures of the wetting phase ($P_{inj}$ and $P_{prod}$), where they are all assumed to be known parameters of the problem under study. 

%----------------------- what is random
An ideal prediction scenario calls for the complete knowledge of the aforementioned parameters. However, data sparsity and heterogeneity of the input definitions render such a perfect framework unattainable. The primary justification to consider the inputs as stochastic variables is lack of knowledge, i.e. having limited measurements which are only available at well locations, for the static properties (porosity, permeability) as well as dynamic properties (saturation, injection/production rates) of the reservoir. The uncertainty can also arise from random initial/boundary conditions, as well as fluid-rock interactions. Unlike rock properties that are inherent to the specific subsurface of evaluation, and therefore may not be quantified deterministically, the fluid properties are usually accessible through lab experiments and are often taken as deterministic values. To this end, we focus on the uncertainty in porosity field $\phi(x)$ and total Darcy flux $q(t)$ (the coefficients of the PDE are stochastic) in this study, whereas the rest are assumed to be deterministically known parameters. More precisely, we describe these random parameters using a prescribed PDF, i.e. by assigning a specific probabilistic description $p_{\varphi}(\phi)$ and $p_q(Q)$ to them, fully characterized by their known mean ($\mu$), standard deviation $(\sigma^2)$, and correlation structure $(Cr)$. Once such a probabilistic setting for the input parameters is established, we can propagate the uncertainty from inputs to output quantities of interest, essentially by solving a coupled system of nonlinear transport equation for saturation which satisfies mass conservation, along with the Darcy equation for pressure field. This comprehensive framework enables us to identify a probabilistic spatio-temporal description for the water saturation.

The nonlinear hyperbolic conservation law with a non-convex flux is conventionally studied by employing the Buckley Leverett (BL) equation which is the result of combining mass balance for each phase $\alpha=w,o$ with Darcy's law by \cite{buckley1942mechanism}. 
%------------------------------ starting a "Brief" Derivation of BL
The two-phase extension of the single-phase Darcy's law for wetting and non-wetting phases leads to,

\begin{align} \label{eq:Darcy for uw uo}
	& \textbf{u}_w = -\dfrac{k k_{rw}}{\mu_w} \nabla\varphi_w = -\dfrac{k k_{rw}}{\mu_w}\nabla(P_w + \rho_w g h) = -\dfrac{k k_{rw}}{\mu_w} \nabla (P_o - P_c + \rho_w g h)\\
	& \textbf{u}_o = -\dfrac{k k_{ro}}{\mu_o} \nabla\varphi_o = -\dfrac{k k_{ro}}{\mu_o}\nabla(P_o + \rho_o g h) = -\dfrac{k k_{ro}}{\mu_o} \nabla (P_o + \rho_o g h)
\end{align}

Where capillary pressure relating the pressures of two phases is $P_c(S_w) = P_o - P_w$. In the above-mentioned equations, $\textbf{u}_{\alpha} (\alpha=\{w,o\})$ are the phase Darcy velocities of the wetting (water) and nonwetting (oil) phases. $k_{r\alpha}$, $\mu_{\alpha}$ and $\rho_{\alpha}$ $(\alpha=\{w,o\})$ represent relative permeability, viscosity and density of each phase, respectively. In the case of vertical reservoirs, $h$ is the vertical distance travelled by each phase.
Applying mass conservation on a control volume results in the transport equation (saturation equation),

\begin{align}
&\phi(\textbf{x}) \dfrac{\partial S_{\alpha}}{\partial t} +\nabla\cdot\textbf{u}_{\alpha} = 0\quad\text{for $\alpha=\{w,o\}$} \label{water_sat_before_fw}
% &\phi(\textbf{x}) \dfrac{\partial S_o}{\partial t} +\nabla\cdot\textbf{u}_o = 0
\end{align}

Where the phase saturations, $S_w$ and $S_o$ are the volume fraction of the pore space occupied by the corresponding fluid phase. Consequently, $S_w + S_o =1$ simplifies the two mass conservation equations for the wetting and nonwetting phases, to the incompressiblity condition of $\nabla\cdot\textbf{u}_T = 0$ for total Darcy flux. 
% Accordingly, four unknowns of the problem, i.e. pressures and saturations fo each phase, are completely defined 
Subsequently, we obtain an equation in terms of pressures,
%
% \begin{align}\label{qt_divided_by_A}
% & \textbf{u}_T = \dfrac{\textbf{q}_T}{A} = \textbf{u}_w + \textbf{u}_o = -\dfrac{k k_{rw}}{\mu_w} \nabla (P_o - P_c + \rho_w g h) -\dfrac{k k_{ro}}{\mu_o} \nabla (P_o + \rho_o g h)\nonumber\\
% & = -\lambda_T\nabla P_o + \lambda_w\nabla P_c - (\lambda_o P_o +\lambda_w P_w) g \nabla h
% \end{align}

\begin{align}\label{eq:u_t}
& \textbf{u}_T = \textbf{u}_w + \textbf{u}_o = -\lambda_T\nabla P_o + \lambda_w\nabla P_c - (\lambda_o P_o +\lambda_w P_w) g \nabla h
\end{align}

Where the relative mobilities are defined as $\lambda_{\alpha} = \dfrac{k k _{r\alpha}}{\mu_{\alpha}}$ $(\alpha=\{w,o\})$ for each phase and the total mobility reads $\lambda_T = \lambda_o + \lambda_w$. The three terms on the right-hand side of Eq.~(\ref{eq:u_t}) represent viscous, capillary and buoyancy forces, respectively.
% Imposing the incompressibility condition on this $\textbf{u}_T$ results in,
% %
% \begin{align}
% \nabla.(\lambda_T \nabla P_o - \lambda_w \nabla P_c + (\lambda_o \rho_o + \lambda_w \rho_w) g\nabla h )=0
% \end{align}
% %
The dimensionless flux function of the wetting phase, referred to as the fractional flow of water is defined as the wetting phase velocity divided by the total velocity,
%
% \begin{align}
% f_w = \dfrac{q_w}{q_w+q_o} = \dfrac{\textbf{u}_w}{\textbf{q}_T/A} = \dfrac{\lambda_w}{\lambda_T} \left[1 + \dfrac{\lambda_o A}{\textbf{q}_T} \left(\nabla P_c - \Delta \rho g \nabla h\right)\right]
% \end{align}

\begin{align}\label{eq:flux function uw/ut }
f_w = \dfrac{\textbf{u}_w}{\textbf{u}_T} = \dfrac{\lambda_w}{\lambda_T} \bigg(1 + \dfrac{\lambda_o}{\textbf{u}_T} \left(\nabla P_c - \Delta \rho g \nabla h\right)\bigg)
\end{align}

%-------------------------------- more details on flux function
% 
In the absence of capillarity, the two phase pressures are equal and identical to the global pressure. For such a scenario, we consider one-dimensional flow in an inclined reservoir with a constant dip angle $\theta$, where $0\leq \theta\leq 90$, ranging from 0 for the horizontal reservoirs to 1 for the vertical case studies, resulting in $\nabla h=sin \theta$. The dimensionless gravity number $N_g$ is defined as the ratio of buoyancy to viscous forces, 

\begin{align}
   N_g = \dfrac{k g (\rho_w - \rho_o)}{\mu_o \textbf{u}_T} 
\end{align}
 
Subsequently, defining the viscosity ratio $m$ as $ m = \dfrac{\mu_w}{\mu_o}$, while employing the definitions for mobilities, the fractional flow of water can be recast in the following form,

\begin{align}\label{fractional flow equation}
    f_w(S_w, \theta) = \dfrac{k_{rw}}{k_{rw}+ m k_{ro}}\bigg(1- N_g k_{ro} sin\theta \bigg)
\end{align}

Rewriting the mass conservation of water saturation (Eq.\eqref{water_sat_before_fw}) using the fractional flow $f_w$, while  applying the chain rule and incompressibility condition leads to,
%
% \begin{align}
% &\phi(\textbf{x}) \dfrac{\partial S_w}{\partial t} +\nabla\cdot(\textbf{u}_w) = \phi(\textbf{x}) \dfrac{\partial S_w}{\partial t} +\nabla\cdot(f_w \textbf{u}_T) = 0 
% \end{align}
%
% Applying the chain rule and incompressibility condition leads to,
%
% \begin{align}
%  \dfrac{\partial S_w}{\partial t} + (\dfrac{\textbf{u}_T}{\phi(\textbf{x}) })\nabla f_w = 0
% \end{align}

\begin{align}\label{sat_after_fw}
\dfrac{\partial S_w}{\partial t} + \textbf{v}_T(x,t)\cdot\nabla f_w(S_w) = 0
\end{align}

Where $\textbf{v}_T(x,t)$ is defined as the interstitial velocity (total seepage velocity field) $\textbf{v}_T(\textbf{x},t) = \dfrac{q_T(t)}{\phi(\textbf{x})}$. Where $q_T$ is the total volumetric flow rate, obtained by multiplying the total Darcy velocity by the area A (A is assumed to be 1 in this work). 
we assume the total flux is time-dependent. This is a feasible assumption because incompressibility condition imposes $q_T(x,t)$ to be x-independent and hence equal to the injection flow rate at the inlet boundary which could be time-dependent. That is,  $q_{T}(x,t)= q_{T}(x=0,t) = q_{T_{inlet}}(t)$. In this study, we assume the total velocity field $\textbf{v}_T(x,t)$ is a random variable, where its uncertainty arises from two different sources/physical phenomena; the spatially-varying porosity field is uncertain in space, whereas the temporally-varying total flux is a random variable in time. We will investigate scenarios in which either or both of these phenomena are taken to be stochastic. \\
It should be noted that fractional flow $f(S_w)$ is a continuous smooth, and hence differentiable function. To this end, $\nabla f_w(S_w) = f_w^{\prime}(S_w)\nabla S_w$.
% Also, by defining $\textbf{v}(S_w) :=  (\dfrac{\textbf{u}_T}{\phi(\textbf{x})}) f^\prime(S_w) = \textbf{v}_T f^\prime(S_w) $,
Thereupon, supplemented with the following initial/boundary conditions, Buckley Leverett equation defines an IBVP for water saturation which reads,

\begin{align}\label{eq:saturation_equation_with_BC}
& \dfrac{\partial S_w}{\partial t} + \textbf{v}_T(x,t)f^{\prime}_w(S_w)\cdot\nabla S_w = 0,\qquad \qquad \text{where}\quad \textbf{v}_T = \dfrac{q_T(t)}{\phi(x)}\nonumber\\
& S_w(\textbf{x},t) = S_B = 1- S_{oi}, \quad \textbf{x} \in \Gamma_i, t>0,\nonumber\\
& S_w(\textbf{x},t=0) = S_{wi}, \quad \textbf{x}\in \Omega
\end{align}

Where $S_{wi}$ is the initial water saturation in the heterogeneous reservoir domain $\Omega$. In other words, the reservoir is assumed to be initially oil-saturated with a small irreducible water saturation $S_{wi}$. $S_B$ is water saturation at the injection boundary $\Gamma_i$, i.e. $x=0$, where the wetting phase is being injected. $S_{oi}$ is the irreducible oil saturation. $\Gamma_o$ is the boundary along which the flow outlet happens. Although initial/boundary conditions may be in general random variables, we assume they are both deterministic constants. Also, in order to guarantee self-similarity in solutions of saturation, we impose $S_{wi}$ and $S_{B}$ as uniform in space and time-independent quantities. For the test cases with gravitational force included, the initial condition is non-uniform. Consequently, this equation admits analytical solutions in terms of $x/t$. Moreover, the following Dirichlet boundary conditions for the pressure field (pressure control) is considered,

\begin{align}\label{pressure_BC}
& P(\textbf{x}) = p_{inj},\qquad \textbf{x}\in\Gamma_i\nonumber\\
& P(\textbf{x}) = p_{prod},\qquad \textbf{x}\in\Gamma_o
\end{align}
%------------------------------------------ END of Derivation of BL

Therefore, predictions of two-phase flow entails solving the first-order nonlinear hyperbolic Buckley Leverett equation with a nonconvex flux (the flux function changes its convexity within the given saturation interval). Solving the BL equation provides the distribution of saturation as a function of space and time. The evolution of the saturation field is described in terms of nonlinear kinematic waves, where different complex wave propagation scenarios happen depending on the initial/boundary conditions as well as behavior of the flux function (i.e. whether the flux function includes gravity or not, and if it does, how the convexity of flux function changes throughout the domain). These scenarios include formation of shocks and rarefaction zones as well as reflection of the shocks from endpoints of the domain. The latter case which is more involved, is the physical phenomena behind the so called pure gravity segregation in a sealed gravity column, in which gravitational forces are the only driving mechanism, as will be elaborated more in the upcoming sections.
 
 %------- more elaboration on the flux function: explain three scenarios (horizontal/updip/downdip)
Illustrating behavior of the flux function requires the dip angle as well as relative permeabilities to be fully specified. According to Eq.~(\ref{fractional flow equation}), the fractional flow is indeed a function of saturation and the dip angle, and hence can result in three different scenarios depending on the inclination angle $\theta$, 
\begin{itemize}
    \item $N_g sin\theta=0$ turns Eq.~(\ref{fractional flow equation}) to that of a horizontal reservoir, meaning gravity plays no role and viscous forces are the only driving mechanism.
    \item $N_g sin\theta >0$ represents an updip flooding, in which gravity retards the water flow, thereby reducing velocity of the front movement. It is noteworthy that $sin \theta>0$ can result in a fractional flow smaller than 0. Assuming that the wetting phase is heavier than the nonwetting one, $f_w<0$ represents counter-current flow of the heavier wetting phase moving downdip, while the lighter nonwetting phase is moving updip. The $0<f_w<1$ region indicates the co-current flow of water and oil moving updip.
    \item $N_g sin\theta <0$ results in a downdip flooding scenario, where gravity increases the water flow, and hence causes a faster movement of the saturation front. This scenario corresponds to the counter-current flow of oil flowing updip, while water is flowing downdip in a tilted reservoir. $sin \theta<0$ leads to a fractional flow larger than 1, thereby causing the $f_w>1$ region its fractional flow curve. In contrast, in the $0<f_w<1$ region, the two phases are moving downdip co-currently.
\end{itemize}
%-------------------------------  PLOTTING flux functions
%[htbp] 
\begin{figure}[t!] 
	\centering
	%	\label{fig:a}
% 	\includegraphics[width=57mm]{figures/horizontal-flux-imbibition-drainage.png}
% 		\includegraphics[width=57mm]{figures/1_new.png}
		\includegraphics[width=55mm]{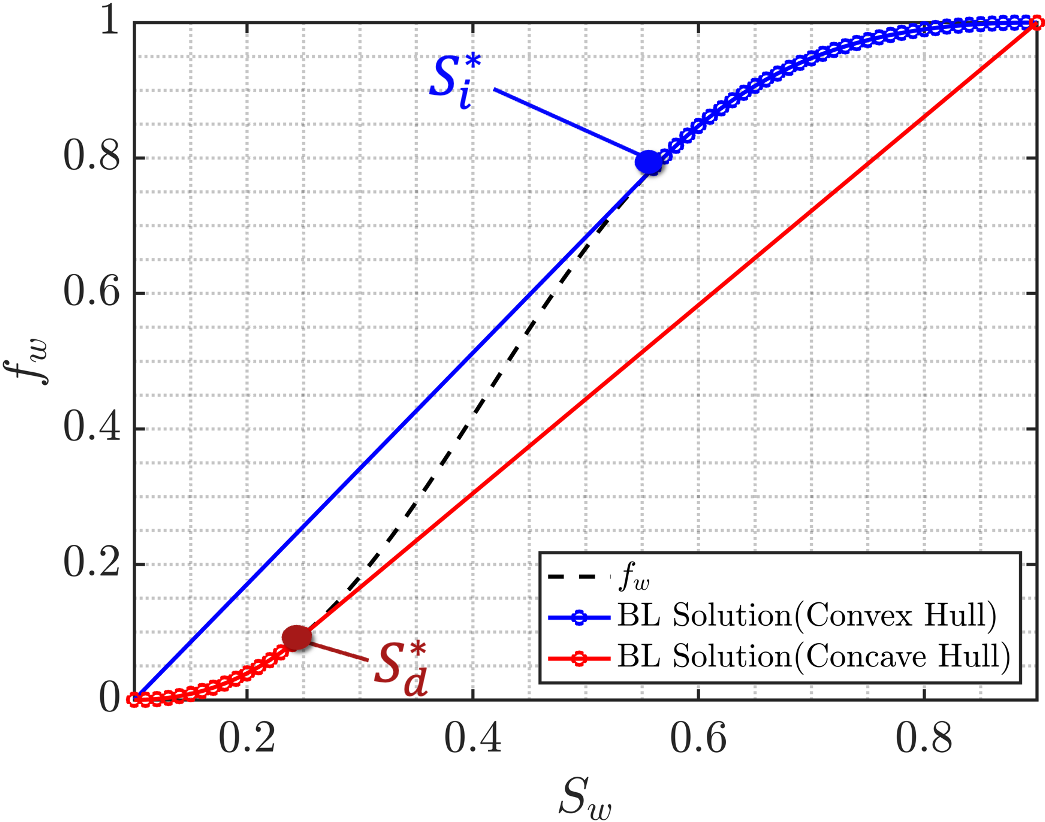}
		\includegraphics[width=55mm]{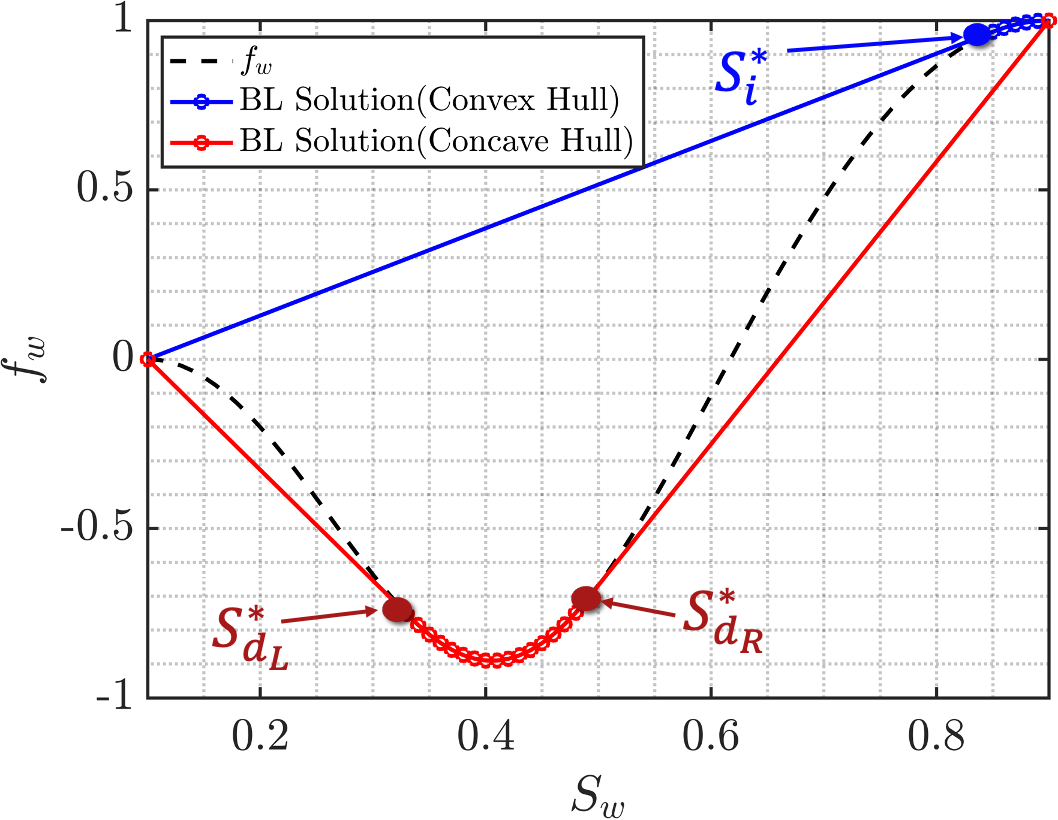}\\
	 \hspace*{20pt} (a) Horizontal flooding\hspace*{68pt}(b) Upslope background drift\\
	%	\label{fig:b}
% 	\includegraphics[width=57mm]{figures/downdip-flux-imbibition-drainage.png}
% 	\includegraphics[width=57mm]{figures/downdip_flux.png}
		\includegraphics[width=55mm]{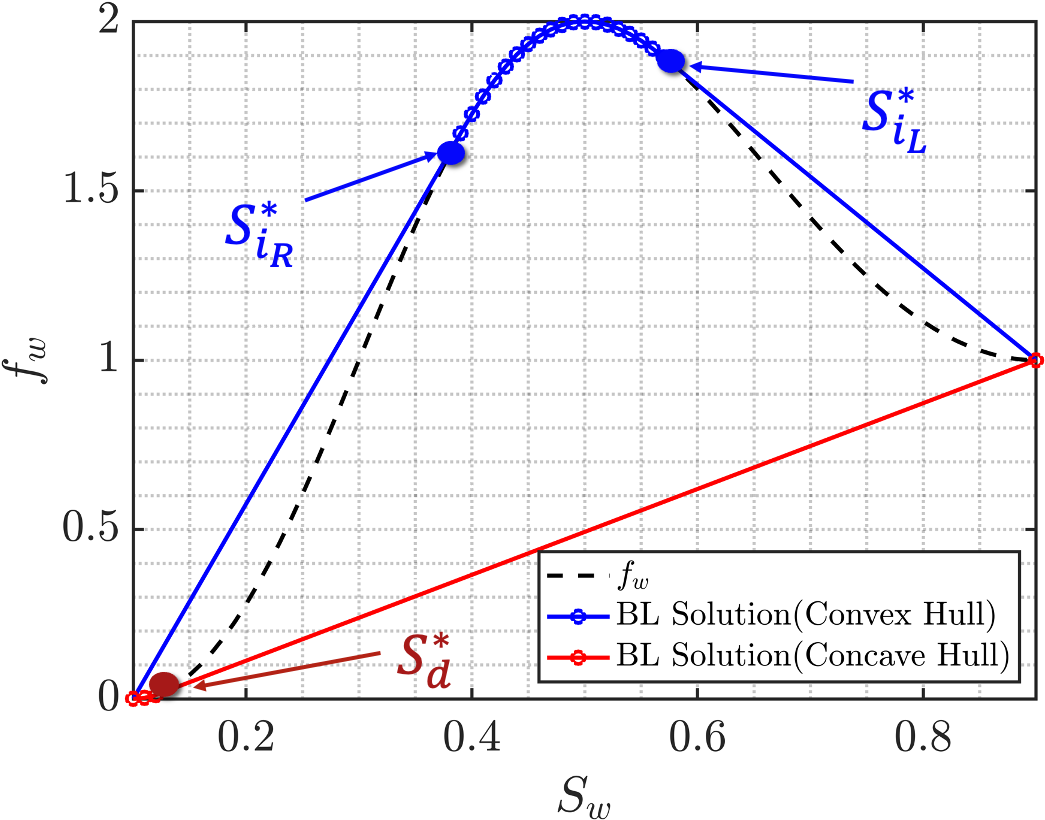}
		\includegraphics[width=55mm]{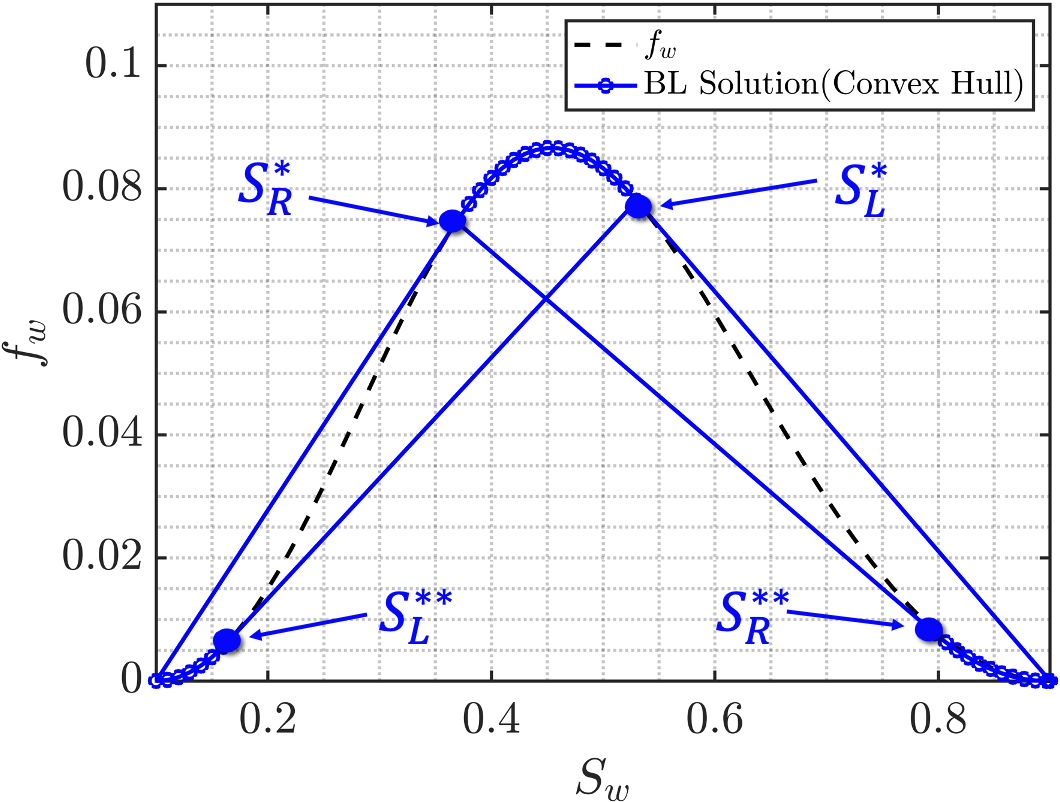}\\

	(c) Downslope background drift \hspace*{38pt}  (d) Gravity column

	\caption{Continuous (analytical) fractional flow curves for the four cases studies of our problem (horizontal, upslope background drift, downslope background drift and pure gravity segregation). The dashed curves are the flux function $f_w(S_w)$.
		In cases (b) and (c) $f(S_w)$ accounts for a combination of gravitational forces and viscous background drift. The blue lines show the concave envelopes corresponding to imbibition at the left edge of the domain, while the red lines are the convex envelopes corresponding to drainage at the right edge. In case (d), the inner envelope manifest reflection of waves from boundaries. All cases have adopted Brooks and Corey relative permeabilities, and have a viscosity ratio of m=0.5. }
	\label{fig:frac_flow_curves_with_concave_convex_hulls}
\end{figure}
The flux function corresponding to these scenarios are schematically represented in Fig.~(\ref{fig:frac_flow_curves_with_concave_convex_hulls}). Horizontal case ($N_g=0$) with an S-shaped curved is characterized by a shock and a trailing rarefaction zone where viscous forces are the only driving mechanism, whereas the vertical updip injection ($N_g=10$) is characterized by a shock and a rarefaction zone, and vertical downdip injection ($N_g=-10$) case is characterized by two shocks and a rarefaction zone in between. The two former cases represent the combined effect of viscous and buoyancy forces. The case in sub-figure (d) is physically similar to the vertical case, except that it has closed boundaries at the top and bottom. Therefore, unlike the vertical flooding in which we inject at one boundary, and recover oil from the other boundary, we assume that in this case, buoyancy leads the whole physics with no injection/viscous forces helping it. To this end, we consider a gravity column in a heterogeneous domain $[0,3]$ confined by a sealing medium at the top and bottom. We define an initial condition with a heavy fluid on top of a lighter fluid, described by setting $S_L=1$ in the top half of the domain and $S_R = 0$ in the bottom half of the domain. The fluids are separated by a sharp interface, and initially the interface is at $Z=1$. Then, the two fluids start segregating, where the heavier fluid would like to move downward (positive $z$ direction) and reside below the lighter, and hence the lighter fluid moves upward (negative $z$ direction). While the shock regions are representative of a single-phase fluid, the rarefaction area is indeed two-phase.
Therefore, the outer envelope shows the two waves before hitting the boundaries for the first time. These waves move in the opposite directions and will be eventually reflected from each boundary, schematically shown by the inner envelope. Note that $m \neq 1$, and the corresponding flux function is asymmetric, and hence, the two waves move asynchronously, i.e. one may hit its boundary much earlier than the other, depending on $m$. Once reflected from both boundaries, the waves follow a very small rarefaction region. Afterwards, at $S^{\star\star}_{R}$ and $S^{\star\star}_{L}$ two shocks form again, start moving towards each other and meet at a point $z_j$, and then reach to the equilibrium at $t_{eq}$, where the initial condition has been completely flipped.

The current study only investigates the imbibition process (indicated by blue lines in Fig.~(\ref{fig:frac_flow_curves_with_concave_convex_hulls})). Nonetheless, extending the manipulations of this work to the drainage process (depicted by red lines) is unambiguously straightforward; the horizontal flooding case would still entail inspection of a rarefaction followed by a shock. While the shock-rarefaction scenarios for updip and downdip would be flipped in the drainage process. That is, contrary to the imbibition scenario, the downdip case gives rise to one shock, while its updip counterpart consists of two shocks with a rarefaction fan in between.

%%-----------------------------------------  Pure segregation
To shed more light on the special case of pure gravity segregation, we will elaborate more on the distinctions between this scenario and the formerly explained downdip flooding, where we assumed injection is happening at the inlet. Strictly speaking, we assume a vertical column in a heterogeneous domain $[0,3]$ with sealing top and bottom boundaries, so $q_T=0$ in pure gravity segregation. Assuming the heavier fluid on top and the lighter fluid beneath gives rise to an unstable situation and needs to reach a vertical equilibrium. Therefore, even though $\textbf{u}_T=0$, we have $\textbf{u}_o \neq 0, \textbf{u}_w\neq 0$, where the continuity condition results in $\textbf{u}_o +\textbf{u}_w=\text{constant}$. Assuming boundary conditions $\textbf{u}_o=\textbf{u}_w=0$ at the top and bottom of the domain, we conclude the constant is zero, and hence $\textbf{u}_o= -\textbf{u}_w$. Consequently, $\textbf{u}_T=0$ entails a revisited definition for $f_w$, as we can no longer use the conventional definition in Eq.~(\ref{eq:flux function uw/ut }). Note that we are using $\textbf{u}$ for the vertical velocity with a small abuse of notation, however, we keep in mind $\textbf{u}$ is referring to a horizontal or vertical velocity depending on the case study.\\
In order to find the corresponding flux function, we start from Eq.~(\ref{eq:Darcy for uw uo}) for the vertical velocity of water and oil. After rearranging and subtracting them, and applying $\text{u}_o = -\text{u}_w$ in the absence of capillarity, we end up with,
%
% \begin{align}
%     \textbf{u}_w \bigg( \dfrac{\mu_w}{k k_{rw}} + \dfrac{\mu_o}{k k_{ro}}\bigg) = g (\rho_o -\rho_w)
% \end{align}
% After some manipulations, this equation can be recast as,

\begin{align}
    \textbf{u}_w = \dfrac{g (\rho_o -\rho_w) k }{\mu_w} \dfrac{k_{rw}}{\bigg( 1+\dfrac{k_{rw} \mu_o}{k_{ro} \mu_w}\bigg)} \equiv \textbf{u}_c G(S_w)
\end{align}
Where the first fraction is assumed to be a characteristic velocity resulted from gravity $\textbf{u}_c$, and the second fraction is the dimensionless flux function $G(S_w)$. Hence, fractional flow for pure gravity segregation is defined as $G(S_w) = \dfrac{\textbf{u}_w}{\textbf{u}_c}$, and is equal to,

\begin{align}
    G(S_w) = \dfrac{k_{rw}}{\bigg( 1+\dfrac{k_{rw} \mu_o}{k_{ro} \mu_w}\bigg)}
\end{align}

Therefore, the transport equation is recast as,

\begin{align}\label{eq: pure segregation-----saturation_equation_with_BC}
& \dfrac{\partial S_w}{\partial t} + \dfrac{\textbf{u}_c}{\phi(z)} G^{\prime}_w(S_w)\cdot\nabla S_w = 0\nonumber\\
& S_w(\textbf{z},t) = S_B = 1- S_{oi}, \quad \textbf{z} \in \Gamma_i, t>0,\nonumber\\
& S_w(\textbf{z},t=0) = 1-S_{oi} , \quad \textbf{z}\in \Omega^{-} \equiv [-1,0]\nonumber\\
& S_w(\textbf{z},t=0) = S_{wi}, \quad \textbf{z}\in \Omega^{+} \equiv [0,+1]
\end{align}

%-------------------- Relative permeabilities/ Brooks& Corey
Numerical treatment of first order hyperbolic conservation laws poses additional challenges, mainly due to their nonlinearity lying solely in the flux function. Such a nonlinear behavior in fractional flow is essentially introduced by relative permeabilities of each phase. There are several empirical models for defining the relative permeabilities as a function of water saturation. Brooks and Corey \cite{brooks1964hydraulic}, and Van Genuchten \cite{van1980closed} are the most commonly employed empirical models. We employ the Brooks and Corey model to define relative permeabilities given by the following expressions,

\begin{align}
& k_{rw}(S_w) = \left(\dfrac{S_w - S_{wi}}{S_B - S_{wi}}\right)^2, \qquad k_{ro}(S_w) = \left(\dfrac{S_B- S_w }{S_B - S_{wi}}\right)^2 \label{eqn:Brooks-Corey-rel_perms}
\end{align}

%-----------------------------------  v_s advection speed
A closer inspection of Eq.~(\ref{eq:saturation_equation_with_BC}) immediately reveals the fact that hyperbolic Buckley Leverett equation is indeed in the form of an advection equation with velocity $\textbf{v}_s(x,t,s)\equiv \dfrac{q(t)}{\phi(x)} f^{\prime}(S_w)$. Combining this relation with Eq.~(\ref{fractional flow equation}) and Eq.~(\ref{eqn:Brooks-Corey-rel_perms}) leads to an expression for the advection speed (for simplicity, we refer to $S_w$ as $S$),

\begin{align} \label{eq: advection velocity horizontal-up/downdip}
\textbf{v}_s(x,t,S) =\dfrac{q(t)}{\phi(x)} \dfrac{\partial f_w(S)}{\partial S} &= \dfrac{q(t)}{\phi(x)} \dfrac{2m (S-S_{wi}) (1-S - S_{oi}) (1-S_{oi} -S_{wi})} {\bigg[ (S-S_{wi})^2 + m (1-S -S_{oi})^2 \bigg]^2}\notag\\
&+N_g \dfrac{q(t)}{\phi(x)}  \dfrac{2 (1-S - S_{oi})(S - S_{wi}) \bigg[ (S - S_{wi})^3 + m(1-S - S_{oi})^3\bigg] }{(1- S_{wi} - S_{oi})^2  \bigg[ (S-S_{wi})^2 + m (1-S -S_{oi})^2 \bigg]^2} 
\end{align}

Where the second term drops for the horizontal cases with a gravity number of $(N_g = 0)$. 
 Similarly, for the pure segregation scenario,

\begin{align} \label{eq: advection velocity pure segregation}
    \textbf{v}_s(z,S) =\dfrac{\textbf{u}_c}{\phi(z)} \dfrac{\partial G(S)}{\partial S}= \dfrac{\textbf{u}_c}{\phi(z)} \dfrac{2 m(1-S - S_{oi})(S - S_{wi}) \bigg[ (S - S_{wi})^3 + m(1-S - S_{oi})^3\bigg] }{(1- S_{wi} - S_{oi})^2  \bigg[ (S-S_{wi})^2 + m (1-S -S_{oi})^2 \bigg]^2} 
\end{align}

%-------------------------- Riemann problem/ analytical solutions MOC
% According to the theory of hyperbolic conservation laws, a conservation equation with constant left and right states denoted by $S_L$ and $S_R$ forms a Riemann problem. If $S_L > S_R$, the Riemann problem has a self-similar solution in the form of.
A hyperbolic conservation law with piece-wise constant initial condition $S(z,t=0)=S_0(z)=\begin{cases} S_L \quad z\leq z_d\notag\\
S_R \quad z > z_d
\end{cases}$ forms a Riemann problem, where $S_L$ and $S_R$ represent the left and right values of the initial condition, separated at a discontinuity location $z_d$. While in general, $S_L$ and $S_R$ can be random, we assume they are deterministic constants. As the flux function changes its convexity throughout the saturation domain, the solution of this Riemann problem can be comprised of shocks and rarefaction zones, depending on whether the flux function is convex or concave over that specific interval. The exact analytical solution to Eq.~(\ref{eq:saturation_equation_with_BC}) subject to the given boundary/initial conditions, in one-dimensional domain of [a,b], and a final simulation time T, for a given realization of the total velocity $\textbf{v}_T(x,t)$, can be expresses as following (\cite{aziz1979petroleum}, \cite{orr2007theory}),
% Exact analytical solution for water saturation can be obtained using frontal advance equation $\dfrac{dx}{dt} = \dfrac{q_t}{\varphi} f^{\prime}(s)$

\begin{align}\label{eq:saturation analytical solution}
S_w(x,t;v)= \begin{cases}
S_B, & x<X_1\\
S_{\text{r}} = (f_w^{\prime})^{(-1)}\Bigg[x \bigg(\dfrac{\bigintss_{ 0}^{t} q(t^{\prime}) dt^{\prime}}{\varphi(x)}\bigg) ^{-1}\Bigg], & X_1\leq x < X_2\\
S_{wi} & x \geq X_2
\end{cases}
\end{align}

Where $X_1=a$ and $X_2=x_f(t;v)$ for the no-gravity case. Whereas the solution for downdip flooding and gravity column before reflection of the waves takes into account the presence of two shocks, resulting in the same solution above, however, $X_1=x_{f_2}(t;v)$ and $X_2=x_{f_1}(t;v)$,
%
% \begin{align}\label{eq:saturation analytical solution-downdip}
% S_w(x,t;v)= \begin{cases}
% s_B, & x < x_{f_2}(t;v)\\
% s_{\text{r}} = (f_w^{\prime})^{(-1)}\Bigg[x \bigg(\dfrac{\bigintss_{ 0}^{t} q(t)^{\prime} dt^{\prime}}{\varphi(x)}\bigg) ^{-1}\Bigg], & x_{f_2}(t;v)\leq x < x_{f_1}(t;v)\\
% s_{wi} & x> x_{f_1}(t;v)
% \end{cases}
% \end{align}
%
where $x_{f_i}(t;v)$ for $ i=1,2$ describes the front location at time t for the given realization of total velocity. This quantity can be found by satisfying the entropy constraint through Rankine-Hugoniot condition at the location of discontinuities (shocks), 
% \begin{align}
%     & x_{f_1}(t;V) = \dfrac{f^{\prime}(S^{\star}_1)}{\phi(x)} \bigintss _0^{t} q(t^{\prime}) dt^{\prime}, \qquad\qquad  x_{f_2}(t;V) = \dfrac{f^{\prime}(S^{\star}_2)}{\phi(x)} \bigintss _0^{t} q(t^{\prime}) dt^{\prime}
% \end{align}

\begin{align}
    \dfrac{dx_{f_i}}{dt} = \dfrac{q(t)}{\phi(x)} \dfrac{f_w(S^{\star}_i) - f_w(S_{wi})}{S^{\star}_i - S_{wi}} 
\end{align}
Where the continuity condition imposed on the speed of the shocks determines $S^{\star}_i$,
\begin{align}
    \dfrac{f_w(S^{\star}_i) - f_w(S_{wi})}{S^{\star}_i - S_{wi}} = f^{\prime}_w(S^{\star}_i)
\end{align}

For the special case of horizontal reservoir domain (with fractional flow $f_w=\dfrac{k_{rw}}{k_{rw} + m k_{ro}}$ supplemented by Brooks and Corey relative permeabilities), $S^{\star}$ has an analytical expression (Eq. 4.7 in \cite{ibrahima2017multipoint}). Nevertheless, we resort to a general numerical root finding scheme which handles $N_g\neq 0 $ cases as well. 
% 
% Examples of saturation profiles when there are either one or two shocks present are depicted using Eq.~(\ref{eq:saturation analytical solution}) in Fig.~(\ref{fig:example saturation profiles}). Uncertainty in $\phi(x)$ (with a high variance) gives rise to roughness in the rarefaction zone.
% While one Monte Carlo sampling clearly displays this roughness, later results from Monte Carlo approach will not manifest any roughness as the result of being averaged out through a large number of Monte Carlo trials.
%
% add this: saturation gradually decreases from ... at ... to .... The discontinuity in the rarefaction region corresponds to the zero-velocity point in the flux function
%
%-------------- plot an example of saturation profiles for horizontal (1 shock) and downdip (2 shocks)
% \begin{figure}[htbp]
% 	\centering
% 	\includegraphics[width=60mm]{figures/satprofile_horizontal.png}
% % 		\includegraphics[width=83mm]{figures/SatProfileStochastic.png}
% 	\includegraphics[width=60mm]{figures/satprofile_vertical.png}
% 	\caption{Examples of a saturation profile at time $t^{\prime}$ for one realization of random $\textbf{v}_T(x,t)$ obtained from Eq.~(\ref{eq:saturation analytical solution}) for horizontal (left) and downdip displacement (right)}
% 	\label{fig:example saturation profiles}
% \end{figure}
%-------------------------------- Nondimensionalization OF BL
\subsection{Nondimensionalization}
We will investigate the impact of uncertainty in the inputs on the output saturation field, first by assuming that $\phi(x)$ is random and $q$ is a deterministic constant, second by assuming that $q_T(t)$ is random and $\phi$ is a deterministic constant, and third by assuming both are stochastic fields. For the case in which $q_T(t)$ is random with the given mean $\mu_q$ and correlation length $\tau_q$, we can make the transport Eq.~(\ref{eq:saturation_equation_with_BC}) dimensionless by defining,
% \begin{align}
%     x_D=\dfrac{x}{L}, \qquad t_D=\dfrac{t}{\tau_q}, \qquad q^{\star}_T(t)=\dfrac{q_T(t)}{\mu_q},\qquad L=\mu_q\tau_q
% \end{align}

 \begin{align}
    x_D=\dfrac{x}{L}, \qquad t_D=\dfrac{1}{L} \bigintssss_0^t q(t^\prime) dt^\prime
\end{align}

This rescaling will recast Eq.~(\ref{eq:saturation_equation_with_BC}) as the dimensionless form in the one-dimensional domain $\Omega$,
% \begin{align}
%     \phi(x)\dfrac{\partial S_w}{\partial t_D} + q^{\star}_T(t)\dfrac{\partial f(S_w)}{\partial x_D} = 0
% \end{align}
 %%%%%%%%%%%%%%%%%%%%%
 
\begin{align}
    \phi(x)\dfrac{\partial S_w}{\partial t_D} + \dfrac{\partial f(S_w)}{\partial x_D} = 0
\end{align}

 %%%%%%%%%%%%%%%%%%%%
For the cases in which $q$ is a deterministic constant, we can simply use the dimensionless quantities below to make Eq.~(\ref{eq:saturation_equation_with_BC}) dimensionless,

\begin{align}
    x_D=\dfrac{x}{L}, \qquad t_D=\dfrac{q t}{L}
\end{align}

This definitions lead to the dimensionless form of Eq.~(\ref{eq:saturation_equation_with_BC}),

\begin{align}
    \phi(x)\dfrac{\partial S_w}{\partial t_D} + \dfrac{\partial f(S_w)}{\partial x_D} = 0
\end{align}

%
% Note that, we would like to preserve $\phi(x)$ and $q(t)$ explicitly in the dimensionless form of transport equation whenever they are stochastic, hence the numerical processing of the stochastic fields can be done straightforwardly.
%-------------- pure segregation non-dimensionalization
In order to make transport Eq.~(\ref{eq: pure segregation-----saturation_equation_with_BC}) for pure gravity segregation dimensionless, since there is no injection flux for this scenario, we will only investigate the effect of uncertainty in porosity field on the water saturation. Therefore, we introduce the dimensionless time and distance parameters,

\begin{align}
    t_D = t\dfrac{g (\rho_o - \rho_w) k}{\mu_w H} =\dfrac{t}{\tau}, \qquad \qquad z_D = \dfrac{z}{H}
\end{align}

Where $H$ is the total length of the vertical domain and $\tau = \dfrac{\mu_w H}{g (\rho_o -\rho_w) k}$ is the characteristic time scale for gravity segregation. That is, if $\tau=1$, the particle has enough time to travel along the whole $H$.
Such definitions allow us to work with the dimensionless form of Eq.~(\ref{eq: pure segregation-----saturation_equation_with_BC}) as,

\begin{align}\label{eq: dimensionless pure segregation-----saturation_equation_with_BC}
& \phi(z)\dfrac{\partial S_w}{\partial t_D} + \dfrac{\partial G(S_w)}{\partial z_D} = 0
\end{align}

%%%%%%%%%%%%%%%%%%%%%%%%%%%%%%%%%%%%%%%%%%%%%%%%%%%%%%%%%%%%%%%%%%%%%%%%%%%%%%%%%%%
%%%                            METHOD OF CHARACTERISTICS
%%%%%%%%%%%%%%%%%%%%%%%%%%%%%%%%%%%%%%%%%%%%%%%%%%%%%%%%%%%%%%%%%%%%%%%%%%%%%%%%%%%
% \section{Method of Characteristics}

%--------------------------------------- Plotting characteristics lines of
\begin{figure}[t!]
 	\centering
	%	\label{fig:a}
\includegraphics[width=33mm]{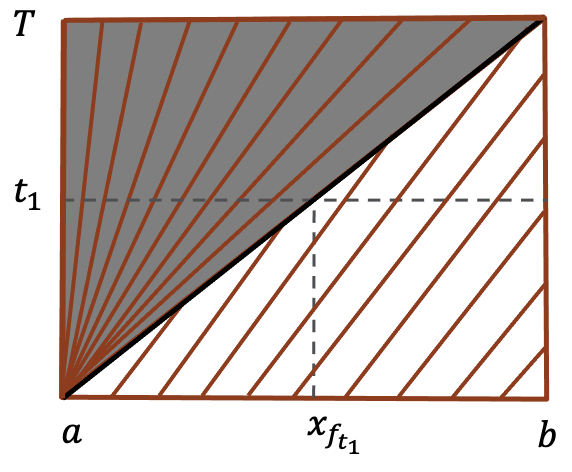}
\includegraphics[width=33mm]{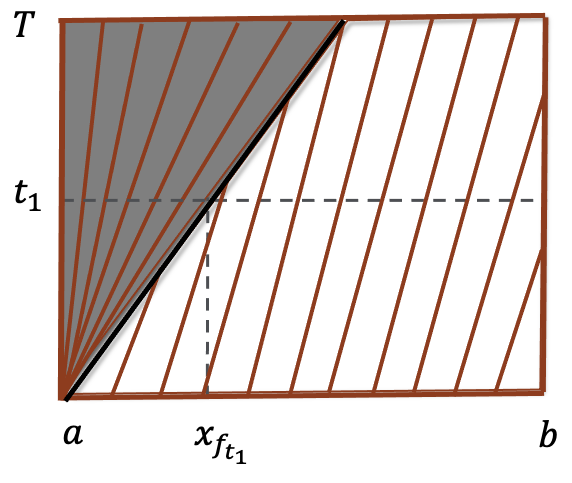}
\includegraphics[width=33mm]{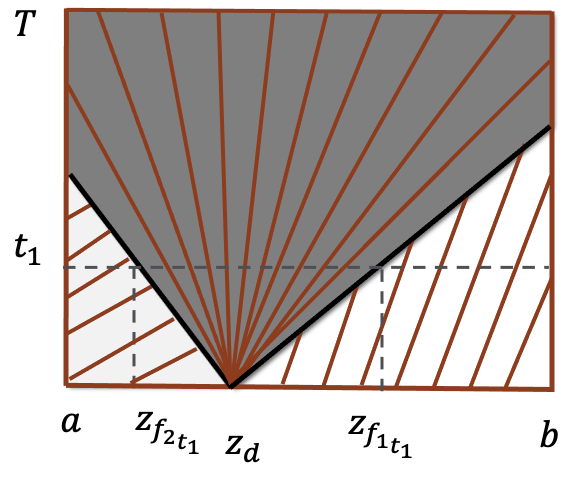}
\caption{Characteristics plots for horizontal, updip and downdip injection (from left to right). }
	\label{fig:characteristics plots for horizontal, updip and downdip injection }
\end{figure}
\begin{figure}[t!]
	\centering
\includegraphics[width=40mm]{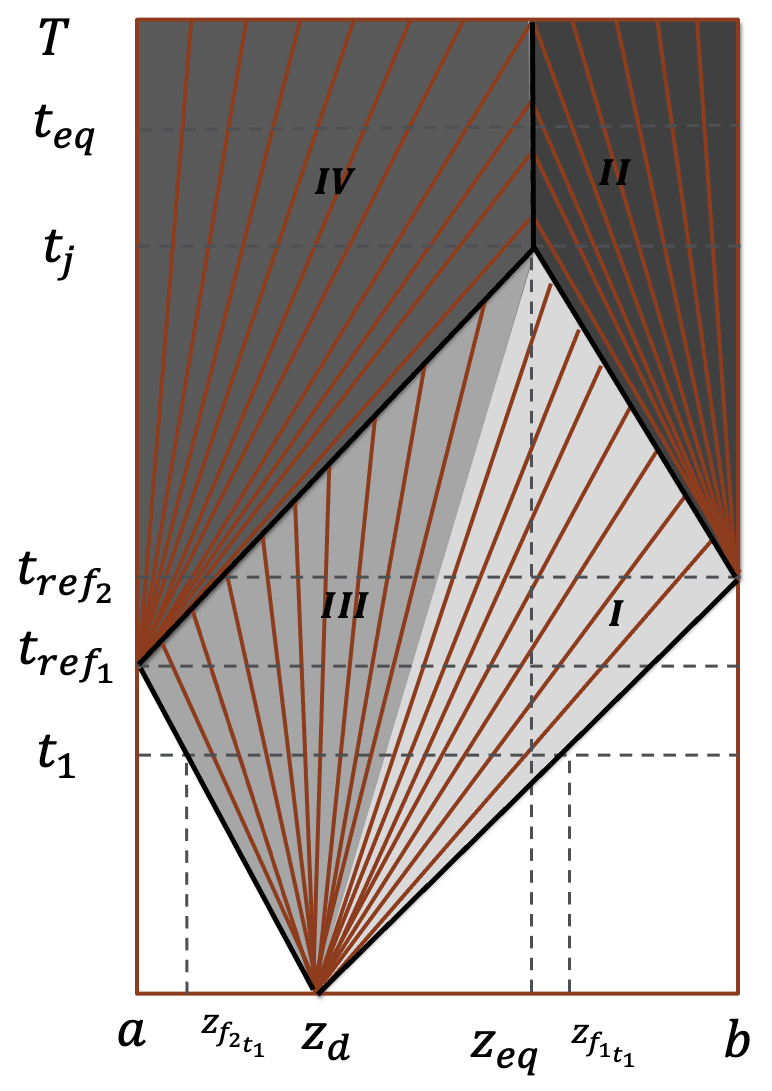}
\includegraphics[width=72mm]{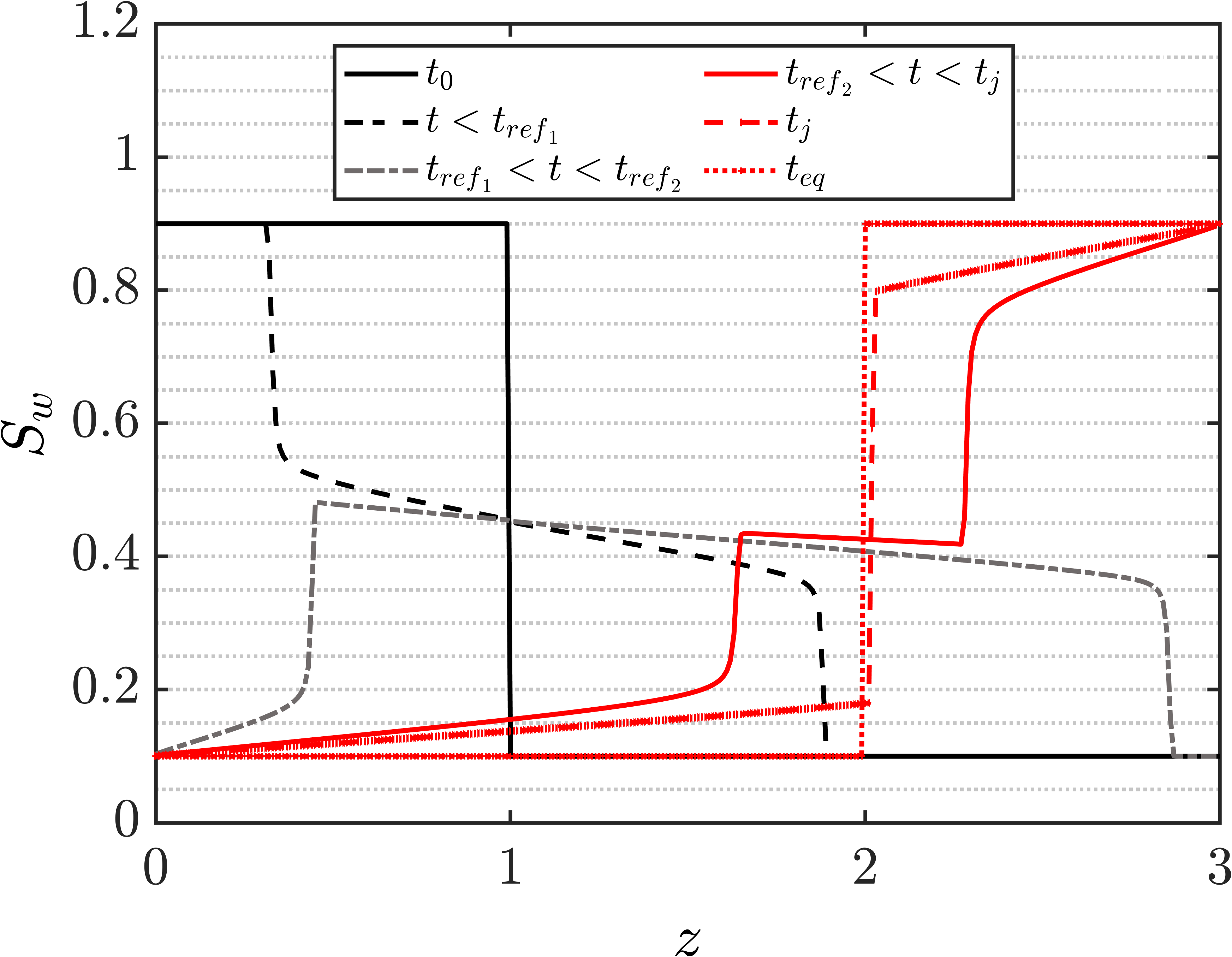}
		\caption{(left) Characteristic plot in the $z-t$ plane, (right) the corresponding finite volume solution of the saturation profiles $S_w(z,t)$ for a gravity column with a heavy fluid on top of a light fluid initially separated by a sharp interface at $z_d$. After reflection from boundaries the two shocks join each other at an equilibrium point $z_{j}$. Asymmetric pattern of the right and left branches arises from $m\neq 1$ and the asymmetric non-uniform initial condition. Solid black lines on the characteristics plot correspond to the shocks. Regions I, II, III, IV represent the right-moving shock before and after reflection, left-moving shock before and after reflection, respectively.}
	\label{fig:characteristics of downdip}
\end{figure}
%------------------------------------------------------

% \begin{figure}[htbp]
%  	\centering
% %  \includegraphics[width=50mm]{figures/characteristics_gravity_column3.png}
% \includegraphics[width=45mm]{figures/characteristics_gravity_column2.png}
% 	\includegraphics[width=95mm]{figures/downdip_timeplots.png}
% % 		\includegraphics[width=60mm]{figures/MRST1.png}
%   		% \includegraphics[width=40mm]{figures/characteristics_horizontal.png}
% 	\caption{Solution for a gravity column with heavy fluid on top of a light fluid computed by front tracking for a model with Corey relative permeabilities with nw = no = 2 and equal viscosities. The left column shows the wave pattern in the (z,t) plane, where each blue line corresponds to the propagation of a discontinuous state. The right column shows the solution S(z,t) at four different times.}
% 	\label{fig:characteristics of downdip}
% \end{figure}

%%%%%%%%%%%%%%%%%%%%%%%%%%%%%%%%%%%%%%%%%%%%%%%%%%%%%%%%%%%%%%%%%%%%%%%%%%%%%%%%%%%
%%%                            CDF METHOD
%%%%%%%%%%%%%%%%%%%%%%%%%%%%%%%%%%%%%%%%%%%%%%%%%%%%%%%%%%%%%%%%%%%%%%%%%%%%%%%%%%%%
\section{Single-point CDF equation for Saturation} \label{section: Single-point CDF equation for Saturation}
In order to provide full probabilistic distribution of water saturation, we will extend the idea of \cite{wang2013cdf} to solve a deterministic equation that governs the space-time evolution of the CDF of water saturation $F_s(S_w;\textbf{x},t)$ for more general cases of gravitational forces taken into consideration. This approach overcomes the limitations of solving the nonlinear hyperbolic Eq.~(\ref{eq:saturation_equation_with_BC}), by introducing a linear hyperbolic equation for the so-called random ``raw'' CDF function of water saturation,

\begin{align}
\Pi(\theta,S_w;\textbf{x},t) = \mathcal{H}(\theta - S_w(\textbf{x},t))
\end{align}

where $\mathcal{H}(.)$ is the Heaviside step function and $\theta$ is a deterministic value (outcome) that the random water saturation $S_w$ takes at a space-time point (\textbf{x}, t).

For continuous solutions of Eq.~(\ref{eq:saturation_equation_with_BC}), the linear stochastic hyperbolic equation for the raw CDF reads \cite{wang2013cdf},

\begin{align}\label{eq: pi equation with BC}
& \dfrac{\partial \Pi}{\partial t} + \textbf{v}_s(\theta, x, t)\cdot \nabla\Pi = 0\qquad x\in\Omega, t>0\nonumber\\
& \Pi(\theta, S_w;x,t=0) = \mathcal{H}(\theta - S_{wi})\nonumber\\
& \Pi(\theta, S_w;x=0,t) = \mathcal{H}(\theta - 1+ S_{oi})
\end{align}

Where the boundary condition corresponds to the injection of wetting phase fluid, i.e. imbibition process at the inlet $x=0$. The advection velocity is defined using Eq.~(\ref{eq: advection velocity horizontal-up/downdip}) for the first three cases, and it follows Eq.~(\ref{eq: advection velocity pure segregation}) for pure gravity segregation.

% \begin{align}
% \textbf{v}_s(\theta, x,t) =\dfrac{q(t)}{\phi(x)} \dfrac{\partial f_w(\theta)}{\partial \theta} = & \dfrac{q(t)}{\varphi(x)} \dfrac{2m (\theta-s_{wi}) (1-\theta - s_{oi}) (1-s_{oi} -s_{wi})} {\bigg[ (\theta-s_{wi})^2 + m (1-\theta -s_{oi})^2 \bigg]^2}+\notag\\
% &+N_g \dfrac{q(t)}{\phi(x)}  \dfrac{2 (1-\theta - s_{oi})(\theta - s_{wi}) \bigg[ (\theta-s_{wi})^3 - m(1-\theta - s_{oi})^3\bigg] }{(1- s_{wi} -s_{oi})^2  \bigg[ (\theta-s_{wi})^2 + m (1-\theta -s_{oi})^2 \bigg]^2} 
% \end{align}
% Where the second term drops for the horizontal cases with a gravity number of $(N_g = 0)$. 
% For the pure gravity segregation we will follow Eq.~(\ref{eq: advection velocity pure segregation}),
% %
% \begin{align} 
%     \textbf{v}_s(\theta, z) =\dfrac{\textbf{u}_c}{\phi(z)} \dfrac{\partial G(\theta)}{\partial \theta}= \dfrac{\textbf{u}_c}{\phi(z)} \dfrac{2 m(1-\theta-S_{oi})(\theta - S_{wi}) \bigg[ (\theta - S_{wi})^3 + m(1- \theta- S_{oi})^3\bigg] }{(1- S_{wi} - S_{oi})^2  \bigg[ (\theta-S_{wi})^2 + m (1-\theta -S_{oi})^2 \bigg]^2} 
% \end{align}
%
As mentioned before, Eq.~(\ref{eq: pi equation with BC}) holds only for the continuous solutions of the raw CDF, in which we are indeed excluding the treatment of shocks (discontinuities). A general form of Eq.~(\ref{eq: pi equation with BC}) comprises of imposing the non-smooth solutions to fulfill entropy conditions at the location of discontinuities, allowing for a physically meaningful solution. This is primarily done by adding a kinetic defect term $\mathcal{M}(\theta, x,t)$ to Eq.~(\ref{eq: pi equation with BC}), making it 
$\dfrac{\partial \Pi}{\partial t} + \textbf{v}_s(\theta, x, t)\cdot \nabla\Pi = \mathcal{M}(\theta, x,t)$,
which essentially results in a solution valid for the the entire domain of saturation. Although, the higher dimensional problems require solving the later equation as an alternative to Eq.~(\ref{eq: pi equation with BC}) , their one-dimensional counterparts do not suffer from additional complications emerged from interaction of waves at higher dimensions, and hence, the problem in one-dimension could be studied by employing Eq.~(\ref{eq: pi equation with BC}) for the separate smooth parts of the domain. To this end, for horizontal displacement and updip flooding, we divide the solution of Eq.~(\ref{eq: pi equation with BC}), into two sub-regions similarly to \cite{wang2013cdf},

\begin{align}
\Pi(\theta, x, t) = \begin{cases}
\Pi_a = \mathcal{H}(\theta - S_{wi}), \qquad\hfill & S_{wi} <\theta < S^\star, \qquad\hfill x_{f}(t;v)< x < b\\
\Pi_b = \mathcal{H}(\theta - S_r), \qquad\hfill & S^\star <\theta < 1-S_{oi} , \qquad\hfill a < x < x_f(t;v)
\end{cases}
\end{align}

Whereas for the downdip flooding case as well as the pure segregation scenario before the fastest-moving wave reflects from its boundary, we deal with three sub-regions. That is, for $0<t<t_{ref_1}$, 

\begin{align}\label{eq: pi_b pure seg, before reflection}
\Pi(\theta, x, t) = \begin{cases}
\Pi_a = \mathcal{H}(\theta - S_{wi}), \qquad\hfill & S_{wi} <\theta < S_R^\star,  \qquad\hfill x_{f_R}(t;v)< x < b\\
\Pi_b = \mathcal{H}(\theta - S_r), \qquad\hfill & S_R^\star < \theta < S_L^\star,  \qquad\hfill x_{f_L}(t;v) < x <  x_{f_R}(t;v)\\
\Pi_c = \mathcal{H}(S_{B} - \theta), \qquad\hfill & S_L^\star <\theta  < 1-S_{oi},  \qquad\hfill a < x < x_{f_L}(t;v)
\end{cases}
\end{align}

It should be noted that $\Pi_a$ and $\Pi_c$ are independent of $x$ and $t$.
After the waves reflect from boundaries in the pure segregation case, three separate rarefaction fans along with two shocks in between develop and move towards each other. As is represented in Fig~\ref{fig:characteristics of downdip}, the rarefaction region in the middle is indeed the continuation of the characteristics in the rarefaction fan of the region $\pi_b$ in Eq.~\ref{eq: pi_b pure seg, before reflection}, up to the time $t_{j}$. Eventually, when the two shocks meet at time $t_{j}$, they form one shock along with two rarefaction zones. Therefore, at $t=t_{j}$, the middle rarefaction area has totally disappeared and we have $ x_{f_L}(t;v) = x_{f_R}(t;v) = x_{eq}$.
That being said, after the slowest-moving wave has reflected from its boundary, i.e. for $t_{ref_2}<t<t_{j}$, 

\begin{align}\label{eq: pi_b pure seg, after reflection}
\Pi(\theta, x, t) = \begin{cases}
    \Pi_a = \mathcal{H}(\theta - S_{r_L}), \quad  S_{wi} <\theta  < S_L^{\star\star}, \quad a < x < x_{f_L}(t;v)\\
    \Pi_b = \mathcal{H}(\theta - S_{r_m})\mathcal{H}(t_{j} - t)+0\cdot \mathcal{H}(t-t_{j}), \quad S_R^\star < \theta < S_L^\star,  \quad\ x_{f_L}(t;v) < x <  x_{f_R}(t;v)\\
    \Pi_c = \mathcal{H}(\theta - S_{r_R}), \qquad S_R^{\star\star} <\theta < 1 - S_{oi},  \qquad x_{f_R}(t;v)< x < b
\end{cases}  
\end{align}

% The time interval between $t_{ref_1}<t<t_{ref_2}$ has a mix character of relations Eq.~(\ref{eq: pi_b pure seg, before reflection}), Eq.~(\ref{eq: pi_b pure seg, after reflection}).

%%%%%--------------------------
 The ensemble average of $\Pi$ over all possible values of $S(x,t)$ yields $F_s(S_w;x,t)$ which is a deterministic quantity. Using the single-point PDF of water saturation, $p_s(S_w;x,t)$ we have,

\begin{align}
F_s(\theta;x,t) = \big\langle \Pi(\theta, S_w(x,t))\big\rangle = \int_{-\infty}^{\infty} \mathcal{H}(\theta-S_w^\prime) p_s(S_w^\prime;x,t) dS_w^\prime 
\end{align}

Therefore, by solving a linear equation for $\pi$, followed by finding only the first ensemble moment of $\pi$, we find spatio-temporal evolution of the full single-point CDF of saturation. This is a more straightforward approach than first numerically solving the nonlinear Eq.~(\ref{eq:saturation_equation_with_BC}) for an exhaustive number of Monte Carlo realizations of the random variables, and subsequently discovering the numerical CDF and PDF using methods like \textit{kernel density estimation} (\cite{botev2010kernel}). 

It is noteworthy that similarly to the past literature (\cite{wang2013cdf}, \cite{ibrahima2017multipoint}), we prefer employing a CDF framework rather than working with PDF equations, due to the smoothness of CDF as well as the trivial formulation of boundary conditions for CDF methodology, i.e. $F_s(\theta=S_{wi};x,t) = 0 $ and $F_s(\theta=1-S_{oi};x,t) = 1$. The boundary conditions can not be uniquely defined for the PDF scheme. Therefore, using CDF obviates the need for dealing with closure approximations inherent to the PDF scheme. 

The method of characteristics can be utilized to form an analytical solution for the one-dimensional linear hyperbolic Eq.~(\ref{eq: pi equation with BC}), specifically in the continuous rarefaction zone where there is no discontinuity present. While the analytical solution for a horizontal reservoir has been studied by \cite{wang2013cdf}, the more general case studies comprising of buoyancy forces demand manipulations for handling two shocks introduced in this work. The raw CDF solution for the horizontal reservoirs reads \cite{wang2013cdf},

%--------------------------------------
\begin{align}
&\Pi_b = \mathcal{H}(\theta-S_{wi})\mathcal{H}(x-C) + \mathcal{H}(\theta -1+S_{oi})\mathcal{H}(C-x)  
\end{align}

%------------------------------------- derive pi_b for vertical-pure segregation
In order to define a family of characteristics $x=x(t;x_0)$ along which the original hyperbolic PDE becomes an ODE $\dfrac{d\pi}{dt} = 0$, we start by using equation $\dfrac{dx}{dt} = \textbf{v}(\theta,x(t),t), \qquad x(t=0)=x_0$, where $x_0$ specifies where the characteristic line has originated from. Thus $x=\bigintssss_0^t \textbf{v}(\theta,x(t^\prime), t^\prime) dt^\prime +x_0$, 
%
% \begin{align}
%     x=\bigintssss_0^t \textbf{v}(\theta,x(t^\prime), t^\prime) dt^\prime +x_0
% \end{align}
%
For the aforementioned setting of our problem, i.e. the initial and boundary conditions of the gravity column, the characteristic solution is depicted in Fig~(\ref{fig:characteristics of downdip}). The non-unity viscosity ratio causes an asymmetry and non-synchronism in the movement of two waves. To this end, the solution will be discussed in four distinct regions, i.e. for the right-moving branch before being reflected from right boundary and after being reflected, and Similarly, for the left-moving branch before and after being reflected from left boundary.   

As Fig.~(\ref{fig:characteristics of downdip}) displays, before the faster wave (left-moving wave in $(S_D, S^\star_L)$) hits into its boundary ($t<t_{{ref}_1}$), all the characteristics are emanating from $z$ axis where $t=0$. Therefore, initial condition is used to find the solution. Whereas, at $t = t_{{ref}_1}$, the left-moving wave gets reflected from left boundary and the characteristics start originating from t axis where $z=0$, while the direction of characteristics has been reversed by the virtue of the reflection from boundary. Thus, after reflection of the left-moving wave, and before the two waves meet each other at time $t_{j}$ and stabilize at $t_{eq}$ somewhere in the middle of domain identified by $z_{eq}$, we use the boundary condition at $z=a$ to find the solution. The analysis of the right-moving wave $(S^\star_R,S_D)$ follows a similar trend. Before hitting the right boundary $t < t_{{ref}_2}$, characteristics originate from $z$ axis, and initial condition is leveraged to find the solution. However, during $t_{{ref}_2} < t < t_{j}$, characteristics develop from $t$ axis at the right boundary $(z=L)$ and continues up to $t_{j}$ and location $z_{eq}$, where it meets the other branch coming towards it. 

%%%%%%%%%%%%%%%%%% from presentation
\begin{align}
\Pi_b =
\begin{cases}
    \Pi_{b_L} = 
        \begin{cases}
\mathcal{H}(\theta - S_B) \mathcal{H}(C - x) \mathcal{H}(x_d - x) + \mathcal{H}(\theta - S_D)\mathcal{H}(x - C) \mathcal{H}(x_d - x), \quad 0< t < t_{{ref}_1}\\\\
\mathcal{H}(\theta - S_{wi}) \mathcal{H}(C - x) \mathcal{H}(x_d - x)+  \mathcal{H}(\theta - S^{\star\star}_L) \mathcal{H}(x - C) \mathcal{H}(x_d - x), \quad t_{{ref}_1} < t < t_j
        \end{cases}\\\\
    \Pi_{b_R} = 
        \begin{cases}
\mathcal{H}(\theta - S_{wi}) \mathcal{H}(x - C) \mathcal{H}(x - x_d) + \mathcal{H}(\theta - S_D) \mathcal{H}(C - x) \mathcal{H}(x - x_d), \quad 0 < t < t_{{ref}_2}\\\\
        % \Pi(x=0) \mathcal{H}(x - C) \mathcal{H}(x - x_{d}) = 
        \mathcal{H}(\theta-S_{B}) \mathcal{H}( x - C) \mathcal{H}(x - x_d) + \mathcal{H}(\theta - S^{\star\star}_R) \mathcal{H}(C - x) \mathcal{H}(x - x_d), \quad t_{{ref}_2} < t < t_j
        \end{cases}
\end{cases} 
\end{align}

%%%%%%%%%%%%%%%%%%%%%%%%%%%%%%%%%%%%%%%%%%%%%
% \begin{align}
% \Pi_b =
% \begin{cases}
%     \Pi_b_L = 
%         \begin{cases}
%         \Pi(x=0) \mathcal{H}(C - x) \mathcal{H}(x_d - x) + \Pi(x=x_d) \mathcal{H}(x - C) \mathcal{H}(x_d - x), \qquad \hfill 0< t < t_{{ref}_1}\\
%         % \Pi(t=0) \mathcal{H}(C - x) \mathcal{H}(x_{d} - x) =
%           \mathcal{H}(\theta - S_{wi}) \mathcal{H}(C - x) \mathcal{H}(x_d - x)+  \mathcal{H}(\theta - S^{\star\star}_L) \mathcal{H}(x - C) \mathcal{H}(x_d - x), \qquad \hfill t_{{ref}_1} < t < t_j
%         \end{cases}\\
%     \Pi_b_R = 
%         \begin{cases}
%         \Pi(t=0) \mathcal{H}(x - C) \mathcal{H}(x - x_d) + \Pi(x=x_d) \mathcal{H}(C - x) \mathcal{H}(x - x_d), \qquad\hfill 0 < t < t_{{ref}_2}\\
%         % \Pi(x=0) \mathcal{H}(x - C) \mathcal{H}(x - x_{d}) = 
%         \mathcal{H}(\theta-S_{B}) \mathcal{H}( x - C) \mathcal{H}(x - x_d) + \mathcal{H}(\theta - S^{\star\star}_R) \mathcal{H}(C - x) \mathcal{H}(x - x_d), \qquad\hfill t_{{ref}_2} < t < t_j
%         \end{cases}
% \end{cases}
% \end{align}
%
Where $\Pi_{b_R}$ and $\Pi_{b_L}$ manifest development of characteristics in the right and left branches, respectively. Also, we use $\Pi(t=0)= \mathcal{H}(\theta - S_{wi}), \Pi(x=0) = \mathcal{H}(\theta - S_{B})= \mathcal{H}(\theta-1+ S_{oi}), \Pi(x=x_d) = \mathcal{H}(\theta - S_{D})$.
%%%%%%%%%%%%%%%%%%%%%%%%%%%%%%%%%%%%%%%%%%%%%%%%%%%%%%%%%%%%%%%%%%%%%
We define $C(\theta,x,t) = \bigintss_0^t v(\theta,x(t^{\prime}),t^\prime) dt^\prime = \dfrac{1}{\phi(x)} \dfrac{\partial f(\theta)}{\partial \theta}\bigintss_0^t q(t^\prime) dt^\prime = \dfrac{\partial f(\theta)}{\partial \theta}\bigintss_0^t v(x,t^\prime) dt^\prime$.

%%%%%%------------------------------ F_s(S_w) horizontal 
We will present the general formulation of CDF of water saturation, with uncertainty in both porosity field and injection flux encapsulated in $ U(x,t)= \dfrac{q(t)}{\phi(x)} $. For the cases with one shock, i.e. a horizontal reservoir as well as an inclined reservoir with updip flooding, the CDF of saturation is then formulated as below (\cite{wang2013cdf}),

\begin{align} \label{eq: CDF horizontal}
F_s(\theta;x,t) =
\begin{cases}
  \bigintss_{0}^{\infty}  \Pi_a \mathcal{H}(x-x_f) P_{U}(u) du, \hfill S_{wi} \leq \theta<S^\star\\
  \bigintss_{0}^{\infty}  \Pi_a \mathcal{H}(x-x_f) P_{U}(u) du + \Pi_b \mathcal{H}(x_f-x) P_{U}(u) du,\quad \hfill S^\star \leq \theta\leq 1-S_{oi}
\end{cases}
\end{align}

We introduce $F_{U(x,t)}$, CDF of the input random velocity field $U(x,t)$, and employ the notation $a\vee b$ and $a\wedge b$ for the $max(a,b)$ and $min(a,b)$ respectively. We expand this relation using a similar approach in \cite{ibrahima2017multipoint} as follows,
%------------------ F_s_a horizontal

\begin{align} \label{eq: F_s_a horizontal with Ut kde generated}
 F_{s_a}(\theta;x,t) & =
  \bigintsss_{0}^{\infty}   \Pi_a \mathcal{H}(x-x_f) P_{U}(u)  du = \mathcal{H}(\theta-S_{wi}) \bigintsss_{0}^{\infty} \mathcal{H} \bigg(x- f^{\prime}(S^{\star}) U(x,t) \bigg) P_{U}(u) du\nonumber\\
& = \mathcal{H}(\theta-S_{wi}) \bigintsss_{0}^{\infty} \mathcal{H}\bigg(\dfrac{x}{f^{\prime}(S^{\star})}- u \bigg) P_{U}(u) du =  \mathcal{H}(\theta-S_{wi}) F_{U(t)}\bigg(\dfrac{x}{f^{\prime}(S^{\star})}\bigg)
\end{align}

%------------------ F_s_b horizontal
\begin{align}\label{eq: F_s_b horizontal with Ut kde generated}
 & F_{s_b}(\theta;x,t) =  F_{s_a}(\theta;x,t)  + \bigintsss_{0}^{\infty}\Pi_b \mathcal{H}(x_f-x) P_{U}(u)  du  \nonumber\\
& = F_{s_a}(\theta;x,t)  + \bigintsss_{0}^{\infty}\Bigg[\mathcal{H}(\theta-S_{wi}) \mathcal{H}(x- C) \mathcal{H}(x_f - x) 
 +\mathcal{H}(\theta-S_B) \mathcal{H}(C - x) \mathcal{H}(x_f - x) \Bigg] P_{U}(u) du \nonumber\\
& = F_{s_a}(\theta;x,t)  + \bigintsss_{0}^{\infty}\Bigg[\mathcal{H}(\theta-s_{wi}) \mathcal{H}\bigg(x- f^{\prime}(\theta) U(x,t)\bigg) \mathcal{H} \bigg(f^{\prime}(S^{\star}) U(x,t) - x \bigg) \nonumber\\
& \qquad \qquad\qquad\qquad\qquad + \mathcal{H}(\theta-S_B) \mathcal{H}\bigg(f^{\prime}(\theta) U(x,t) - x\bigg) \mathcal{H}\bigg(f^{\prime}(S^{\star}) U(x,t) - x \bigg) \Bigg] P_{U} du \nonumber\\
& = F_{s_a}(\theta;x,t)  + \bigintsss_{0}^{\infty}\Bigg[\mathcal{H}(\theta-S_{wi}) \mathcal{H}\bigg(\dfrac{x}{f^{\prime}(\theta)}-  u\bigg) \mathcal{H} \bigg( u - \dfrac{x}{f^{\prime}(S^{\star})} \bigg)  \nonumber\\
& \qquad \qquad\qquad\qquad\qquad + \mathcal{H}(\theta-S_B) \mathcal{H}\bigg( u - \dfrac{x}{f^{\prime}(\theta)}\bigg) \mathcal{H}\bigg( u - \dfrac{x}{f^{\prime}(S^{\star})} \bigg) \bigg] P_{U} du \nonumber\\
& = F_{s_a}(\theta;x,t) + \mathcal{H}(\theta-S_{wi}) \Bigg( F_U \bigg(\dfrac{x}{f^{\prime} (\theta)}\bigg) - F_U \bigg(\dfrac{x}{f^{\prime} (S^{\star})} \bigg)\Bigg) + \mathcal{H}(\theta-S_{B}) \Bigg(1- F_U \bigg(\dfrac{x}{f^{\prime} (\theta)} \vee \dfrac{x}{f^{\prime} (S^{\star})} \bigg)\Bigg)
\end{align}

%-----------------------------------------------------------------------------------
% Due to the parametric uncertainty in porosity, or/and injection flux, the total velocity field $\textbf{v}_T$ is random, and so is water saturation $S_w$ and location of the front $x_f(t)$. Therefore, the joint PDF of saturation and front location $(P_{s,x_{f}})$ are encapsulated in this relation through $P_{\Phi}$. That is, $P_{s,x_f} (s^\prime, x_f^\prime; x,t) ds^\prime dx_f^\prime = P_{\Phi} d\Phi$. 
%-------------- F_s(S_w) before reflection
For the downdip case as well as the pure gravity segregation case before reflection of the waves from boundaries, the CDF equation reads,

\begin{align} \label{eq: CDF gravity segregation, before reflection}
F_s(\theta;x,t) =\begin{cases}
  \bigintss_{0}^{\infty}  \Pi_a \mathcal{H}(x-x_{f_R}) P_U(u) du, \hfill       S_{wi}\leq \theta<S_R^\star\\
 \bigintss_{0}^{\infty} \big[\Pi_a \mathcal{H}(x-x_{f_R}) + \Pi_{b_R}                  \mathcal{H}(x_{f_R}-x) \big] P_U(u) du, \hfill  S_R^\star \leq \theta\leq S_D\\
 \bigintss_{0}^{\infty} \big[ \Pi_{b_L}  \mathcal{H}(x-x_{f_L})\big] P_U(u) du, \hfill  S_D \leq \theta\leq S_L^\star\\
 \bigintss_{0}^{\infty} \big[ \Pi_{b_L}   \mathcal{H}(x-x_{f_L}) + \Pi_c \mathcal{H}(x_{f_L}-x)\big]  P_U(u) du,\quad \hfill S_L^\star<\theta\leq 1-S_{oi}
\end{cases}
\end{align}

Leveraging $F_{U(x,t)}$, this relation expands as follows.
%------------------ F_s_a downdip
For $x > x_D$, $S_{wi}< \theta < S^{\star}_R $ we have,

\begin{align} \label{eq: F_s_a downdip with Ut kde generated}
 F_{s_a}(\theta;x,t) & =
  \bigintsss_{0}^{\infty}   \Pi_a \mathcal{H}(x-x_{f_R}) P_{U}(u)  du = \mathcal{H}(\theta-S_{wi}) \bigintsss_{0}^{\infty} \mathcal{H} \bigg(x- f^{\prime}(S^{\star}_R) U(x,t) \bigg) P_{U}(u) du\nonumber\\
& = \mathcal{H}(\theta-s_{wi}) \bigintsss_{0}^{\infty} \mathcal{H}\bigg(\dfrac{x}{f^{\prime}(S^{\star}_R)}- u \bigg) P_{U}(u) du =  \mathcal{H}(\theta-S_{wi}) F_{U}\bigg(\dfrac{x}{f^{\prime}(S^{\star}_R)}\bigg)
\end{align}

%------------------ F_s_b downdip  right side x>xD
For $x > x_D$, $S^{\star}_R < \theta < S_D$ we have,

\begin{align}\label{eq: F_s_b_R downdip with Ut kde generated}
& F_{s_{b_R}}(\theta;x,t)  =  F_{s_a}(\theta;x,t)  + \bigintsss_{0}^{\infty}\Pi_b \mathcal{H}(x_{f_R}-x) P_{U}(u)  du  \nonumber\\
& = F_{s_a}(\theta;x,t)  + \bigintsss_{0}^{\infty}\Bigg[\mathcal{H}(\theta-S_{wi}) \mathcal{H}(x - C) \mathcal{H}(x_{f_R} - x) 
 +\mathcal{H}(\theta-S_D) \mathcal{H}(C - x) \mathcal{H}(x_{f_R} - x) \Bigg] P_{U}(u) du \nonumber\\
& = F_{s_a}(\theta;x,t) + \mathcal{H}(\theta-S_{wi}) \Bigg( F_U \bigg(\dfrac{x}{f^{\prime} (\theta)}\bigg) - F_U \bigg(\dfrac{x}{f^{\prime} (S^{\star}_R)} \bigg)\Bigg) + \mathcal{H}(\theta-S_{D}) \Bigg(1- F_U \bigg(\dfrac{x}{f^{\prime} (\theta)} \vee \dfrac{x}{f^{\prime} (S^{\star}_R)} \bigg)\Bigg)
\end{align}

%------------------ F_s_b downdip. left side x<xD
And for $x < x_D$, $S_D < \theta < S^{\star}_L$ we use,

\begin{align}\label{eq: F_s_b_L downdip with Ut kde generated}
 F_{s_{b_L}}(\theta;x,t) & =  \bigintsss_{0}^{\infty}\Pi_b \mathcal{H}(x - x_{f_L}) P_{U}(u)  du  \nonumber\\
& =\bigintsss_{0}^{\infty}\Bigg[\mathcal{H}(\theta-S_{B}) \mathcal{H}( C- x) \mathcal{H}(x - x_{f_L}) 
 +\mathcal{H}(\theta-S_D) \mathcal{H}(x - C) \mathcal{H}(x - x_{f_L}) \Bigg] P_{U}(u) du \nonumber\\
& =   \mathcal{H}(\theta-S_{B}) \Bigg(  F_U \bigg(\dfrac{x}{f^{\prime} (S^{\star}_L)} \bigg)- F_U \bigg(\dfrac{x}{f^{\prime} (\theta)}\bigg)\Bigg) + \mathcal{H}(\theta-S_{D}) \Bigg(1- F_U \bigg(\dfrac{x}{f^{\prime} (\theta)} \vee \dfrac{x}{f^{\prime} (S^{\star}_L)} \bigg)\Bigg)
\end{align}

%----------------------- F_s_c downdip
for $x < x_D$, $ S^{\star}_L < \theta < S_B$ we leverage,

\begin{align} \label{eq: F_s_a downdip with Ut kde generated}
 F_{s_c}(\theta;x,t) & = F_{s_{b_L}}(\theta;x,t)+
 \bigintsss_{0}^{\infty}   \Pi_c \mathcal{H}(x_{f_L} - x) P_{U}(u)  du \nonumber\\ & = F_{s_{b_L}}(\theta;x,t) + \mathcal{H}(S_{B} - \theta) \bigintsss_{0}^{\infty} \mathcal{H} \bigg( f^{\prime}(S^{\star}_L)U(x,t) -x \bigg) P_{U}(u) du\nonumber\\
& = F_{s_{b_L}}(\theta;x,t)+ \mathcal{H}(S_{B} - \theta) \bigintsss_{0}^{\infty} \mathcal{H} \bigg( u -\dfrac{x}{f^{\prime}(S^{\star}_L)} \bigg) P_{U}(u) du \nonumber\\
& = F_{s_{b_L}}(\theta;x,t)+ \mathcal{H}(S_{B} - \theta) \Bigg( 1 - F_{U} \bigg(\dfrac{x}{f^{\prime}(S^{\star}_L)} \bigg) \Bigg) 
\end{align}

%----------- F_s(S_w) gravity segregation/ after reflection
Whereas the gravity segregation scenario after the reflection of waves follows the CDF relation below,

\begin{align} \label{eq: CDF gravity segregation, after reflection}
F_s(\theta;x,t) =\begin{cases}
 \bigintss_{0}^{\infty}  \Pi_{b_L} \mathcal{H}(x_{f_L} - x) P_U(u) du, \quad S_{wi}\leq \theta < S_L^{\star\star}\\
 \bigintss_{0}^{\infty} \big[\Pi_{b_L} \mathcal{H}(x_{f_L} - x) + \Pi_b                       \mathcal{H}(x_{f_R}-x)\mathcal{H}(x-x_{f_L}) \big] P_U(u) du, \quad S_R^\star \leq \theta\leq    S_L^\star\\
 \bigintss_{0}^{\infty} \big[\Pi_{b_L} \mathcal{H}(x_{f_L} - x) + \Pi_b                      \mathcal{H}(x_{f_R}-x)\mathcal{H}(x-x_{f_L})  + \Pi_{b_R} \mathcal{H}(x-x_{f_R})\big]  P_U(u) du,\\
  \hfill S_R^{\star\star}<\theta\leq 1-S_{oi}
\end{cases}
\end{align}

This relation is consequently expanded as below.
%------------------ F_s_a gravity segregation/after reflection from boundaries
For $x<x_{S_L^{\star\star}}$, $S_{wi} < \theta < S^{\star\star}_L$ we have,

\begin{align}\label{eq: F_s_a gravity column with Ut kde generated}
& F_{s_{a}}(\theta;x,t)  = 
 \bigintsss_{0}^{\infty}\Pi_{b_L} \mathcal{H}(x_{f_L}-x) P_{U}(u)  du  \nonumber\\
& = \bigintsss_{0}^{\infty}\Bigg[\mathcal{H}(\theta - S_L^{\star\star}) \mathcal{H}(x- C) \mathcal{H}(x_{f_L} - x) 
 +\mathcal{H}(\theta - S_{wi}) \mathcal{H}(C - x) \mathcal{H}(x_{f_L} - x) \Bigg] P_{U}(u) du \nonumber\\
& = \mathcal{H}(\theta - S_L^{\star\star}) \Bigg( F_U \bigg(\dfrac{x}{f^{\prime} (\theta)}\bigg) - F_U \bigg(\dfrac{x}{f^{\prime} (S_L^{\star\star})} \bigg)\Bigg) + \mathcal{H}(\theta-S_{wi}) \Bigg(1- F_U \bigg(\dfrac{x}{f^{\prime} (\theta)} \vee \dfrac{x}{f^{\prime} (S_L^{\star\star})} \bigg)\Bigg)
\end{align}

%------------------ F_s_b gravity segregation/after reflection from boundaries
Analogously to the previous derivations, for $x_{S_L^{\star\star}}<x<x_{S_R^{\star\star}}$, $S^{\star}_R < \theta < S^{\star}_L$ we have,

\begin{align} \label{eq:Fsbr after reflection}
 F_{s_{b_R}}(\theta;x,t) &= F_{s_a}(\theta;x,t) + \mathcal{H}(\theta-S_{wi}) \Bigg( F_U \bigg(\dfrac{x}{f^{\prime} (\theta)}\bigg) - F_U \bigg(\dfrac{x}{f^{\prime} (S_3)} \bigg)\Bigg) \notag\\
& + \mathcal{H}(\theta-S_{D}) \Bigg(1- F_U \bigg(\dfrac{x}{f^{\prime} (\theta)} \vee \dfrac{x}{f^{\prime} (S_3)} \bigg)\Bigg)
 \end{align}
 \begin{align}\label{eq:Fsbl after reflection}
 F_{s_{b_L}}(\theta;x,t)& = F_{s_{a}} + F_{s_{b_R}}+
\mathcal{H}(\theta-S_{B}) \Bigg(  F_U \bigg(\dfrac{x}{f^{\prime} (S_4)} \bigg)- F_U \bigg(\dfrac{x}{f^{\prime} (\theta)}\bigg)\Bigg) \notag\\
& + \mathcal{H}(\theta-S_{D}) \Bigg(1- F_U \bigg(\dfrac{x}{f^{\prime} (\theta)} \vee \dfrac{x}{f^{\prime} (S_4)} \bigg)\Bigg)
\end{align}

%------------------ F_s_c gravity segregation/after reflection from boundaries
For $x > x_{S_R^{\star\star}}$, which corresponds to $ S^{\star\star}_R < \theta < 1-S_{oi}$ we leverage,

\begin{align} \label{eq: F_s_c gravity column with Ut kde generated}
& F_{s_c}(\theta;x,t)  =  F_{s_{a}}(\theta;x,t) +  F_{s_{b}}(\theta;x,t) + 
 \bigintsss_{0}^{\infty}\Pi_{b_R} \mathcal{H}(x - x_{f_R}) P_{U}(u)  du = F_{s_{a}}(\theta;x,t) +  F_{s_{b}}(\theta;x,t)   \nonumber\\
&  +\bigintsss_{0}^{\infty}\Bigg[\mathcal{H}(\theta - S_{B})  \mathcal{H}(x- C) \mathcal{H}(x - x_{f_R}) 
 +\mathcal{H}(\theta - S_R^{\star\star})\mathcal{H}(C - x) \mathcal{H}(x - x_{f_R} ) \Bigg] P_{U}(u) du \nonumber\\
& = F_{s_{a}}(\theta;x,t) + F_{s_{b_R}}(\theta;x,t) + F_{s_{b_L}}(\theta;x,t)\notag\\
& + \mathcal{H}(\theta-S_{B}) \Bigg(1- F_U \bigg(\dfrac{x}{f^{\prime} (\theta)} \vee \dfrac{x}{f^{\prime} (S_R^{\star\star})} \bigg)\Bigg) + \mathcal{H}(\theta - S_R^{\star\star}) \Bigg( F_U \bigg(\dfrac{x}{f^{\prime} (S_R^{\star\star})} \bigg)- F_U \bigg(\dfrac{x}{f^{\prime} (\theta)}\bigg) \Bigg) 
\end{align}

\begin{figure}[htbp]
 	\centering
	%	\label{fig:a}
\includegraphics[width=60mm]{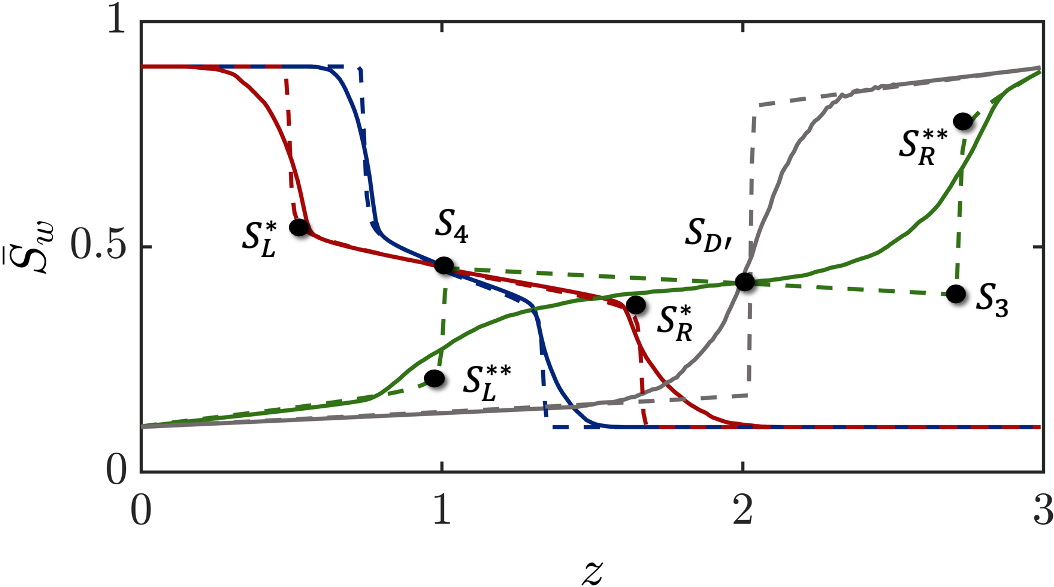}
\caption{Saturation profile for the inverted gravity column, before and after being reflected from the boundaries, solid lines are the Monte Carlo results, while dashed lines represent the corresponding deterministic solutions.}
	\label{fig:saturation annotated }
\end{figure}

Where $F_{s_{b}}(\theta;x,t) $ should be found from the solutions of the before reflection time in Eq. \ref{eq: F_s_b_R downdip with Ut kde generated} and Eq.\ref{eq: F_s_b_L downdip with Ut kde generated}. However, those solutions were developed for $(S_R^{\star}, S_L^{\star})$ rarefaction zone, whereas, the middle rarefaction region here has a different saturation range $(S_3, S_4)$ (Fig.\ref{fig:saturation annotated }), which keeps changing and becoming smaller over time, until it totally fades away at $t_j$ where one shock forms in $z_{eq}$. That is, after reflection from boundaries, the solution is no longer self-similar, and the formula developed in Eq.\ref{eq: F_s_a gravity column with Ut kde generated} and Eq.\ref{eq: F_s_c gravity column with Ut kde generated} using the method of distributions, even though following similar logic to the previous equations, fail to accurately capture the $F_s(s;x,t)$ predicted from Monte Carlo scheme. Fig. \ref{fig:cdfs for gravity column PHI random} represents a reasonable prediction for CDF of saturation at $t_1, t_2$ (before reflection), while at $t_3, t_4$ (after reflection) the CDF of saturation obtained from Eq.\ref{eq: F_s_a gravity column with Ut kde generated} and Eq.\ref{eq: F_s_c gravity column with Ut kde generated}, don't capture the solution of Monte Carlo scheme. 

%--------------------------------------- Plotting sat profile

%%%%%%%%%%%%%%%%%%%%%%%%%%%%%%%%%%%%%%%%%%%%%%%%%%%%%%%%%%%%%%%%%%%%%%%%%%%%%%
% \begin{align} \label{eq: F_s_a gravity seg with Ut kde generated}
%  F_{s_a}(\theta;x,t) & =
%   \bigintsss_{0}^{\infty}   \Pi_a \mathcal{H}(x_{f_L} - x) P_{q}  dQ = ...
% \end{align}
% %------------------ F_s_b gravity segregation/after reflection from boundaries
% We have already found the solution to this region as the rarefaction fan of $t<t_{eq}$.
% \begin{align}\label{eq: F_s_b gravity seg with Ut kde generated}
%  F_{s_b}(\theta;x,t) & =  F_{s_a}(\theta;x,t)  + \bigintsss_{0}^{\infty}\Pi_b \mathcal{H}(x_{f_R}-x)\mathcal{H}(x-x_{f_L}) P_{q}  dQ  \nonumber\\
% & = F_{s_a}(\theta;x,t)  + \bigintsss_{0}^{\infty}\Bigg[ \Bigg] P_{q} dQ \nonumber\\
% & ...
% \end{align}
% %----------------------- F_s_c gravity segregation/after reflection from boundaries
% \begin{align} \label{eq: F_s_a gravity seg with Ut kde generated}
%  F_{s_c}(\theta;x,t) & =
%   F_{s_a}(\theta;x,t)  + F_{s_b}(\theta;x,t)  +\bigintsss_{0}^{\infty}   \Pi_c \mathcal{H}(x - x_{f_R} ) P_{q}  dQ = ...
% \end{align}
%-------------------------------------------------------

Note that CDF is a cumulative quantity, and hence we need to add the CDF of all the previous saturation intervals, starting from the lowest saturation range onward. 
Note that $F_{U(x,t)}(x,t)$ is only a function of time as $F_{q(t)}(t)$ when the injection flux $q(t)$ is the sole source of uncertainty, and is only a function of space $F_{\phi(x)}(x)$ when porosity field $\phi(x)$ is the sole source of uncertainty. It is a space-time dependent function for cases when both fields are random. We use Kernel Density Estimation (KDE) to find $F_{U(x,t)}$.
% Also, as we have assumed the injection flux and porosity are two independent physical process, we will use $P_{\varphi, q} = P_{\varphi} P_{q}$.
%
% While Eqs.~(\ref{eq: CDF horizontal})-(\ref{eq: CDF gravity segregation, after reflection}), formulate the CDF of saturation considering both porosity and injection flux as random sources, the extension to other cases is straightforward. That is, for scenarios in which porosity is the sole source of uncertainty, $P_{\varphi, q} d\Phi dQ$ will be replaced with $P_{\varphi} d\phi$, and for case studies in which injection flux is the sole source of randomness, we replace $P_{\varphi, q} d\Phi dQ$ with $P_{q} dQ$.
%
%--------------------------------  find C
% In order to find $C(\theta,x,t) = \dfrac{1}{\phi(x)} \dfrac{\partial f(\theta)}{\partial \theta}\bigintss_0^t q(t^\prime) dt^\prime$, similar to \cite{wang2013cdf}, we can come up with some simplifying approximations for $\bigintss_0^t q(t^\prime) dt^\prime$. To this end, we can estimate $C(\theta,x,t)$ using the expressions below, depending on the correlation time $\tau_{\phi}$,
%
% For $x \ll \tau_{q}$, we assume $\bigintsss_0^t q(t^\prime)dt^\prime \approx \mu_q t$. 
%
% For $x \gg \tau_{q}$, we use a multiplicative temporal Gaussian white noise assumption for estimating the integral, i.e. $\bigintsss_0^t q(t^\prime)dt^\prime \approx \mathcal{N}(t\mu_q, 2t\sigma_q^2)$.
%
% For intermediate times, we use central limit theorem to approximate the integral.
%

We will validate and verify these assumptions by comparing the $F_s(\theta;x,t)$ obtained from Eqs.~(\ref{eq: CDF horizontal})-(\ref{eq: CDF gravity segregation, after reflection}) with their corresponding $F_s(S,x,t)$ obtained from kernel density estimation applied on the Monte Carlo simulation of Eq.~(\ref{eq:saturation_equation_with_BC}) and Eq.~(\ref{eq: pure segregation-----saturation_equation_with_BC}).
It's worth mentioning that in general $U(x,t)=\dfrac{q(t)}{\phi(x)}$, and hence when the injection flux is random, we generate $F_U(t)(t)$ by applying KDE on 10000 samples of $U(t)=q(t)$ while keeping the porosity constant. On the other hand, when porosity is the only random variable, we generate $F_U(x)(x)$ by applying KDE on $U(x)=\dfrac{1}{\phi(x)}$ by generating 10000 samples of $\phi(x)$ while keeping the injection flux constant. Similarly, when both fields are random, 10000 samples of each random variable is generated and then KDE is applied on $U(x,t)=\dfrac{q(t)}{\phi(x)}$. In all three cases the bandwidth is selected to be $n=10^{10}$ in order to guarantee the accuracy of matching with the corresponding Monte Carlo solution of the finite volume Godunov scheme.

%%%%%%%%%%%%%%%%%%%%%%%%%%%%%%%%%%%%%%%%%%%%%%%%%%%%%%%%%%%%%%%%%%%%%%%%%%%
%%%%%       NUMERICAL SETTING : PHI RANDOM, Q RANDOM, BOTH RANDOM
%%%%%%%%%%%%%%%%%%%%%%%%%%%%%%%%%%%%%%%%%%%%%%%%%%%%%%%%%%%%%%%%%%%%%%%%%%%%
%------------------------------------------------------
\begin{figure}[htbp]
 	\centering
	%	\label{fig:a}
\includegraphics[width=100mm]{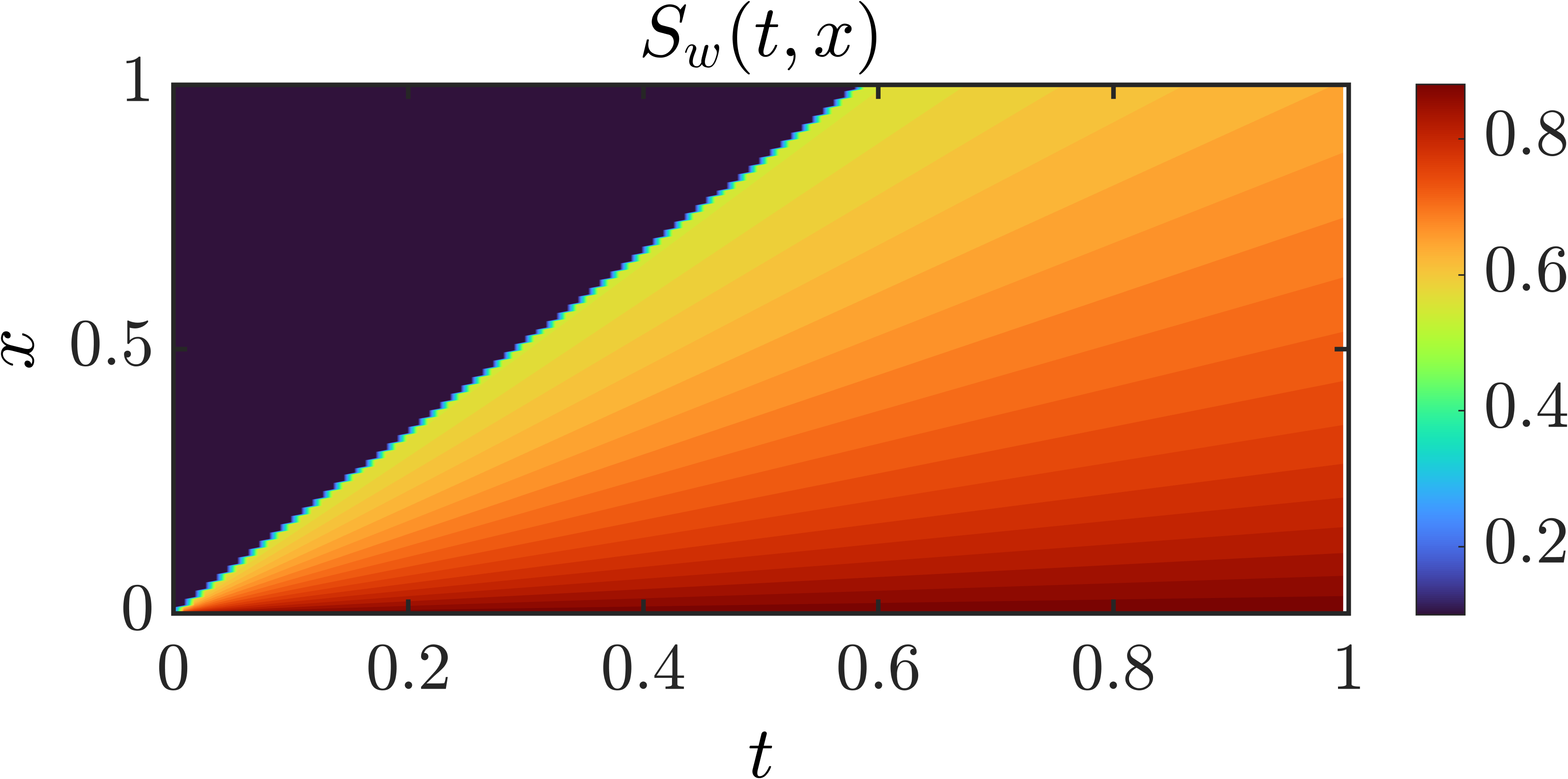}

\vspace{1cm}

\includegraphics[width=98mm]{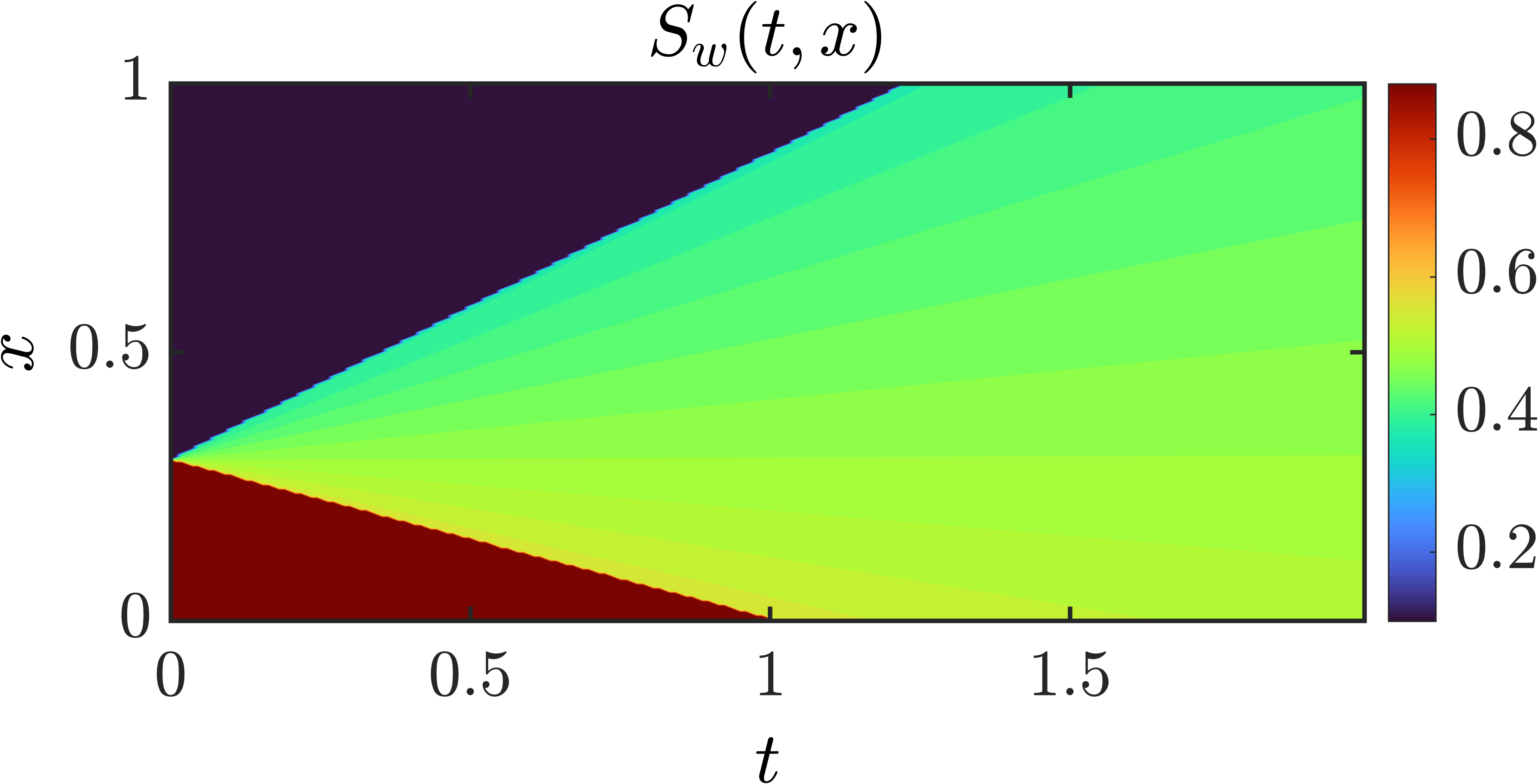}

\vspace{1cm}

\includegraphics[width=100mm]{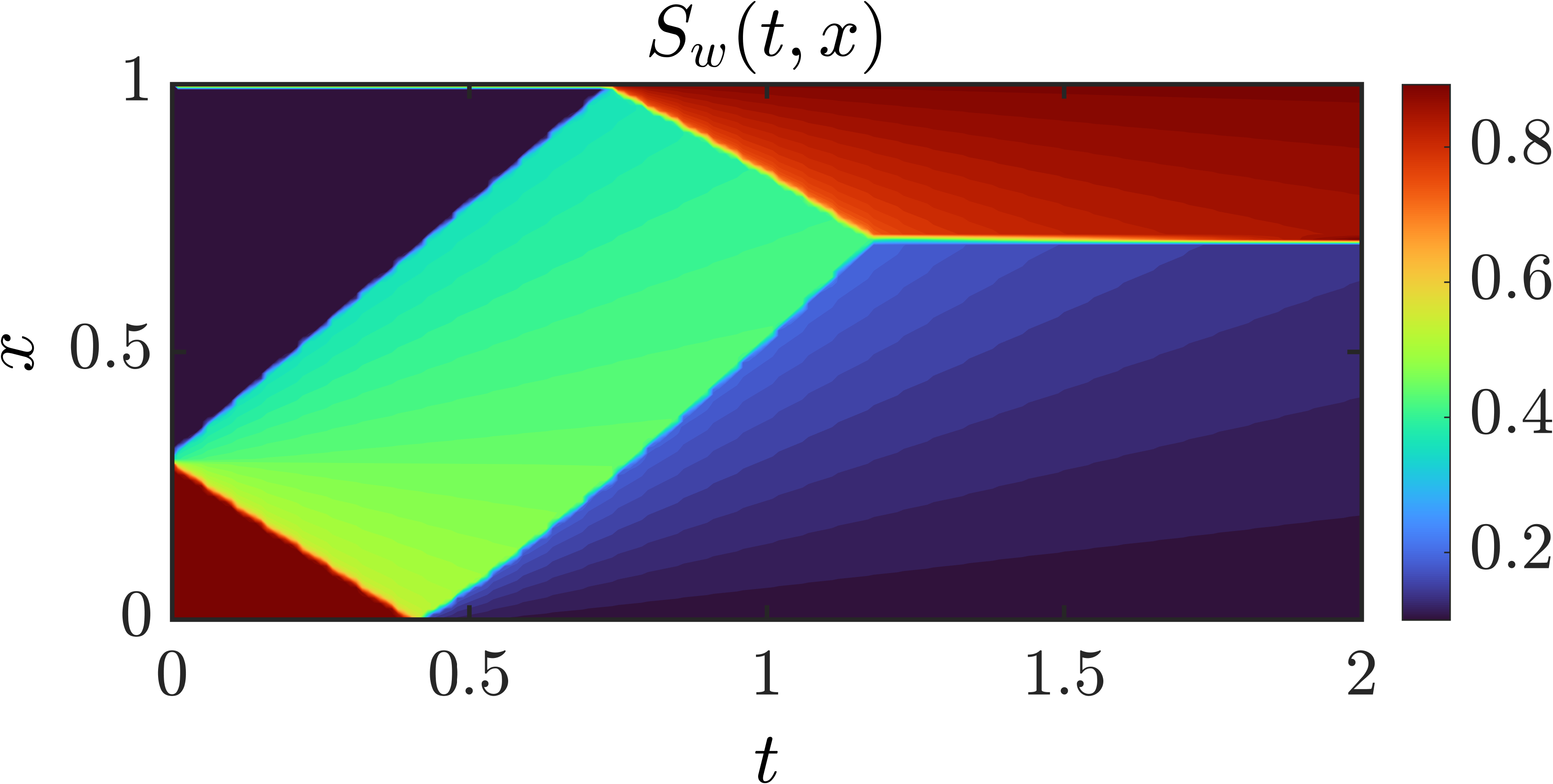}

\caption{Spatio-temporal evolution of the mean saturation field represents development of the shock(s), (top) horizontal flooding with one shock, (middle) downdip flooding with two shocks, (bottom) gravity column with reflection of two shocks from boundaries. Initial discontinuity for the downdip and gravity column cases is located at $Z_d=0.3$ and all the plots are scaled to the same spatial range for better comparison.}
	\label{fig:shock development contours }
\end{figure}

%----------------------------------------------------------------------------

\section{Numerical Experiments} \label{section : Numerical experiments}
\subsection{Setting}
For the Monte Carlo simulation of the nonlinear hyperbolic equation for saturation Eq.~(\ref{eq:saturation_equation_with_BC}) with complicated initial conditions, one must generally resort to numerical schemes like finite element, difference, or volume methods. Accordingly, we have employed finite volume discretization of Godunov scheme to solve the dimensionless form of the nonlinear hyperbolic conservation Eq.~(\ref{eq:saturation_equation_with_BC}) in one-dimension. While previous studies by \cite{ibrahima2015distribution} , \cite{ibrahima2017multipoint}, \cite{ibrahima2018efficient}, have employed a Lagrangian streamline simulation methodology for numerical treatment of the horizontal displacement, we have applied a finite volume strategy to enable numerical manipulation of the multi-shock problems like the downdip displacement in our study.

The CDF and PDF of water saturation $F_s(S_w)$, and $f_s(S_w)$ are generated using kernel density estimation \cite{botev2010kernel} with 10,000 samples of each random variable, while the bandwidth parameter is taken to be $n=2^{10}$. Intuitively, we would like to choose the smallest bandwidth parameter that data will allow. After experimenting with the influence of this parameter on the resulting CDF and PDF plots, we realized that values smaller than $n=2^{7}-2^{12}$ result in an under-smoothed curve with too much spurious data artifacts, while values greater than n yield an over-smoothed curve in which much of the underlying structure is obscured. CDF plots usually need $n=2^{10}$ to capture the transitions from shock to rarefaction zones accurately, while such a large bandwidth leads to overfitting in the PDF plots. Hence, we use $n=2^{8}$ to generate  the PDF plots.

\begin{figure}[htbp]
	\centering
	\includegraphics[width=60mm]{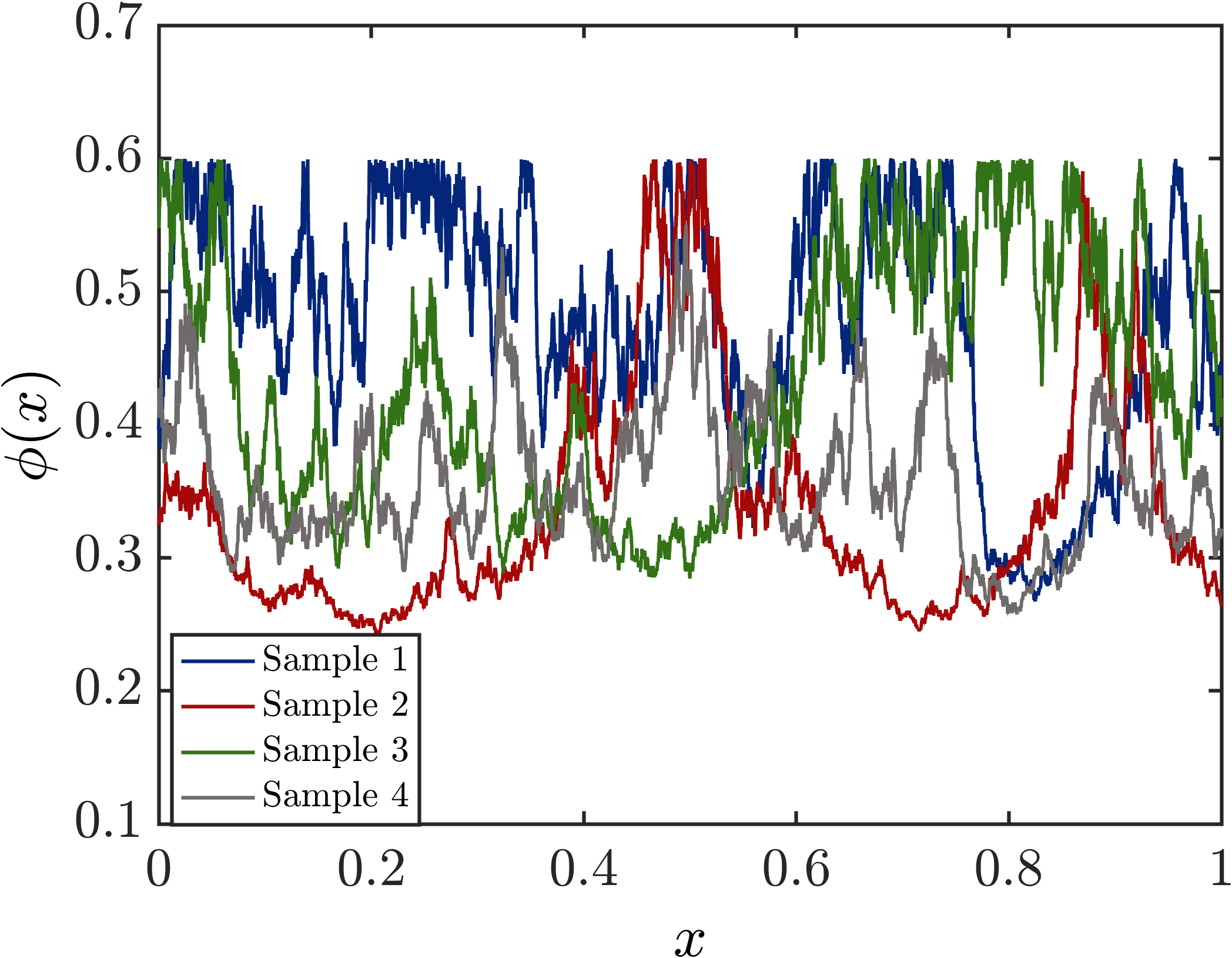}
	\includegraphics[width=60mm]{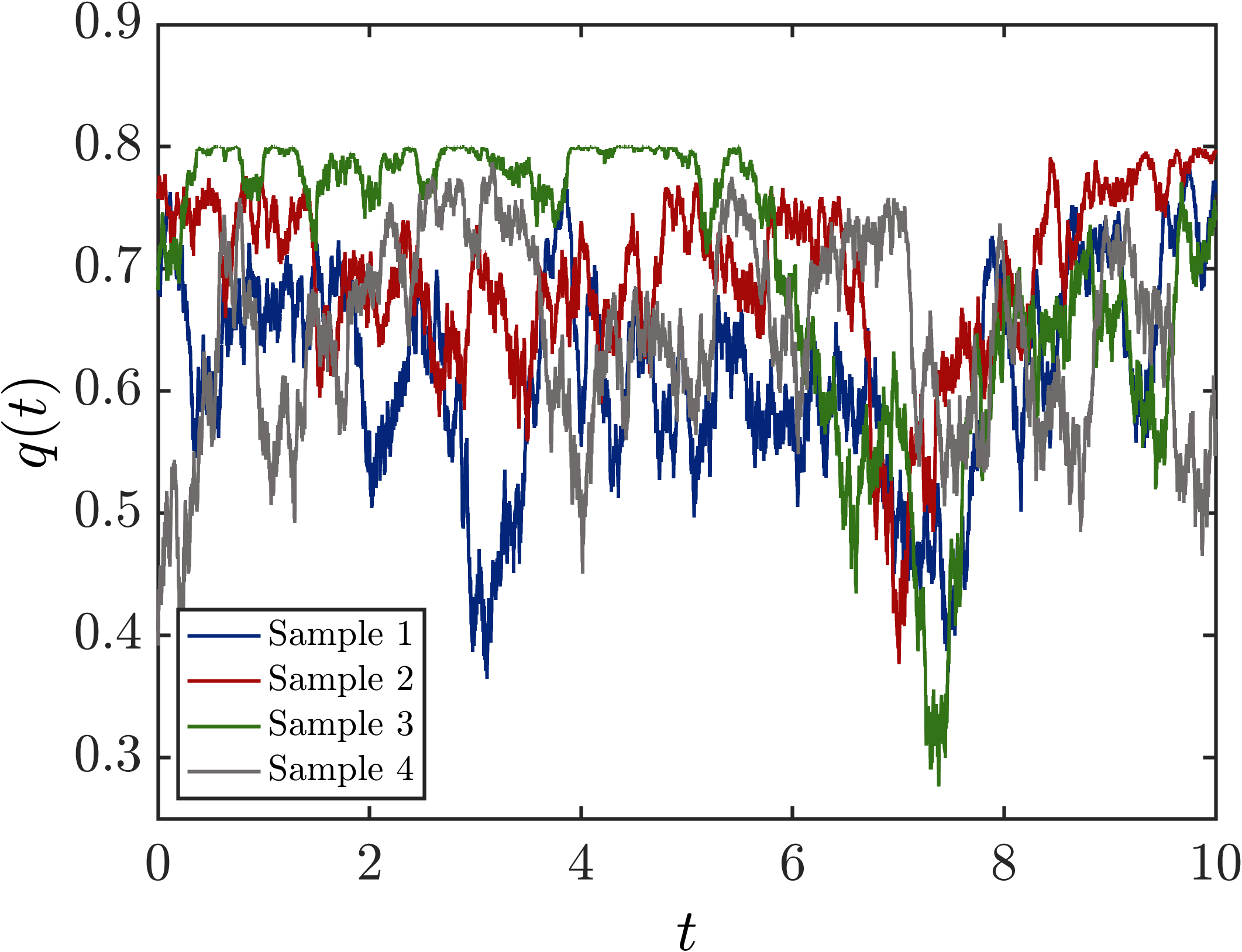}
	\caption{Multiple realizations of the random porosity field (left) and random injection flux (right) with the parameters defined in section~\ref{section:phi and q both random} }
	\label{fig:realizations}
\end{figure}

We will present the influence of parametric uncertainty on the saturation uncertainty by inspecting the mean, variance, temporal evolution of CDF and PDF of water saturation at a specific location, spatial evolution of CDF and PDF of water saturation at a specific point in time and the point-wise average of the CDFs generated at all locations along the domain. The aformentioned profiles are all computed using both Monte Carlo sampling of the Buckley Leverett equation, and the method of distributions. We will study distinct cases for only $\phi(x)$ random, only $q(t)$ random, and when both are stochastic fields. Random porosity field has been restricted to the range $[0.3, 0.6]$, and stochastic injection flux has established to vary between $[0.3,0.8]$ as depicted  in Fig.\ref{fig:realizations}. 
All cases have employed the Brooks and Corey expressions for the relative permeabilities as a function of saturation. We will demonstrate all the experiments for $m=0.5$, which was formerly studied in test cases of \cite{ibrahima2015distribution}, \cite{ibrahima2017multipoint}, \cite{dongxiao1999stochastic}. The irreducible wetting and non-wetting saturations are taken to be $S_{wi}=S_{oi}=0.1$. The injection, if present, always happens at the inlet of domain $x=a$. Final simulation time is $T=1$ for all cases except in the gravity column simulation with $T=10$.

A convergence study at the end of results section will demonstrate the sensitivity of solution to the spatial and temporal grid sizes, as well as number of Monte Carlo trials used to find the results. Also, we will quantify the error between the CDF obtained from Monte Carlo sampling of Godunov equations and the method of distribution. 
Given the CDF of water saturation, the first two statistical moments of saturation read,

\begin{align}
  & \langle S_w\rangle = 1 - \bigintssss_0^1 F_{S_w} (\zeta) d\zeta\\
  & \sigma^2_{S_w} = 1 - \bigintssss_0^1 F_{S_w} (\sqrt{\zeta}) d\zeta - \langle S_w\rangle^2
\end{align}

%-----------------------------------------------
Fig.~(\ref{fig:shock development contours }) represents spatio-temporal evolution of the average saturation fields and development of the shocks for each case study using the Monte Carlo scheme (in all cases the spatial domain is scaled to (0,1) range for easier comparison).

%%%%%%%%%%%%%%%%%%%%%%%%%%%%%%%%%%%%%%%%%% only porosity random
\subsubsection{Numerical setting for stochastic porosity field $\phi(x)$}

In the current scenario, we consider the injection flux $q_T(t)$ at the boundary to be a deterministic constant as $q_{inlet} = 0.3$. Hence, the uncertainty in the porosity field is the only source of randomness. The porosity field $\phi(x)$ is characterized by its know mean, variance, covariance structure, and dimensionless correlation length.
%
% In order to satisfy the general continuity equation for incompressible flow, $\nabla\cdot \boldsymbol{q}=0$, the total flux q has to be constant in x, and is hence equal to the injection flow rate $\boldsymbol{q}(x,t)=q(t)=constant= q_0(t)$. This function can be deterministic or uncertain.

Similarly to \cite{ibrahima2015distribution}, in order to enforce the physical constraint $0<\phi(x)<1$, we first define a random Gaussian field $\gamma(x) \sim \mathcal{N}(\mu_{\gamma}, C_{\gamma}(x))$, with mean $\mu_x$ and exponential covariance structure $C_{\gamma}(x) = \sigma_{\gamma}^2 exp(-\dfrac{x}{\lambda_{\gamma}})$, where $\sigma_{\gamma}^2$ is the variance of $\gamma$, and $\lambda_{\gamma}$ represents the correlation length. Subsequently, we define porosity field as,

\begin{align}
& \phi(x) = 0.01 \dfrac{2}{\pi} \lvert \text{arctan} (\gamma(x) ) \vert + 0.5\bigg(1-\dfrac{2}{\pi} \vert\text{arctan}  (\gamma(x) )\vert\bigg) 
\end{align}

% In consequence, we can choose $\mu_{\gamma}$, we can control mean of the porosity field, i.e. $\langle \phi  (x) \rangle= \mu_{\phi}$. 

In consequence, we can choose $\mu_{\gamma}$ so that $\langle \phi  (x) \rangle= \mu_{\phi}$. 

For a space discretization $\Delta x$ along the domain $[a,b]$, while defining the spatial distance $x=(n-1)\Delta x$, we define the Gaussian vector $(\gamma)^n = (\gamma(j \Delta x ))_{0\leq j\leq n-1}$ by $(\gamma)^n \sim \mathcal{N} (\mu_{\gamma} \mathds{1}_n , (C_{\gamma})^n)$, where $(C_{\gamma})^n = (C_{\gamma} (\vert i-j \vert \Delta x) )_{1\leq i,j\leq n}$, and $\mathds{1}_n$ is a unit vector of size n.

\begin{figure}[htbp]
	\centering
\includegraphics[width=60mm]{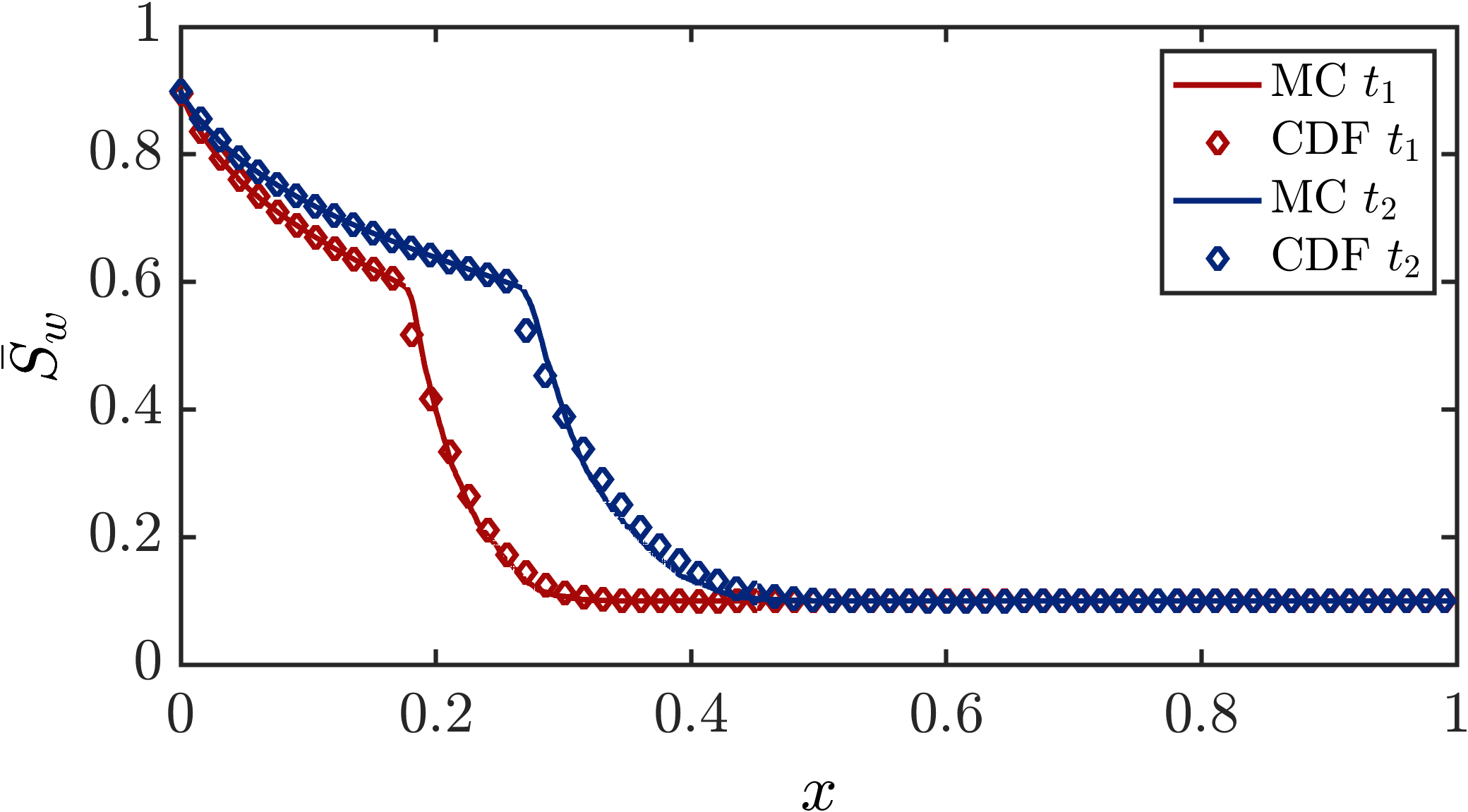}
\includegraphics[width=60mm]{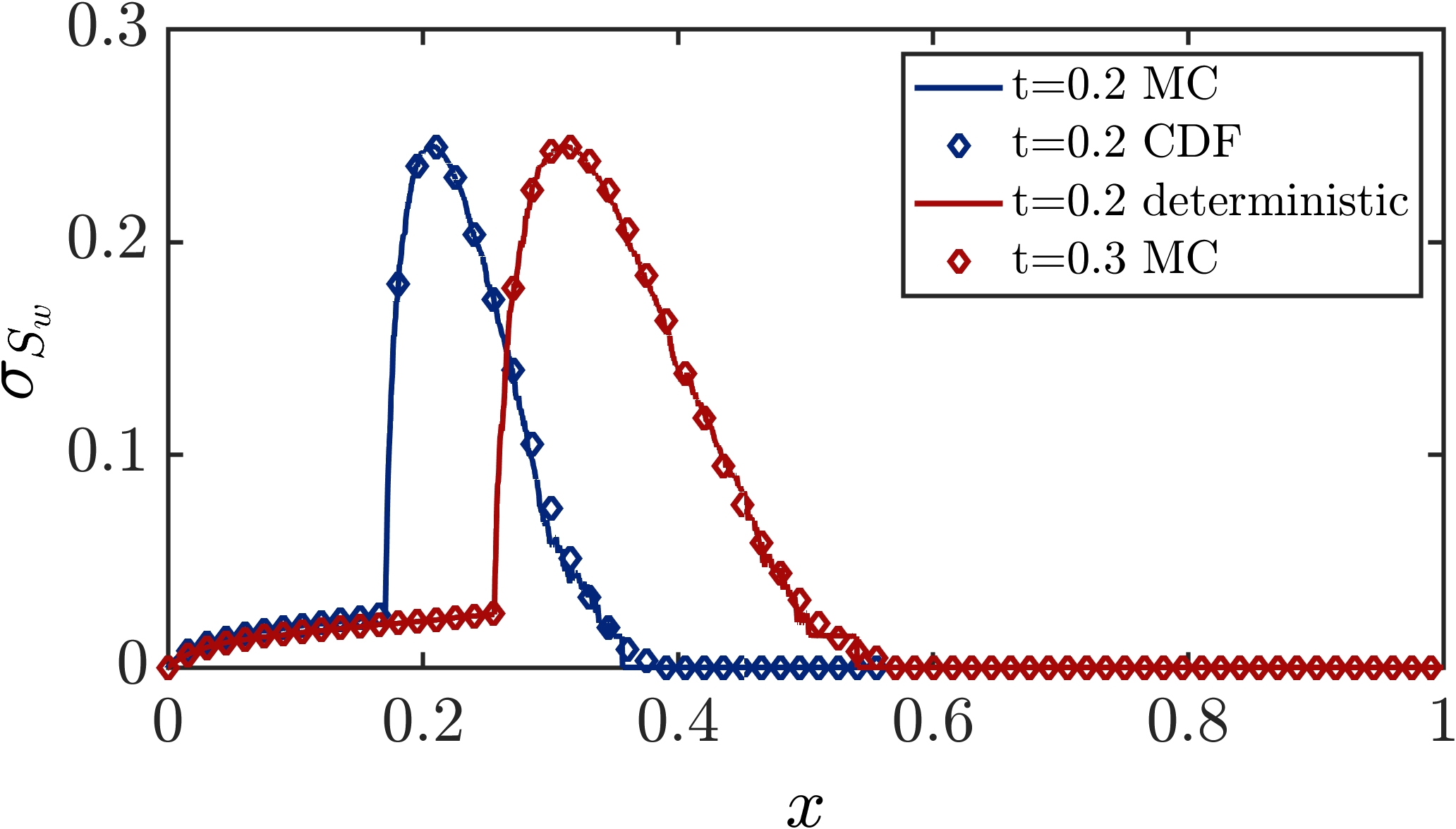}
 \includegraphics[width=60mm]{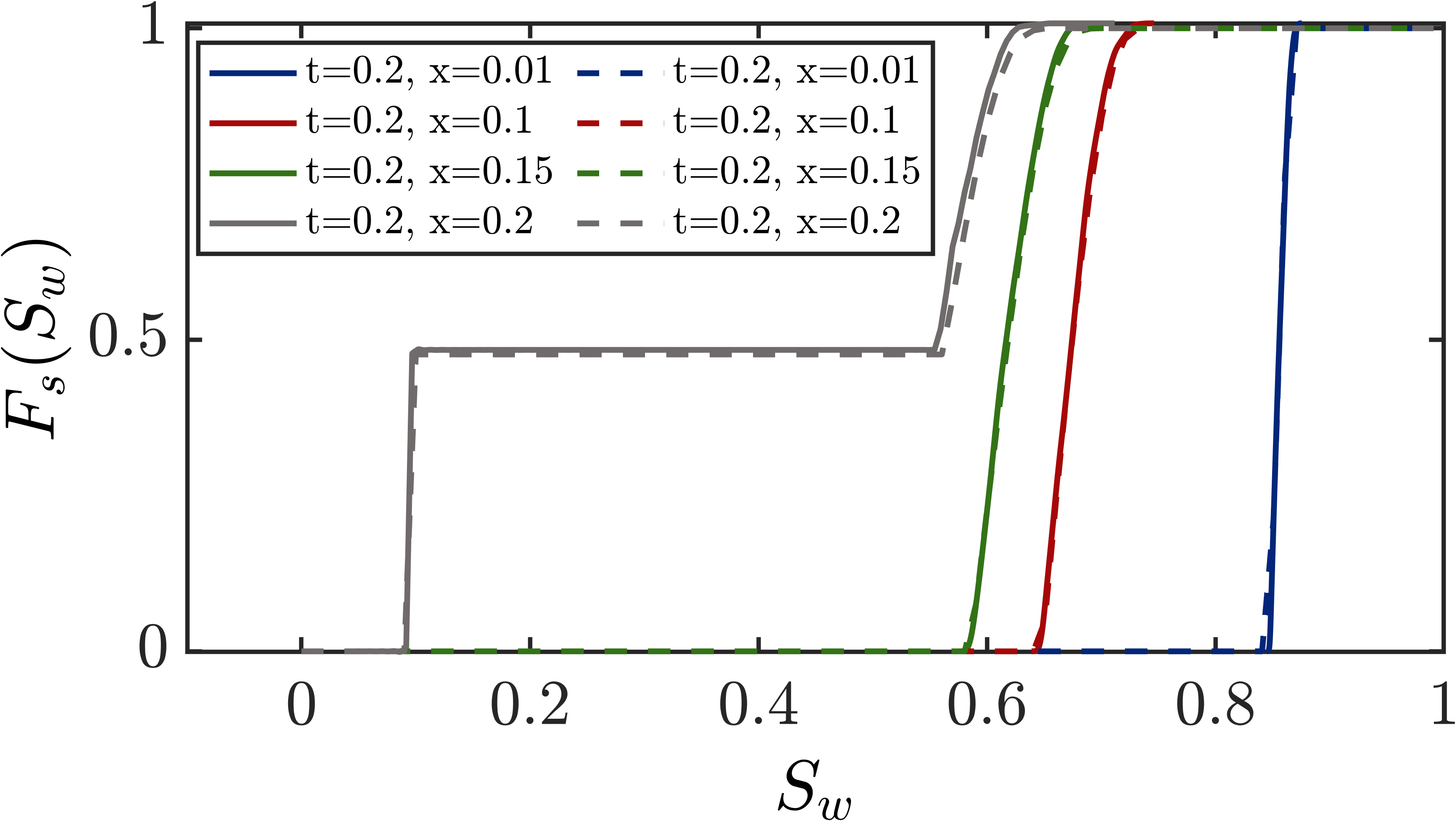}
\includegraphics[width=60mm]{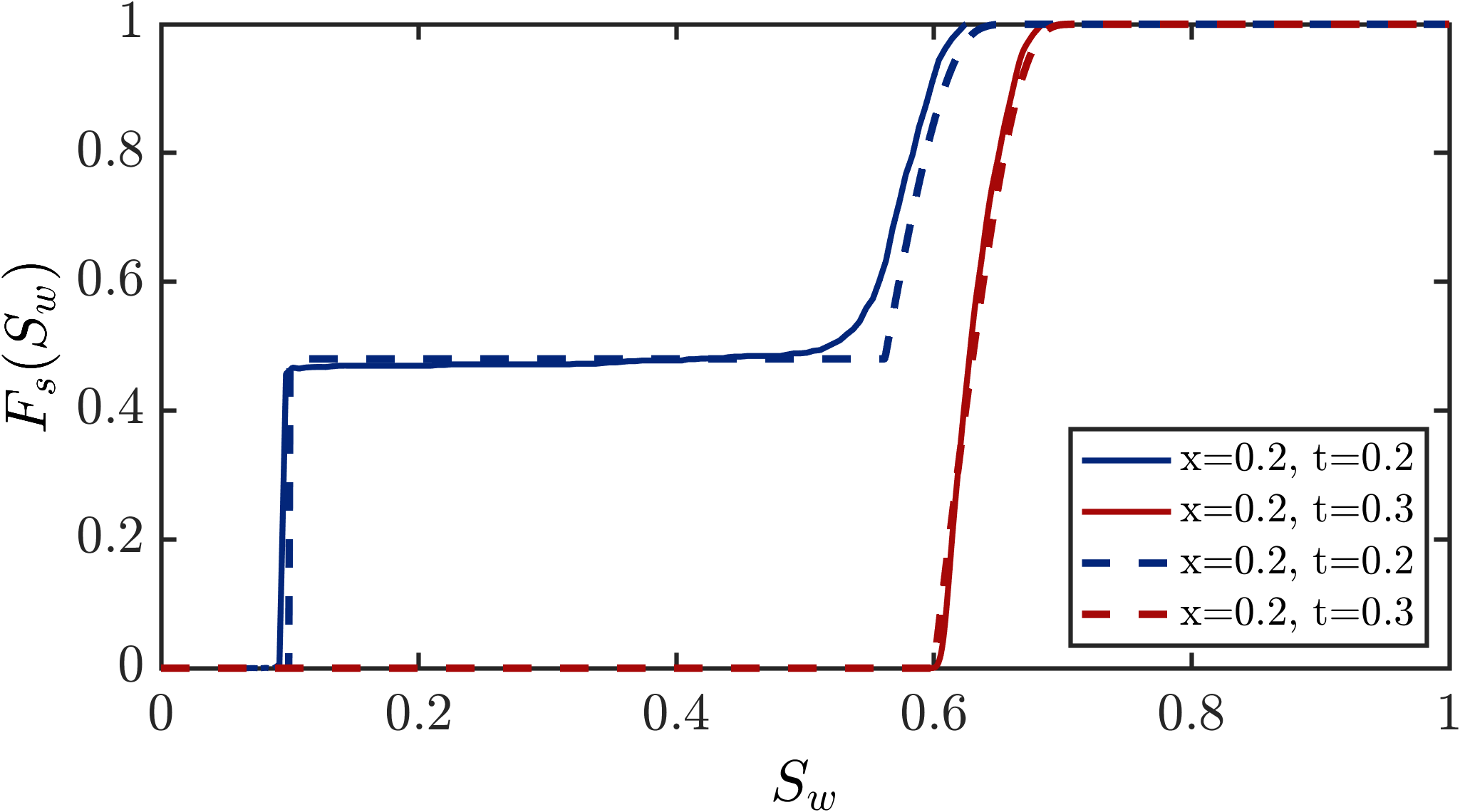}
\includegraphics[width=60mm]{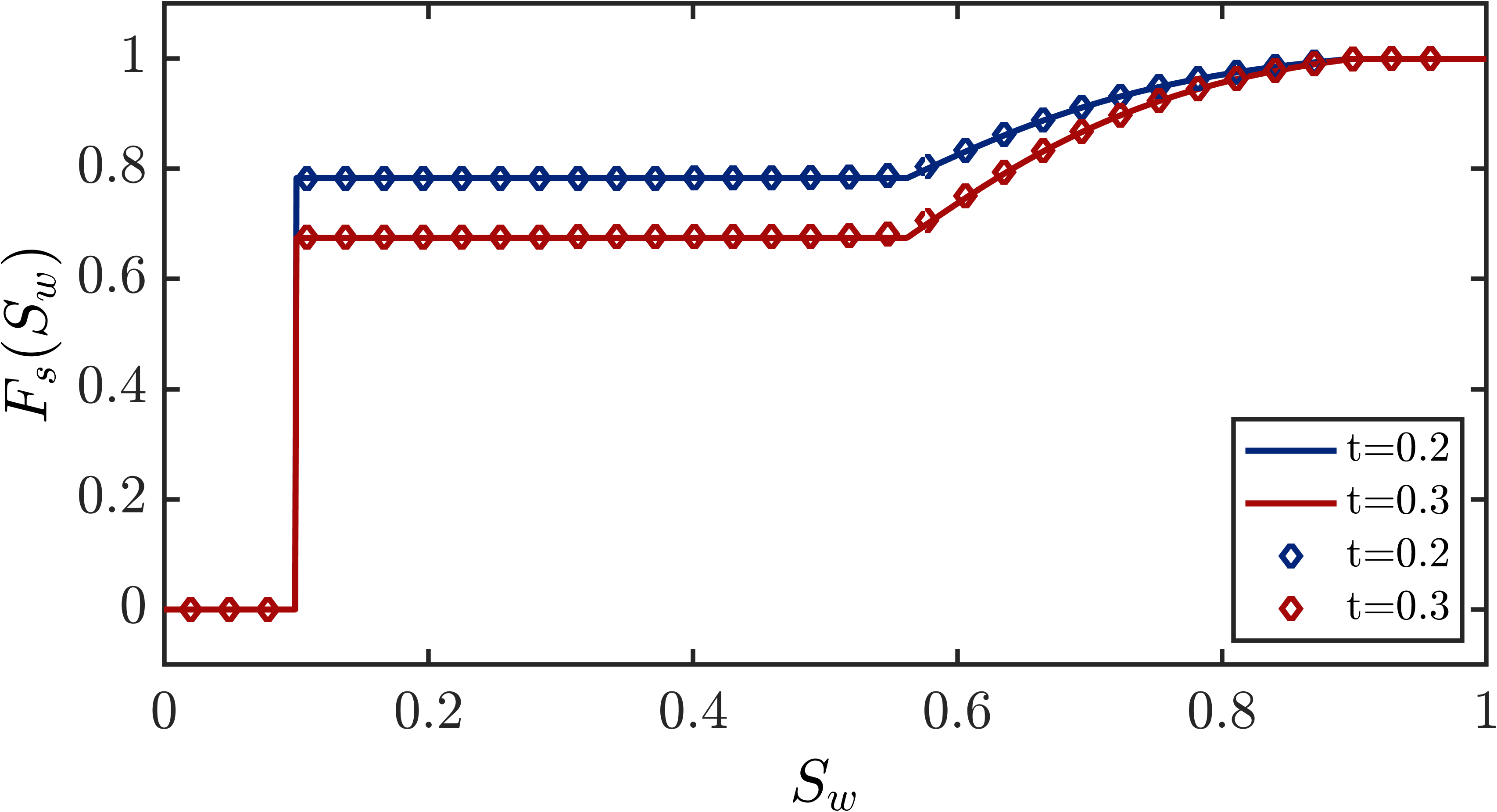}
\includegraphics[width=60mm]{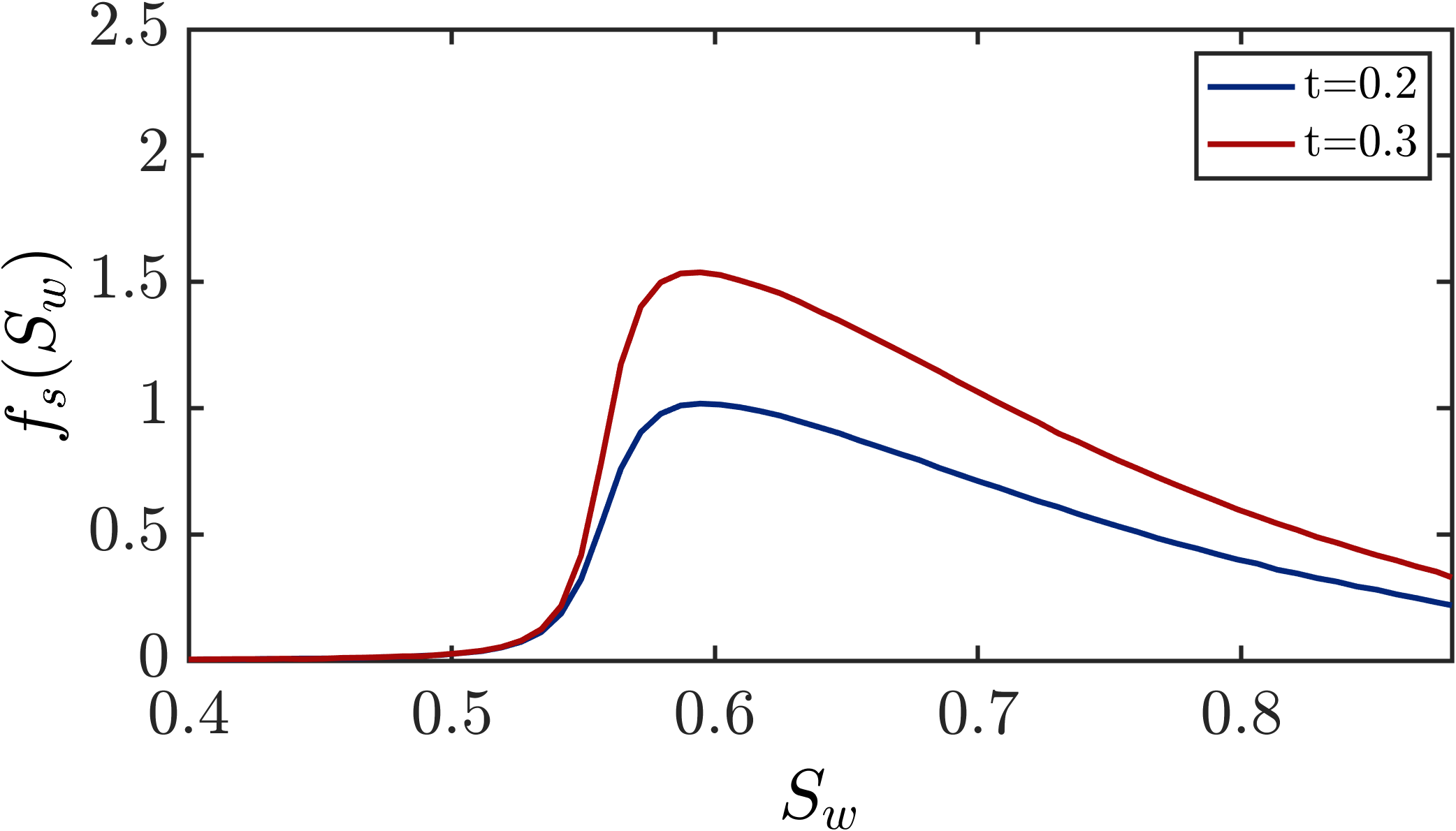}
\includegraphics[width=60mm]{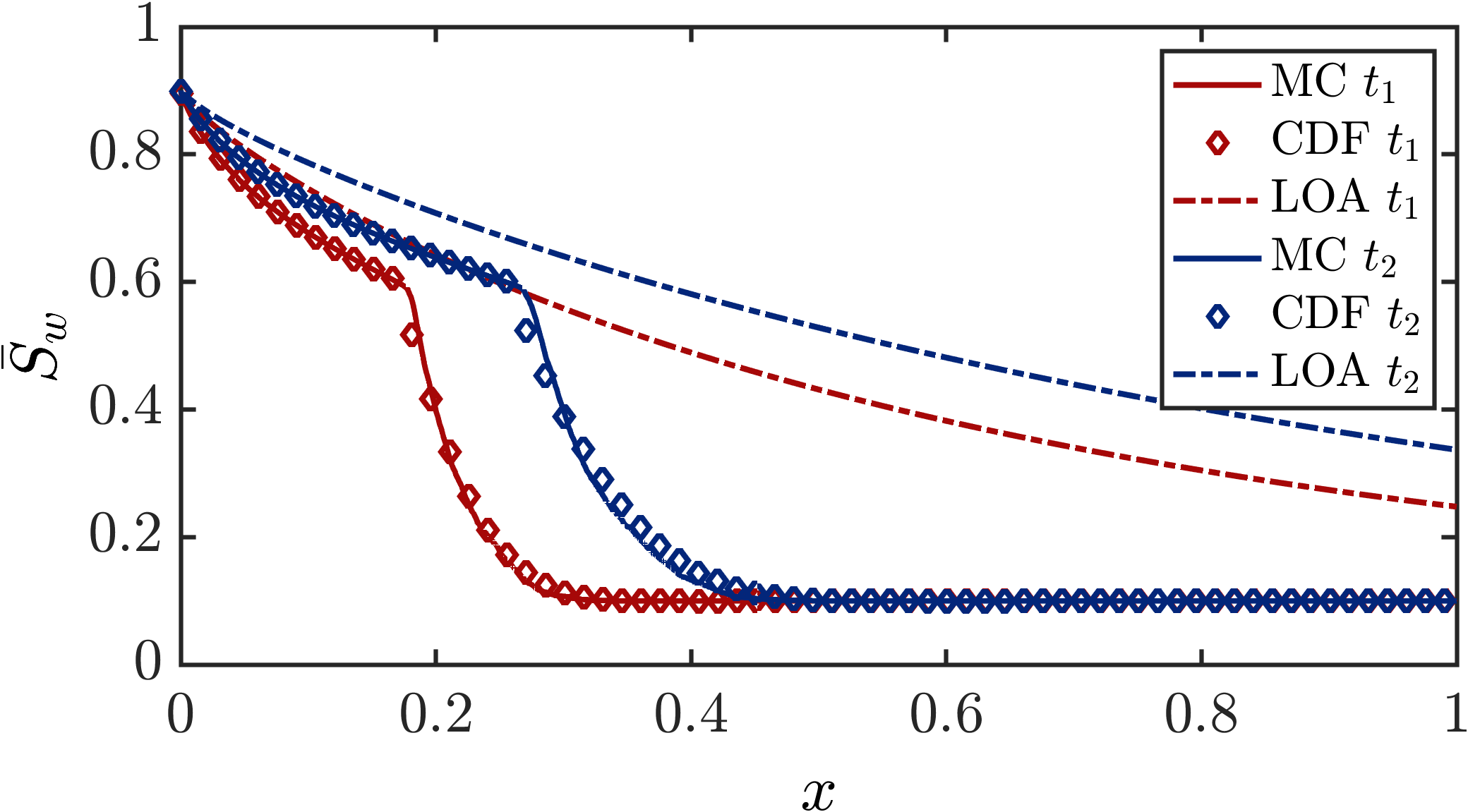}
\includegraphics[width=60mm]{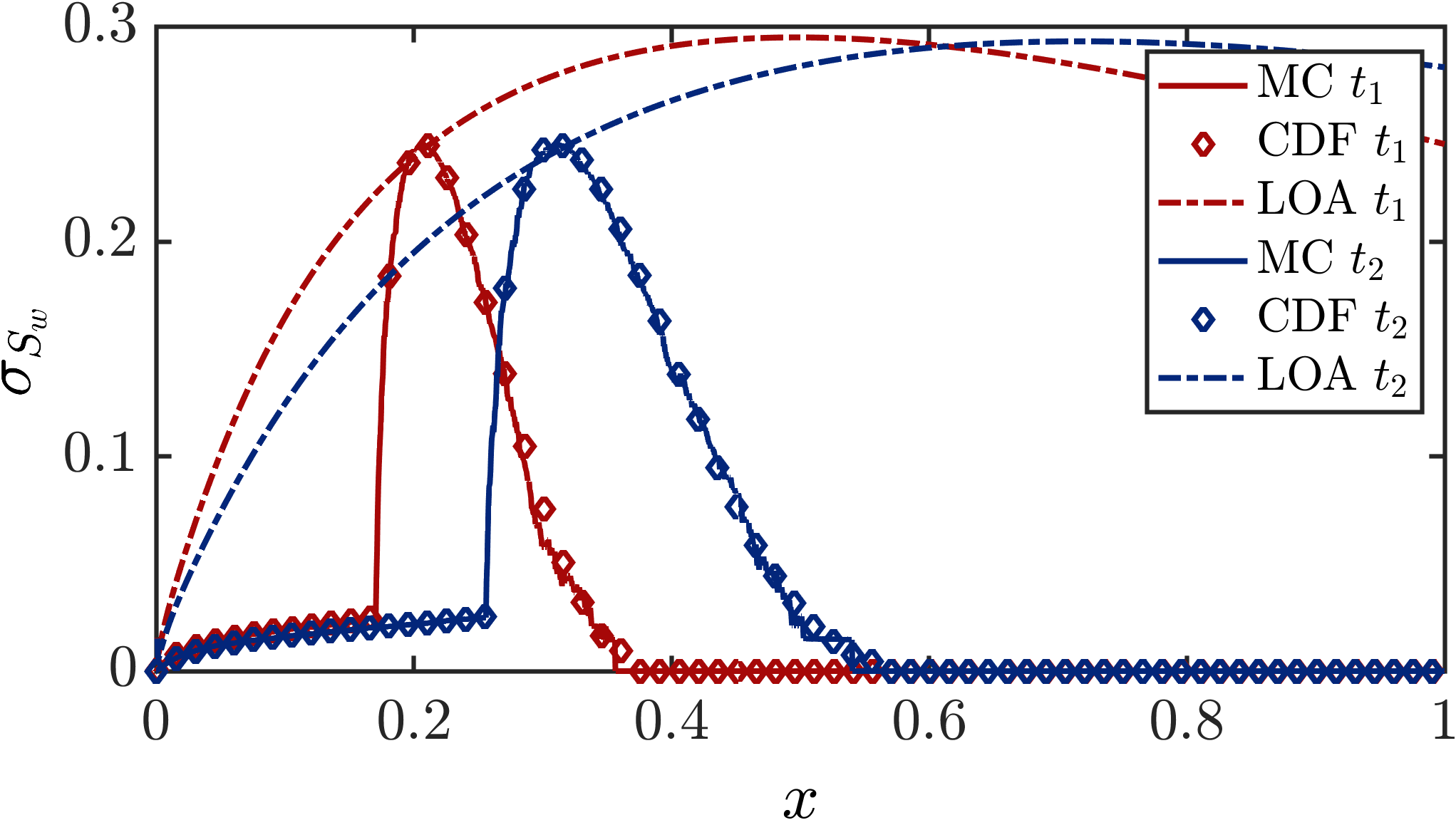}
	\caption{(First row) The first two moments of saturation, (second row) spatial and temporal evolution of the CDF of saturation (dashed lines show the results of method of distributions, while solid lines correspond to Monte Carlo), (third row) point-wise average of the CDF and PDF at all points along the domain, obtained from Monte Carlo and CDF methods, (fourth row) the first two moments obtained using Low-order SME method, vs those of MC and MD. All plots correspond to the horizontal flooding with $\phi(x)$ random, $\mu_{\phi}= 0.3$, $\sigma^2_{\phi}= 0.5$, $\lambda_{\phi}= 0.5 L$, $q=0.3$, at two dimensionless times $t_1=0.2$, $t_2=0.3$. CDF and PDF from Monte Carlo are plotted using a bandwidth of $n=2^{12}$ and $n=2^7$ respectively for the KDE post-processing. }
	\label{fig:cdfs for horizontal PHI random}
\end{figure}

\begin{figure}[htbp]
	\centering
\includegraphics[width=60mm]{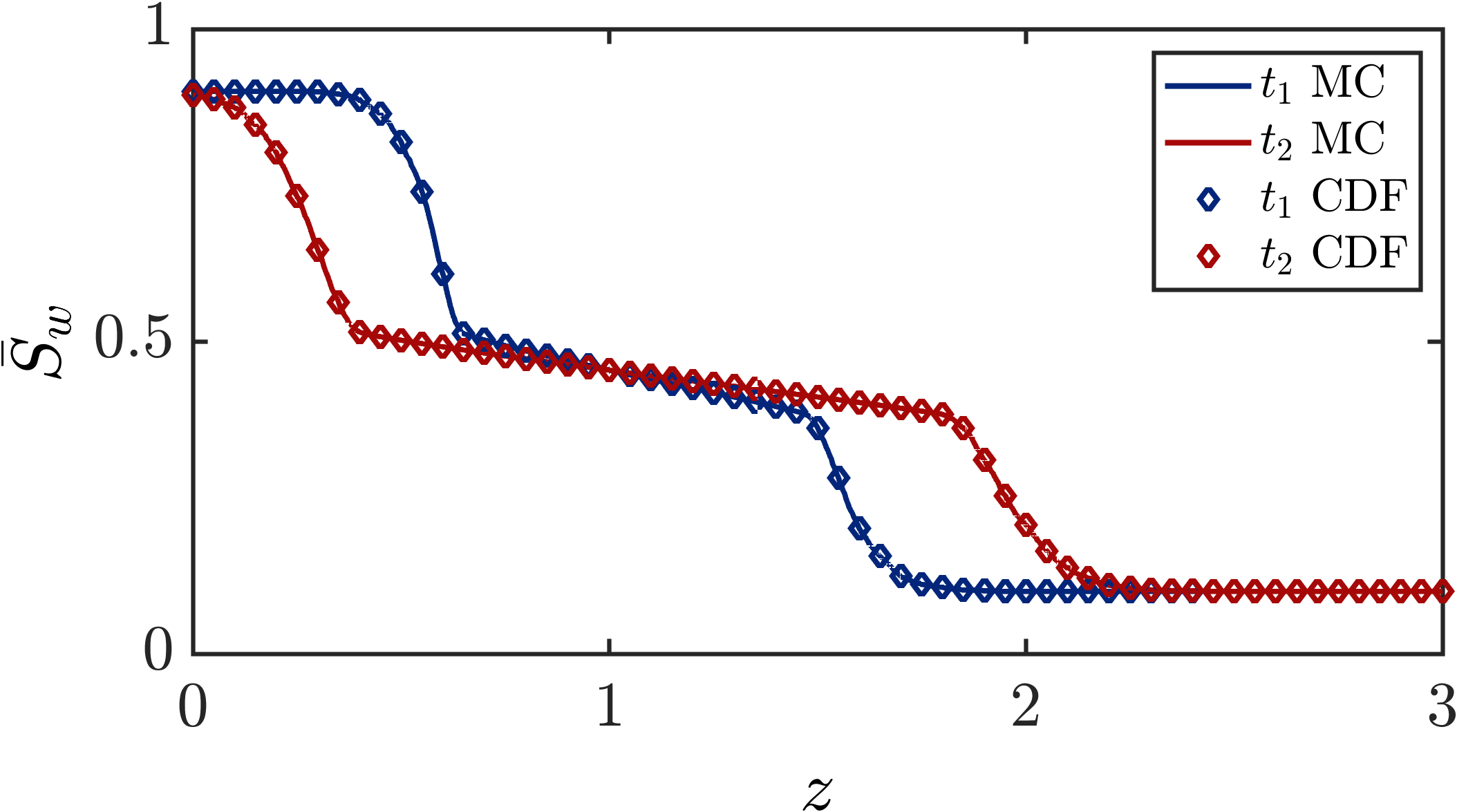}
\includegraphics[width=60mm]{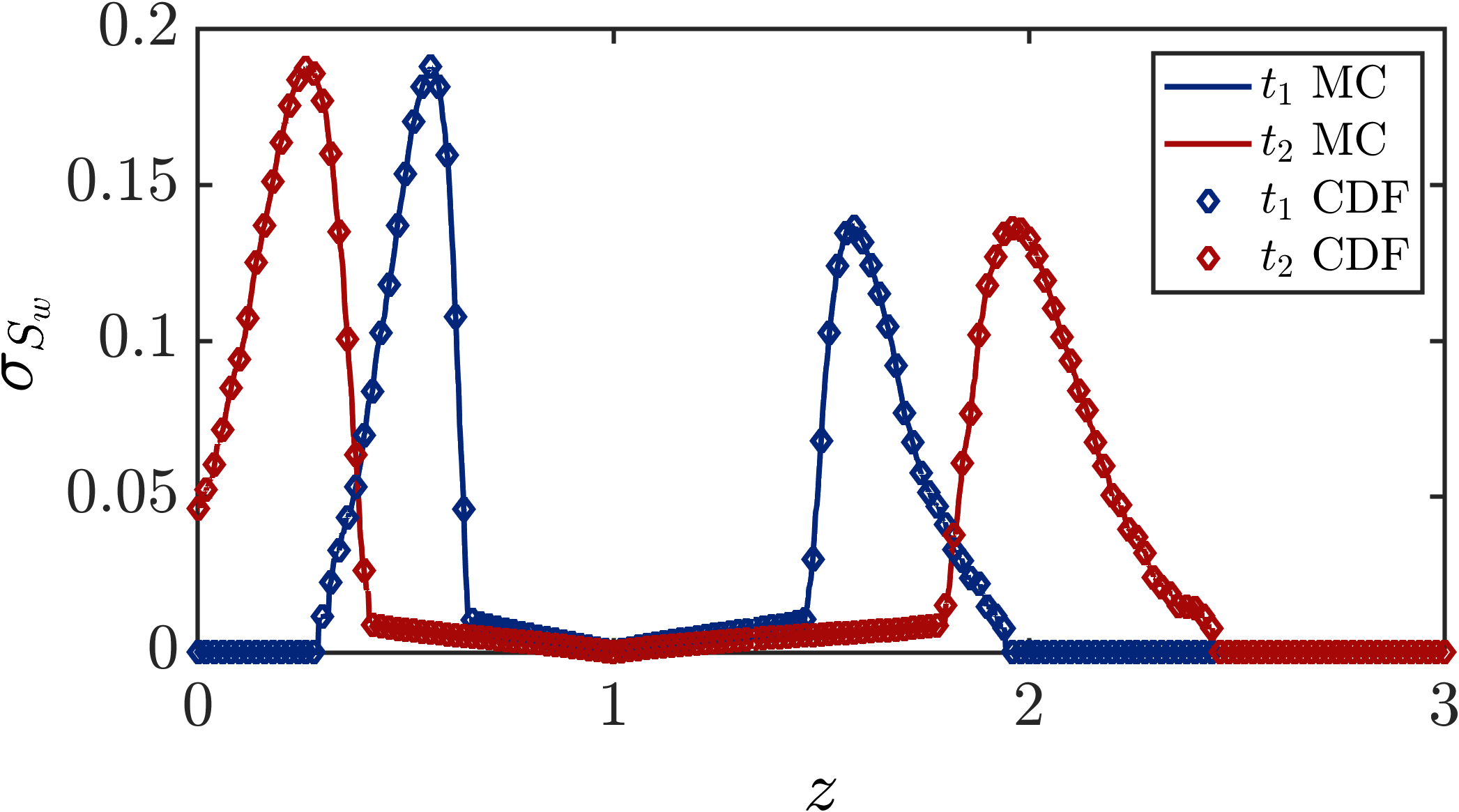}
\includegraphics[width=60mm]{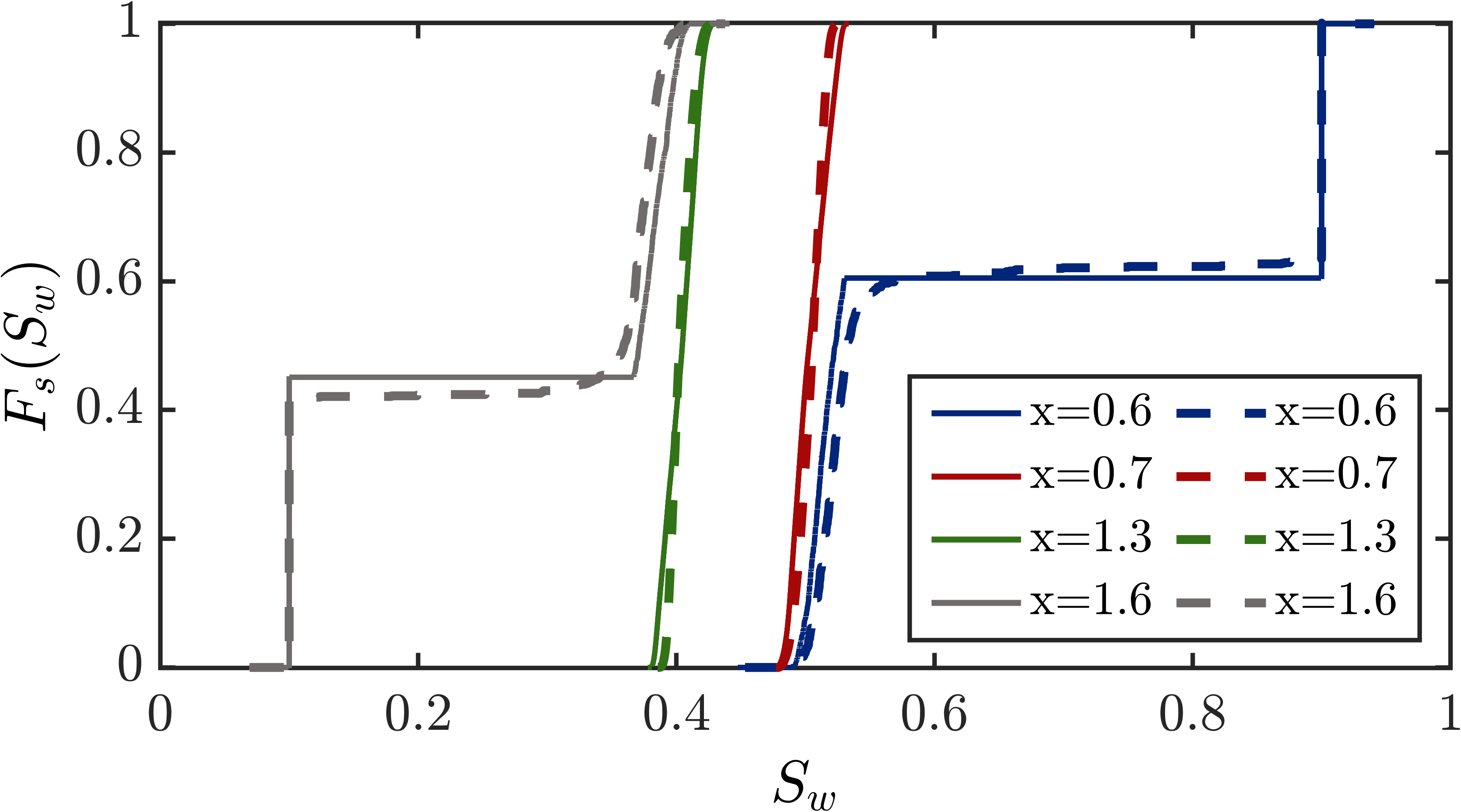}
\includegraphics[width=60mm]{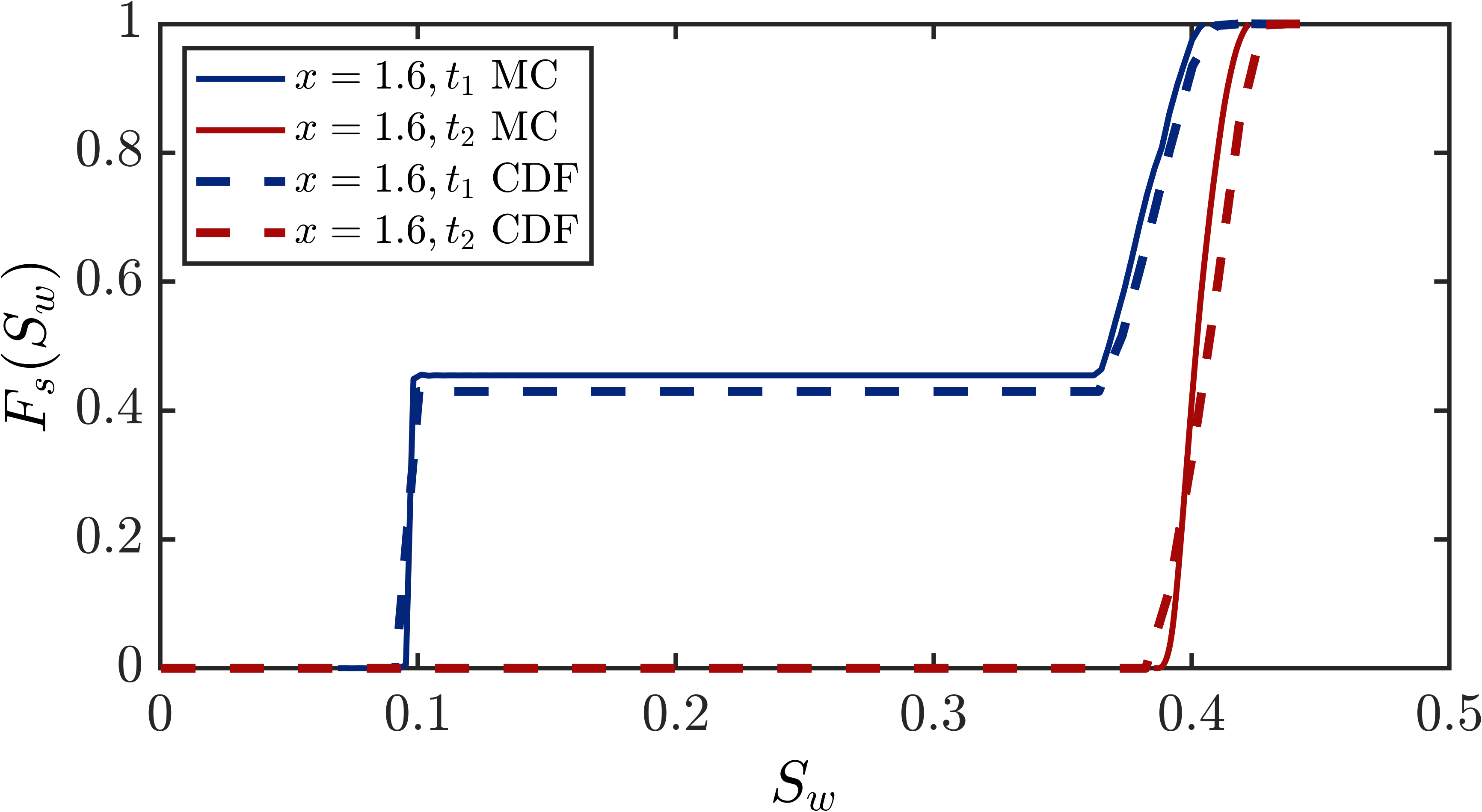}
\includegraphics[width=60mm]{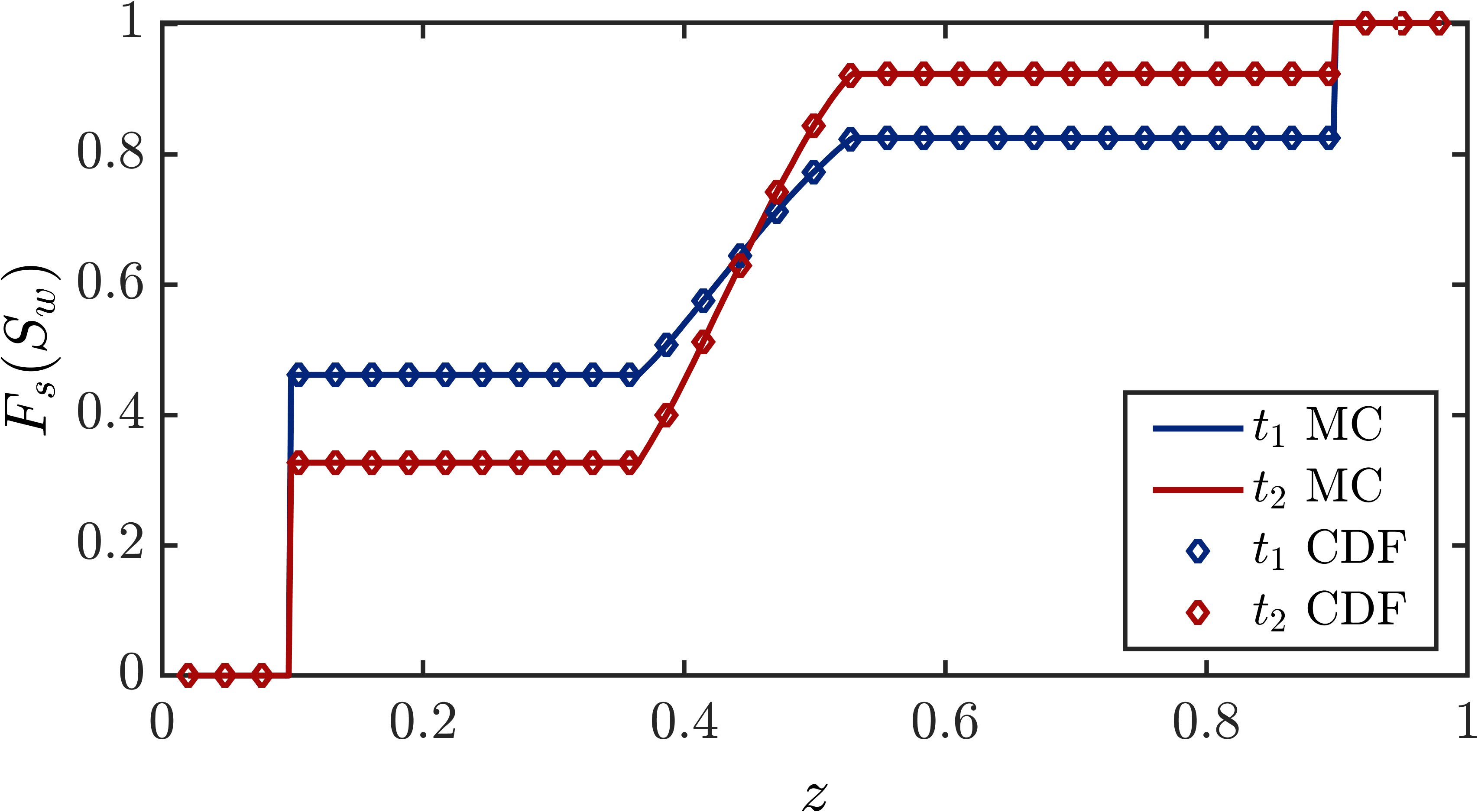}
\includegraphics[width=60mm]{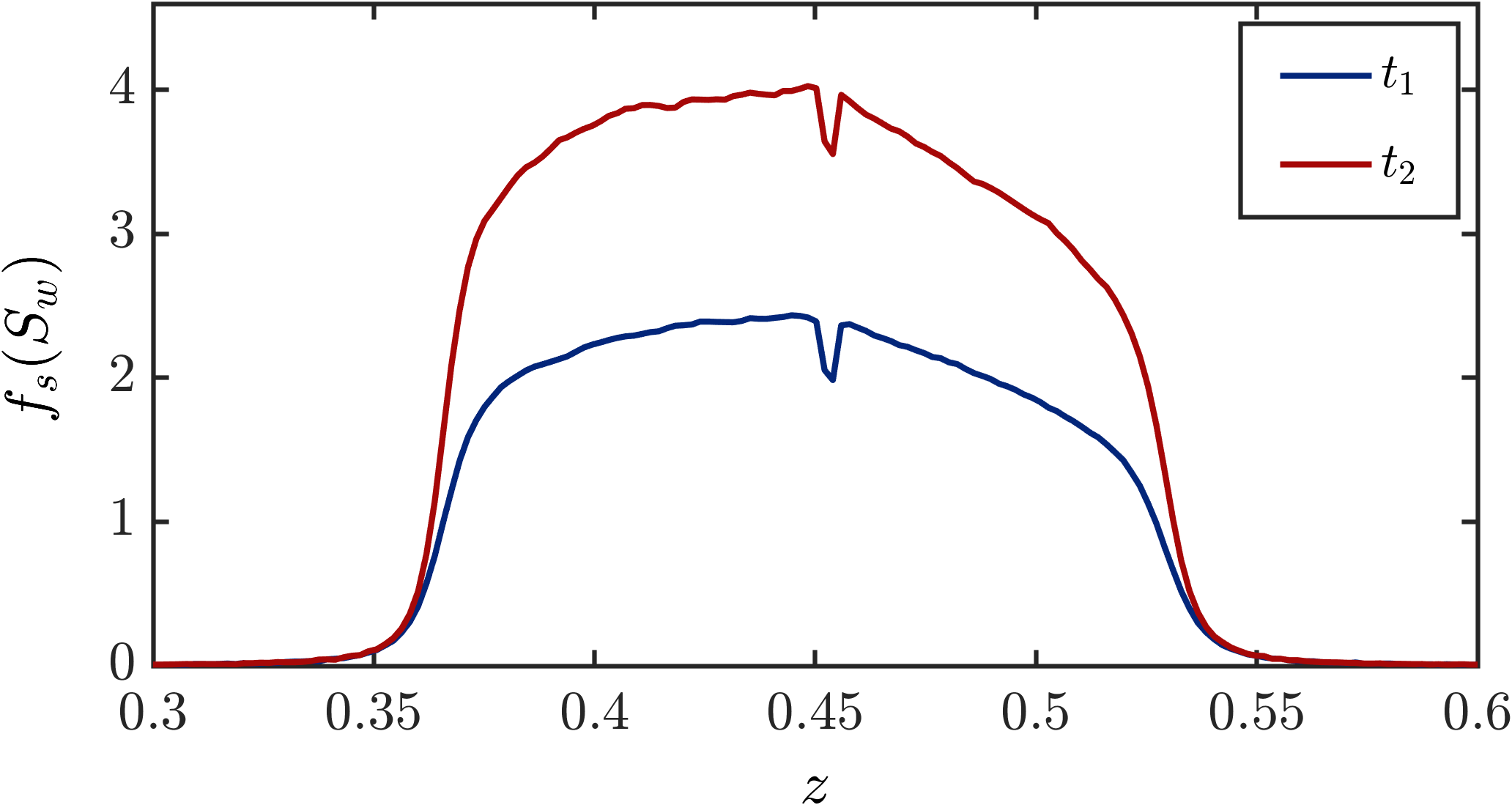}
	\caption{(First row) The first two moments of saturation, (second row) spatial and temporal evolution of the CDF of saturation (dashed lines show the results of method of distributions, while solid lines correspond to Monte Carlo), (third row) point-wise average of the CDF and PDF at all points along the domain, obtained from Monte Carlo and CDF methods. All plots correspond to the downdip flooding with $\phi(x)$ random, $\mu_{\phi}= 0.3$, $\sigma^2_{\phi}= 0.5$, $\lambda_{\phi}= 0.5 L$, $q=0.3$, at two dimensionless times $t_1=0.2$, $t_2=0.3$. CDF and PDF from Monte Carlo are plotted using a bandwidth of $n=2^{12}$ and $n=2^7$ respectively for the KDE post-processing. }
	\label{fig:cdfs for downdip PHI random}
\end{figure}

\begin{figure}[htbp]
	\centering
\includegraphics[width=60mm]{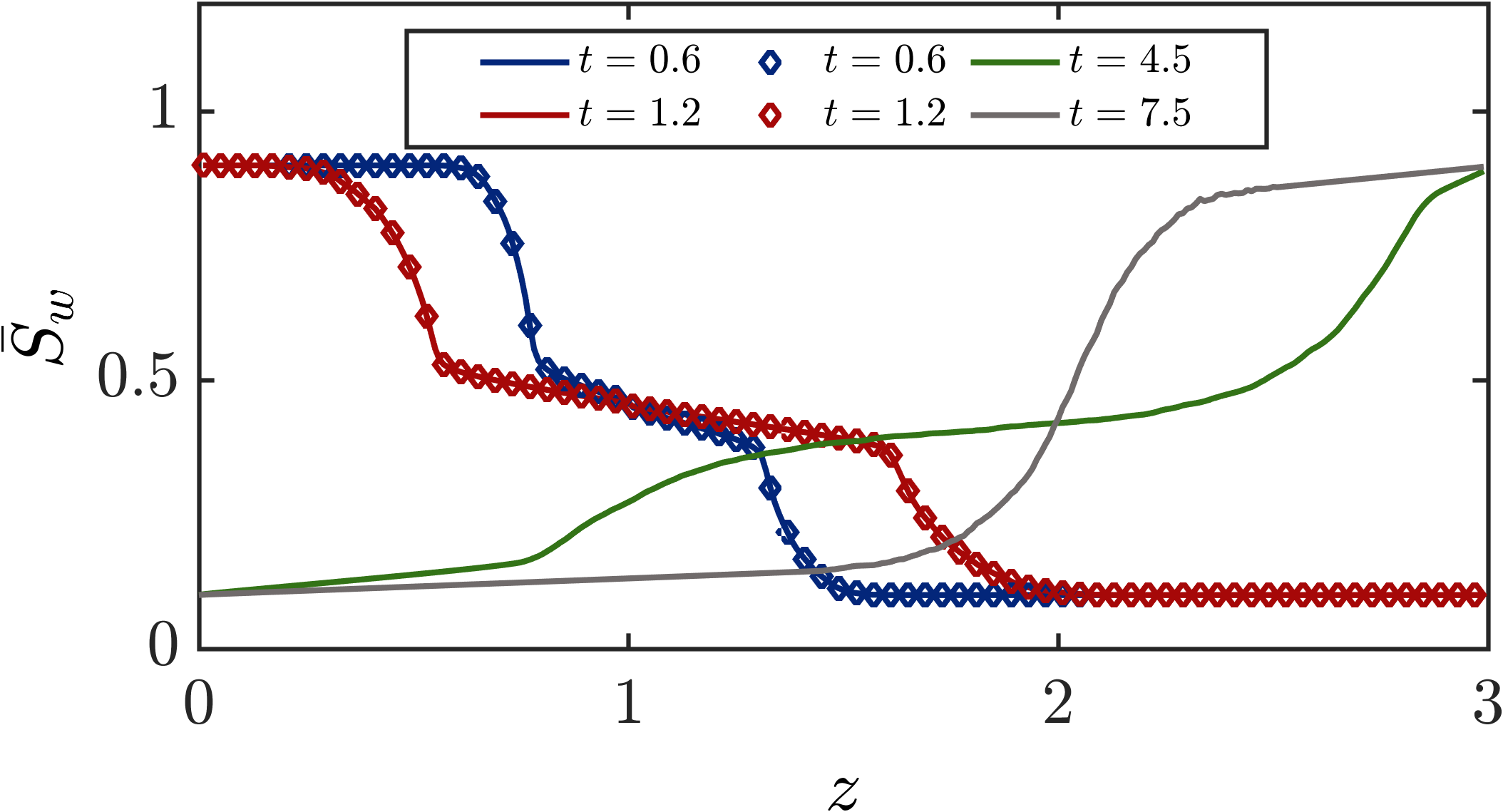}
\includegraphics[width=60mm]{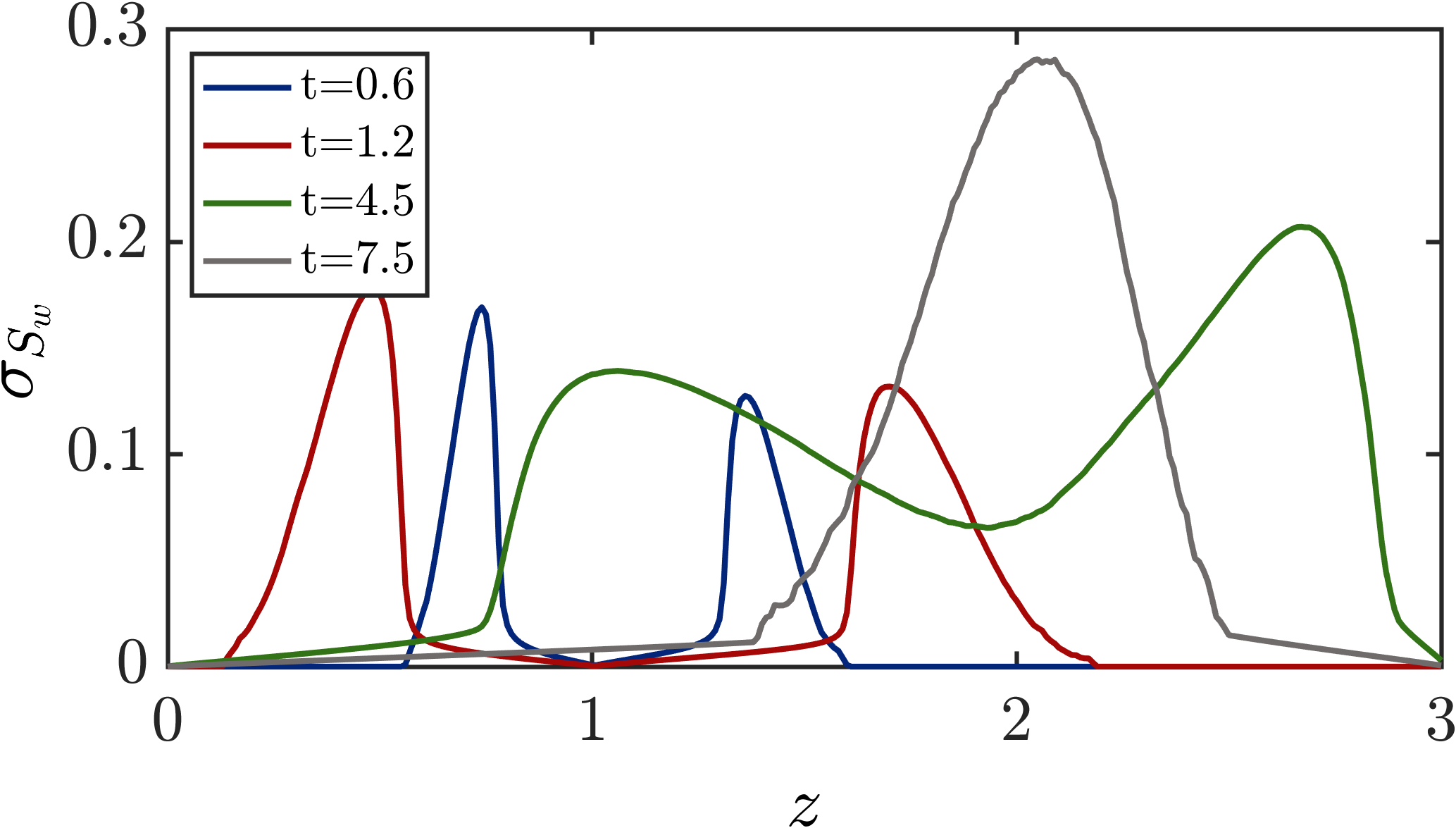}
\includegraphics[width=60mm]{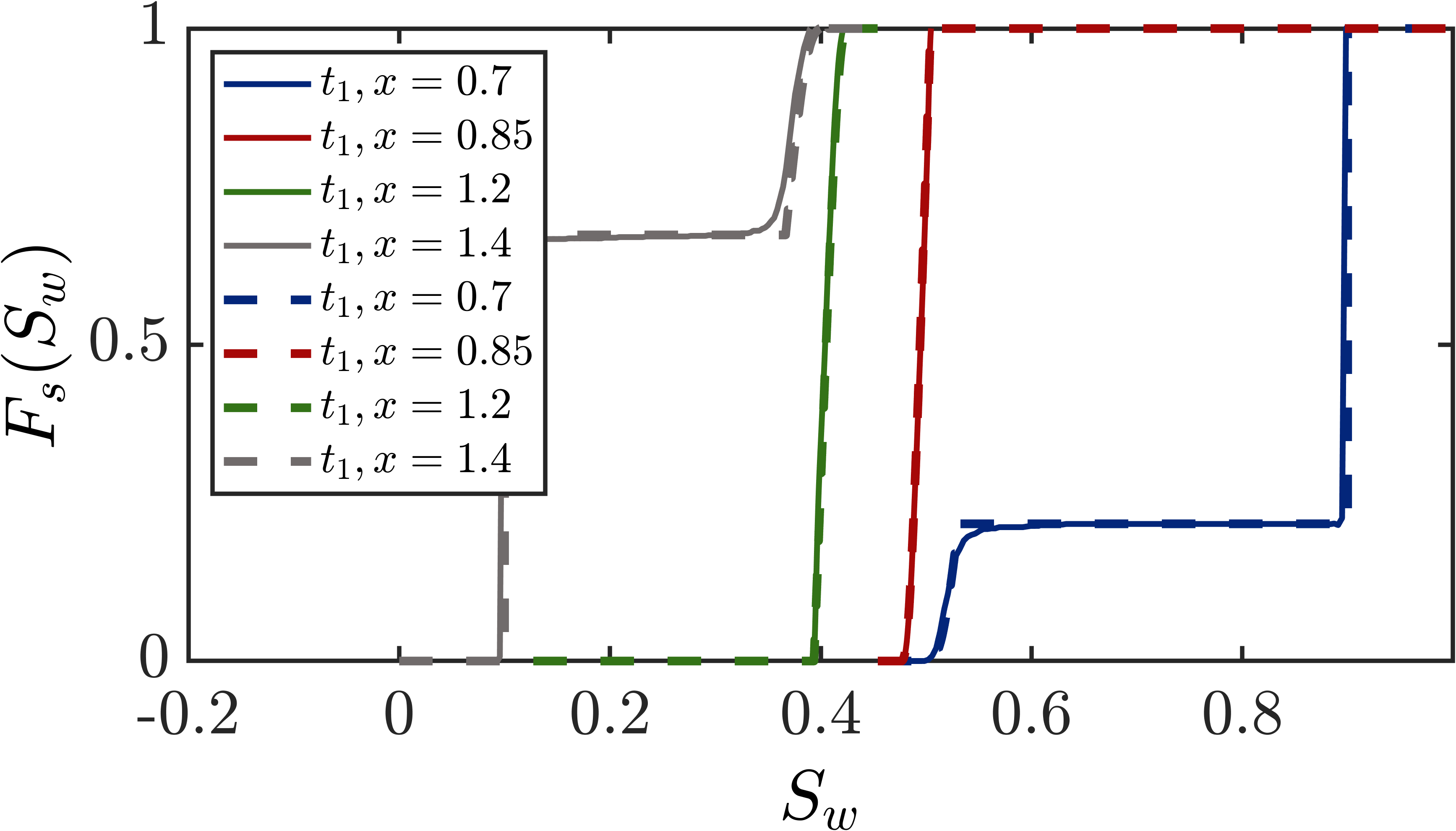}
\includegraphics[width=60mm]{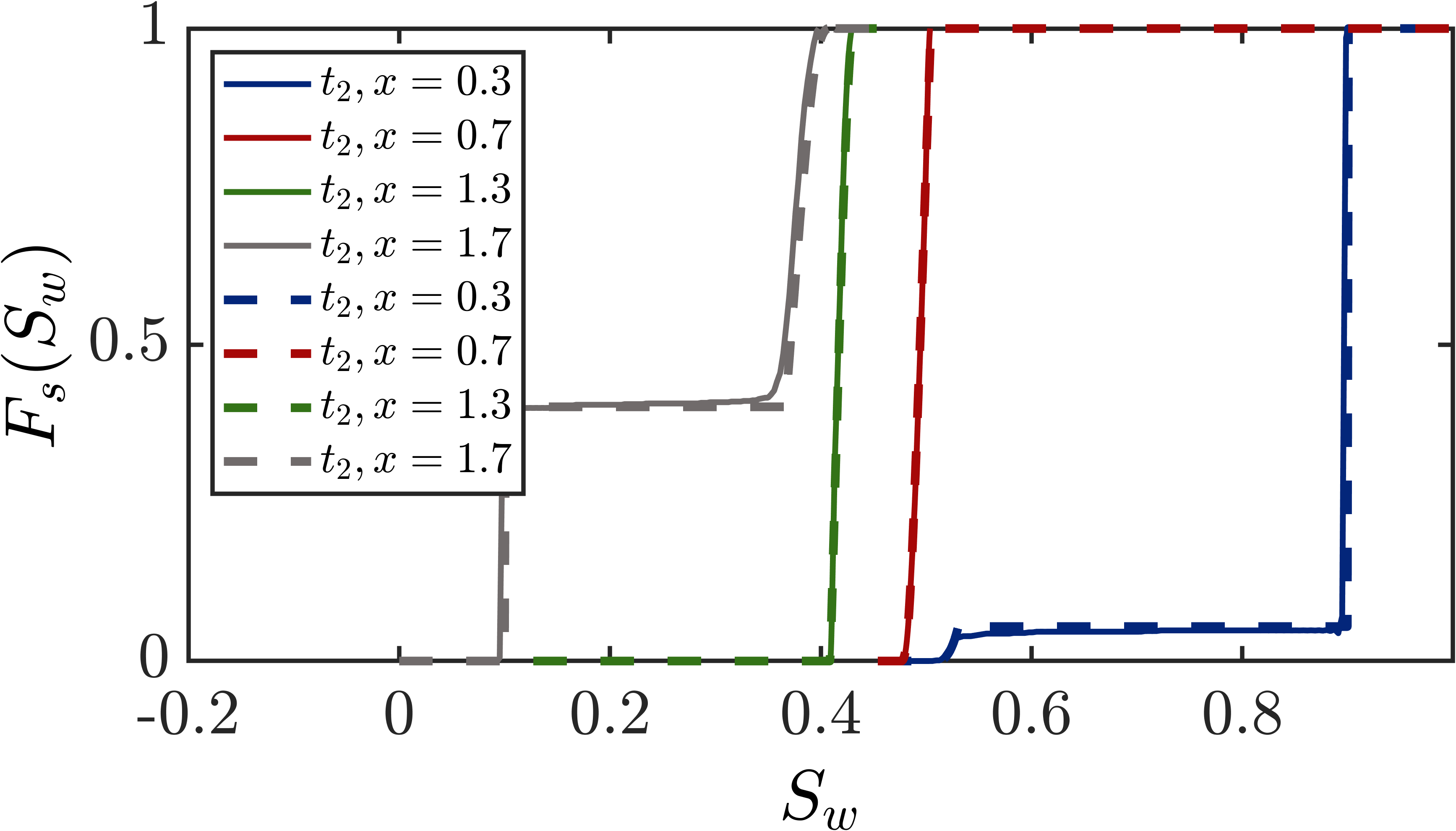}
\includegraphics[width=60mm]{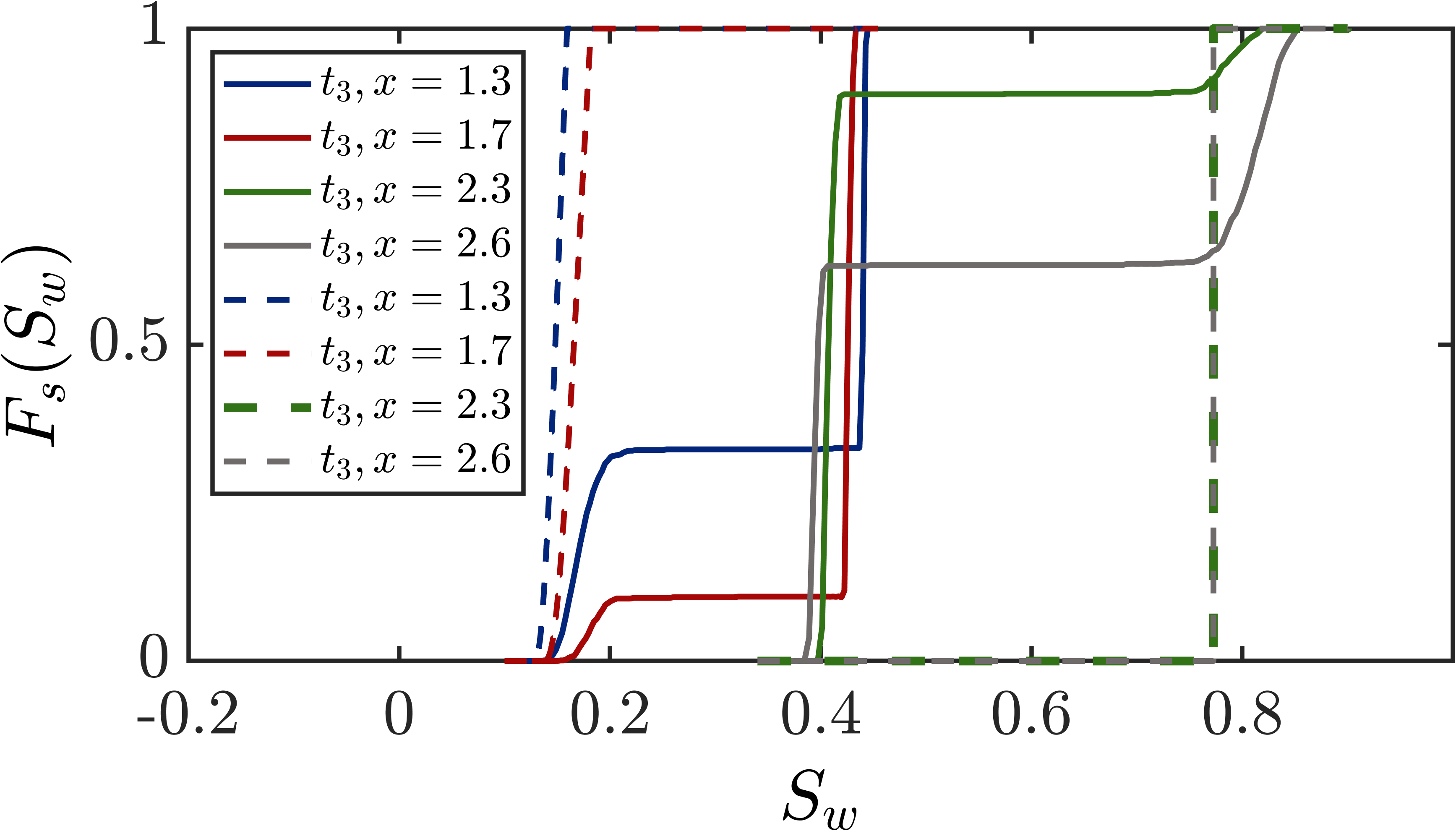}
\includegraphics[width=60mm]{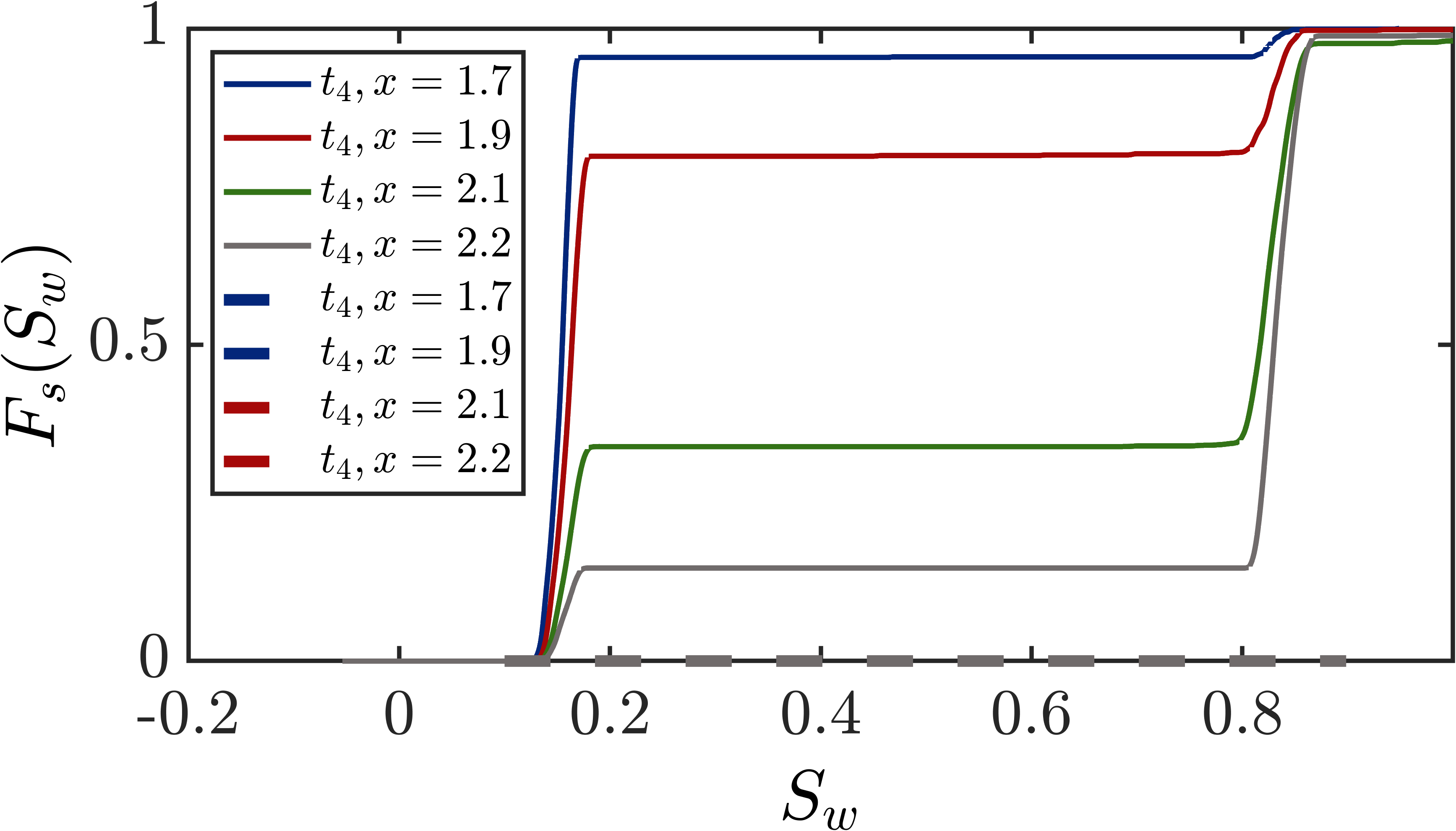}
\includegraphics[width=60mm]{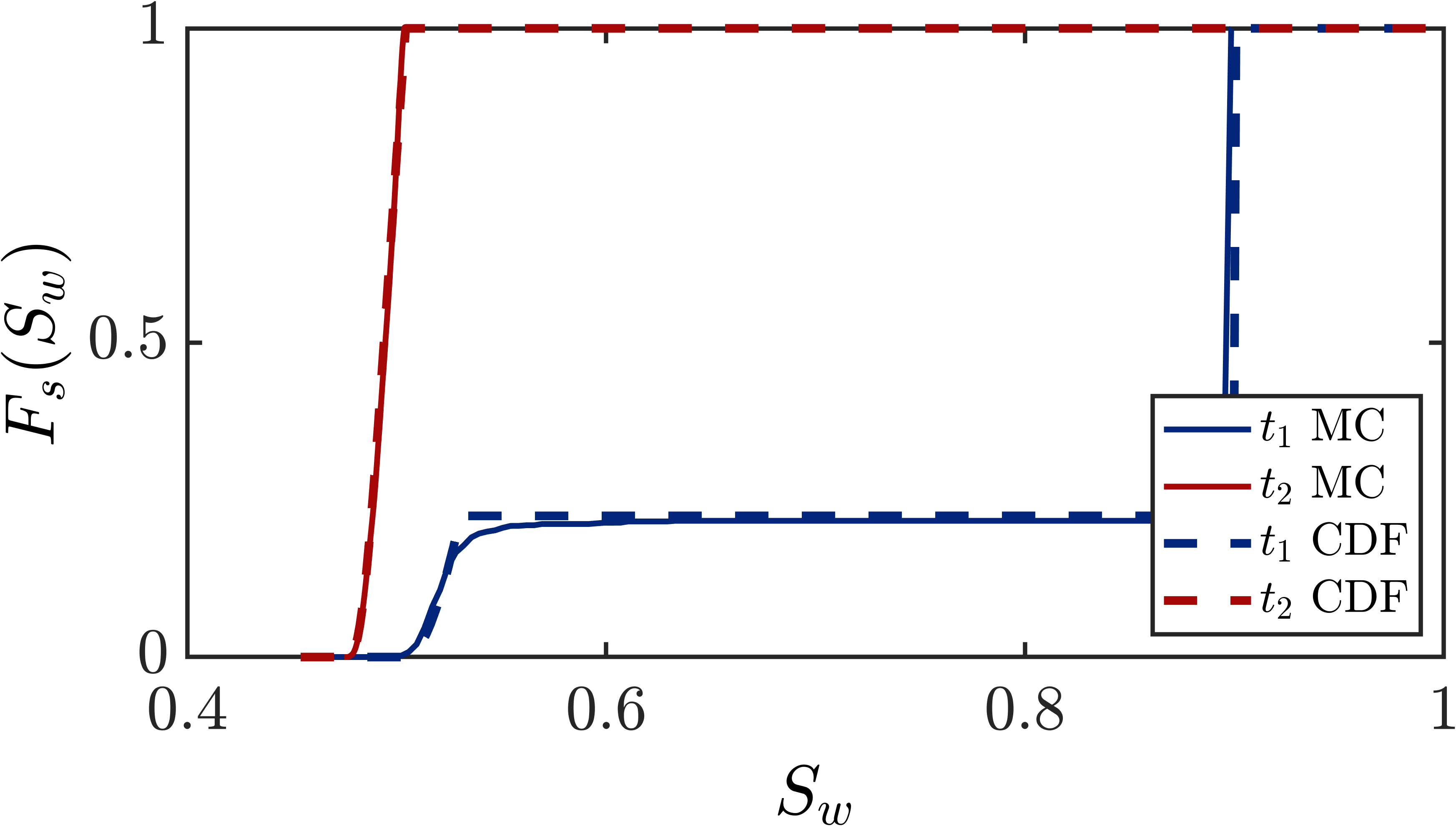}
\includegraphics[width=60mm]{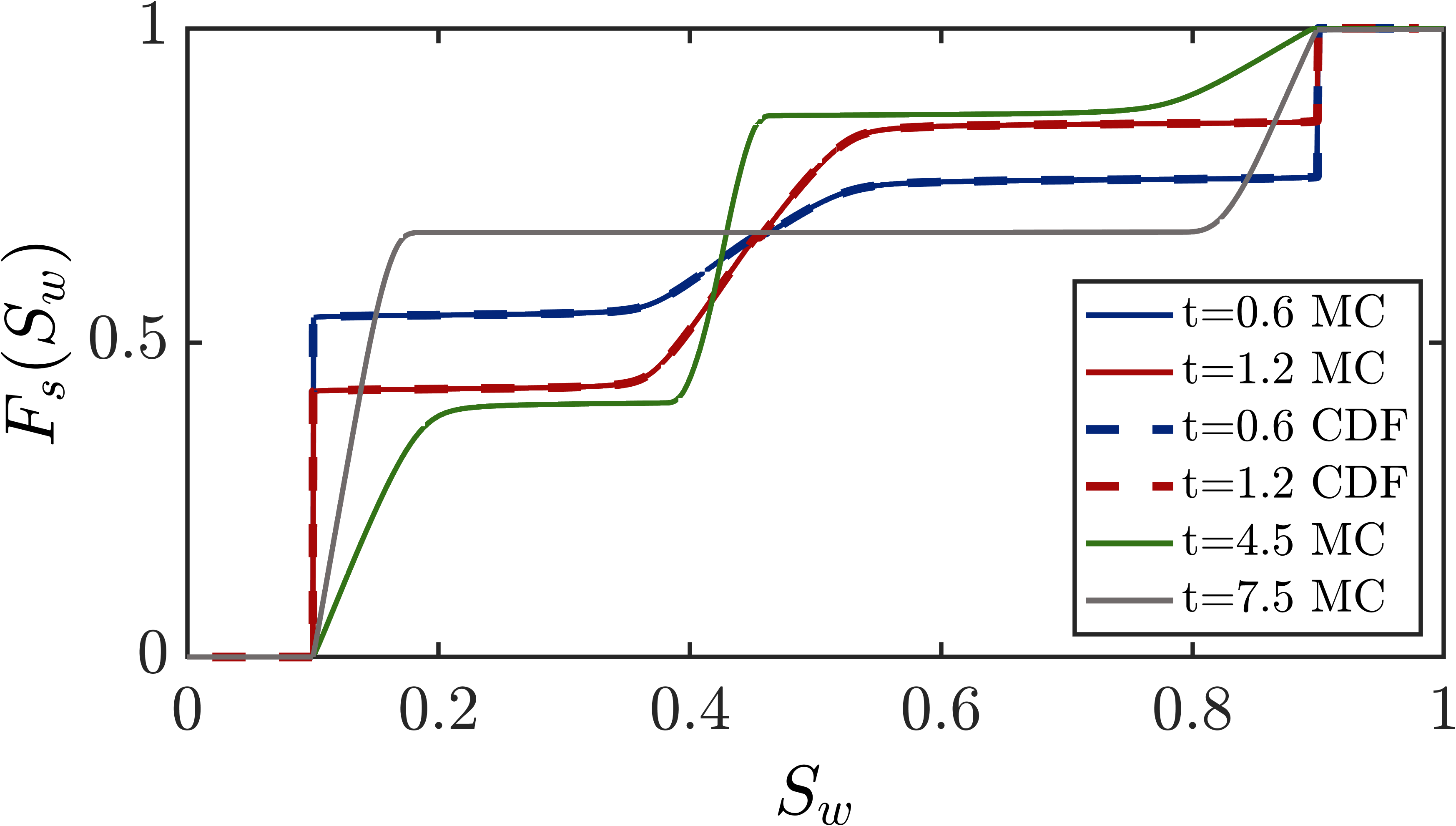}

	\caption{(First row) The first two moments of saturation (solid lines are the result of MC simulation, while markers show the results of method of distributions), (second/third row) spatial evolution of the CDF of saturation at four fixed dimensionless times $t_1=0.6, t_2=1.2, t_3=4.5, t_4 =7.5$ and four dimensionless spatial positions along the domain, temporal evolution of the CDF of saturation at a fixed dimensionless spatial positions along the domain (dashed lines show the results of method of distributions, while solid lines correspond to Monte Carlo), (fourth row) temporal evolution of the CDF of saturation at a fixed dimensionless position $x=0.7 $ and two dimensionless times, as well as point-wise spatial average of CDF of saturation along the domain at four dimensionless times. All simulations represent an inverted gravity column with $\phi(x)$ random, $\mu_{\phi}= 0.3$, $\sigma^2_{\phi}= 0.5$, $\lambda_{\phi}= 0.6 L$, $L=3$, $q=0.3$. Except in the saturation profile, solid lines show the results of MC, while dashed lines are resulted from CDF method.}
	\label{fig:cdfs for gravity column PHI random}
\end{figure}

Throughout our experiments with a stochastic porosity field, unless otherwise stated, we use $\mu_{\phi} = 0.3$, $\sigma_{\gamma}^2 = 0.5$, and $\lambda_{\gamma} =0.5 L$ where $L=1$ and $L=1$ are the dimensionless length of the domain for horizontal and vertical cases, respectively. Nevertheless, in the lower order approximation (SME) section, we keep modifying these parameters and monitoring how the resulting first and second moments of water saturation from Monte Carlo and CDF methods match with those resulting from solving statistical moment equations for the different $\sigma_{\phi}^2$ and $\lambda_{\phi}$ variables.  Figures \ref{fig:cdfs for horizontal PHI random}, \ref{fig:cdfs for downdip PHI random}, \ref{fig:cdfs for gravity column PHI random} show the results for random porosity field as the only source of uncertainty. The mean  saturation  profiles are plotted along with the deterministic solutions, and the smoother solutions of MC and MD methods compared to the sharp solutions of the deterministic case represent the results of averaged solution of a non-zero variance for the underlying stochastic field. Note that the vertical shock regions in the saturation profiles correspond to the horizontal regions in the CDF plots. As these plots represent, our proposed method of distributions is able to predict the behavior of Monte Carlo for mean, standard deviation, and CDF of saturation. The only case for which our method shows limitation is in the gravity column case and only after reflection of waves from boundaries (Fig.~\ref{fig:cdfs for gravity column PHI random}, third row). Unlike to the other cases, this mismatch occurs due to the non-self-similarity of solution in the output state (saturation) domain. That is, as we showed in Eq.~\ref{eq:Fsbl after reflection} and Eq.~\ref{eq:Fsbr after reflection}, we have two non-constant specific saturation values $S_3$ and $S_4$ (Fig.~\ref{fig:saturation annotated }) needed to compute $F_{s_{b_R}}$ and $F_{s_{b_L}}$ respectively, and in general it's hard to keep track of these constantly-changing saturation values. Consequently, solution of the CDF equations is over-complicated and in general does not follow Eq.~\ref{eq:Fsbl after reflection} and Eq.~\ref{eq:Fsbr after reflection}. Therefore, even though the proposed method perfectly captures the two-wave behavior in all the previous examples, it fails only for this specific case.

%%%%%%%%%%----------------------------  horizontal phi random

%%%%%%%%%%----------------------------  downdip phi random

%%%%%%%%%%----------------------------  gravity column phi random
%
%%%%%%%%%%%%%%%%%%%%%%%%%%%%%%%%%%%%%%%%%%%%%%% only q(t) random
\subsubsection{Numerical setting for stochastic injection flux $q(t)$}
Enforcing the general continuity constraint for incompressible flow, $\nabla\cdot q_T(x,t)=0$ makes the total flux $q_T$ to be constant in x, and is therefore equal to the injection flow rate at the left boundary of domain $q(x,t)=q(t)= q_0(t)$. We are interested in finding the solutions to Eq.~(\ref{eq:saturation_equation_with_BC}) while treating $q_0(t)$ as an stochastic field in time with a prescribed CDF. 
Henceforward, we choose to keep porosity as a deterministic constant $\phi=0.3$ for this scenario, thereby only experimenting with the effect of randomness in the injection flux. In order to find the distribution for the random injection flux at the inlet, similarly to the previous case, we assume the mean $\mu_q$, variance $\sigma^2_q$, covariance structure $C_q(t)$, and correlation time $\tau_q$ are all known parameters. To this end, similarly to \cite{ibrahima2015distribution}, we first define a Gaussian random field $\rho_q(t) \sim \mathcal{N} (\mu_{\rho}, C_{\rho}(t))$, with an exponential covariance structure $C_{\rho}(t) = \sigma^2_{\rho} exp(-\dfrac{t}{\tau_q})$, where $\tau_q$ is the correlation parameter for events that are closer in time. Eventually, we define the stochastic injection flux field as,

\begin{align}
& q_T(t) = 0.05 + 0.1 (\rho_q(t))^2
\end{align}

Such a definition allows us to first make sure that $q_T(t)>0$ and second by controlling $\mu_{\rho}$ we can satisfy the criteria $\langle q(t)\rangle =\mu_q$. Any other arbitrary non-negative distribution could be used alternatively.

\begin{figure}[htbp]
	\centering
\includegraphics[width=60mm]{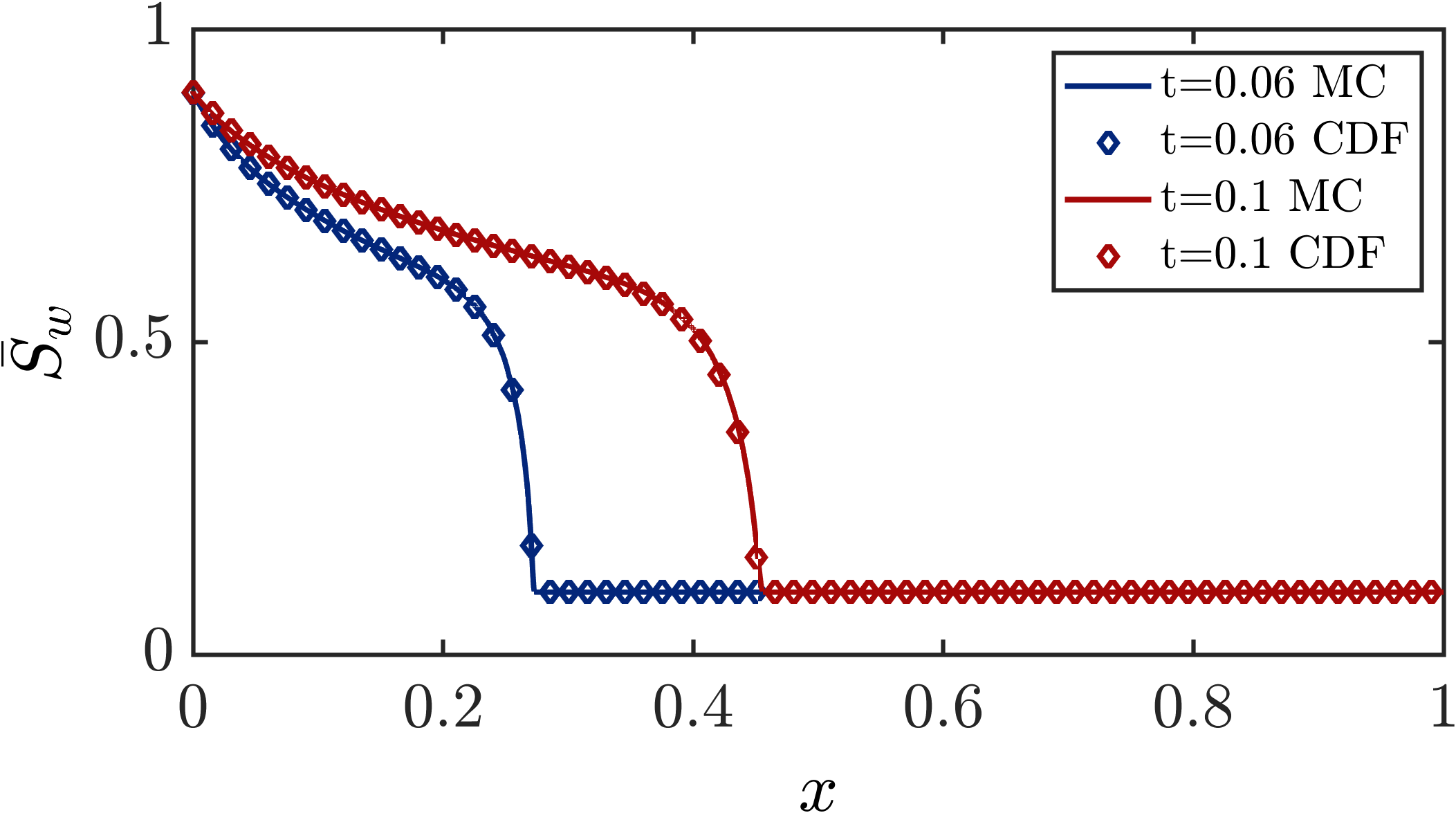}
\includegraphics[width=60mm]{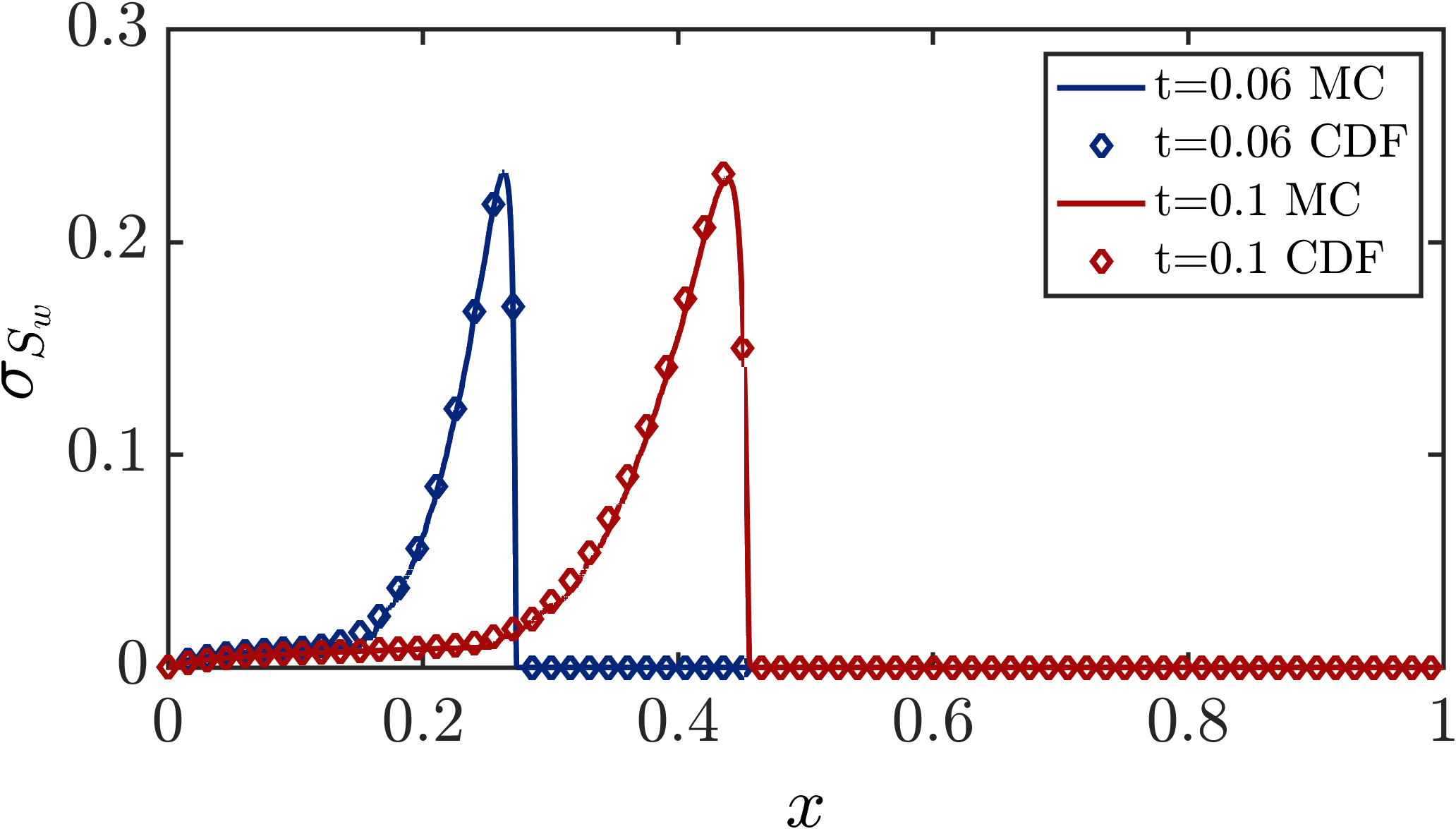}
\includegraphics[width=60mm]{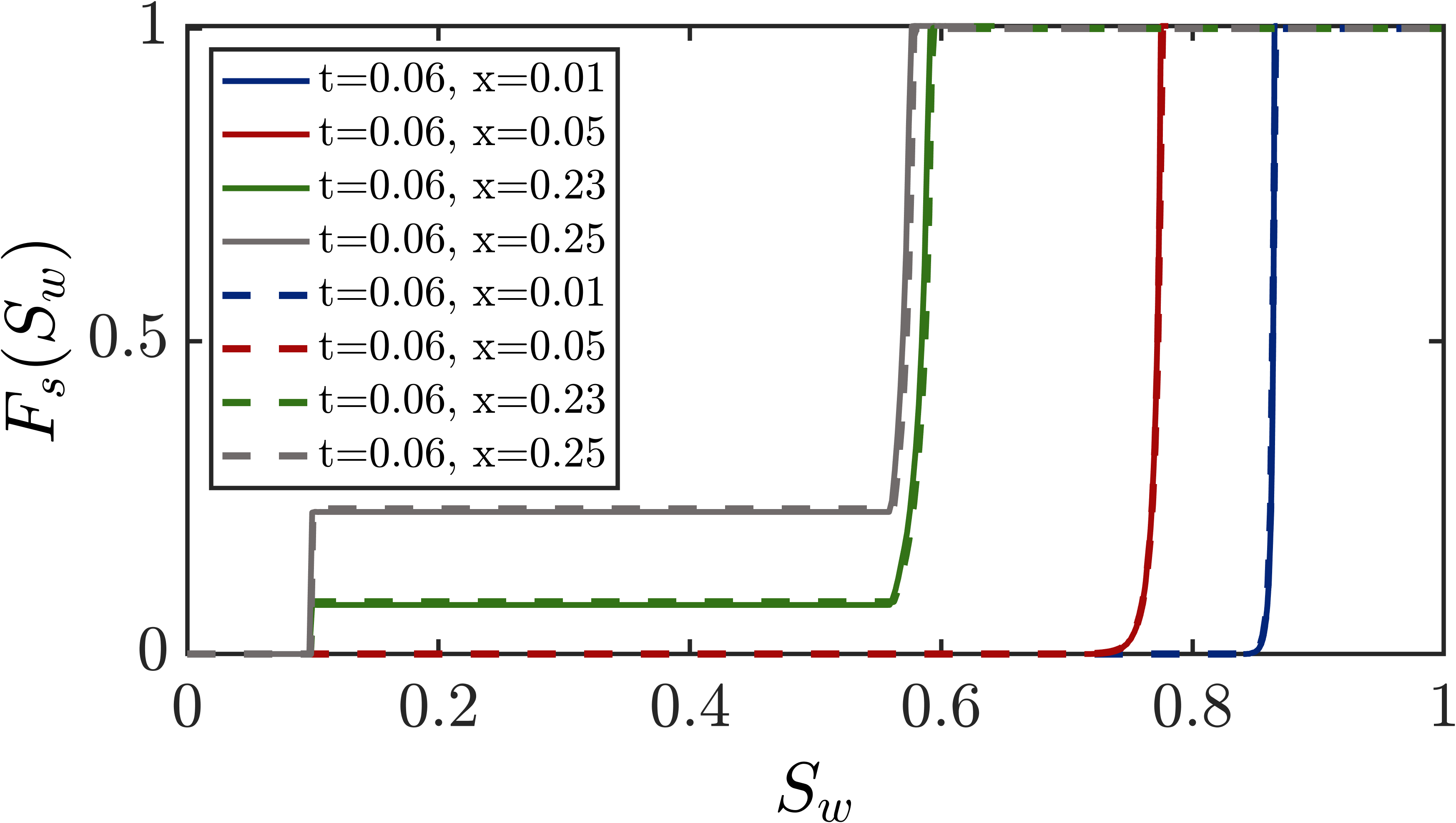}
\includegraphics[width=60mm]{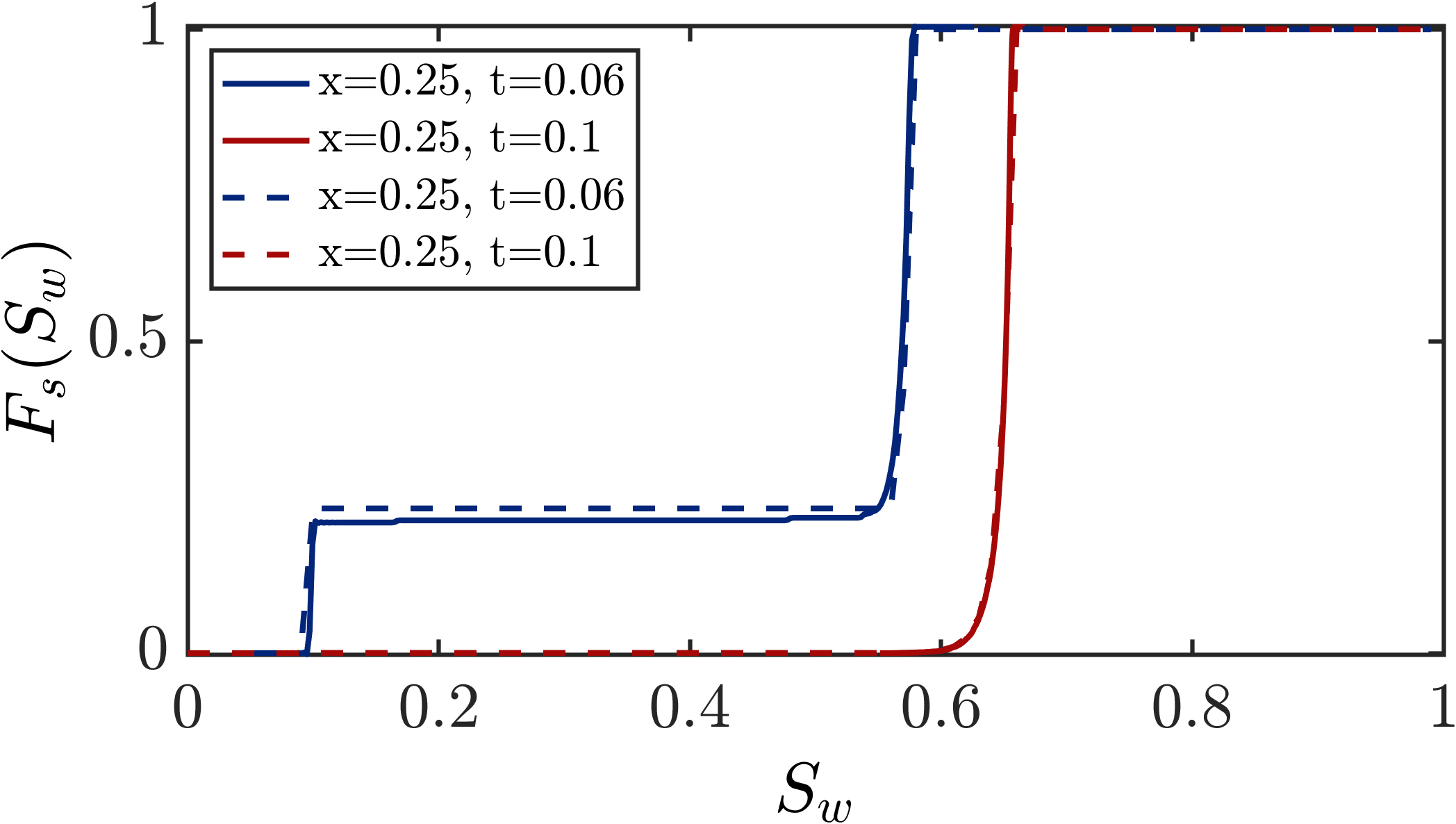}
\includegraphics[width=60mm]{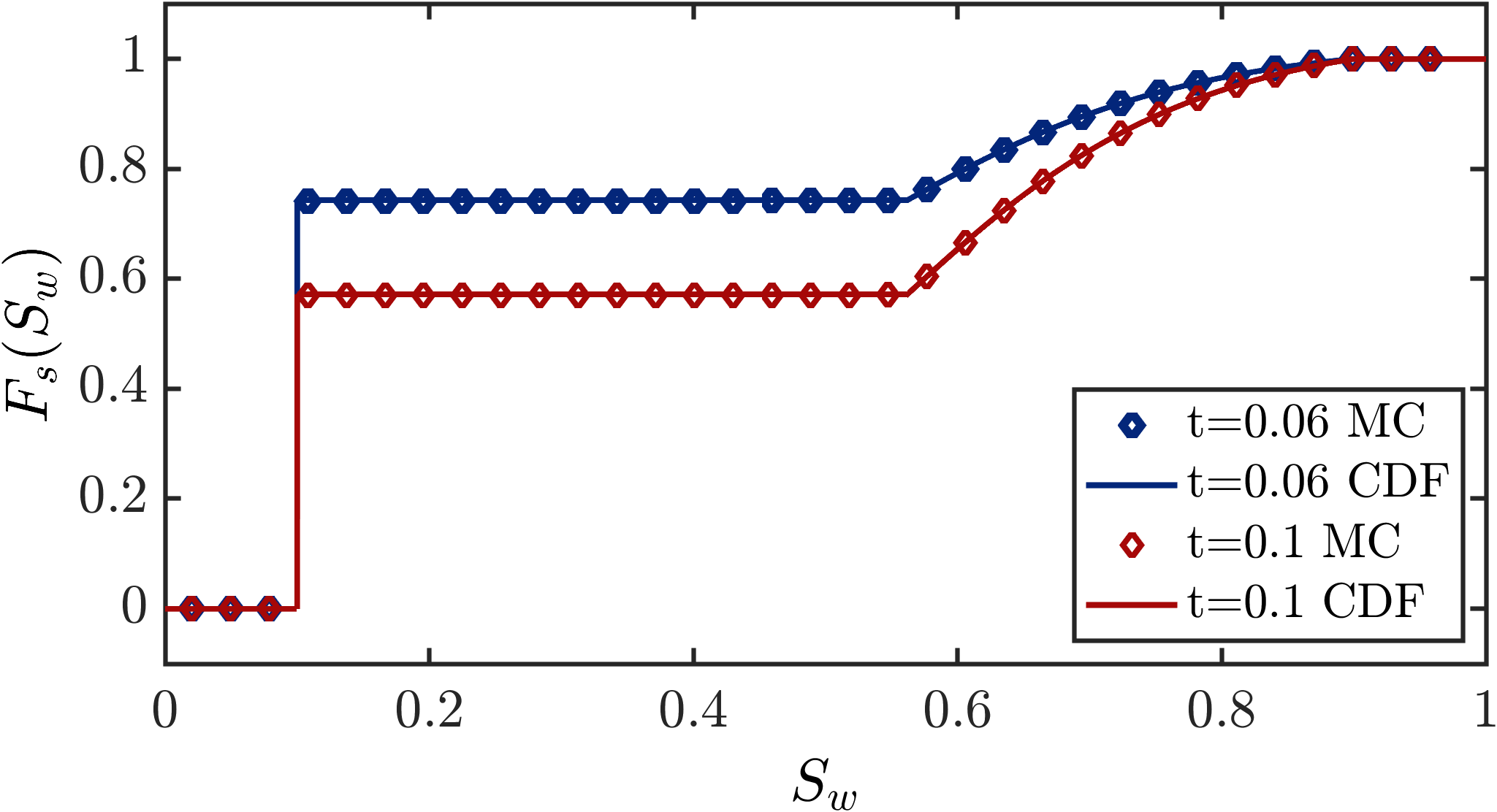}
\includegraphics[width=60mm]{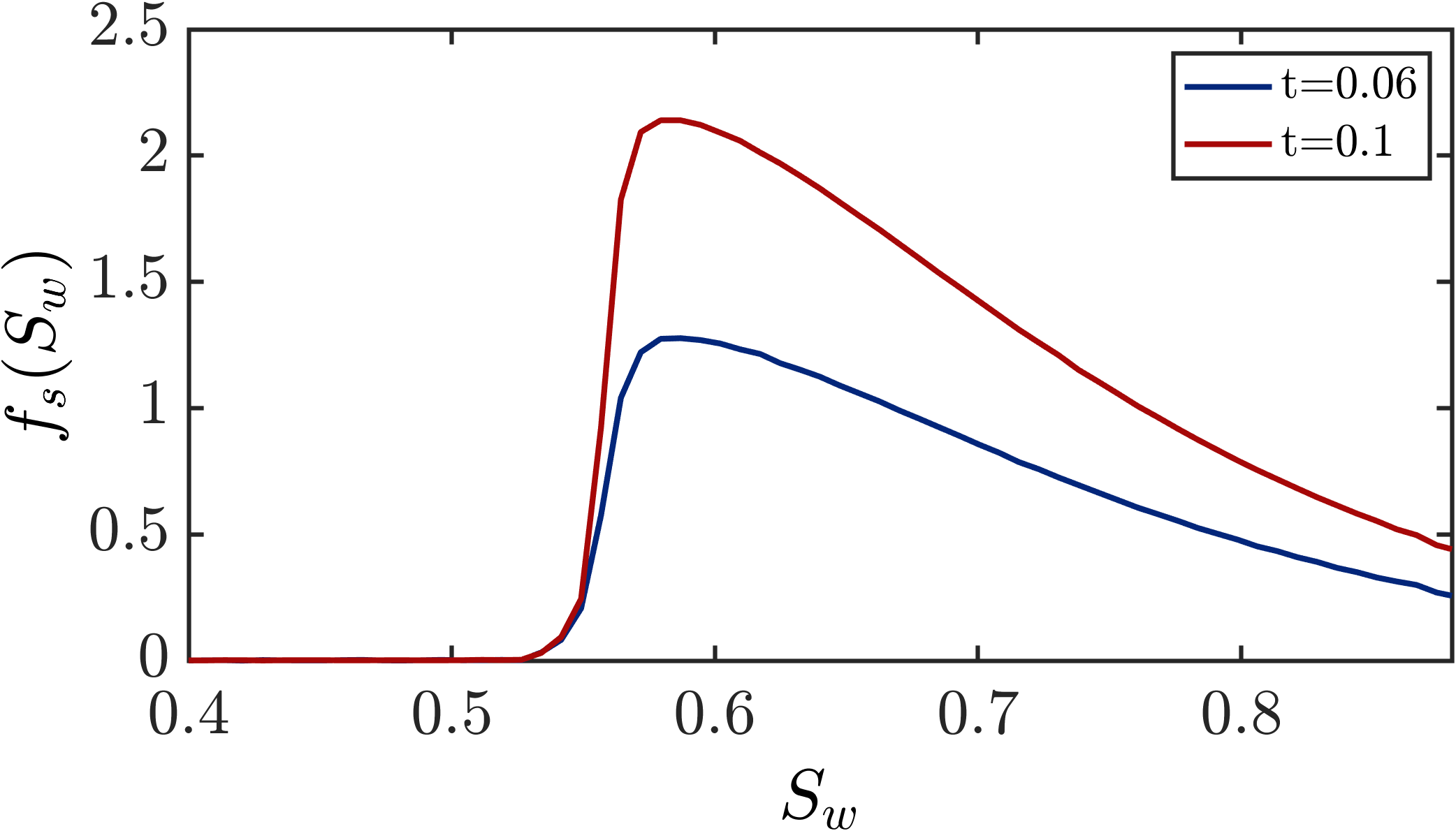}
\includegraphics[width=60mm]{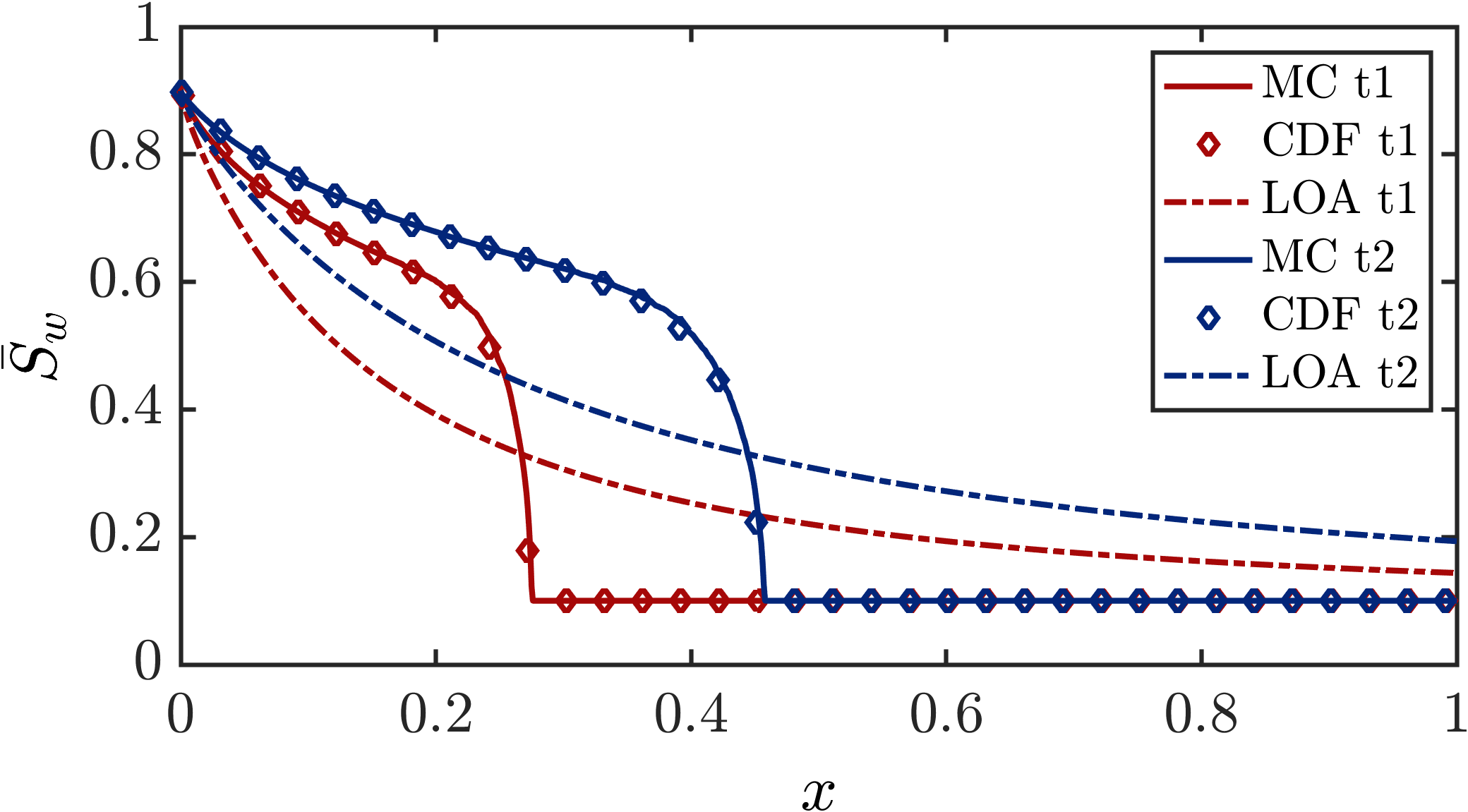}
\includegraphics[width=60mm]{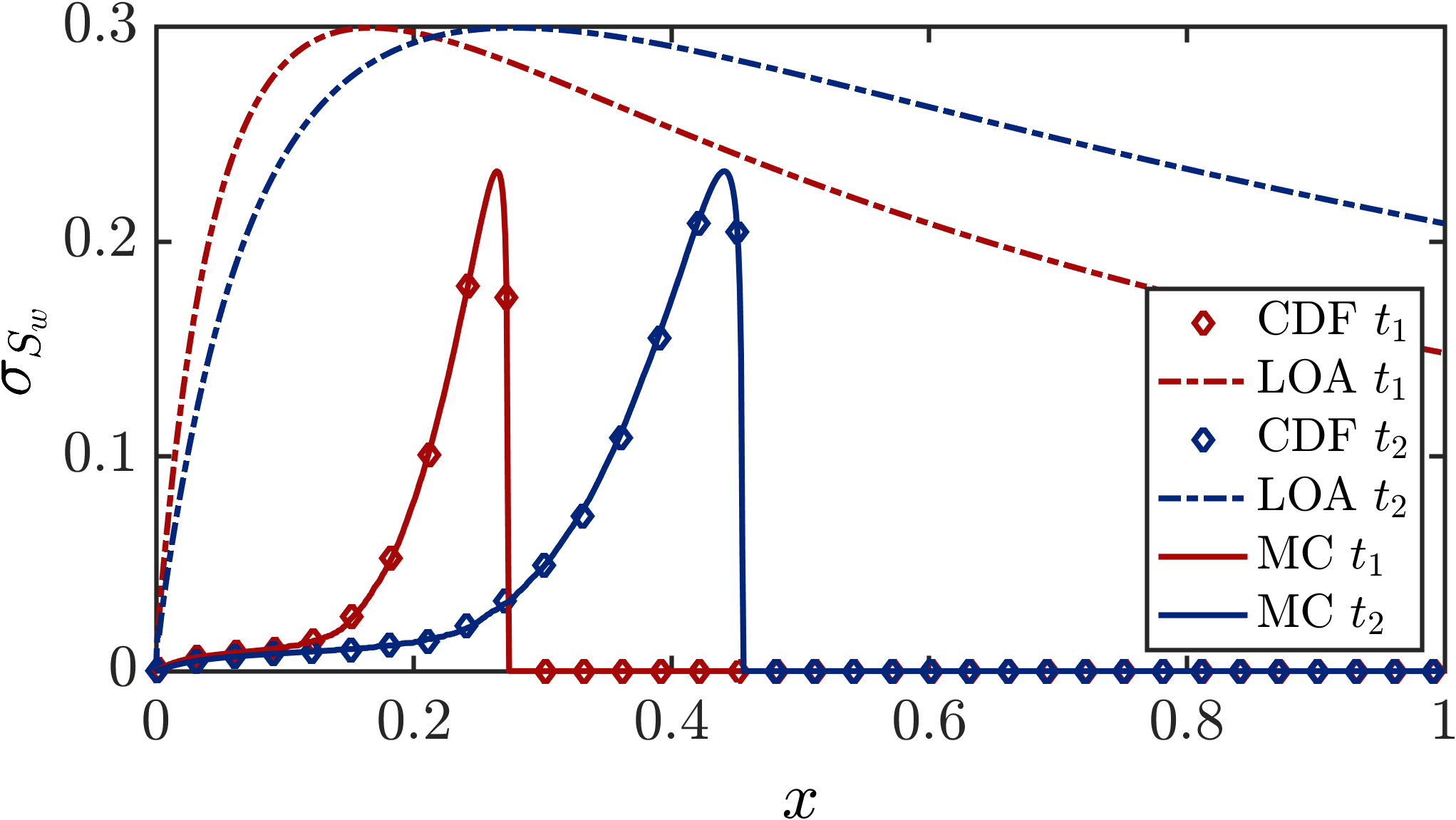}
	\caption{(First row) The first two moments of saturation, (second row) spatial and temporal evolution of the CDF of saturation (dashed lines show the results of method of distributions, while solid lines correspond to Monte Carlo), (third row) point-wise spatial average of the CDF and PDF at all points along the domain, obtained from Monte Carlo and CDF methods, (fourth row) the first two moments obtained using Low-order SME method, vs. those of MC and MD. All simulations represent horizontal flooding, with $q(t)$ random, $\mu_{q}= 0.3$, $\sigma^2_{q}= 0.25$, $\tau_{q}= 0.1 T$, $T=1$, $\phi=0.3$, at two dimensionless times $t_1=0.2$, $t_2=0.3$. CDF and PDF from Monte Carlo are plotted using a bandwidth of $n=2^{12}$ and $n=2^7$ respectively for the KDE post-processing.}
	\label{fig:CDF plots for horizontal Q random}
\end{figure}

\begin{figure}[htbp]
	\centering
\includegraphics[width=60mm]{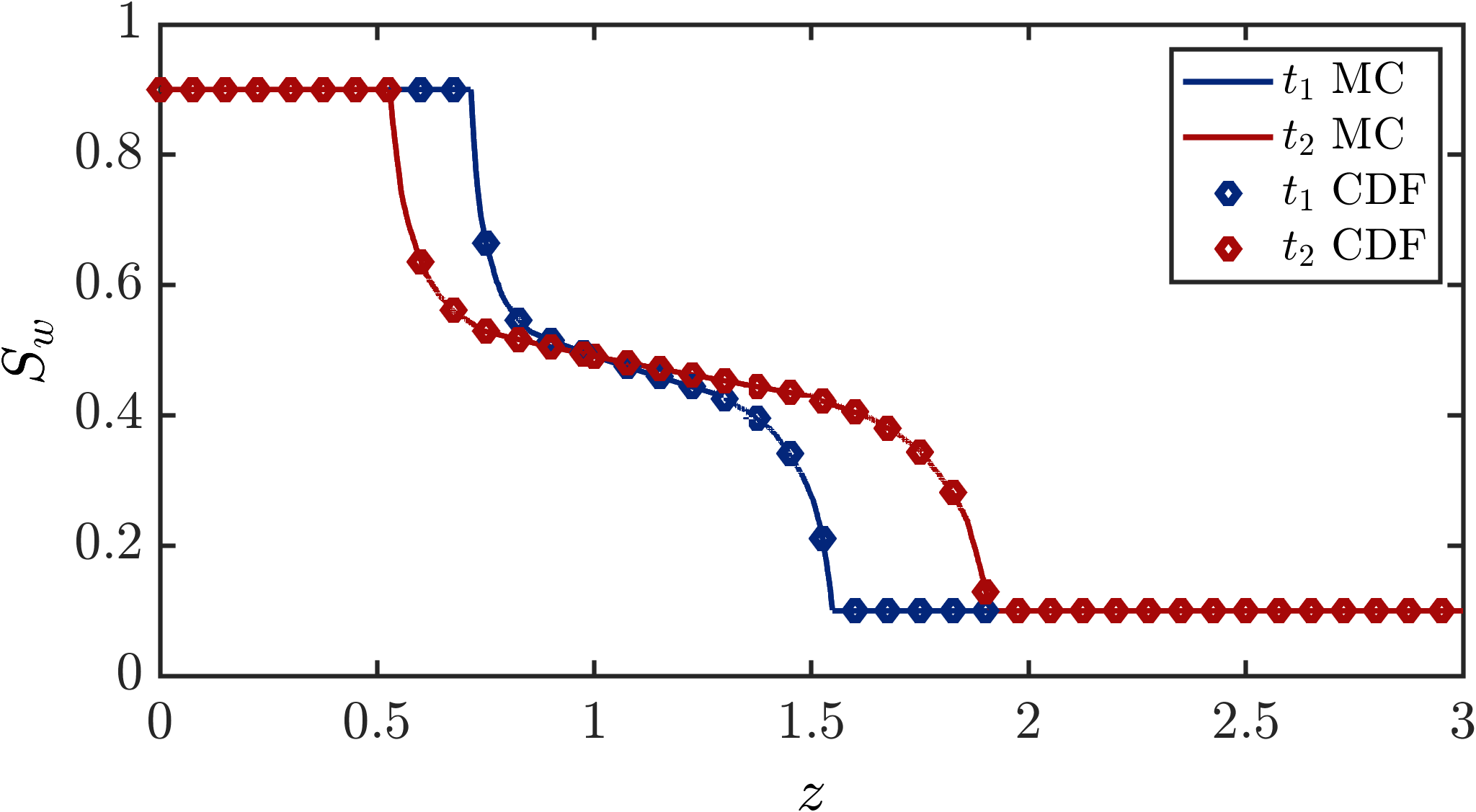}
\includegraphics[width=60mm]{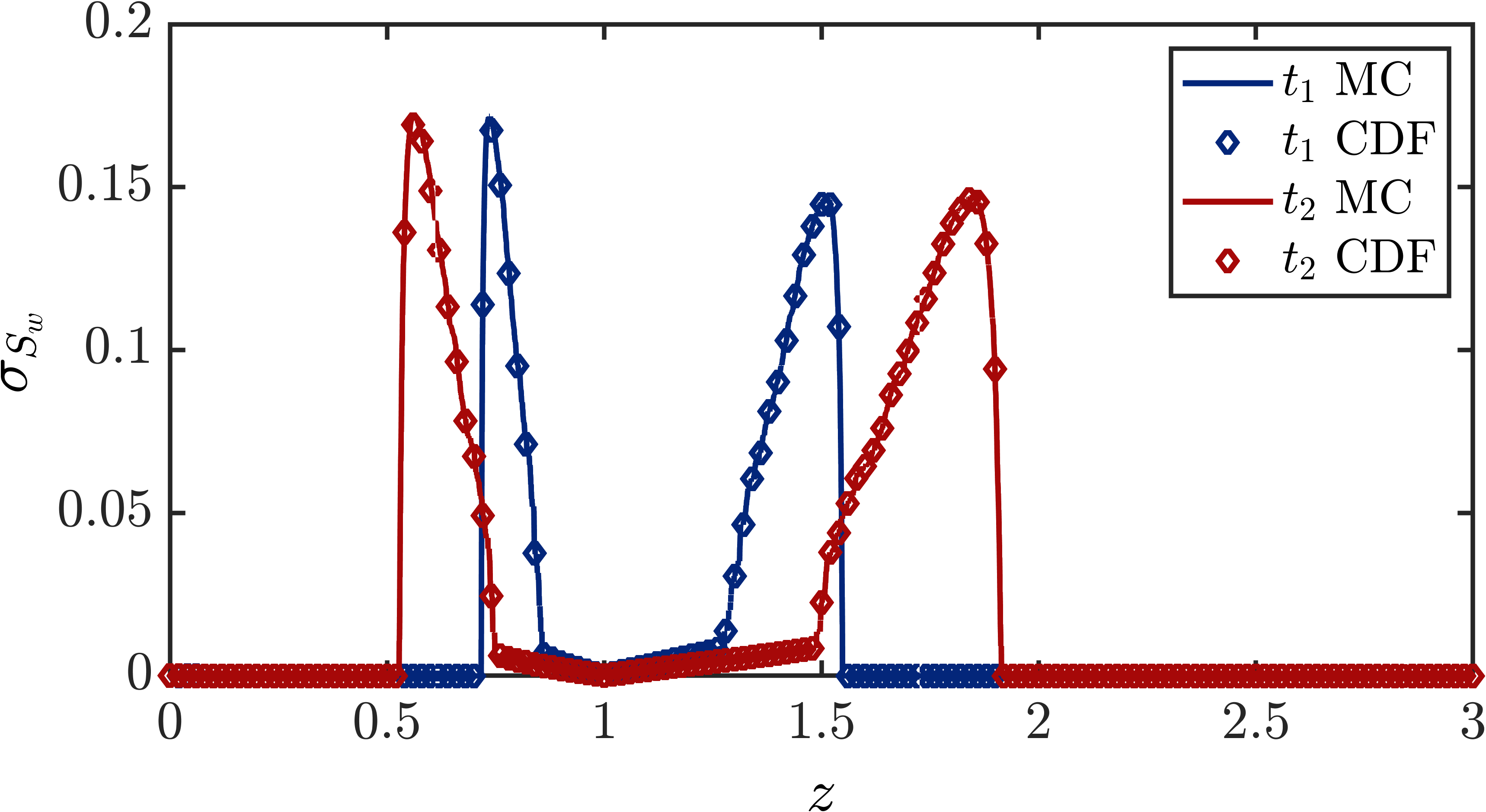}
\includegraphics[width=60mm]{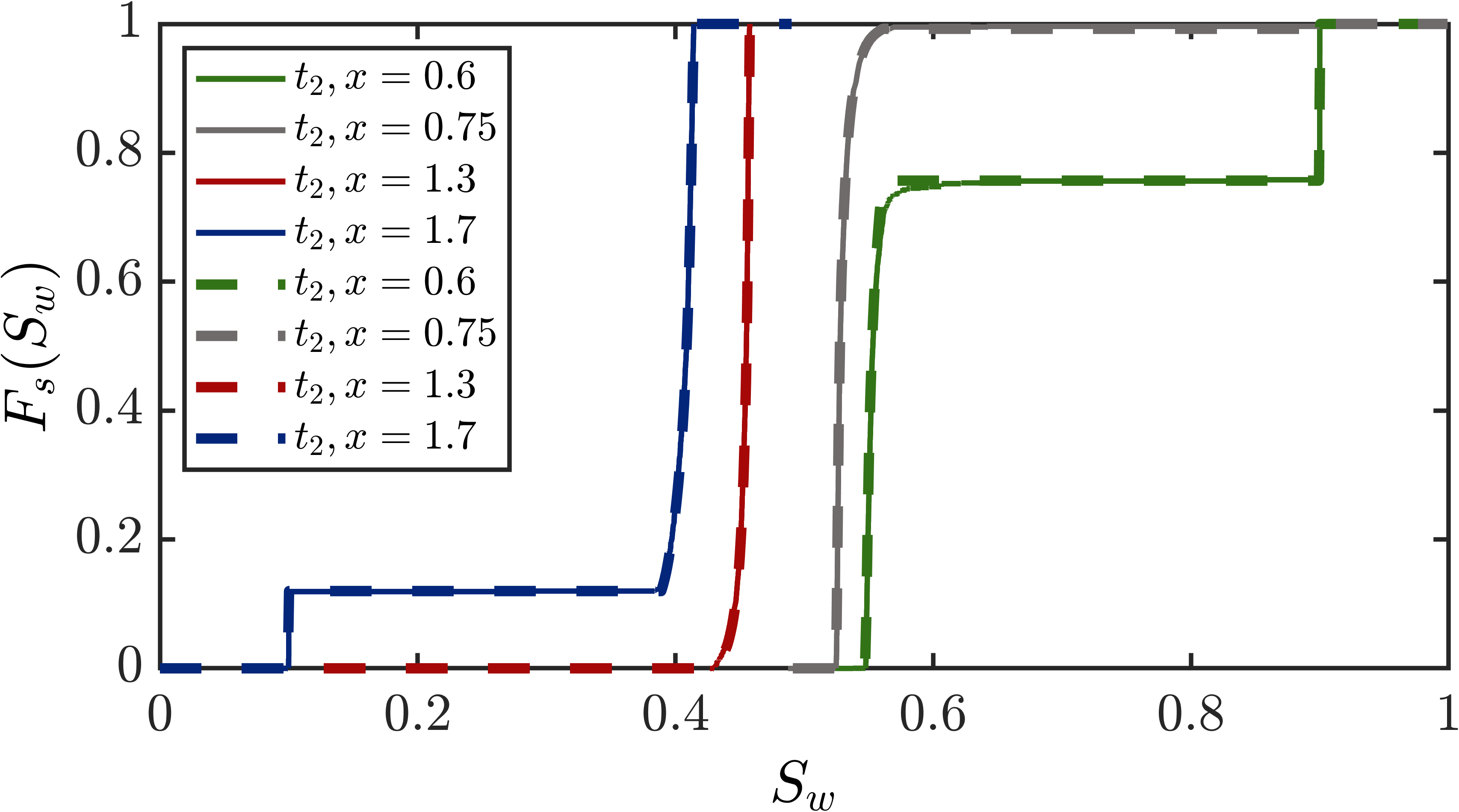}
\includegraphics[width=60mm]{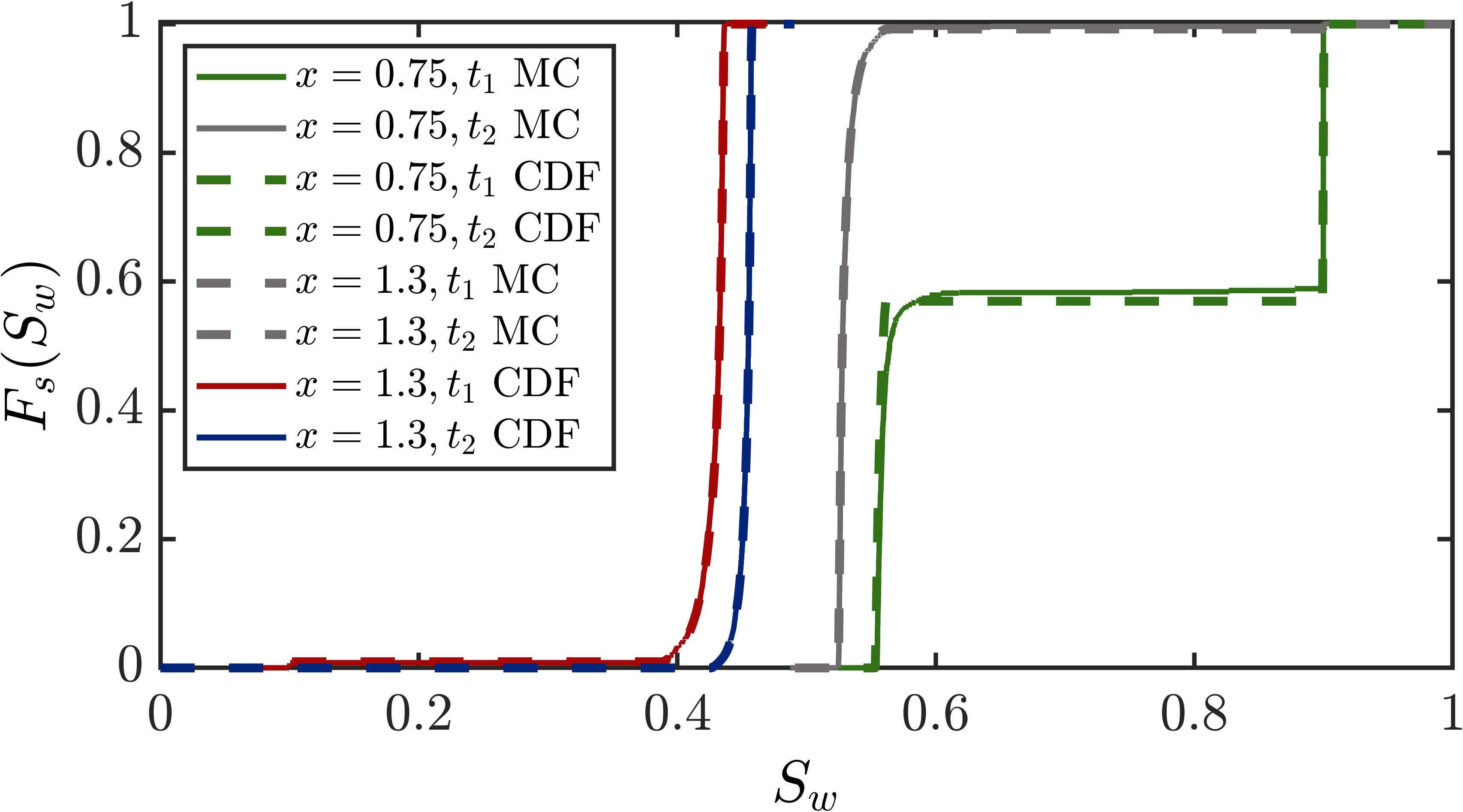}
\includegraphics[width=60mm]{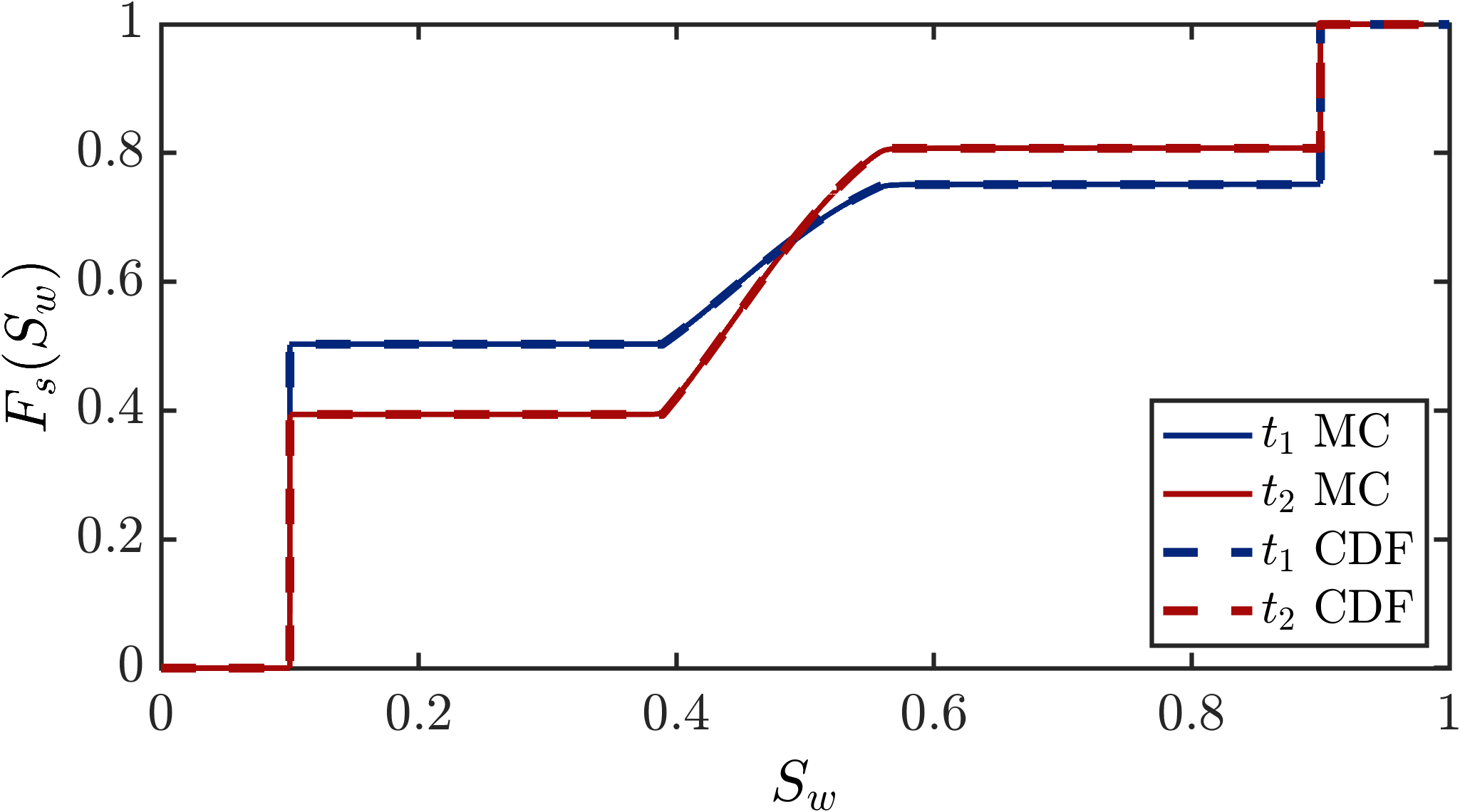}
\includegraphics[width=60mm]{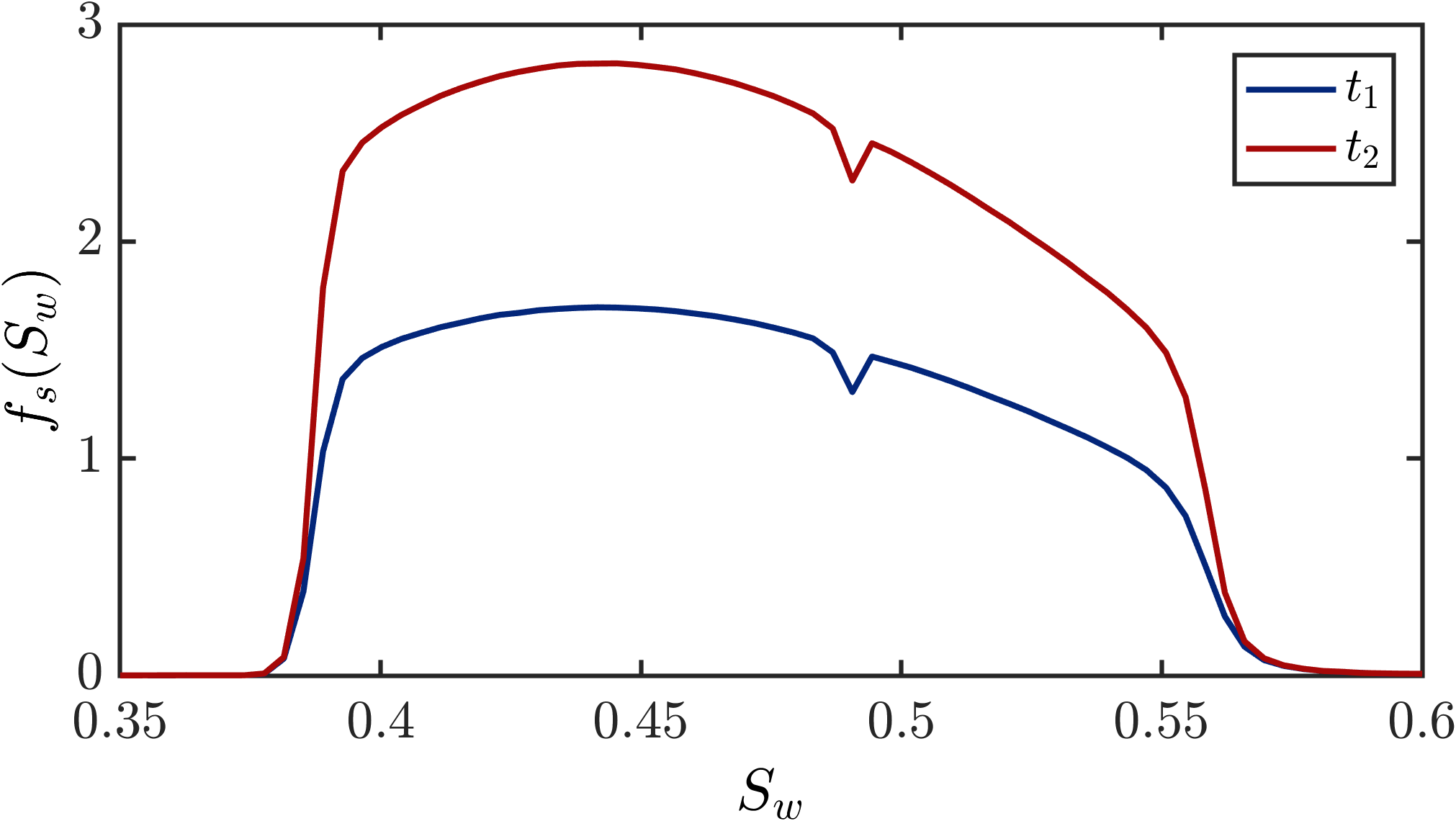}

	\caption{(First row) The first two moments of saturation, (second row) spatial and temporal evolution of the CDF of saturation (dashed lines show the results of method of distributions, while solid lines correspond to Monte Carlo), (third row) point-wise spatial average of the CDF and PDF at all points along the domain, obtained from Monte Carlo and CDF methods. All plots correspond to the downdip flooding with $q(t)$ random, $\mu_q= 0.3$, $\sigma_q= 0.5$, $\tau_q= 0.5 T$, at two dimensionless times $t_1=0.2$, $t_2=0.3$. }
	\label{fig:cdfs for downdip q(t) random}
\end{figure}

Having a time discretization of size $\Delta t$, for any point $t$ in the time interval $[0,T]$, where $t=(n-1)\Delta t$, we define the Gaussian vector $(\rho_q)^n = (\rho_q (j\Delta t))_{0\leq j \leq n-1} $ by $(\rho_q)^n \sim \mathcal{N} (\mu_{\rho} \mathds{1}_n, (C_{\rho})^n)$, where $C_{\rho}^n = ( C_q (\vert i-j\vert \Delta t) )_{1\leq i,j\leq n}$ and $\mathds{1}_n$ is the vector of all ones which is of size n.
Unless otherwise stated, whenever experimenting with uncertainty in $q(t)$, we will use $\mu_q= 0.3$, $\sigma^2_q=0.5$, $\tau_q=0.5 T$, $T=1$, $m=0.5$ . Figures \ref{fig:CDF plots for horizontal Q random} and \ref{fig:cdfs for downdip q(t) random} shows the results for the injection flux as the sole source of randomness.
%%%%%%%%%%%%%%%%%%%%%%%%%%%%%%%%%%. horizontal   q random

%%%%%%%%%%%%%%%%%%%%%%%%%%%%%%%. downdip

%%%%%%%%%%%%%%%%%%%%%%%%%%%%%%%%%%%%%%%%% both phi(x) and q(t) random-correlated
\subsubsection{Numerical setting for uncertainty in both $\phi(x)$ and $q(t)$ } \label{section:phi and q both random}
\begin{figure}[htbp]
	\centering
\includegraphics[width=60mm]{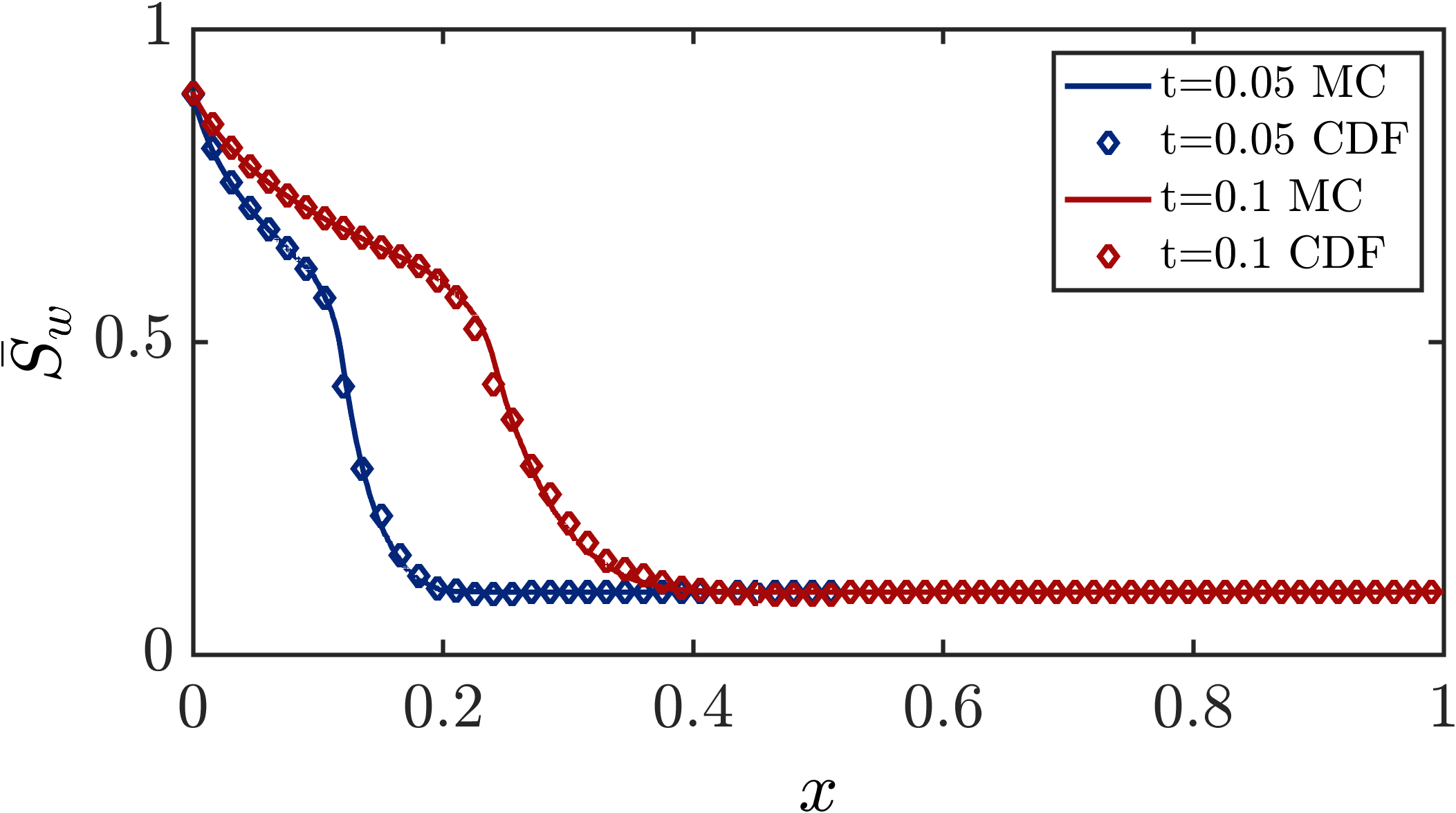}
\includegraphics[width=60mm]{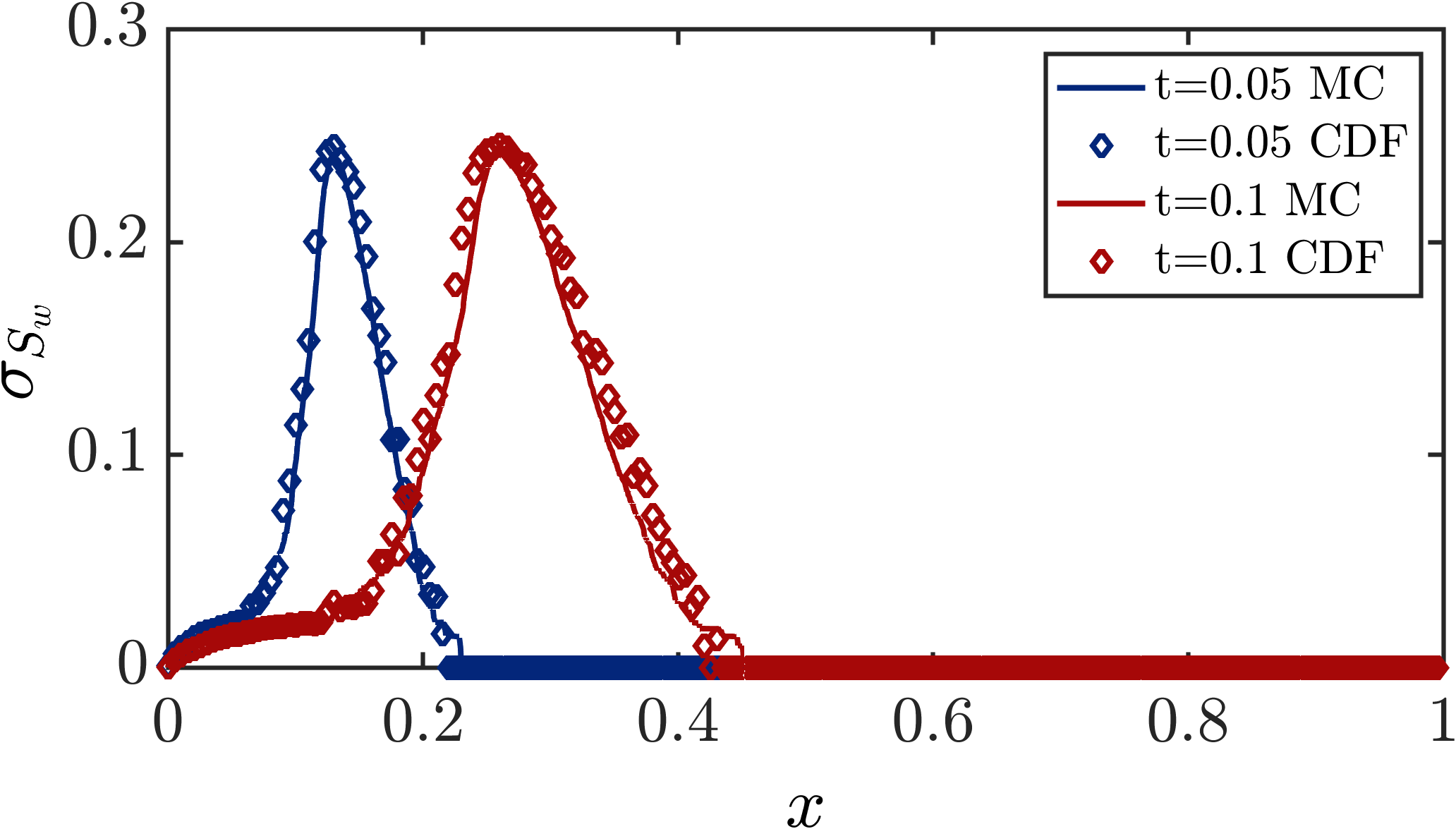}
\includegraphics[width=60mm]{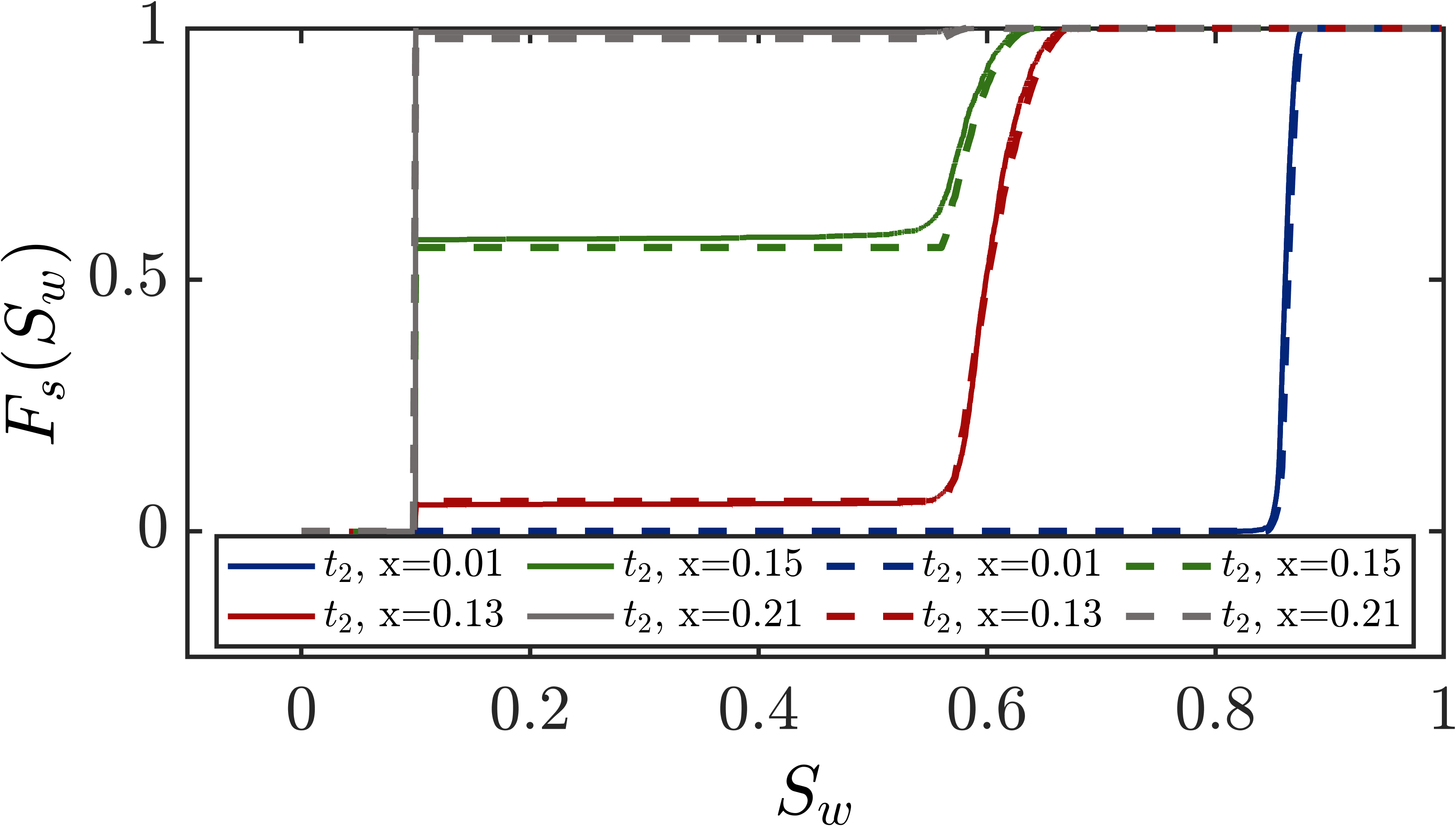}
\includegraphics[width=60mm]{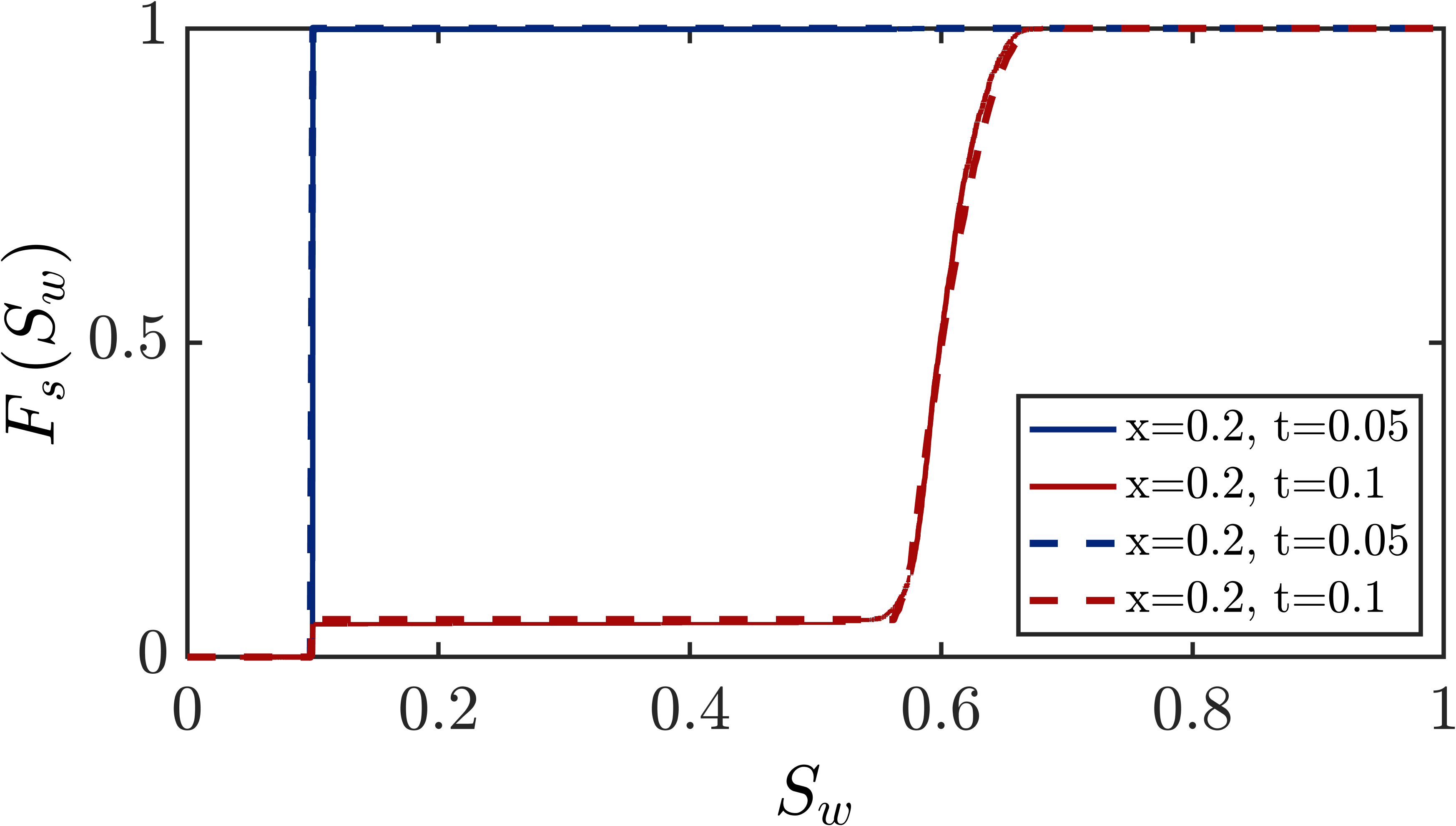}
\includegraphics[width=60mm]{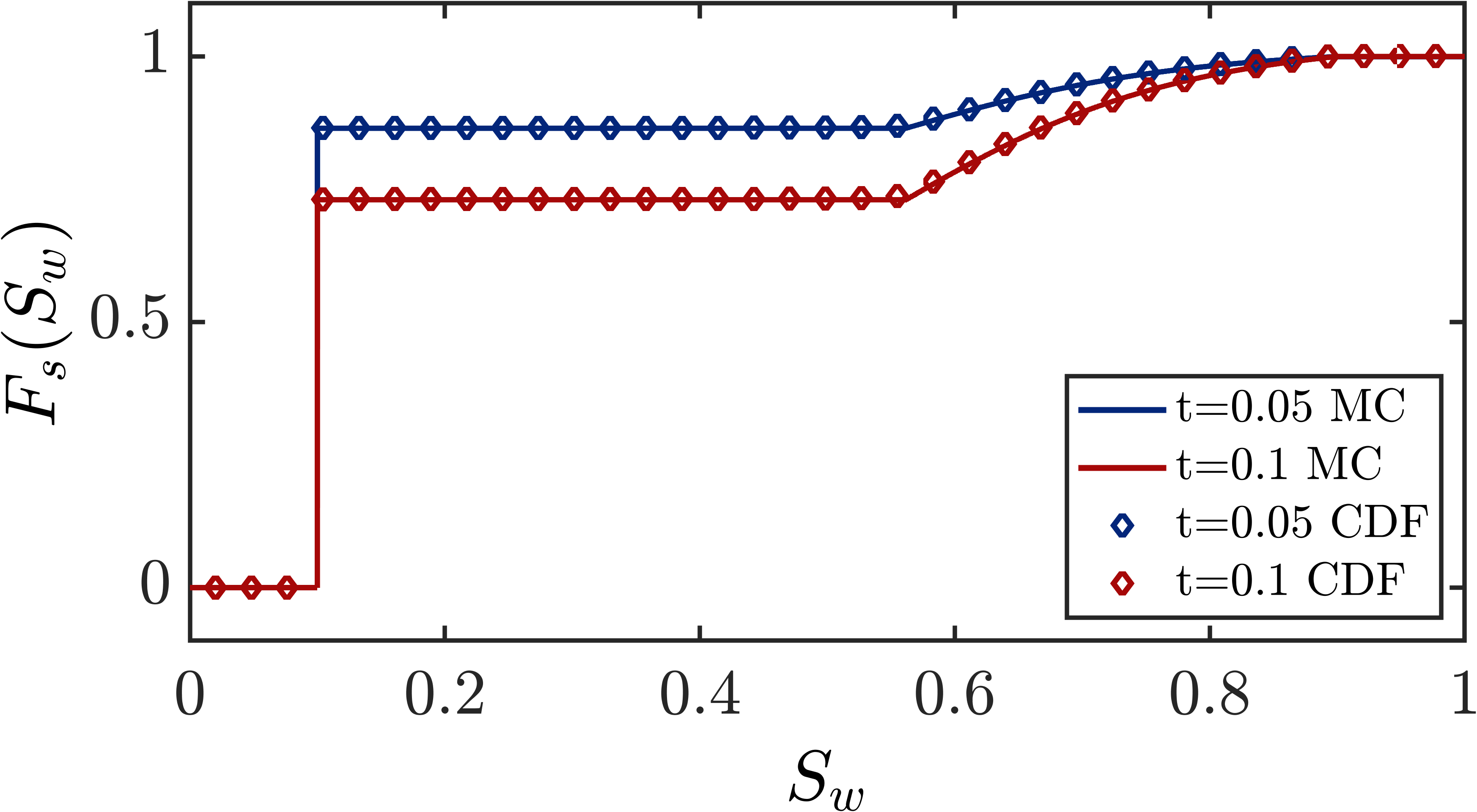}
\includegraphics[width=60mm]{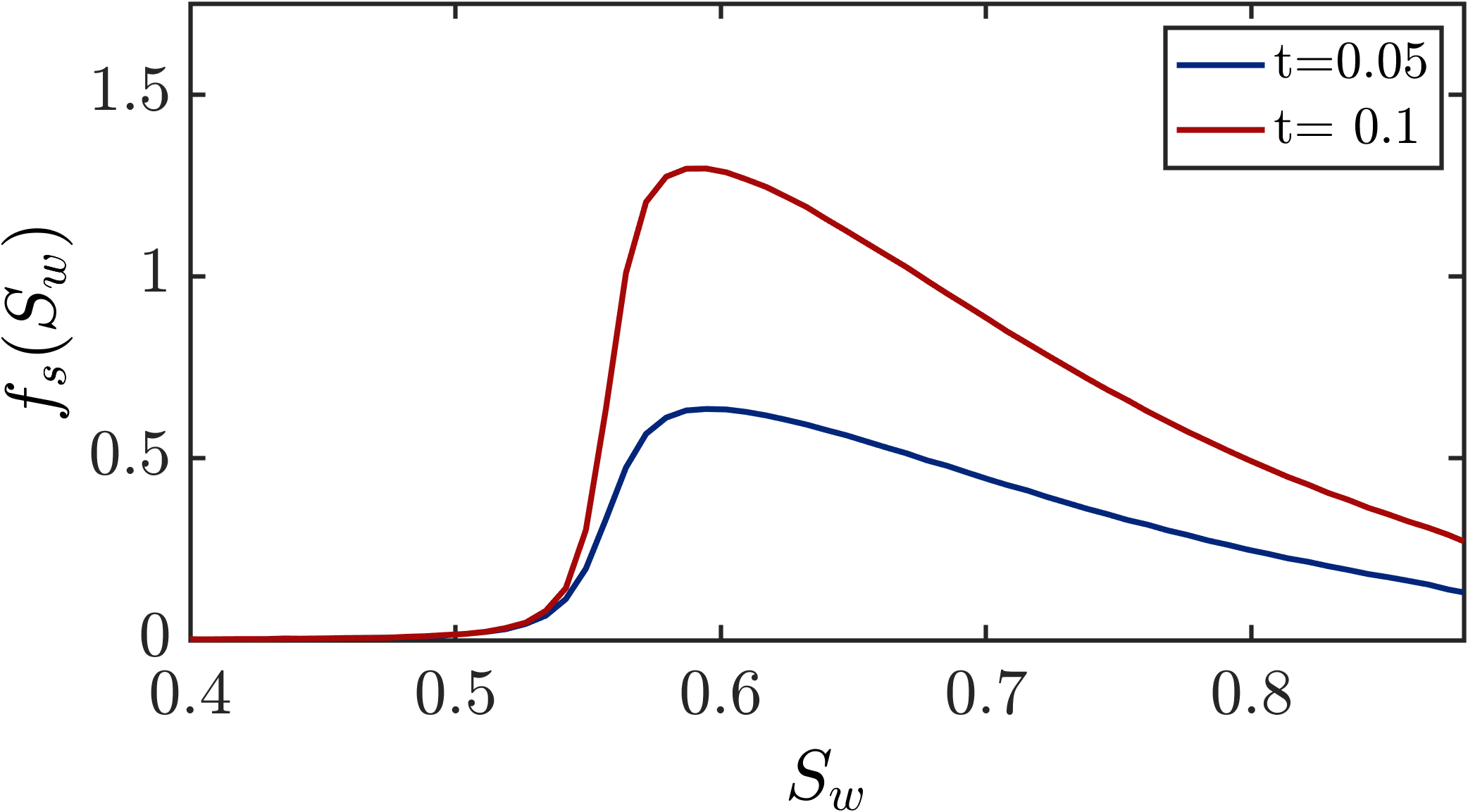}
	\caption{(First row) The first two moments of saturation, (second row) spatial and temporal evolution of the CDF of saturation (dashed lines show the results of method of distributions, while solid lines correspond to Monte Carlo), (third row) point-wise average of the CDF and PDF at all points along the domain, obtained from Monte Carlo and CDF methods, (fourth row) the first two moments obtained using Low-order SME method, vs. those of MC and MD. All simulations represent horizontal case with $\phi(x)$ random, $\mu_{\phi}= 0.3$, $\sigma^2_{\phi}= 0.5$, $\lambda_{\phi}= 0.5 L$, $L=1$, and $q(t)$ random, $\mu_{q}= 0.3$, $\sigma^2_{q}= 0.5$, $\tau_{q}= 0.5 T$, $T=1$. Solid lines are the result of MC simulation, while dashed lines represent the results from CDF method.}
	\label{fig:cdfs for horizontal Q and PHI random}
\end{figure}

In this section, we will illustrate the joint effects of uncertainty in the porosity field as well as the injection flux. We will resort to the same setup as the two previous sections for the construction of $\phi(x)$ and $q_T(t)$ fields. As for the numerical parameters, we will employ $m=0.5$, $\mu_q= 0.3$, $\sigma^2_q=0.5$, $\tau_q=0.5 T$, $T=1$,  $\mu_{\phi}= 0.3$, $\sigma^2_{\phi}=0.5$, $\lambda_{\phi}=0.5 L$. Fig. \ref{fig:realizations} illustrates four random sampling of porosity field as well as injection flux field plotted using the aforementioned values for mean, variance, and correlation length/time for each random variable.
In this section, we will illustrate the joint effects of uncertainty in the porosity field as well as the injection flux. We will encapsulate both uncertainties in the structure of the. stochastic velocity field by representing it as $\textbf{v}_T = \textbf{v}_T(x,t)$.  Fig. \ref{fig:cdfs for horizontal Q and PHI random} shows the results of having both fields as the source of randomness

%%%%%%%%%%----------------------------  horizontal both Q and Phi random

\section{Low-Order Approximation (LOA)} \label{section: LOA}
So far we have demonstrated that estimations of the mean and standard deviation, as well as distribution functions of saturation field from CDF method exhibit a promising match with those obtained from Monte Carlo averaged of a large number of realizations. That is, given an exhaustive number of trials for the random porosity or injection flux followed by solving an upwinding-type problem leads to the solution for saturation field which matches well with the solution from method of distributions. 

The first two moments of water saturation from Monte Carlo and method of distributions were demonstrated to match well. More specifically, in Monte Carlo and MD methods, the saturation moments are obtained through a post-processing step, i.e. after finding the saturation field itself. Alternatively, statistical moment equations (SME) directly solve equations for the first two statistical moments of water saturation, i.e. mean and standard deviation. In this section, we investigate the impact of variance and correlation length of the uncertain input parameters on the first two statistical moments of saturation field. This methodology also referred to as low-order approximation (LOA) involves mathematical manipulations started with the perturbation expansion of the stochastic parameters \cite{dongxiao1999stochastic} and \cite{ibrahima2015distribution}.

%------------------------------------------------
%        FORMULATION OF LOA
%------------------------------------------------
\subsection{Formulation}
We will present the effect of random input on the first two moments of water saturation for different levels of standard deviation and correlation length of the input stochastic field. First we assume that injection flux $q_T$ is a random constant, while porosity $\phi$ is a deterministic constant. In the second scenario, we will assume the opposite setting holds, i.e. we use the equations resulting from perturbation expansion of the porosity field $\phi(x)$, while $q_T$ is a deterministic constant. 

In the first scenario, we assume a deterministic constant value of $\phi=\phi_0=0.3$, and a lognormal distribution for $q$, with $\mu_q=0.3$, while we will experiment with multiple values of variance $\sigma_q^2$ and track the resulting first two moments of the state field $S_w$,

\begin{align}
    & q \sim log \mathcal{N} \bigg(\langle \text{ln}(q)\rangle, \sigma_{\text{ln}(q)}\bigg)\qquad
\langle ln(q)\rangle = ln \bigg(\dfrac{\mu_q^2}{\sqrt{\sigma^2_q+\mu_q^2}}\bigg), \qquad 
\sigma_{ln(q)} = \sqrt{ln \bigg(\dfrac{\sigma_q^2}{\mu_q^2} +1 \bigg)}
\end{align}

Similarly to \cite{dongxiao1999stochastic}, we consider the one-dimensional version of Eq.~(\ref{eq:saturation_equation_with_BC}), and using the idea of displacement along a streamline (\cite{dongxiao1999stochastic}), we define $\tau(x;x_0)$, the travel time (time of flight) of a particle from $x_0=0$ to $x$ in the total velocity field $\textbf{v}_T$ as,

\begin{align}\label{eq: d tau to dx}
    \dfrac{d\tau}{dx} = \dfrac{1}{\textbf{v}_T }= \dfrac{\phi_0}{q}
\end{align}

Which has a solution in the form of $\tau(x;x_0) = (x-x_0)\dfrac{\phi_0}{q}$. Since $\textbf{v}_T$ is random, so is $\tau$, and hence, its statistical moments depend on those of $\textbf{v}_T$. By applying a Reynolds decomposition on $q$, we can estimate the random variable using its ensemble mean (expected value) plus the zero-mean random perturbations (around the mean). Thereupon, Eq.~(\ref{eq: d tau to dx}) will be recast in the following form,

\begin{align}
   \tau(x;x_0) = (x-x_0)\dfrac{\phi_0}{(\mu_q + q^{\prime})} = (x-x_0)\dfrac{\phi_0}{\mu_q } \bigg[ 1 - \dfrac{q^{\prime}}{\mu_q} + \bigg(\dfrac{q^{\prime}}{\mu_q}\bigg)^2 + \cdots\bigg]
\end{align}

Correspondingly, LOA results in the first-order approximation of the first two moments of time of flight as a function of $\mu_q$ and $\sigma^2_q$,

\begin{align}\label{eq:mu tau and sigma tau}
    \langle \tau(x;x_0)\rangle \simeq (x-x_0)\dfrac{\phi_0}{\mu_q},\qquad\qquad\qquad 
    \sigma^2_{\tau} \simeq (x-x_0)^2\phi_0^2 \dfrac{\sigma_q^2}{\mu_q^4}
\end{align}

Furthermore, by prescribing a probability distribution to $\tau$ for the given mean and standard deviation, e.g. lognormal distribution $\tau \sim log \mathcal{N} \bigg(\langle \text{ln}(\tau)\rangle, \sigma_{\text{ln}(\tau)}\bigg)$, where the corresponding moments are defined as (\cite{dongxiao1999stochastic}),

\begin{align}\label{eq:mu ln tau and sigma ln tau}
\big\langle \text{ln}(\tau)\big\rangle = 2 \text{ln} (\langle\tau \rangle) - \dfrac{1}{2} \text{ln}  \bigg[\langle\tau\rangle^2 + \sigma^2_{\tau}\bigg], \qquad \qquad \sigma^2_{\text{ln}(\tau)} = \text{ln} \bigg[\langle\tau\rangle^2 + \sigma^2_{\tau}\bigg] -2 \text{ln}(\langle\tau\rangle)
\end{align}

Therefore, distribution of $\tau$ is fully delineated as a lognormal probability density function, along with the mean $\mu_{\tau}$ and standard deviation $\sigma_{\tau}^2$, which are directly computable using $\mu_q$ and $\sigma_q^2$ in Eq.~(\ref{eq:mu tau and sigma tau}). Therefore, by controlling $\sigma_q^2$ parameter, we can find a low-order estimate of the ensemble mean (expected value) and standard deviation of water saturation using the expressions below,

\begin{align}\label{eq: mean and std of saturation_LOA}
   & \langle S_w (x,t)\rangle = \bigintsss_0^{+\infty} \tilde{S}(\tau, t) p_{\tau}(\tau; x, x_0) d\tau, \notag\\
   & \sigma^2_S(x,t) = \bigintsss_0^{+\infty} \tilde{S}^2(\tau, t) p_{\tau}(\tau; x, x_0) d\tau - \langle S(x,t)\rangle^2
\end{align}

Where where $p_{\tau}(\tau; x,x_0)$ is the probability density function of the (one-particle) travel time $\tau$, and $\tilde{S}(\tau,t)$ can be found by combining Eq.~(\ref{eq:saturation_equation_with_BC}) and Eq.~(\ref{eq: d tau to dx}) resulted in a transformed version of the transport equation for water saturation as,
\begin{align} \label{eq:time of flight transport eqn}
   \dfrac{\partial \tilde{S}}{\partial t} + f_w^{\prime}(S) \dfrac{\partial \tilde{S}(\tau,t)}{d\tau} = 0
\end{align}
as described in \cite{dongxiao1999stochastic}. It should be emphasized that even though \cite{dongxiao1999stochastic} has studies low-order approximation for a horizontal reservoir, we are indeed taking the gravitational effects into account in the flux function while solving Eq.~(\ref{eq:time of flight transport eqn}).

In the second scenario, we assume a constant deterministic injection flux $q=q_0=0.3$ and a random stationary Gaussian porosity field $\phi(x)$ characterized by its mean value of $\mu_{\phi} = 0.3$, and covariance structure $C_{\phi}(x) = \sigma^2_{\phi} exp (-\dfrac{\vert x\vert}{\lambda_{\phi}})$. Given such a definition for the random porosity field, we can leverage the first-order approximation of the first two moments of $\tau$ derived in \cite{dongxiao1999stochastic} as following,

\begin{align}
   & \big\langle \tau(x;x_0) \big\rangle \simeq ( x - x_0 ) \dfrac{\mu_{\phi}}{q_0}, \notag\\ & \sigma^2_{\tau}(x;x_0) \simeq \dfrac{2\sigma^2_{\phi}}{q_0^2} \bigg[\lambda_{\phi}(x-x_0) -\lambda_{\phi}^2 \bigg(1- exp\bigg(-\dfrac{\vert x-x_0\vert}{\lambda_{\phi}} \bigg) \bigg) \bigg]
\end{align}

Furthermore, $\tau$ is assumed log-normally distributed and the same definitions as Eq.~(\ref{eq:mu ln tau and sigma ln tau}) hold for $\langle \text{ln}(\tau)\rangle$ and $\sigma^2_{\text{ln}(\tau)}$. To this end, by controlling $\mu_{\phi}$ and $\lambda_{\phi}$, we will present comparisons of the low-order approximation for the first two moments of water saturation using Eq.~(\ref{eq: mean and std of saturation_LOA}) against those obtained from Monte Carlo and CDF methods.   
%------------------------------------------------
%        NUMERICAL RESULTS OF LOA
%------------------------------------------------
\subsection{Illustrative Examples and Numerical Results}
\begin{figure}[htbp]
	\centering
\includegraphics[width=60mm]{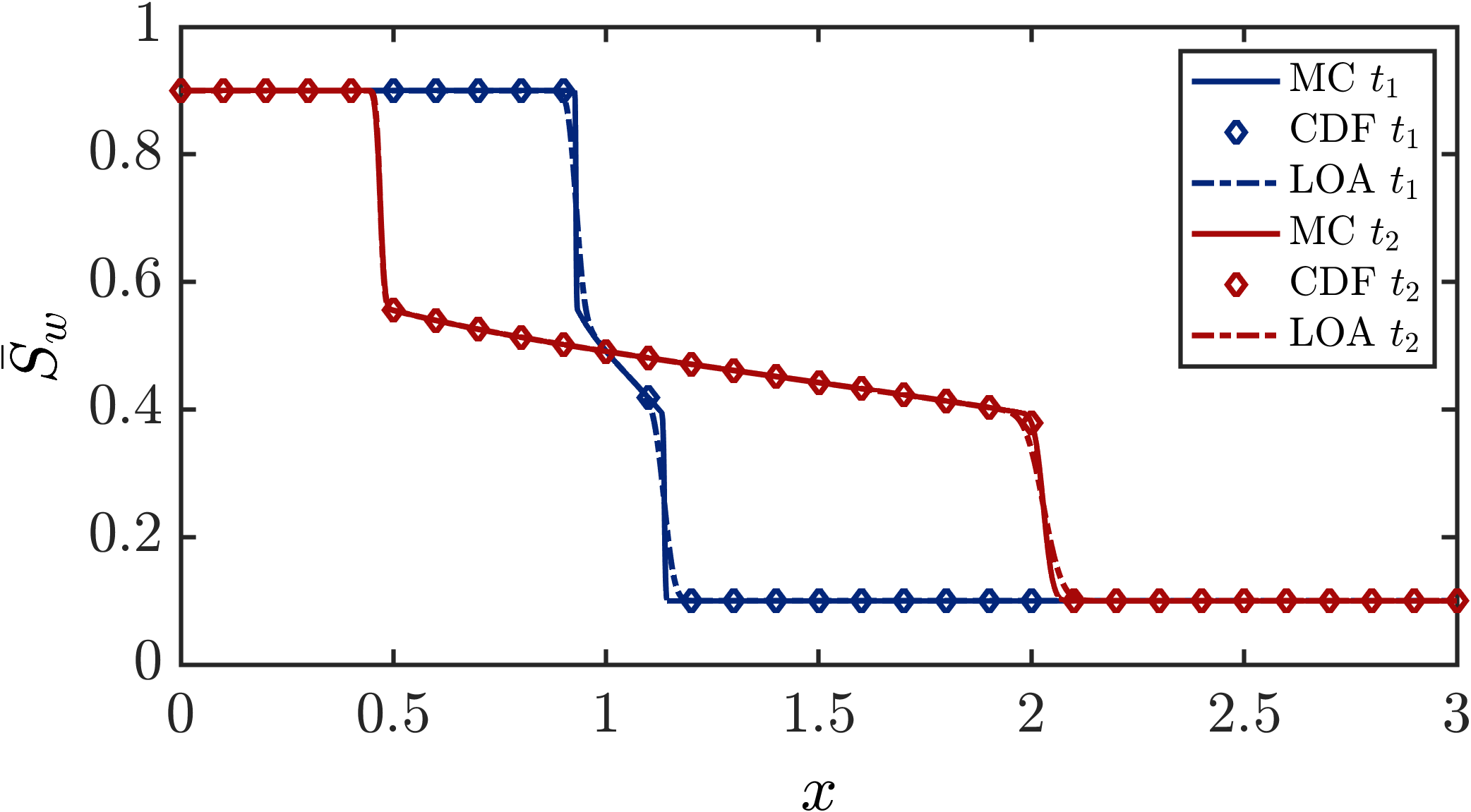}
\includegraphics[width=60mm]{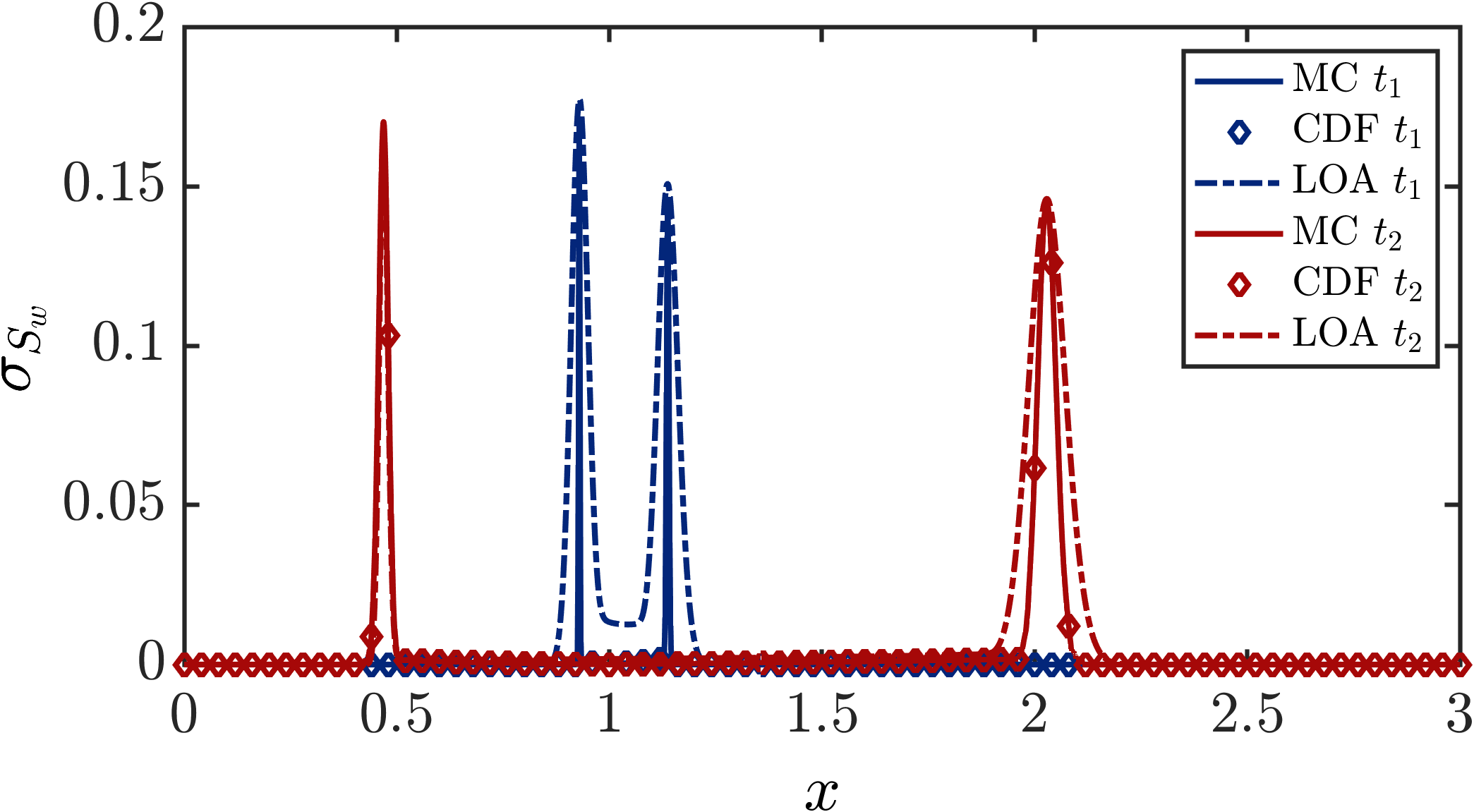}
%%%%%%%%%%%%%%%%%%
\includegraphics[width=60mm]{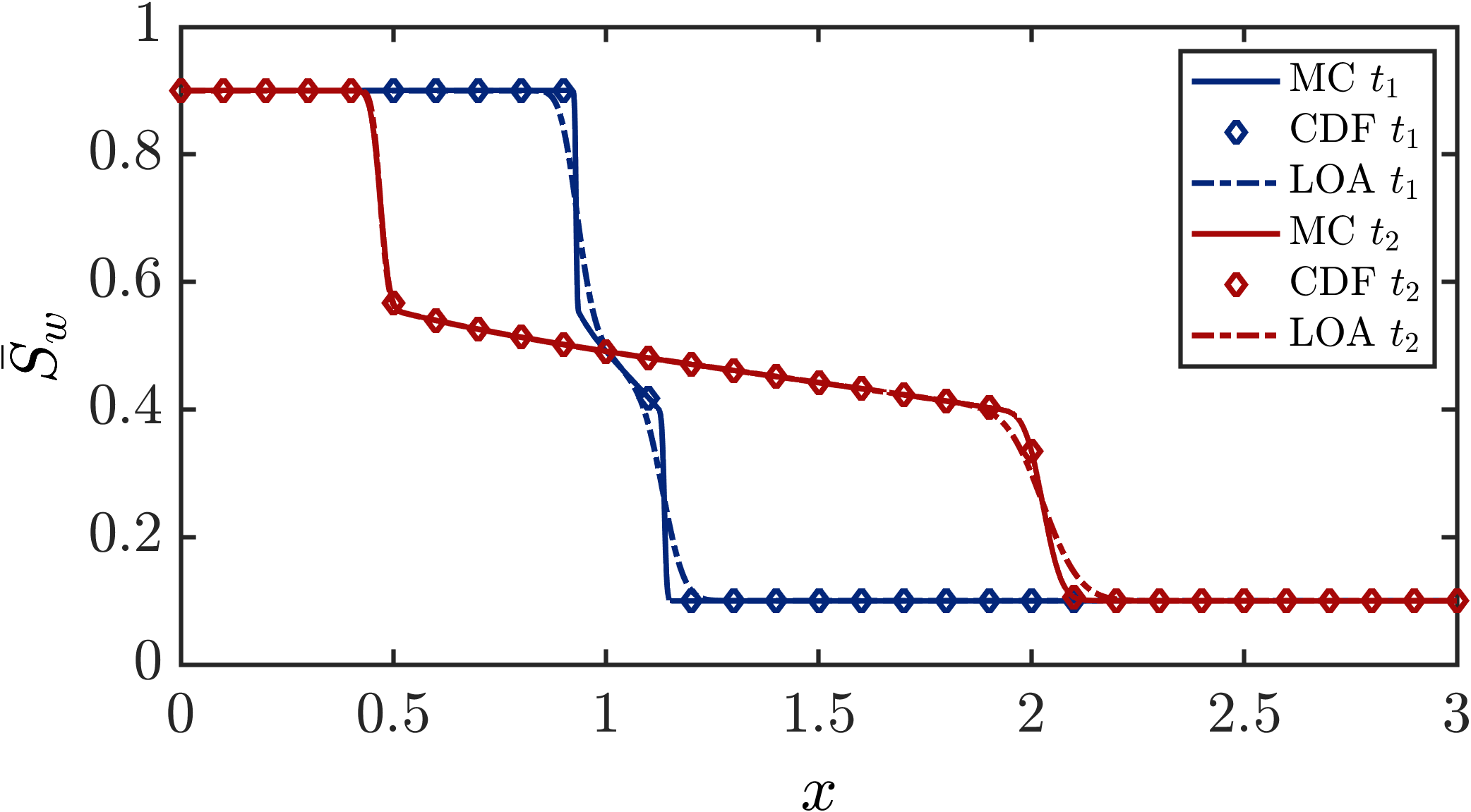}
\includegraphics[width=60mm]{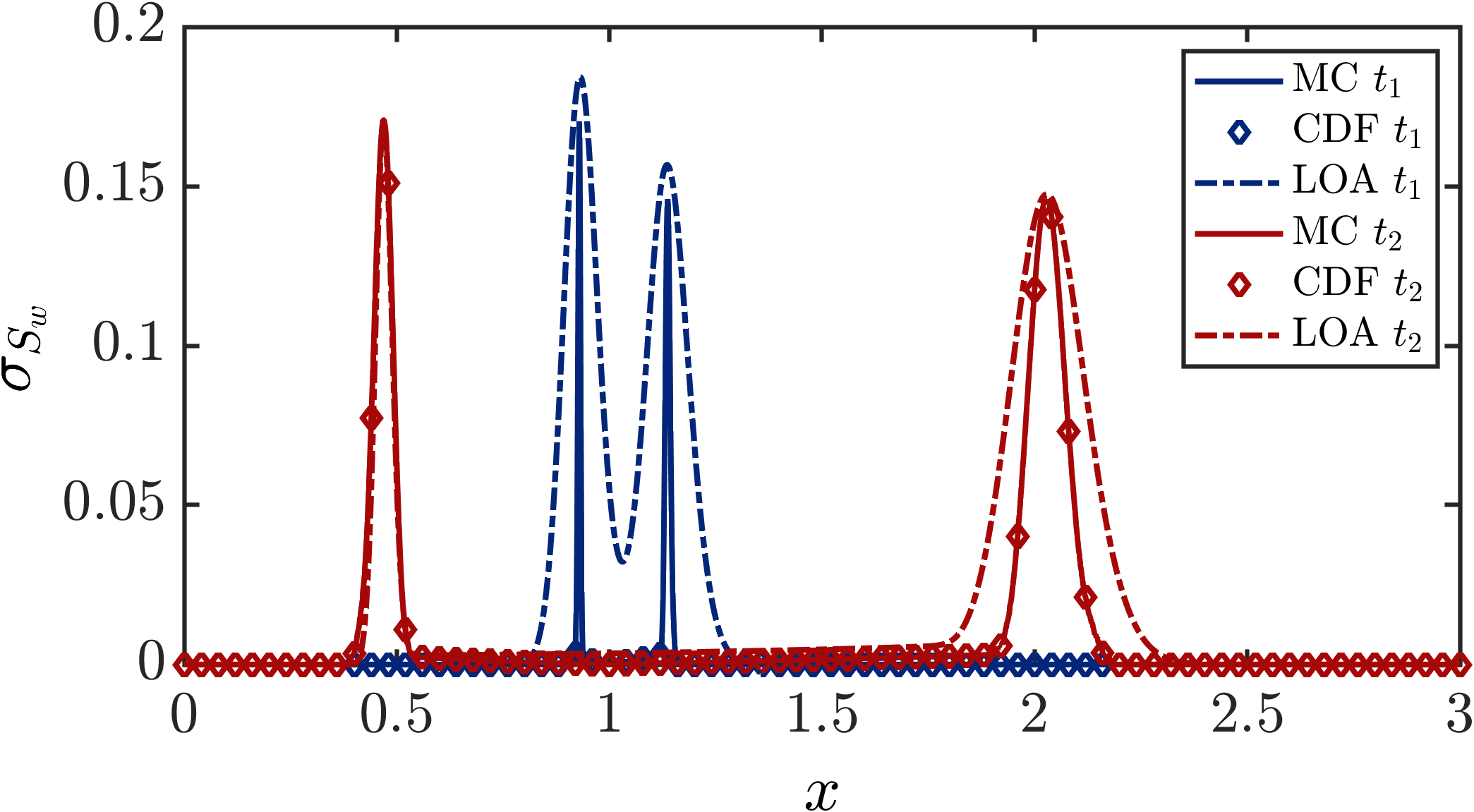}
%%%%%%%%%%%%%%%%%%%%%%%%%%%%%
% \includegraphics[width=60mm]{figures/downdip_LOA_Qrandom_Sat_sigma_0.03_tau0.5.png}
% \includegraphics[width=60mm]{figures/downdip_LOA_Qrandom_STD_sigma_0.03_tau0.5.png}
\includegraphics[width=60mm]{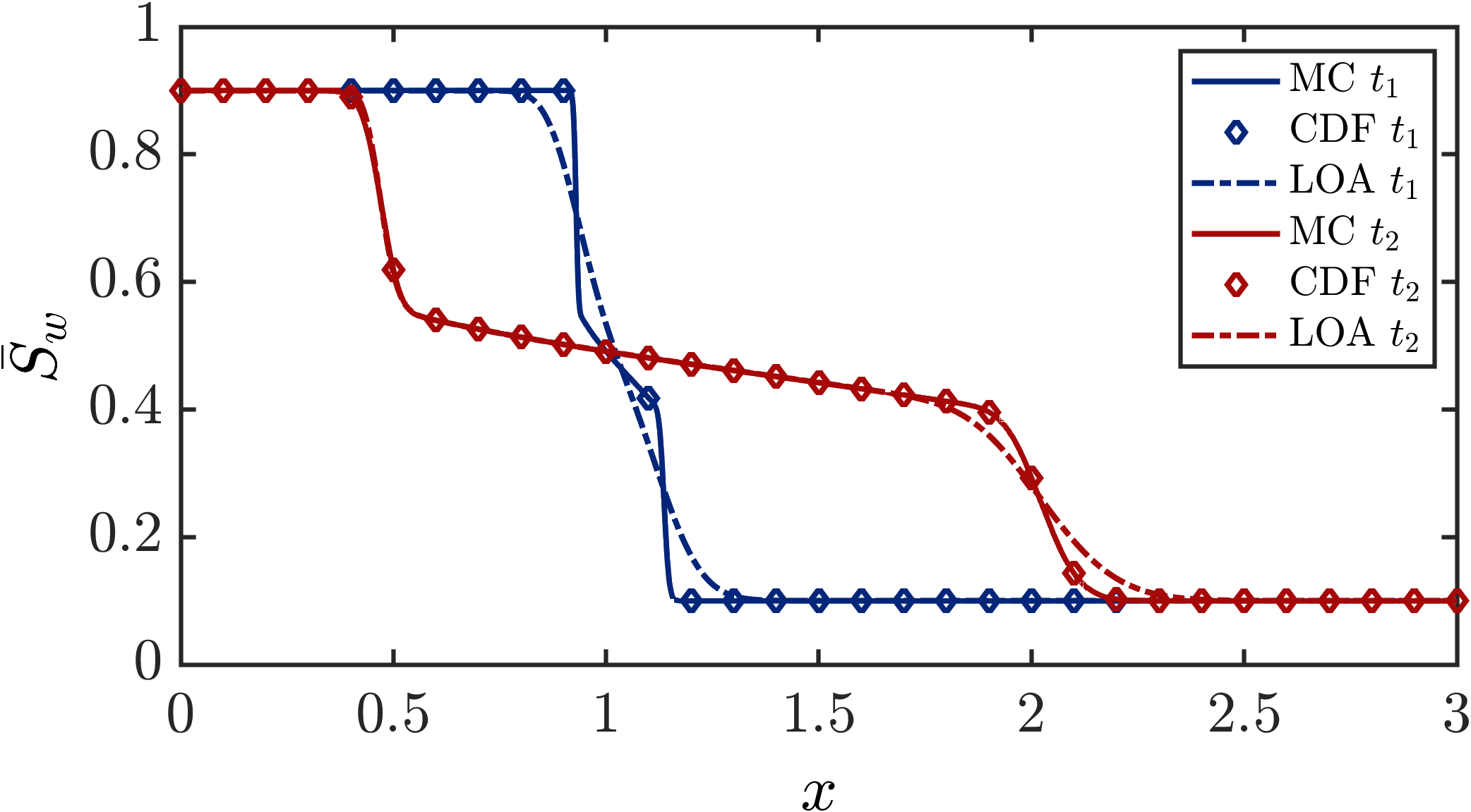}
\includegraphics[width=60mm]{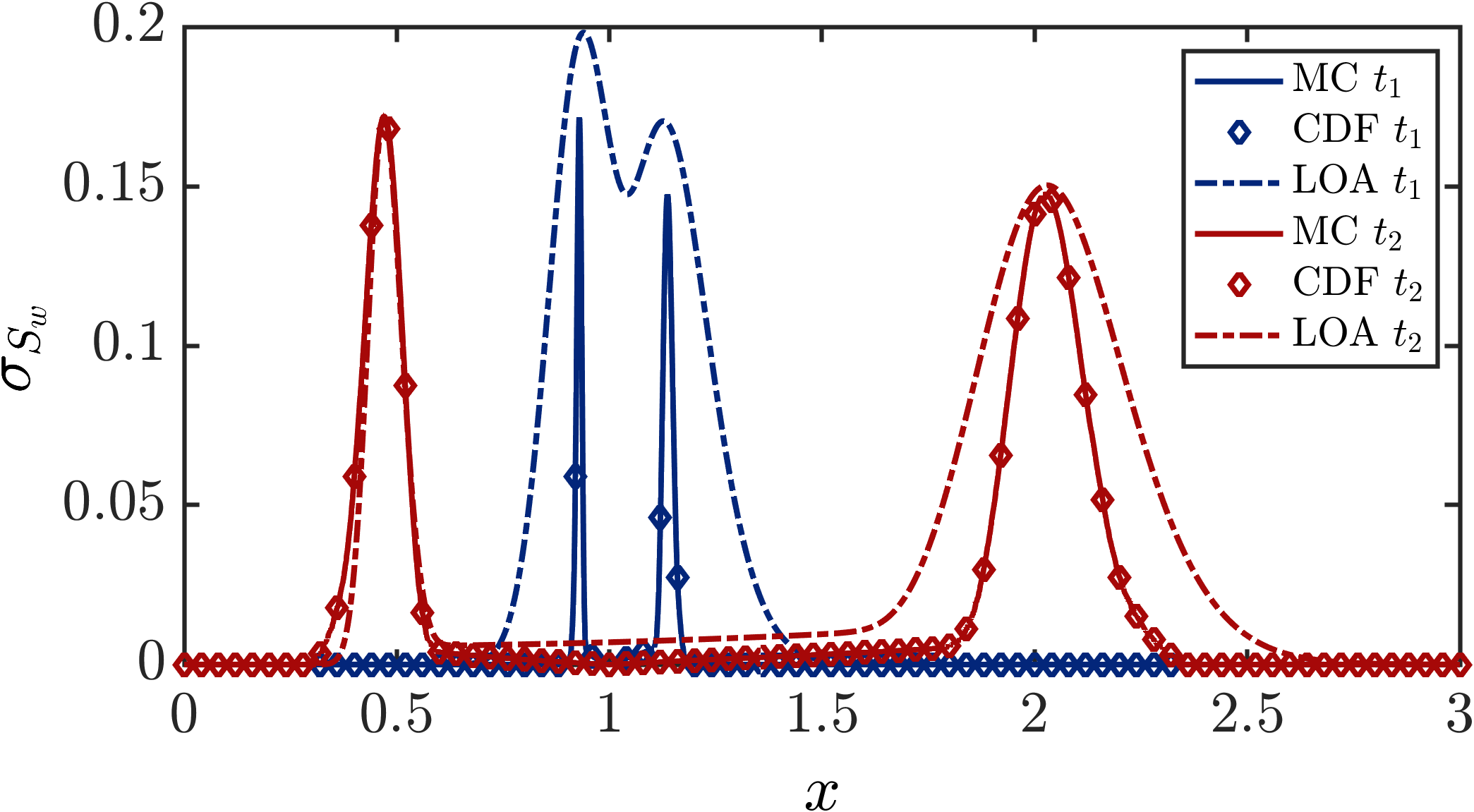}
%%%%%%%%%%%%%%%%%%%%%%%%%%%%%
\includegraphics[width=60mm]{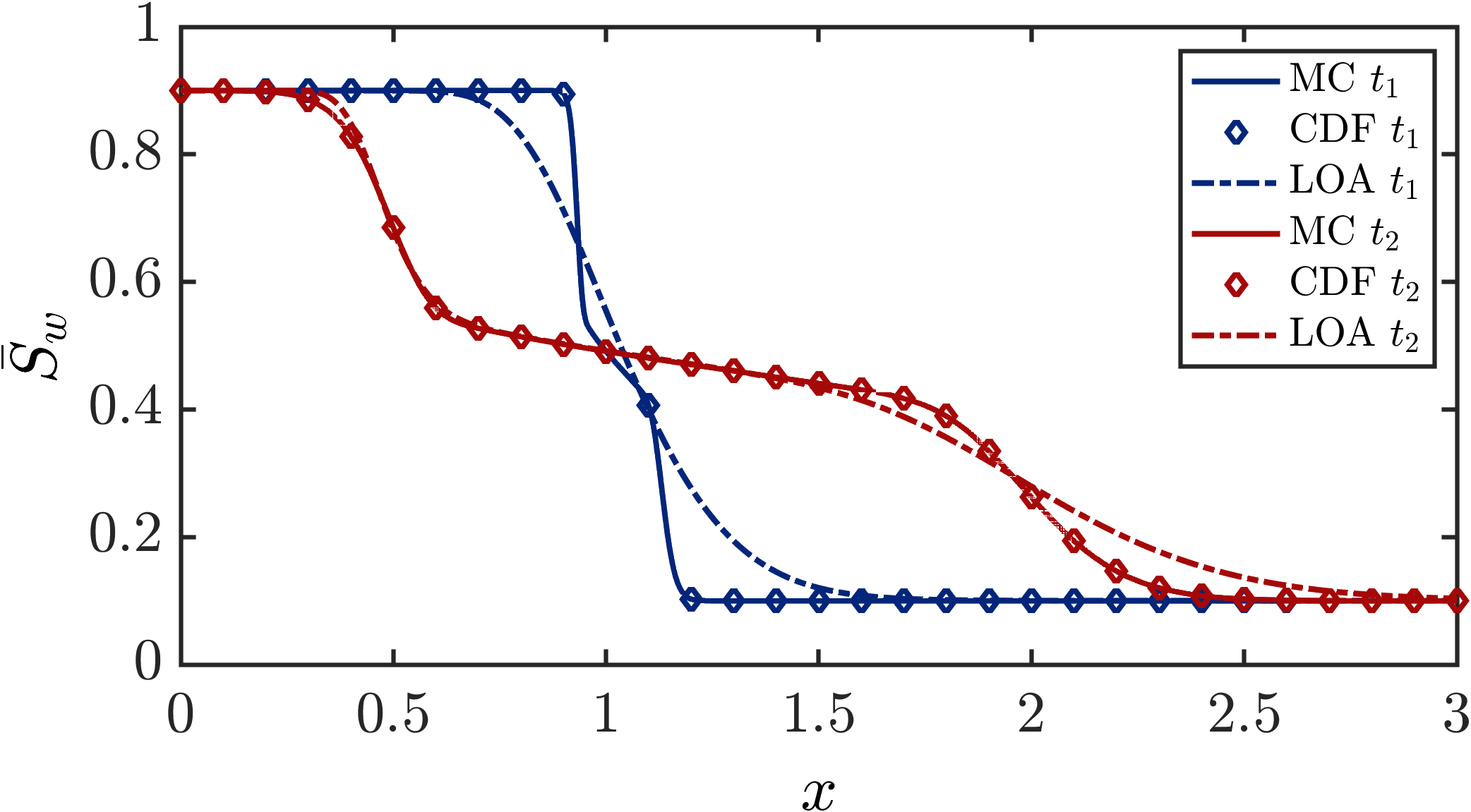}
\includegraphics[width=60mm]{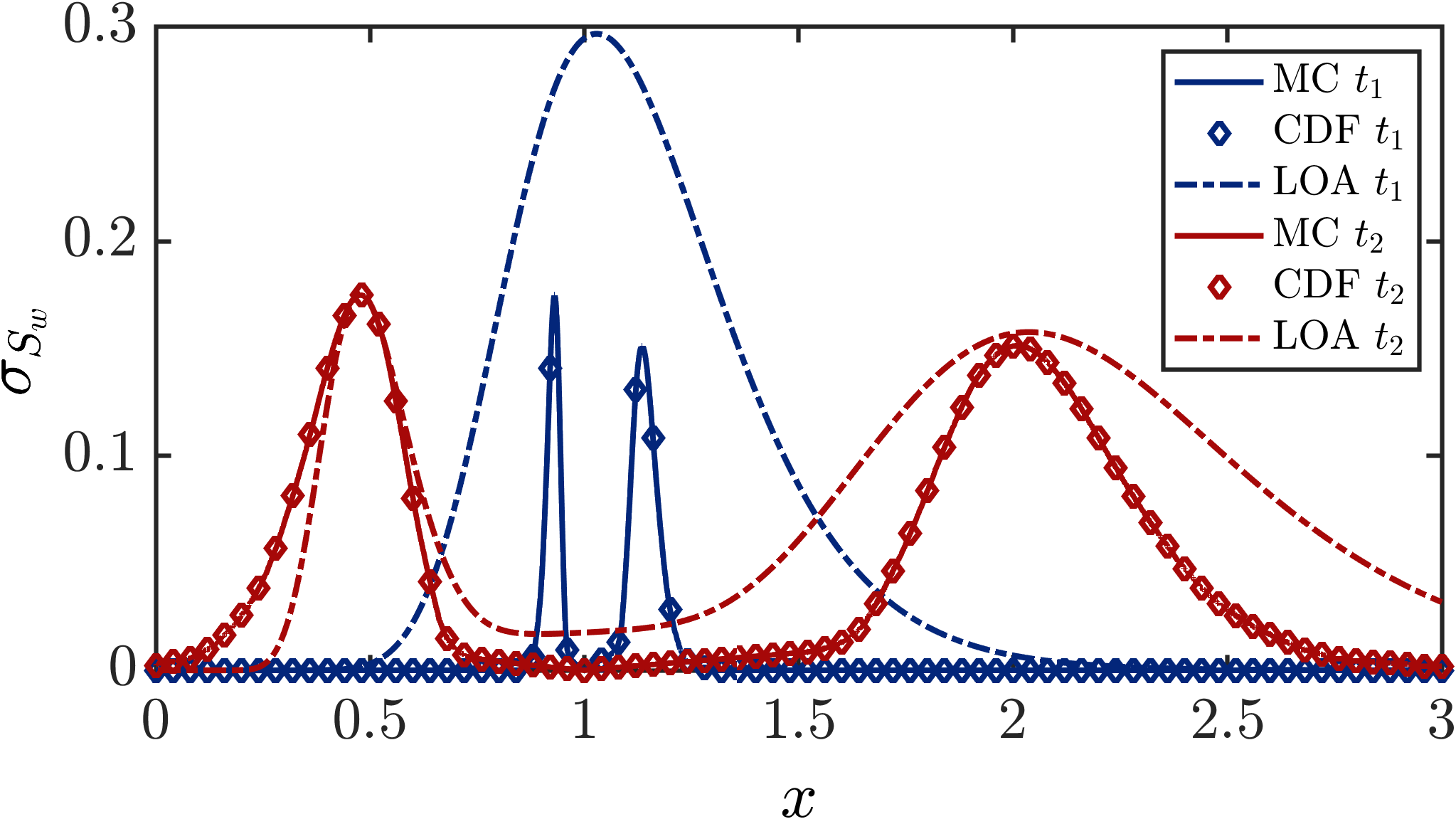}
%%%%%%%%%%%%%%%%%%%%%%%%%%%%%
\includegraphics[width=60mm]{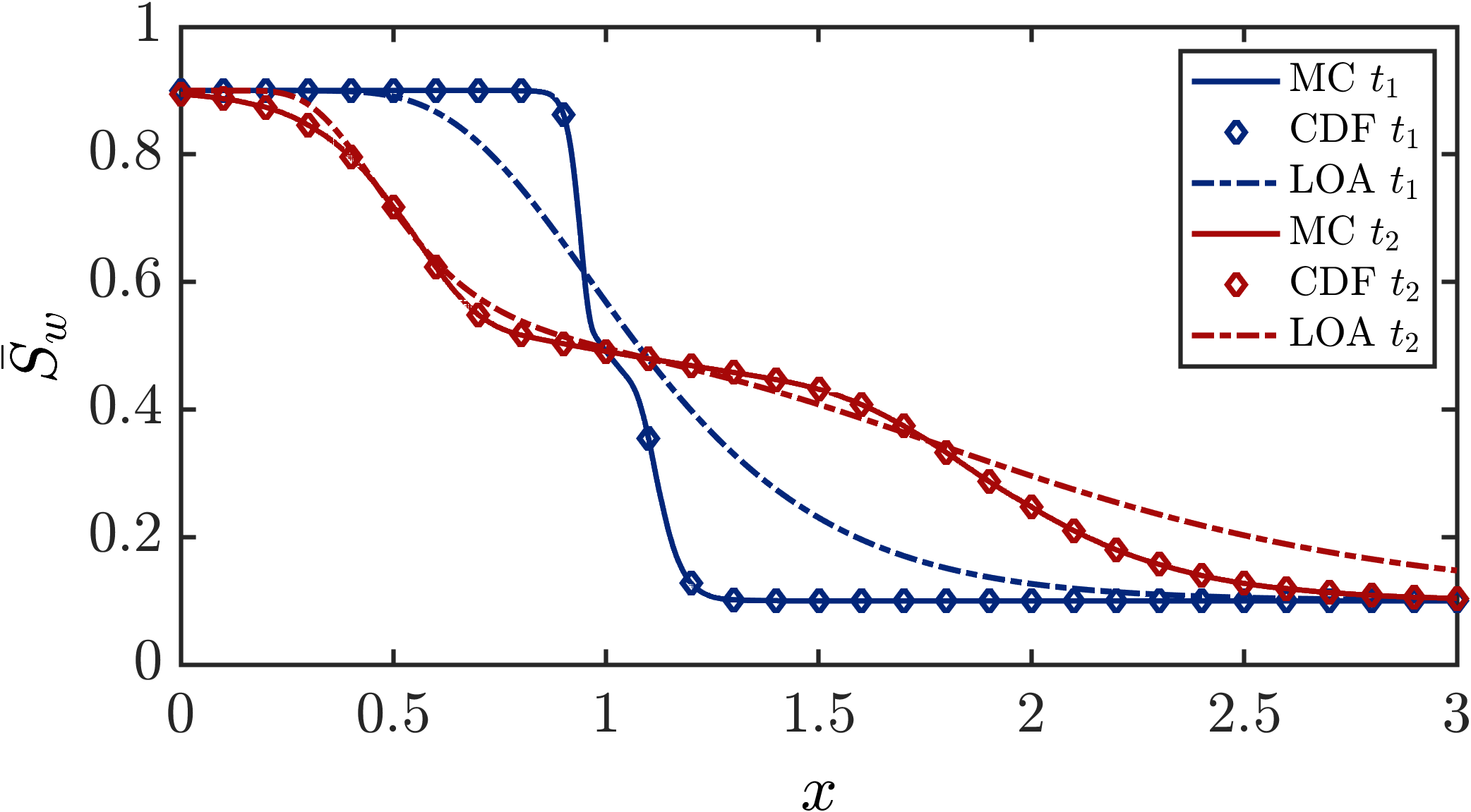}
\includegraphics[width=60mm]{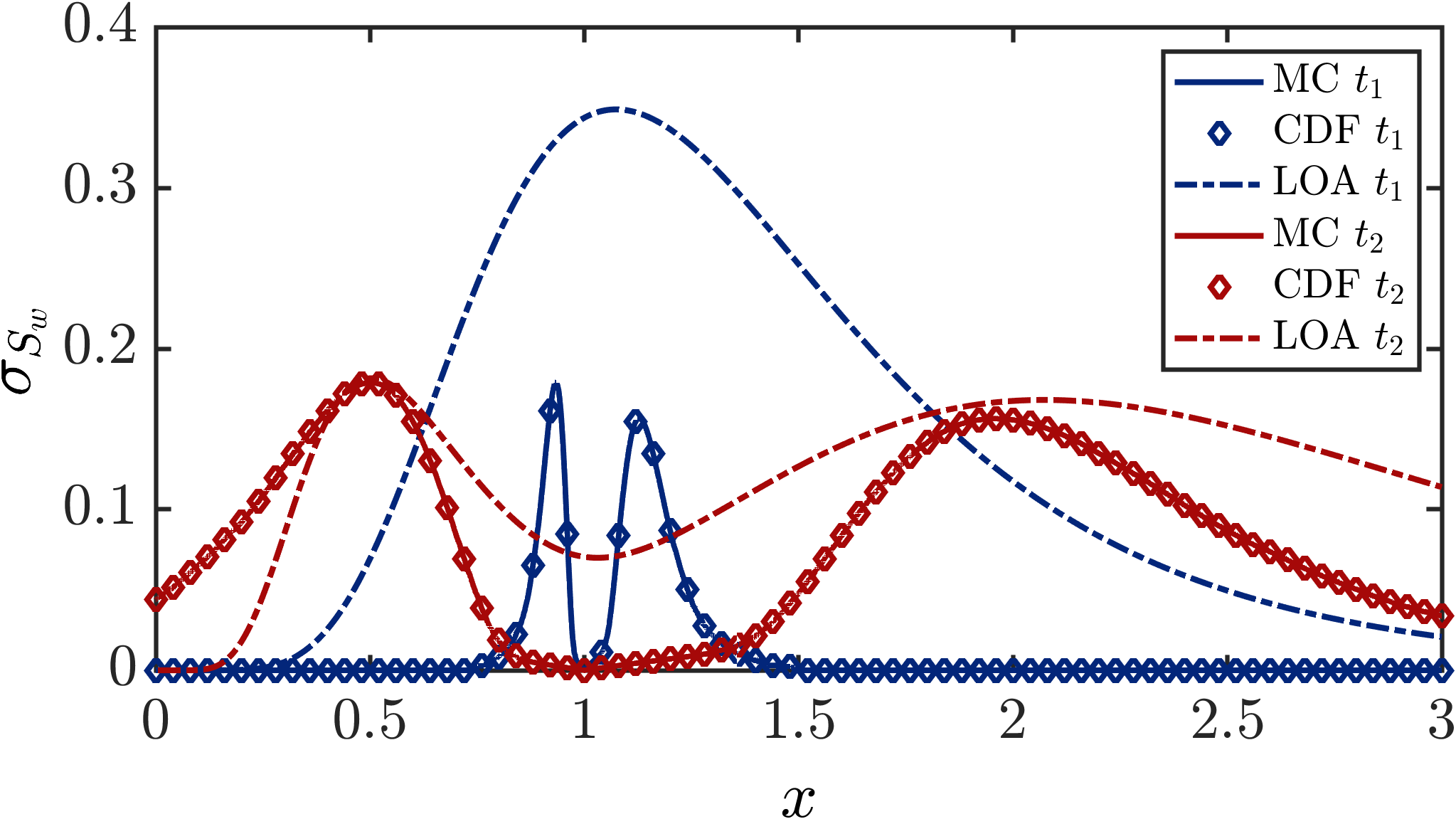}
%%%%%%%%%%%%%%%%%%%%%%%%%%%%%
	\caption{Approximations of the first two moments from LOA, MC and CDF method for the downdip flooding with $q(t)$ random, $\sigma_q = 0.005, 0.01,0.02, 0.05, 0.1$}
% 	, $\Delta x=0.01, \Delta t=0.01$
	\label{fig:LOA_downdip_Q random}
\end{figure}

\begin{figure}[htbp]
	\centering
\includegraphics[width=60mm]{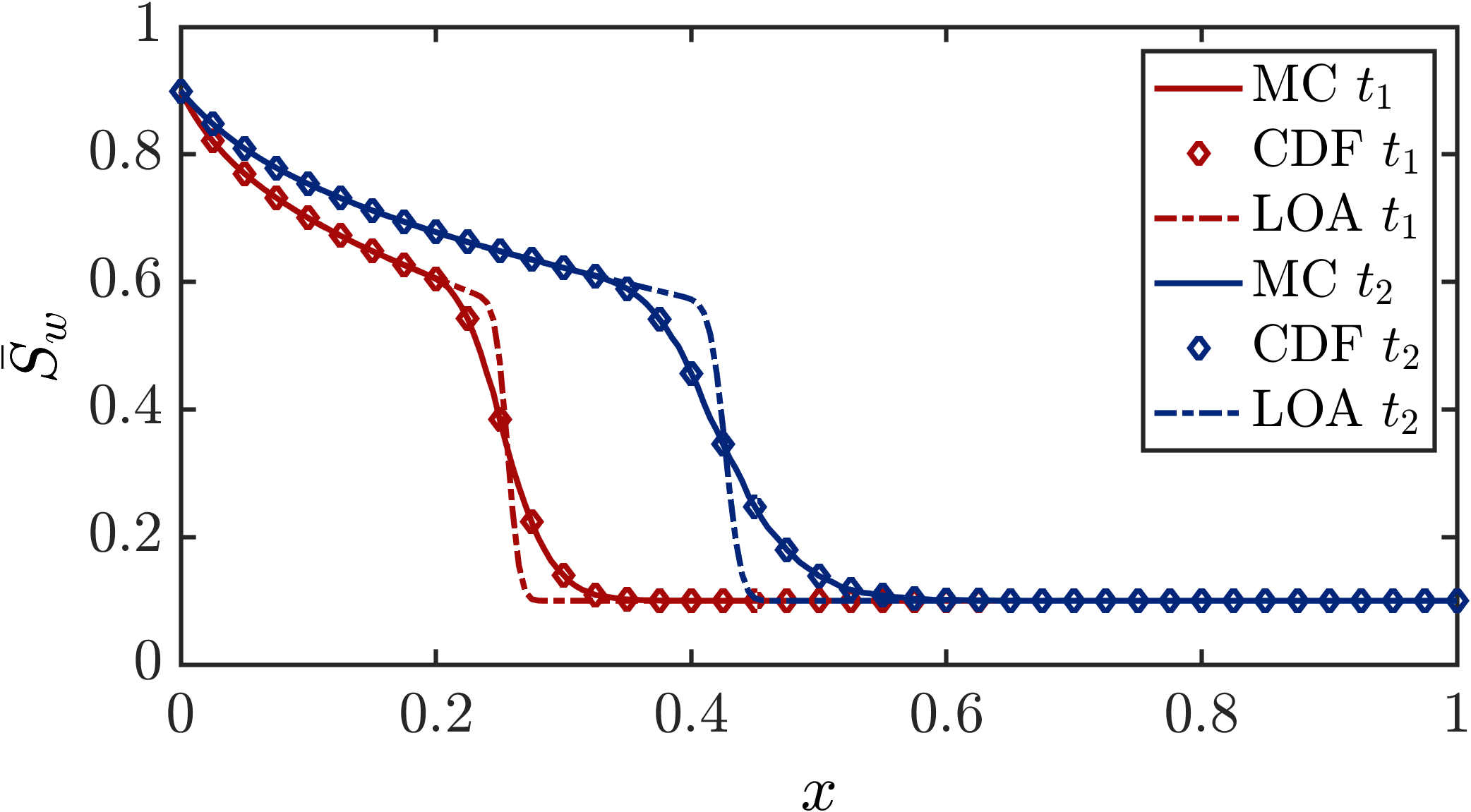}
\includegraphics[width=60mm]{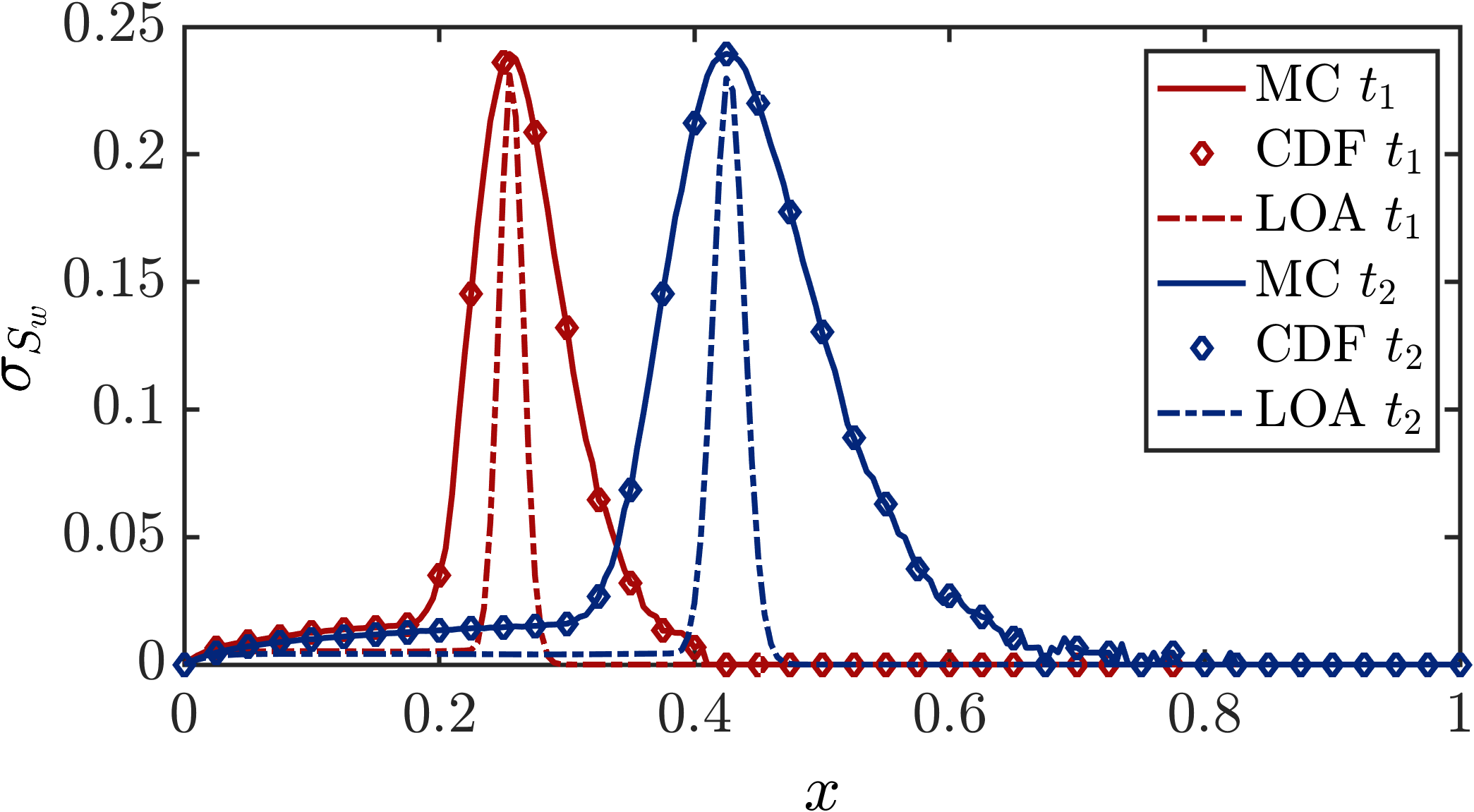}
\includegraphics[width=60mm]{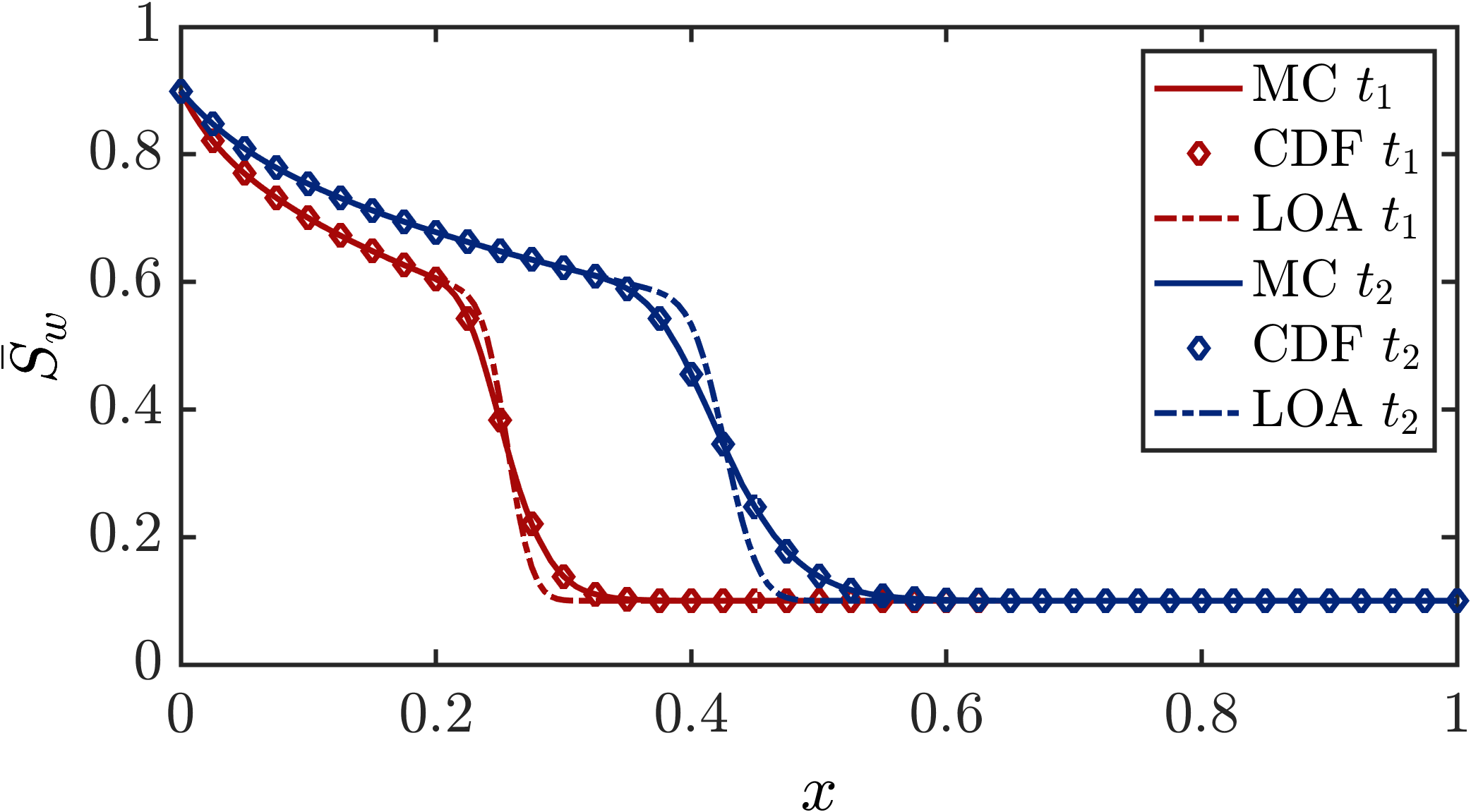}
\includegraphics[width=60mm]{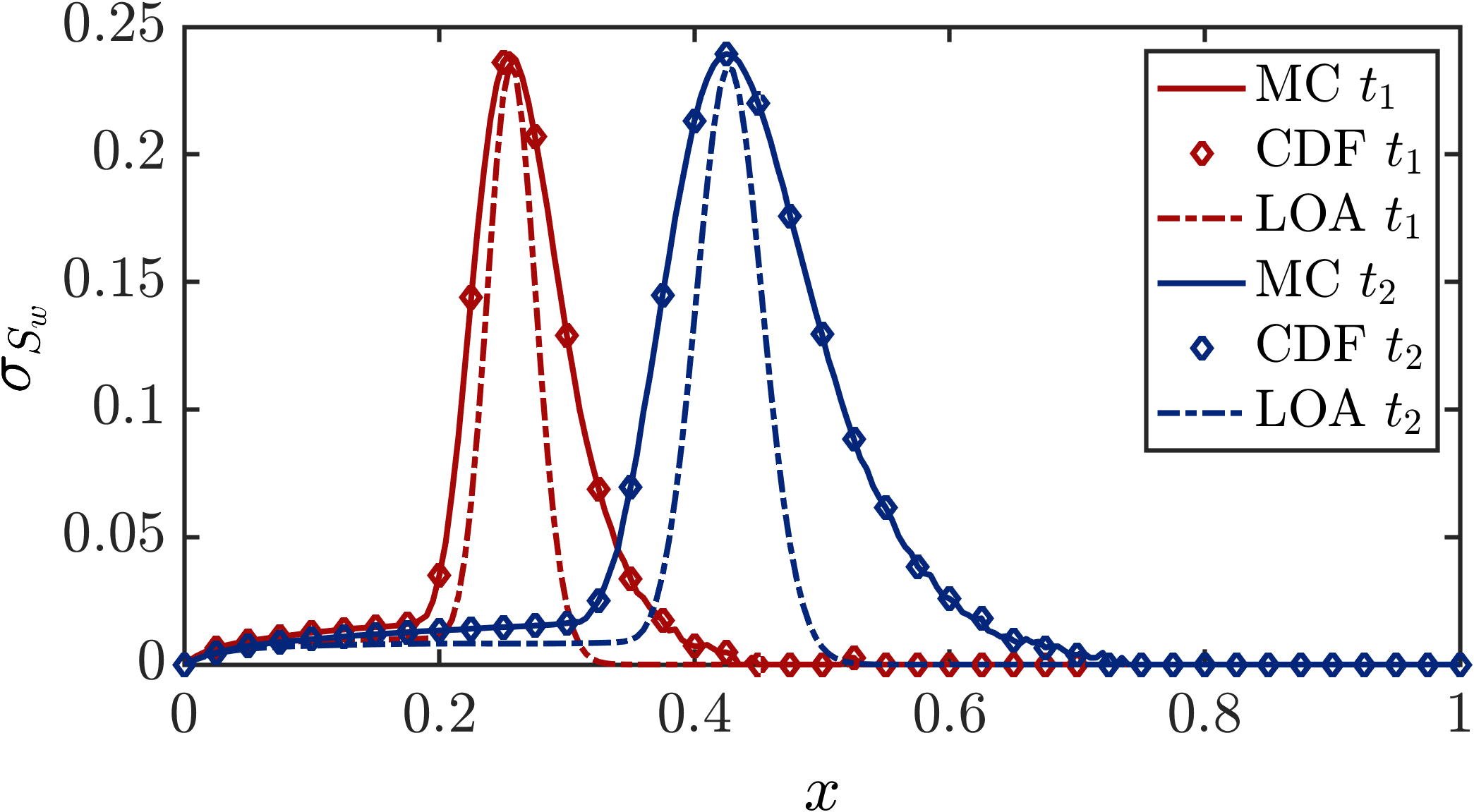}
\includegraphics[width=60mm]{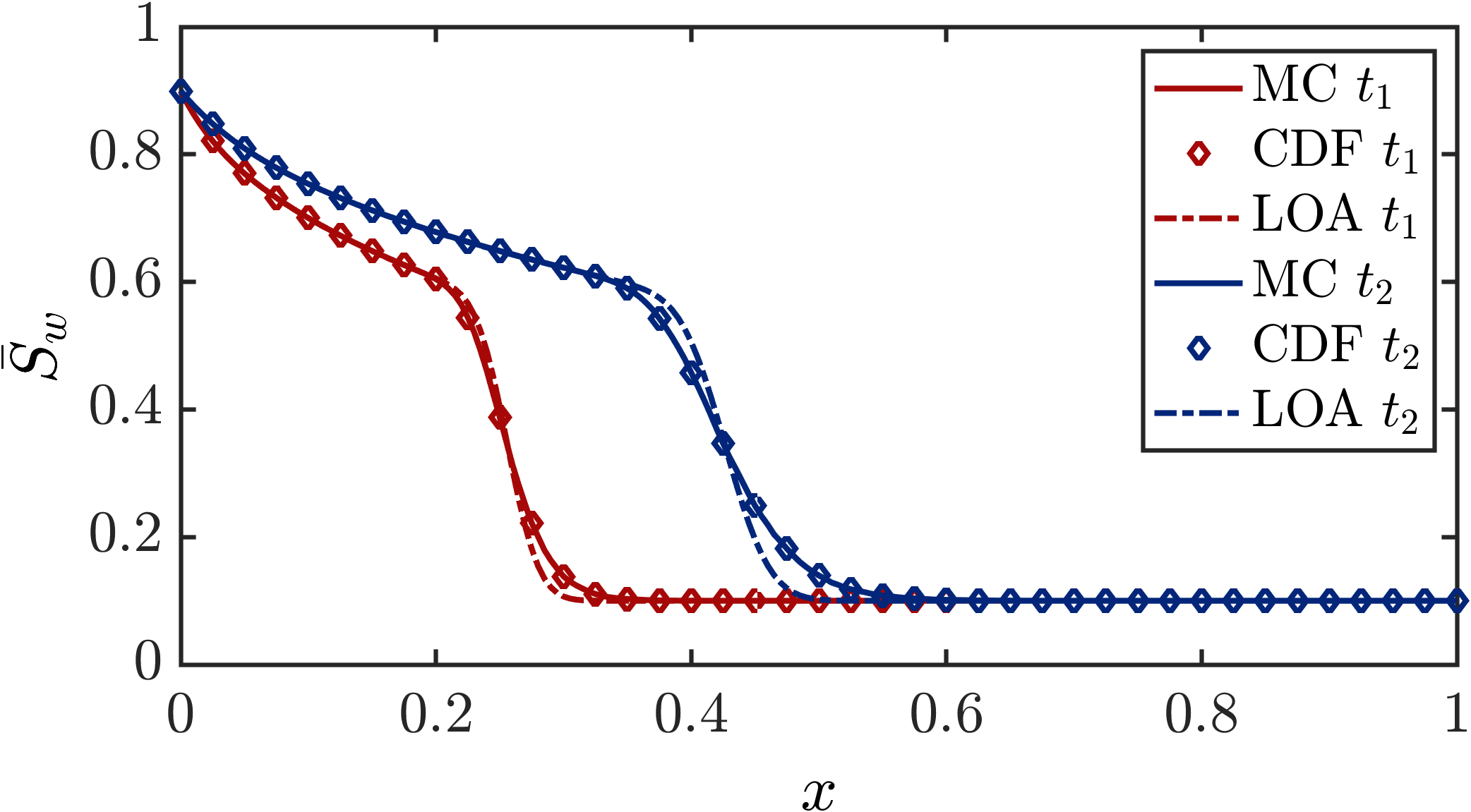}
\includegraphics[width=60mm]{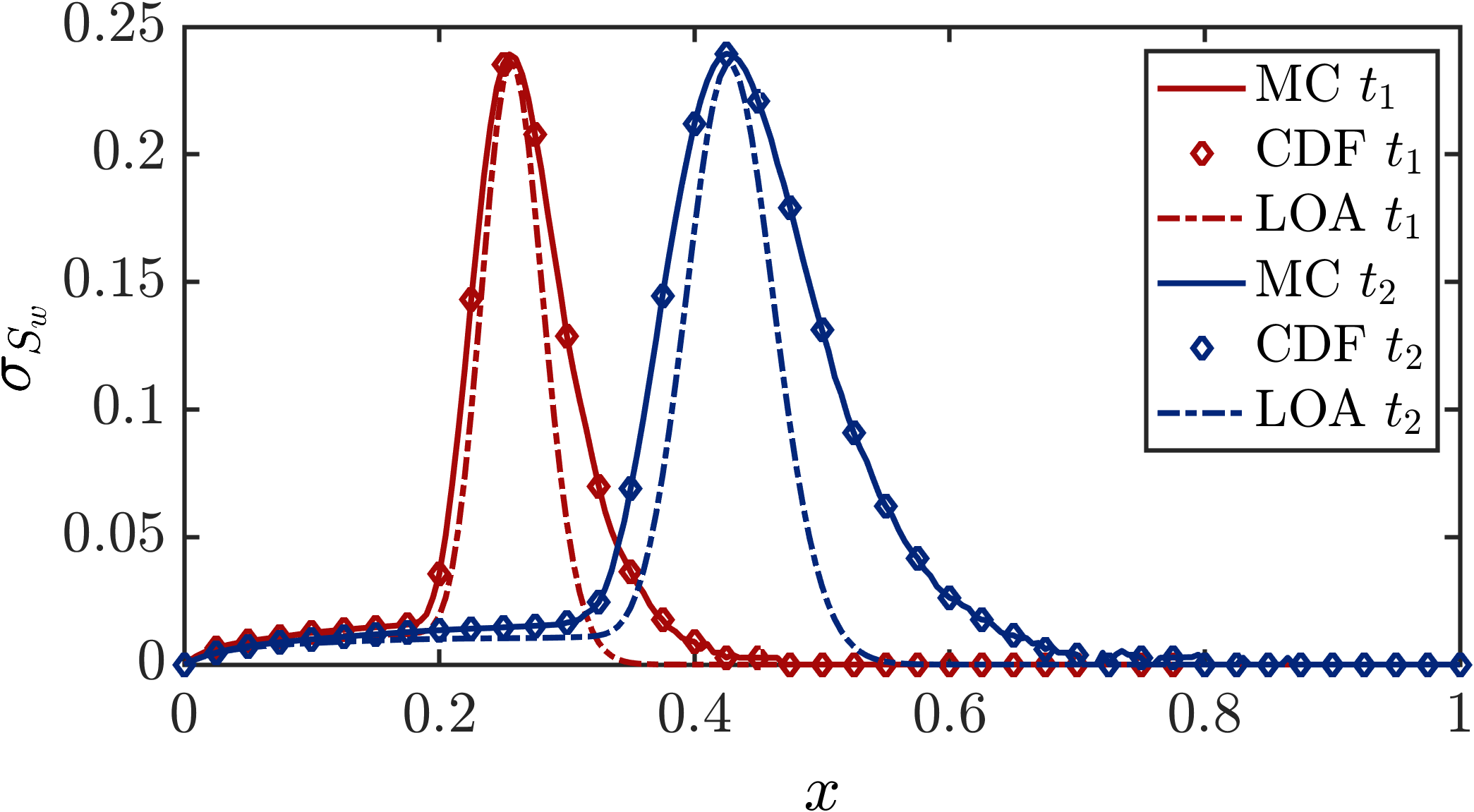}
\includegraphics[width=60mm]{figures/horizontal_LOA_phirandom_Sat_sigma2_1e-3_lambda0.5.png}
\includegraphics[width=60mm]{figures/horizontal_LOA_phirandom_STD_sigma2_1e-3_lambda0.5.png}
\includegraphics[width=60mm]{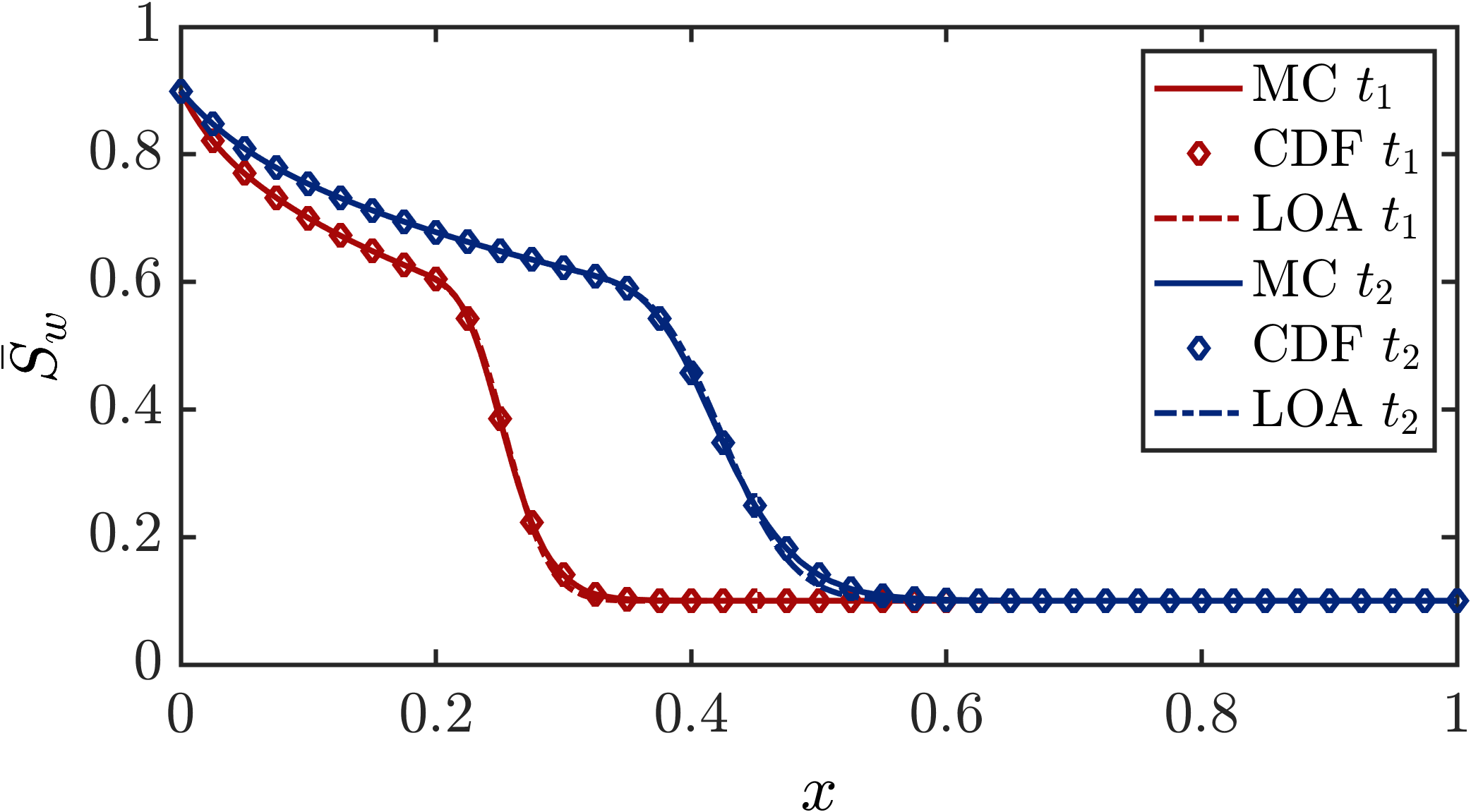}
\includegraphics[width=60mm]{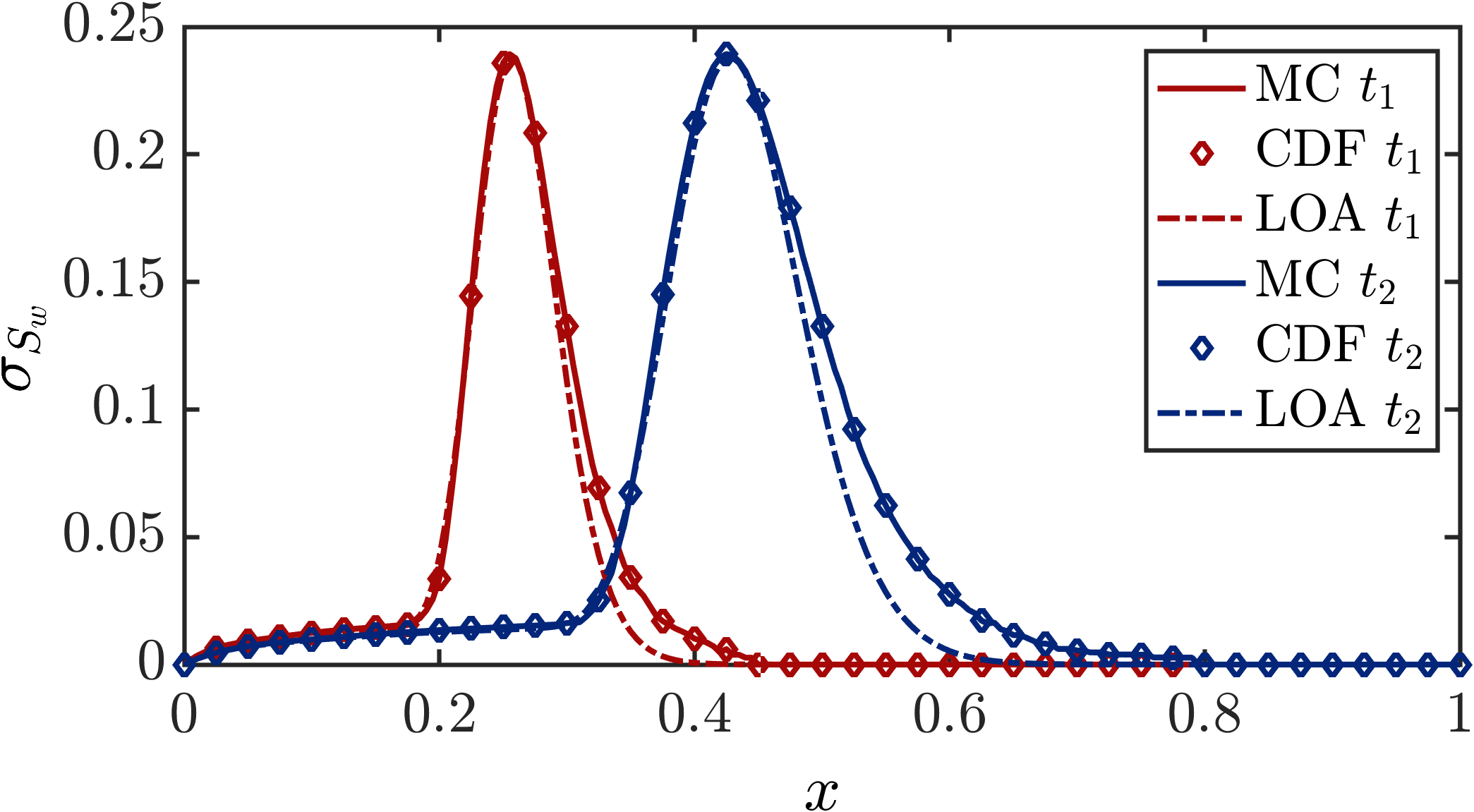}
	\caption{Approximations of the first two moments from LOA, MC and CDF methods for the horizontal flooding with $\phi(x)$ random at dimensionless times $t_1=0.15, t_2=0.25$. For all cases, $\sigma_{\phi}^2=10^{-3}$, $\Delta x=0.005, \Delta t=0.01$, while the correlation length is increasing as $\lambda_{\phi} = 0.01, 0.05, 0.5, 0.7$, respectively.}
	\label{fig:LOA horizontal_phi random___correlation length}
\end{figure}

While the the results from CDF method stay in agreement with those of MC up to large variance and correlation lengths, the accuracy of LOA immediately starts deteriorating even for small variances.  That is because the statistical moments of travel time and velocity are derived as first-order approximations. Therefore, the first-order approximations require the variance of log permeability to be (much) smaller than 1. As depicted in the plots showing sensitivity to the increasing variance, by increasing the variance of input parameters, the saturation variance is increasing, too. Consequently, the saturation profile becomes smoother and deviates from a sharp deterministic profile in the shock region. As the results of sensitivity to the correlation length represents in Fig. \ref{fig:LOA horizontal_phi random___correlation length}, initially the LOA results are far from MC and CDF results. As we increase the correlation length while keeping $\Delta x$ constant, the approximations for both  mean and standard deviation become closer to those of MC. This is because more grid blocks fall within one correlation length, and therefore the underlying domain becomes more homogeneous to resolve. Also, as Fig. \ref{fig:LOA_downdip_Q random} represents, for the cases where there are two shocks, at early times LOA can not discern the two shocks and naively predicts one combined shock, except for very small variances. Whereas, the CDF method accurately captures the solution of MC at both early and later times.  
% \subsubsection{Uncertainty in injection flux $q_T(t)$}
%%%%%%%%%%%%%%%%%%%%%%%%%%%%%%%%%%%%%%%%%%%%%%%%%%%%%%%%%%%%%%%%%%%%%%%%5
%%%%%.                              LOA Q random
%%%%%%%%%%%%%%%%%%%%%%%%%%%%%%%%%%%%%%%%%%%%%%%%%%%%%%%%%%%%%%%%%%%%%%%%%%
%
%%%%%%%%%-------------------------------  LOA horizontal------- Q random
\begin{figure}[htbp]
	\centering
	\includegraphics[width=60mm]{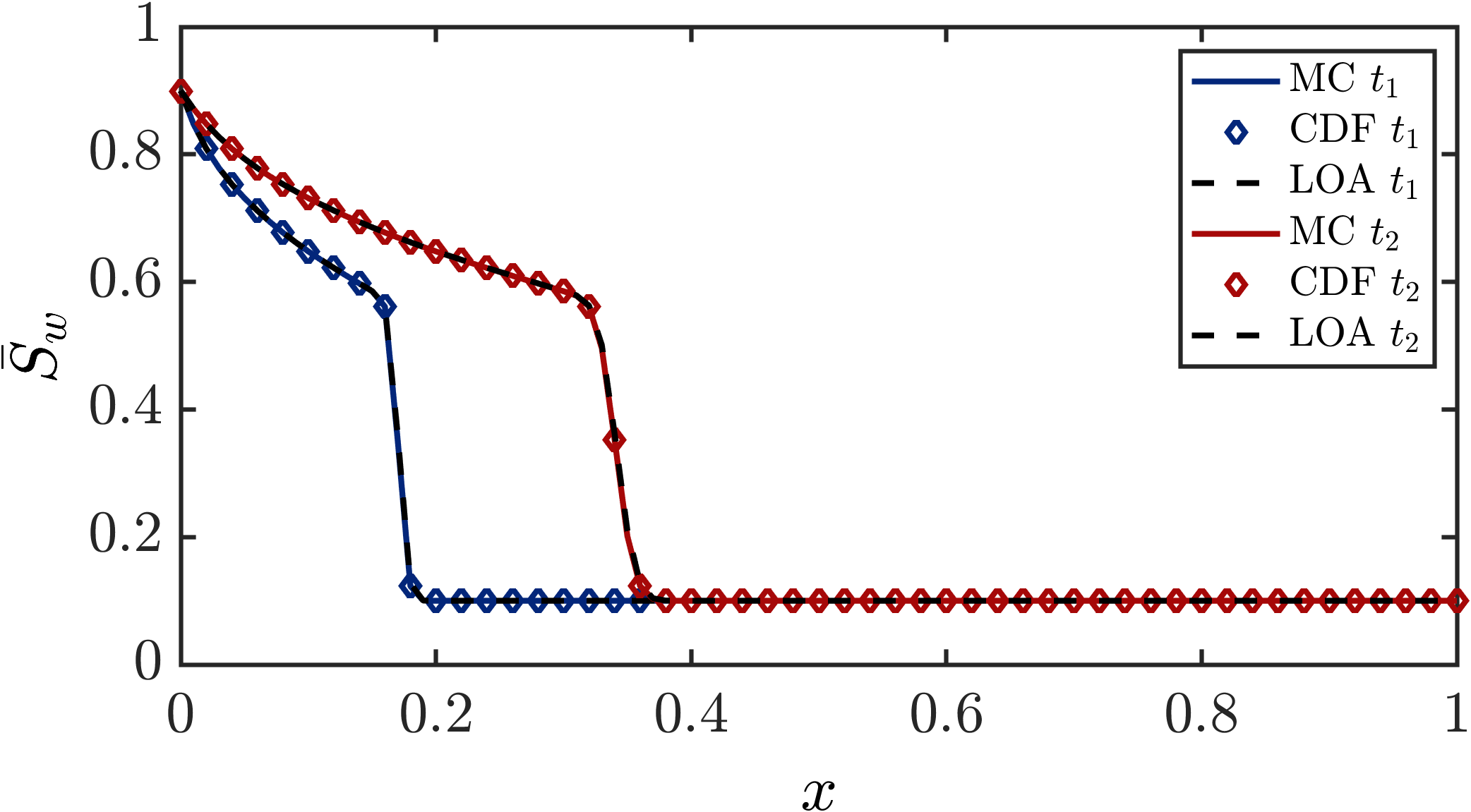}
	\includegraphics[width=60mm]{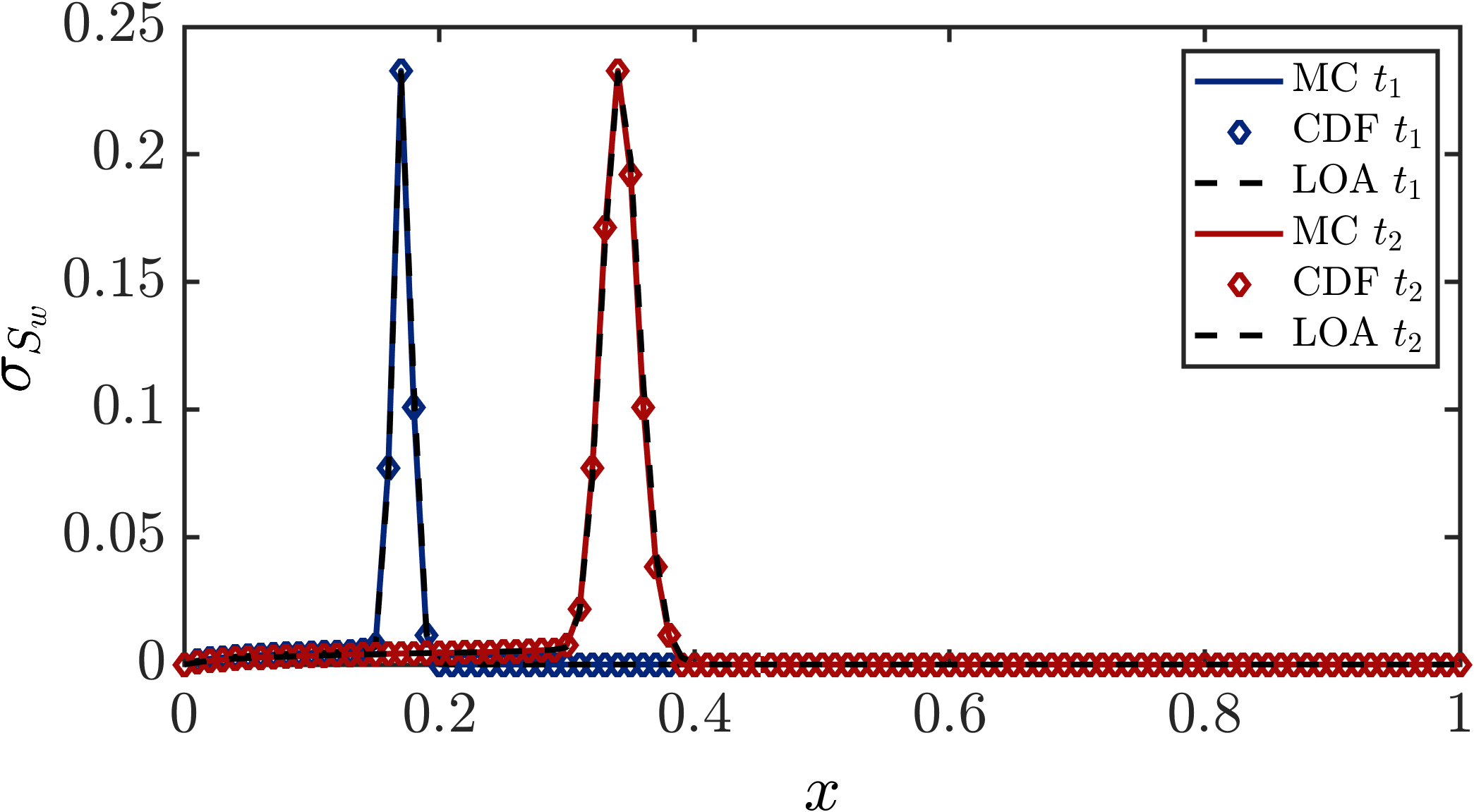}\\
	\includegraphics[width=60mm]{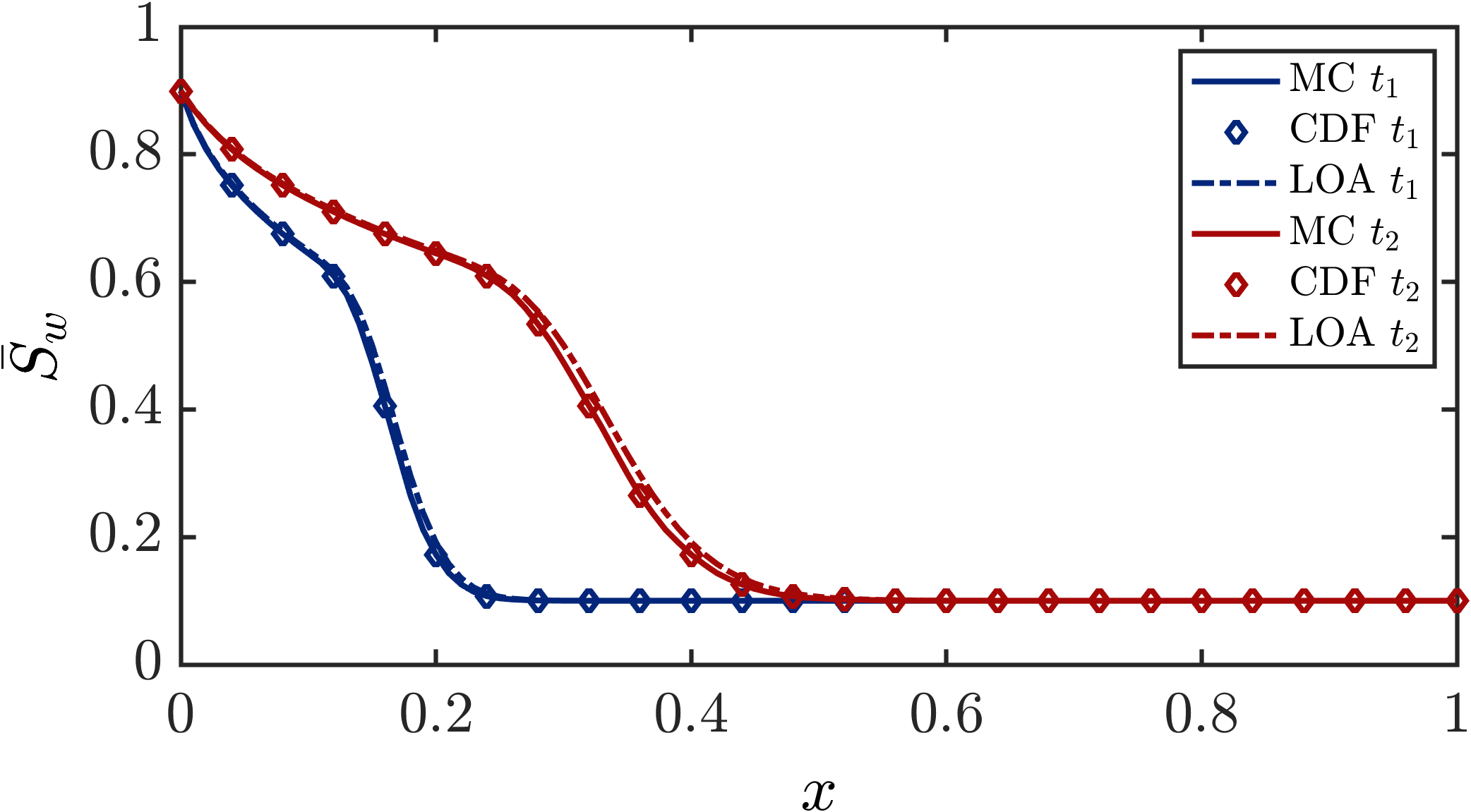}
	\includegraphics[width=60mm]{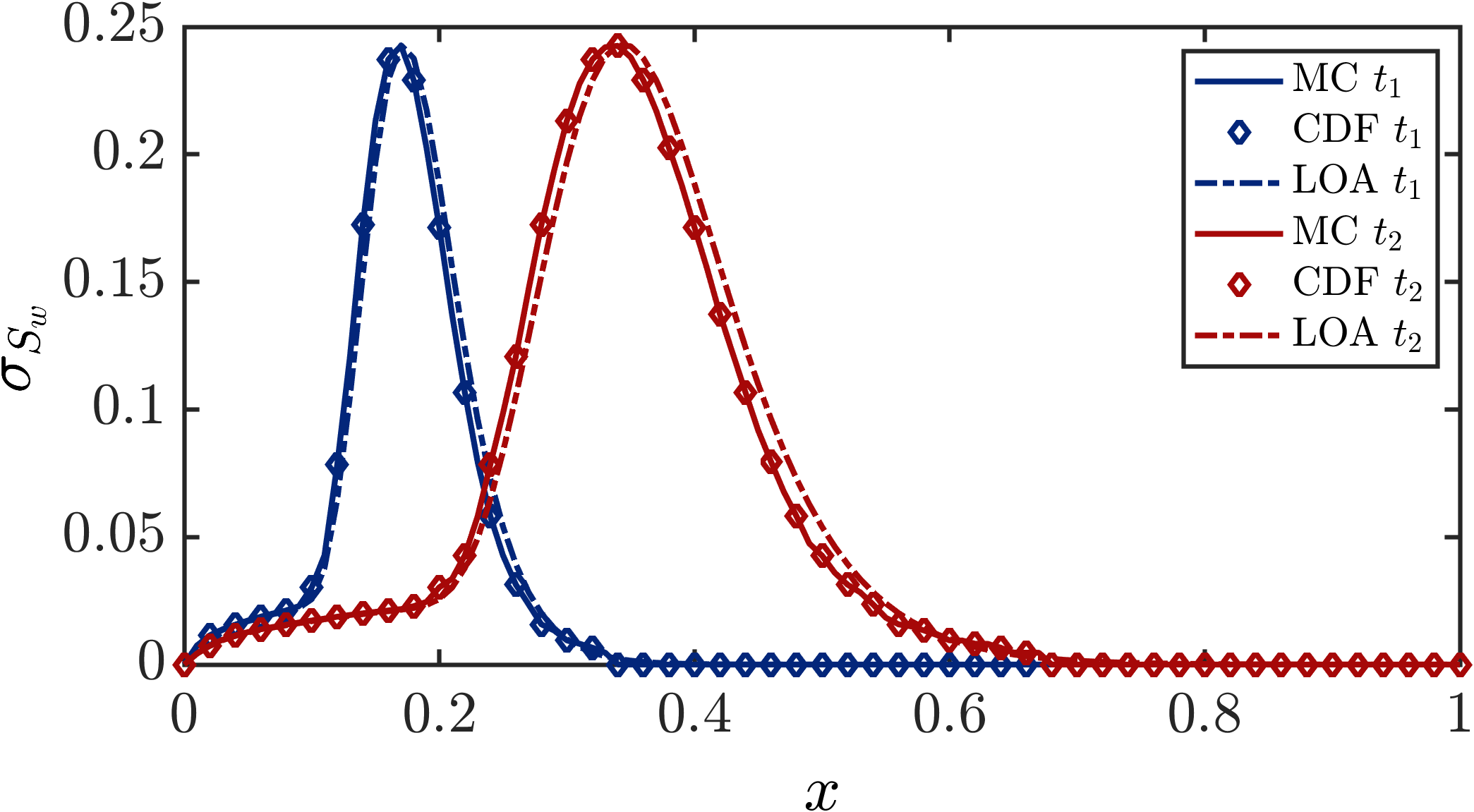}\\
	\includegraphics[width=60mm]{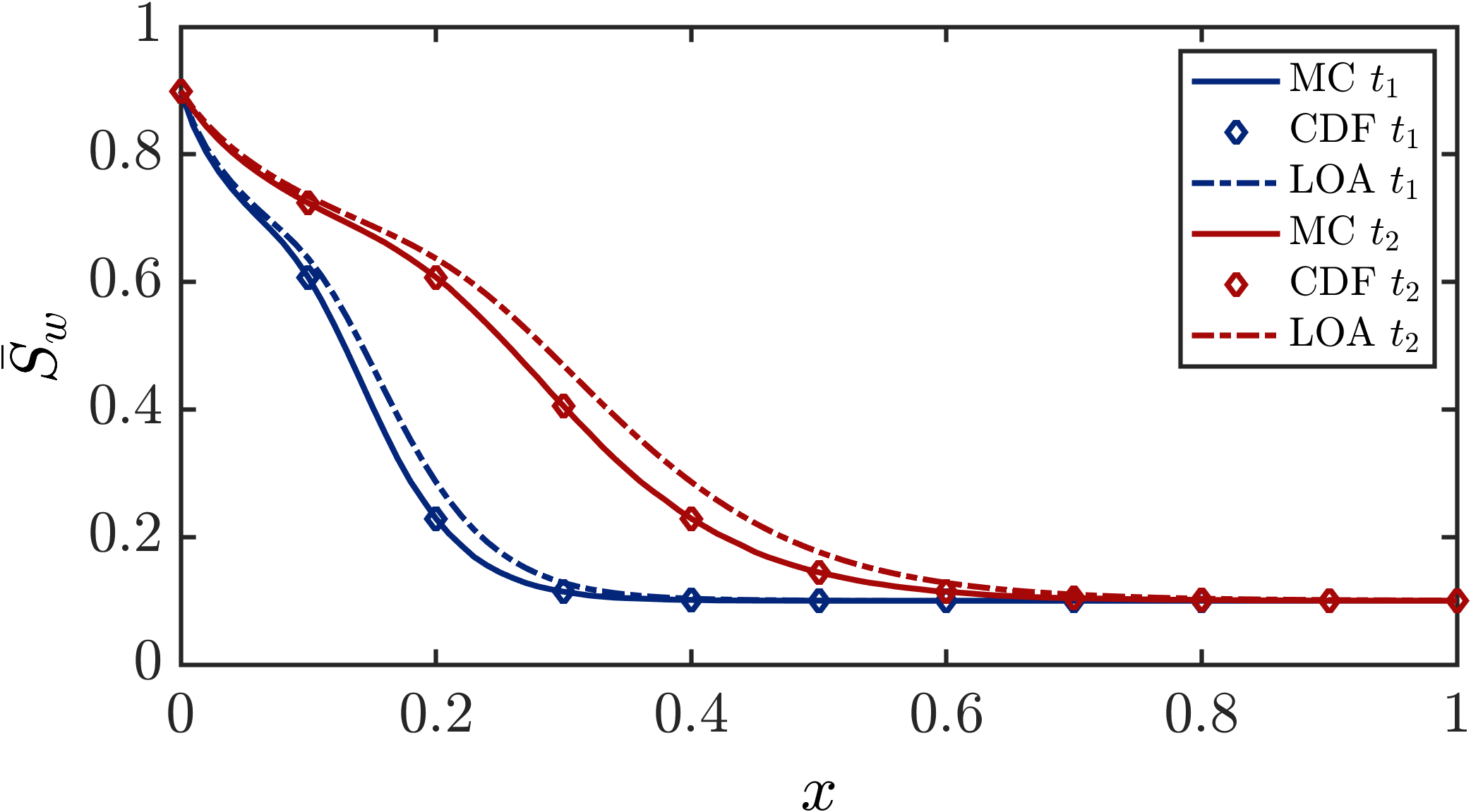}
	\includegraphics[width=60mm]{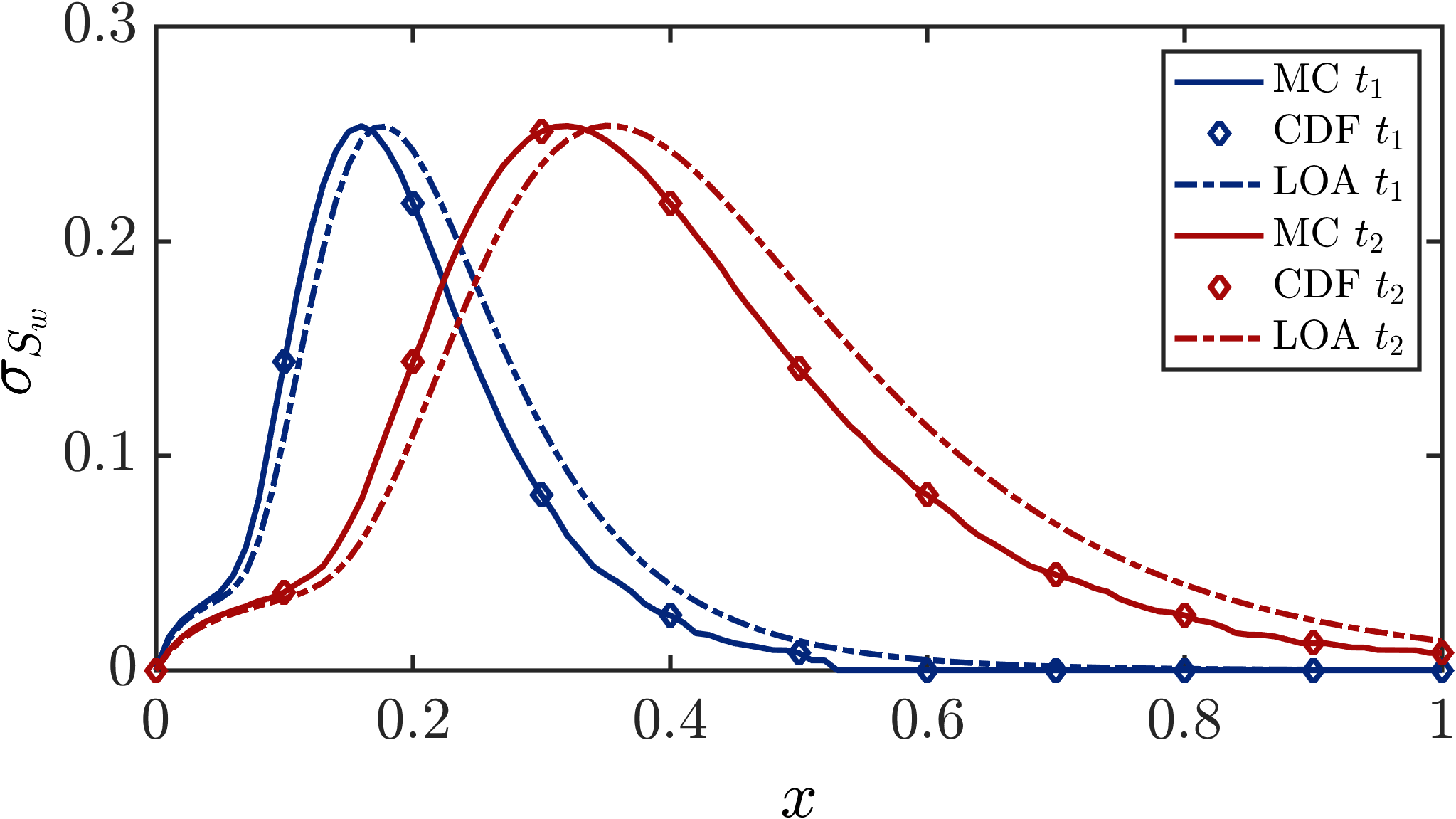}\\
	\includegraphics[width=60mm]{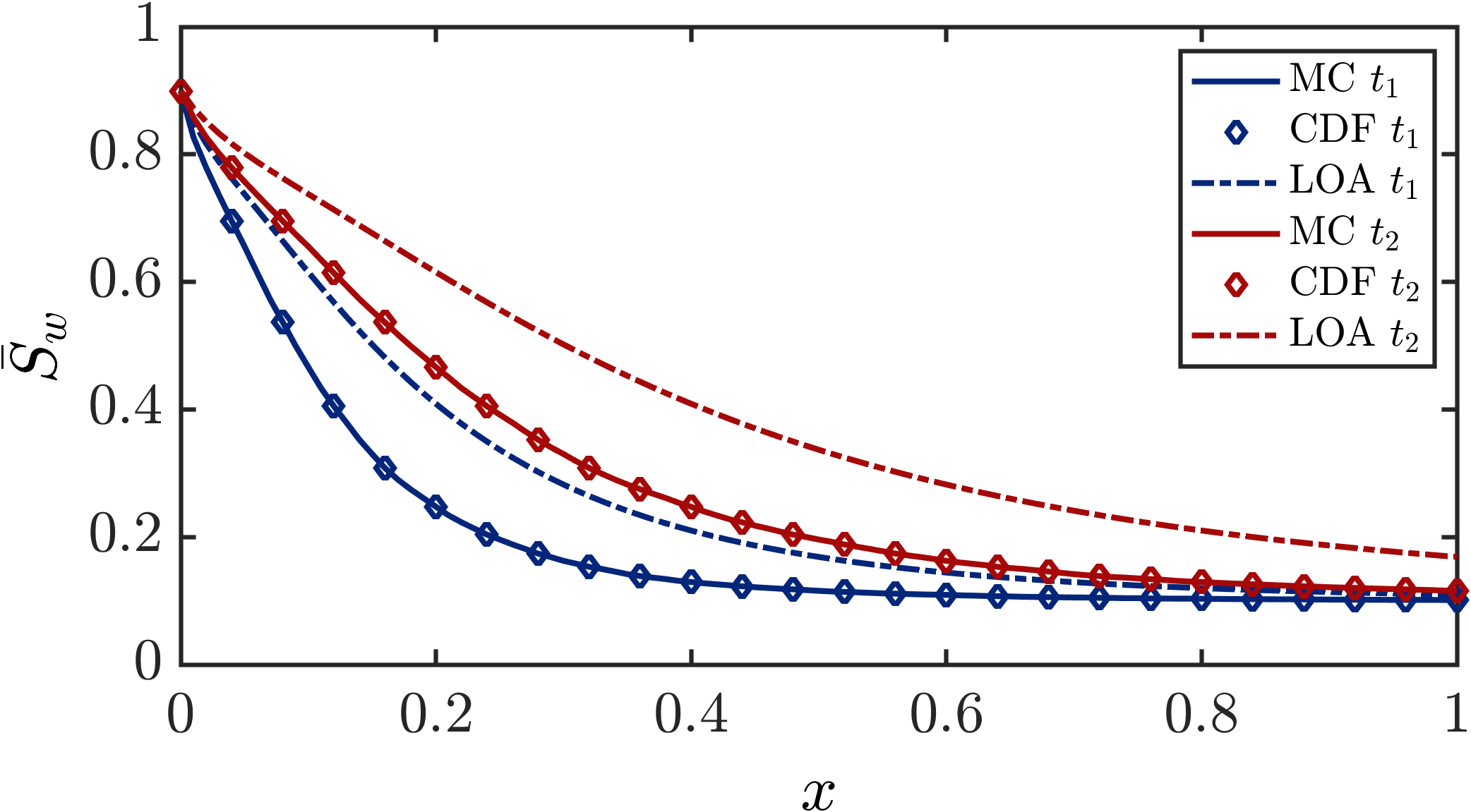}
	\includegraphics[width=60mm]{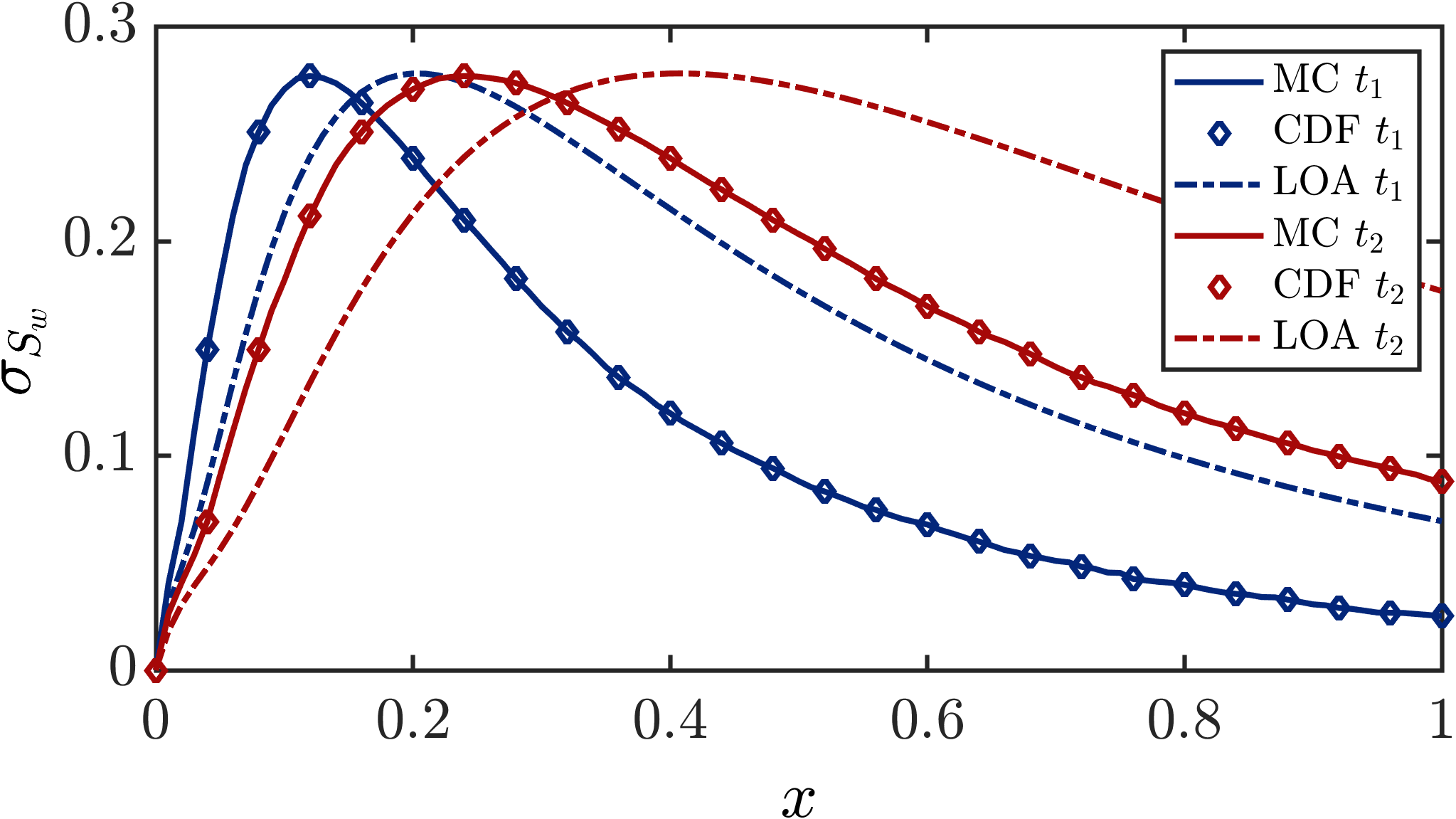}
	\caption{Approximations of the first two moments from LOA, MC and CDF method for the horizontal flooding with $q(t)$ random, $\sigma_q = 0.01, 0.05, 0.1, 0.25$, for all cases $\mu_q = 0.3$ and $\tau_q=0.5T$, $T=1$. As variance increases, LOA approximations of the first two moments deteriorate, whereas CDF results stay in agreement with the corresponding Monte Carlo solutions.}
	\label{fig:LOA_horizontal_Q random}
\end{figure}
%
%%%%%%%%%-------------------------------  LOA downdip------- Q random
%
%%%%%%%%%%%%%%%%%%%%%%%%%%%%%%%%%%%%%%%%%%%%%%%%%%%%%%%%%%%%%%%%%%%%%%%%5
%%%%%.                    LOA phi random
%%%%%%%%%%%%%%%%%%%%%%%%%%%%%%%%%%%%%%%%%%%%%%%%%%%%%%%%%%%%%%%%%%%%%%%%%%
% \subsubsection{Uncertainty in porosity field $\phi(x)$}
%%%%%%%%%-------------------------------  LOA horizontal------- PHI random
\begin{figure}[htbp]
	\centering
\includegraphics[width=60mm]{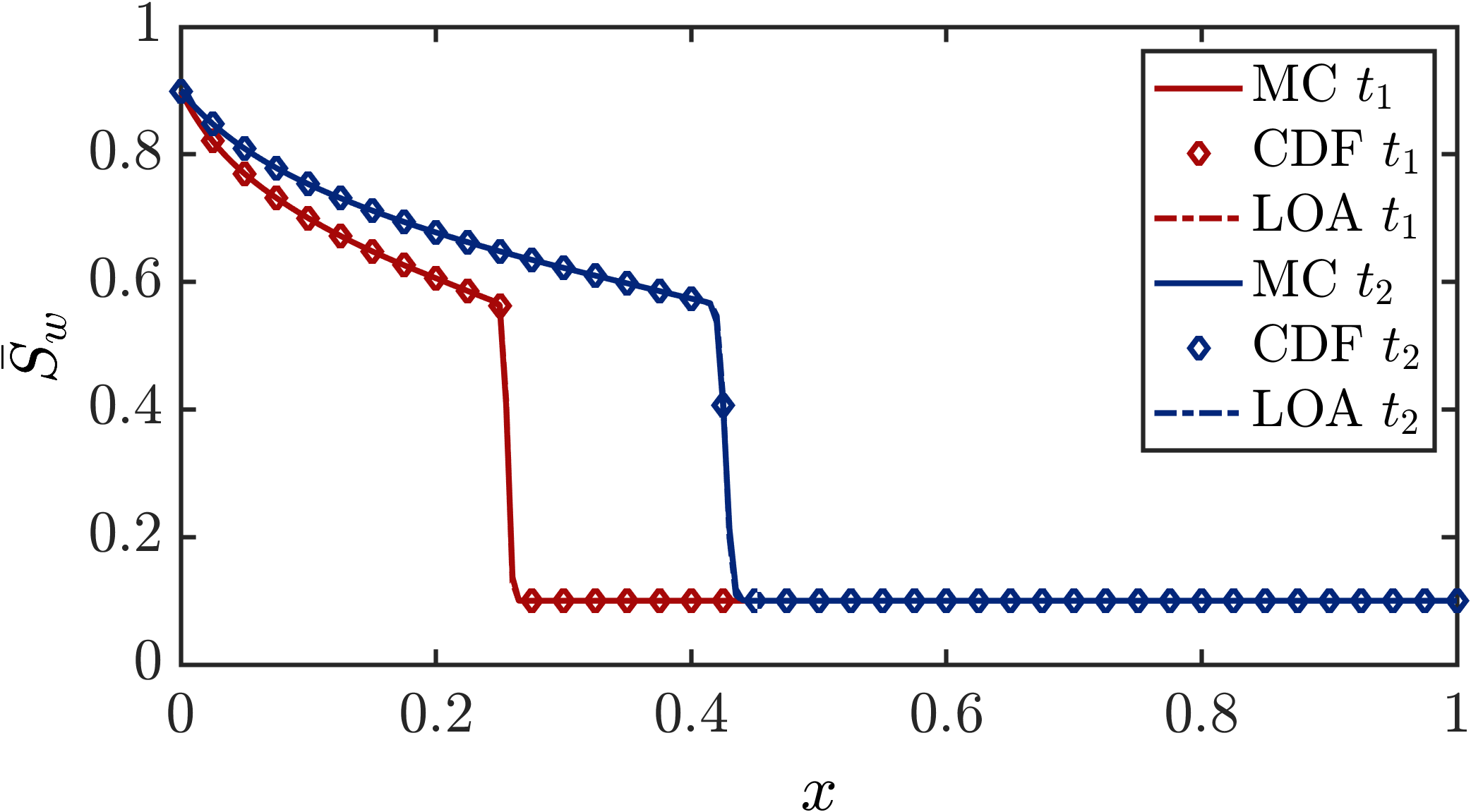}
\includegraphics[width=60mm]{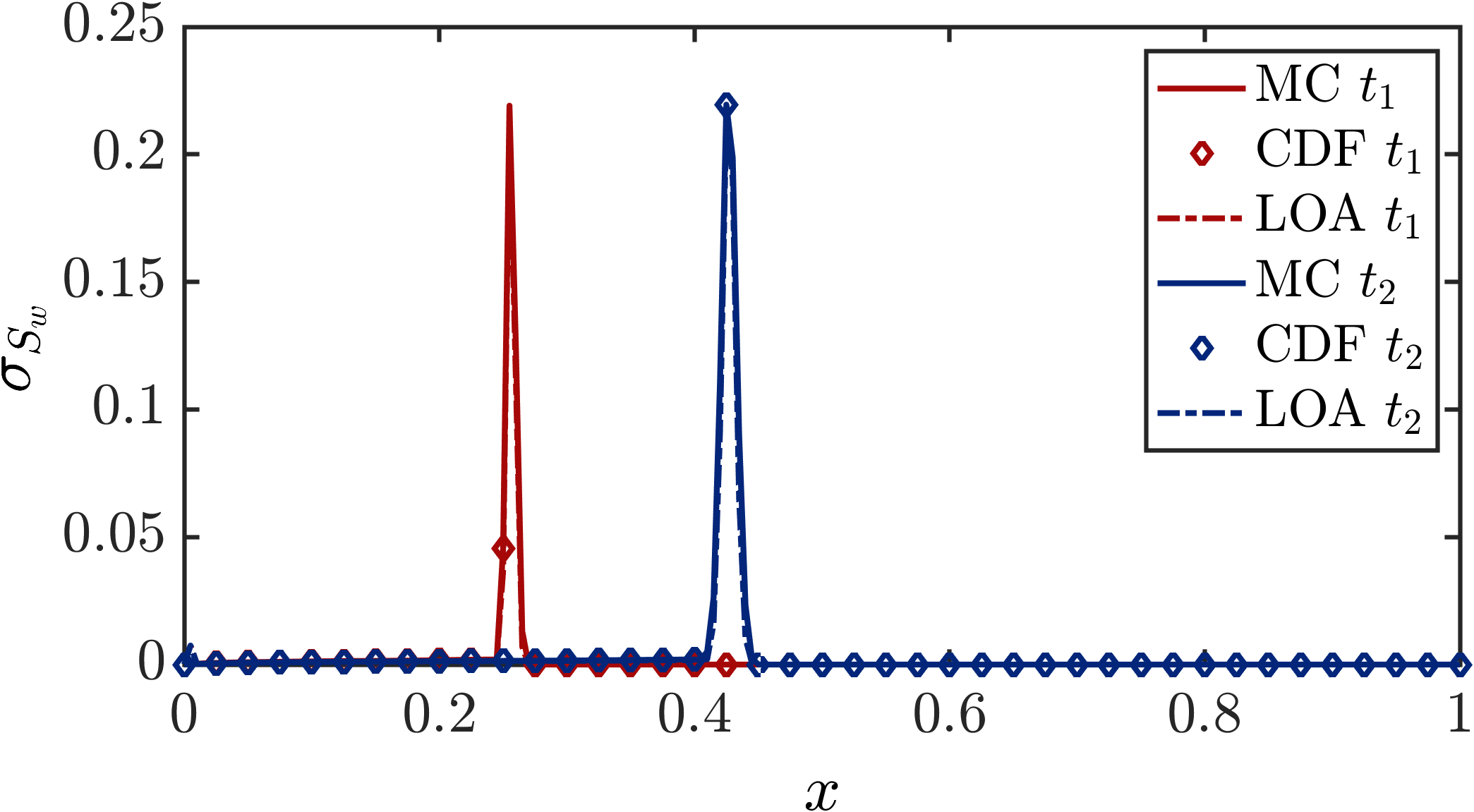}
\includegraphics[width=60mm]{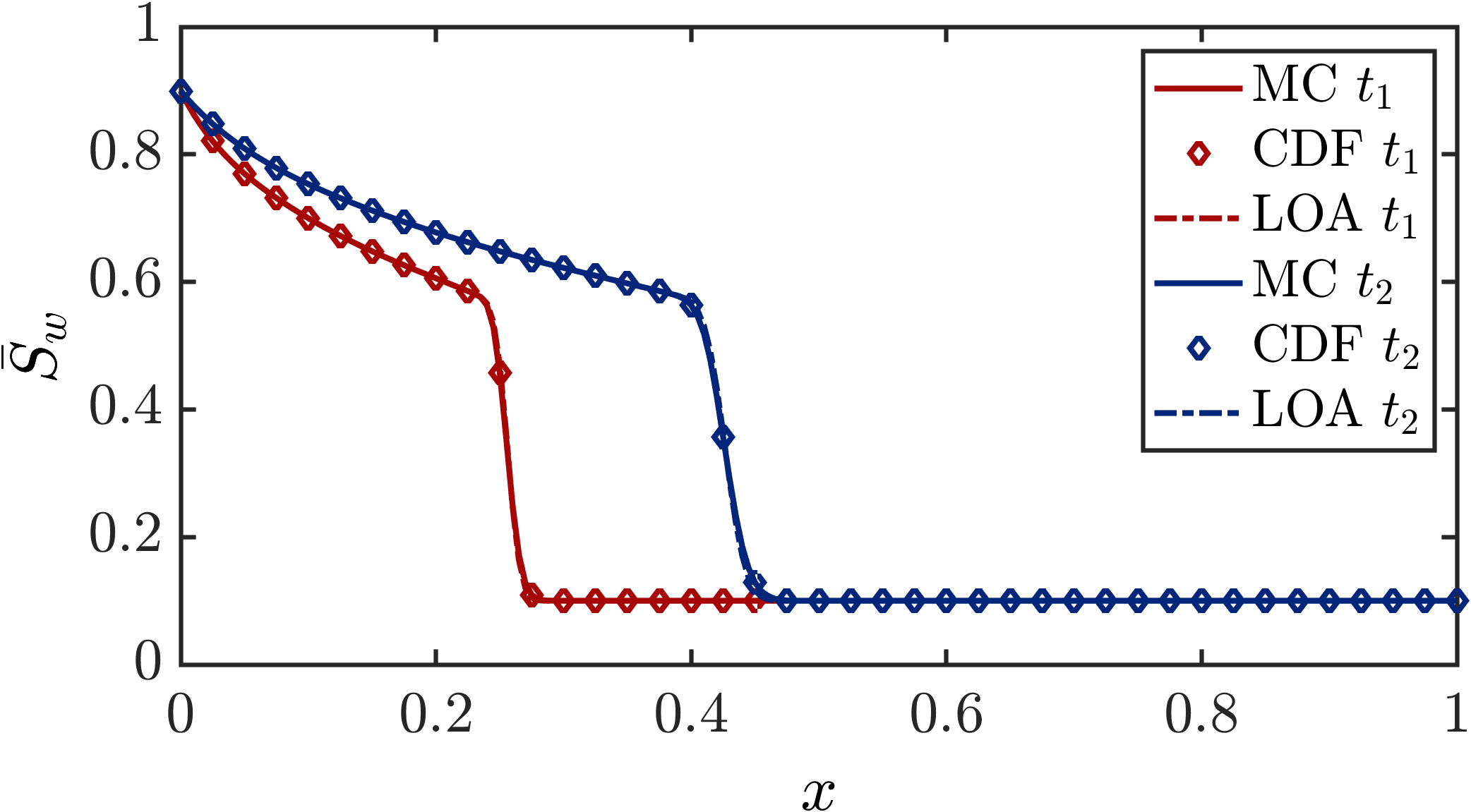}
\includegraphics[width=60mm]{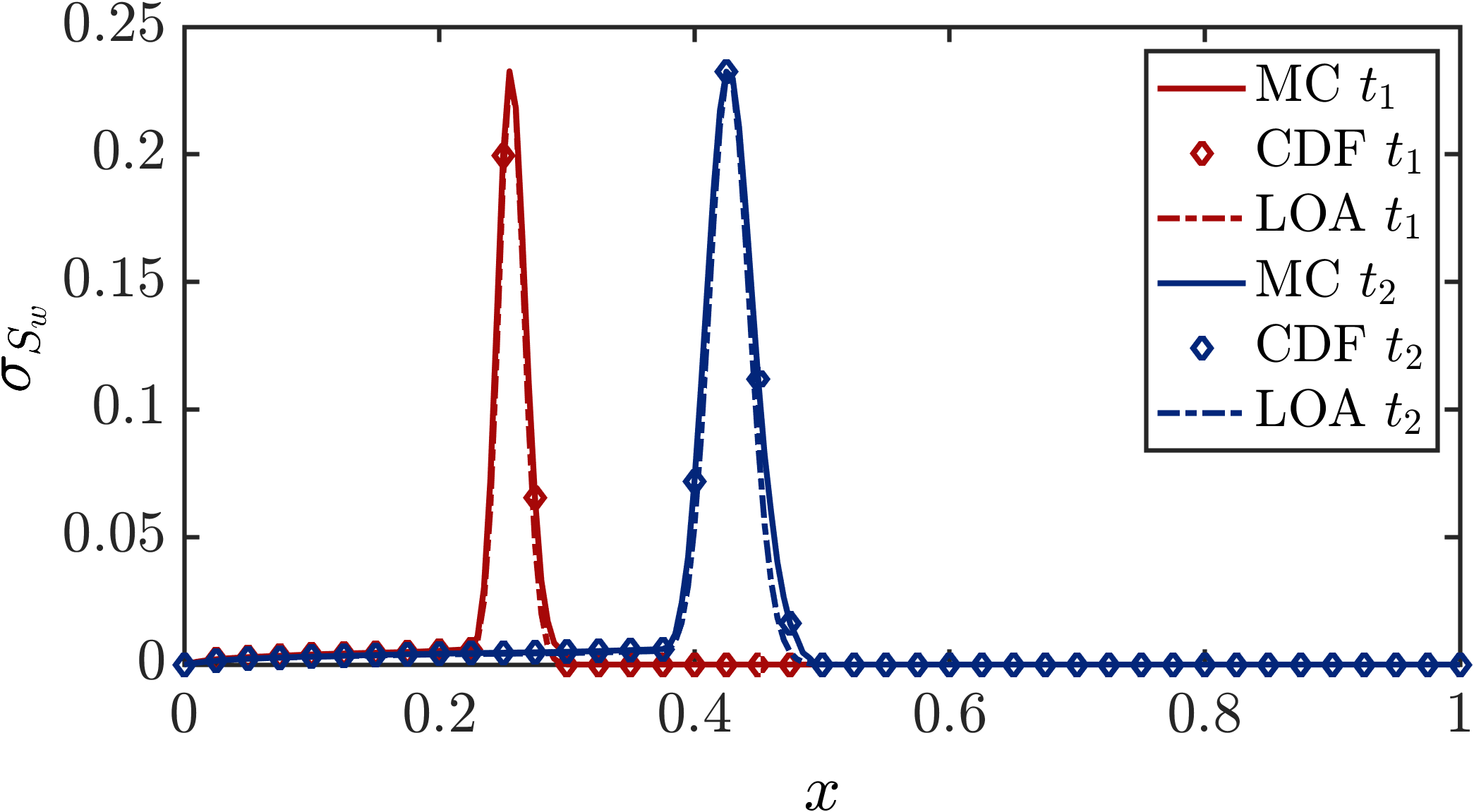}
\includegraphics[width=60mm]{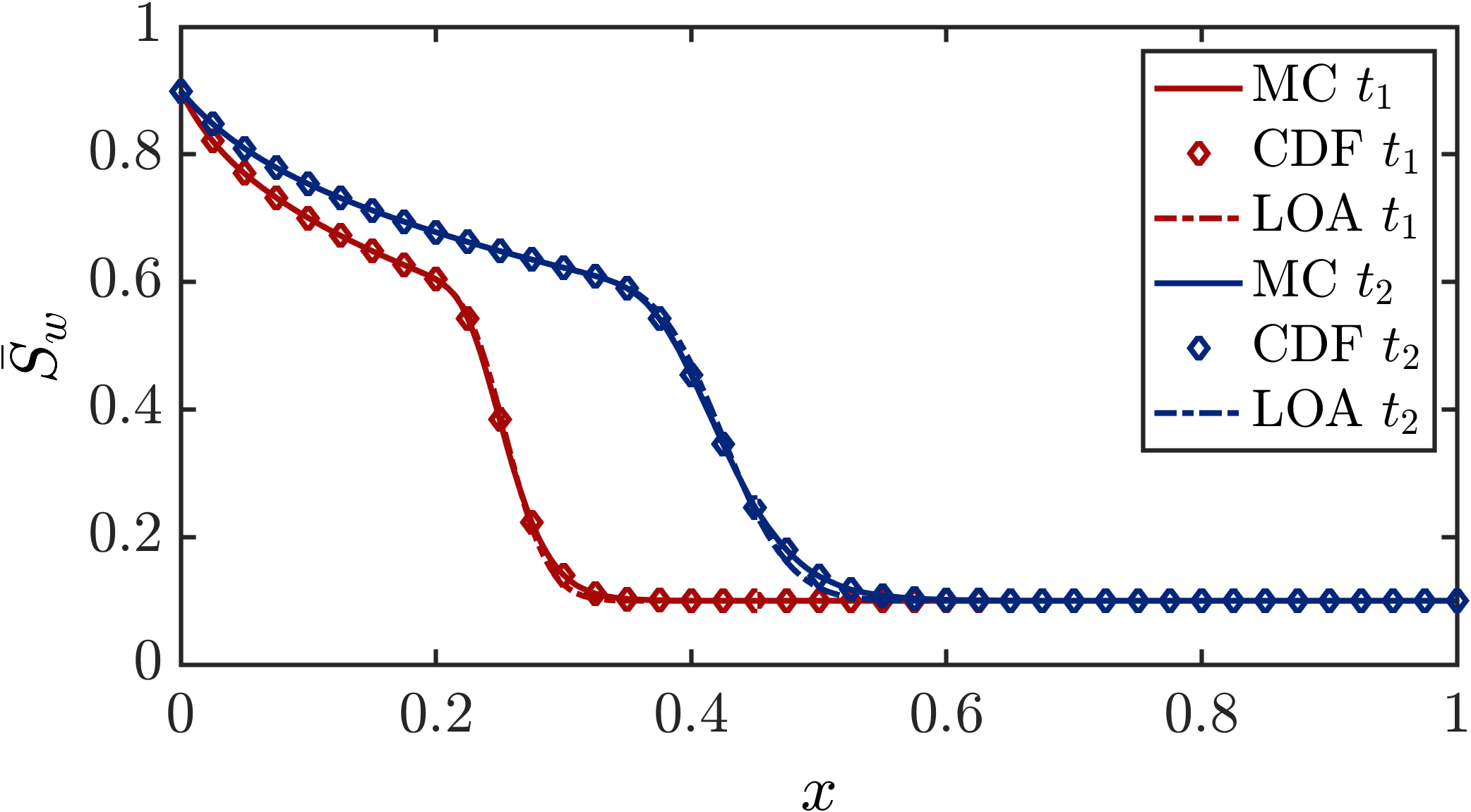}
\includegraphics[width=60mm]{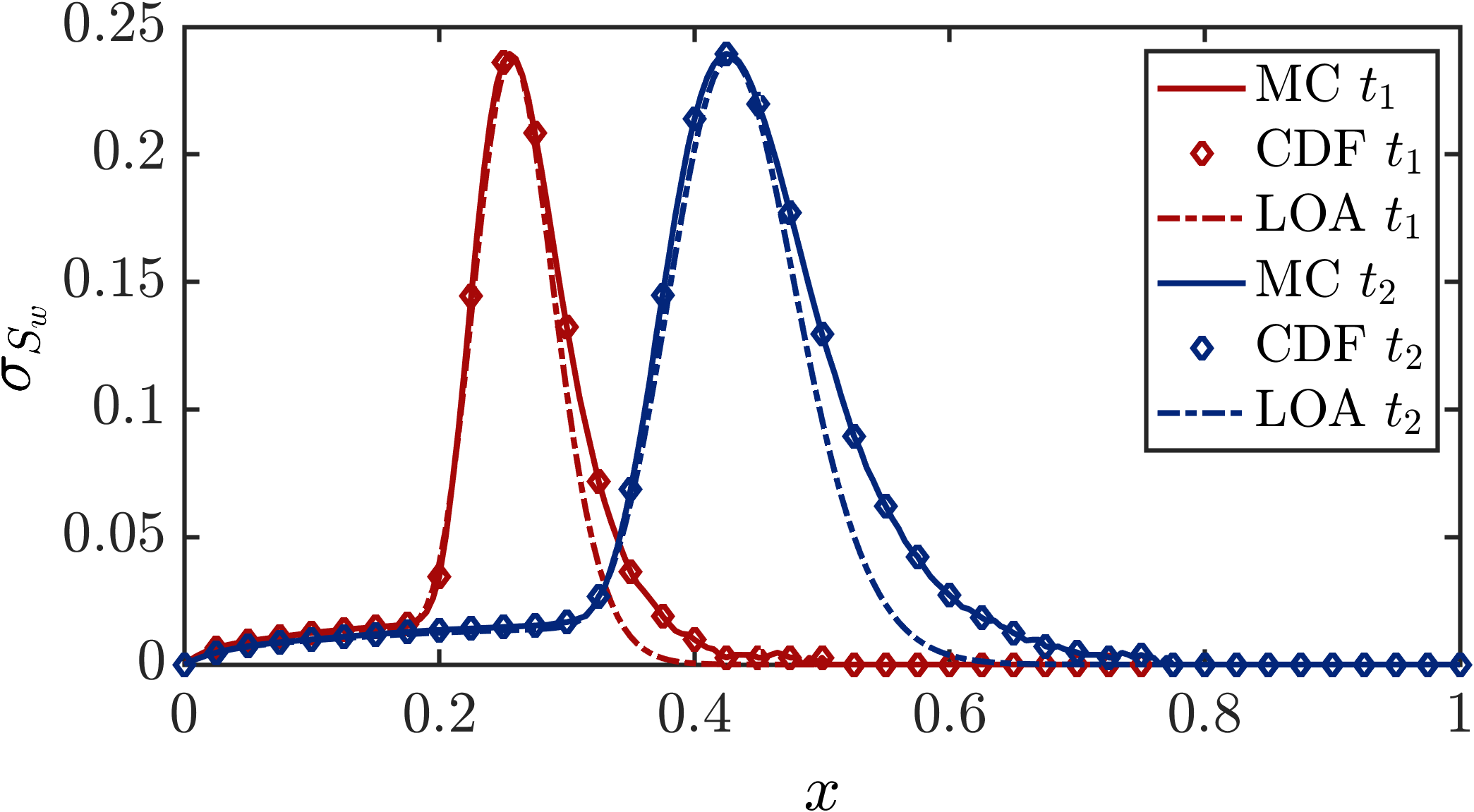}
\includegraphics[width=60mm]{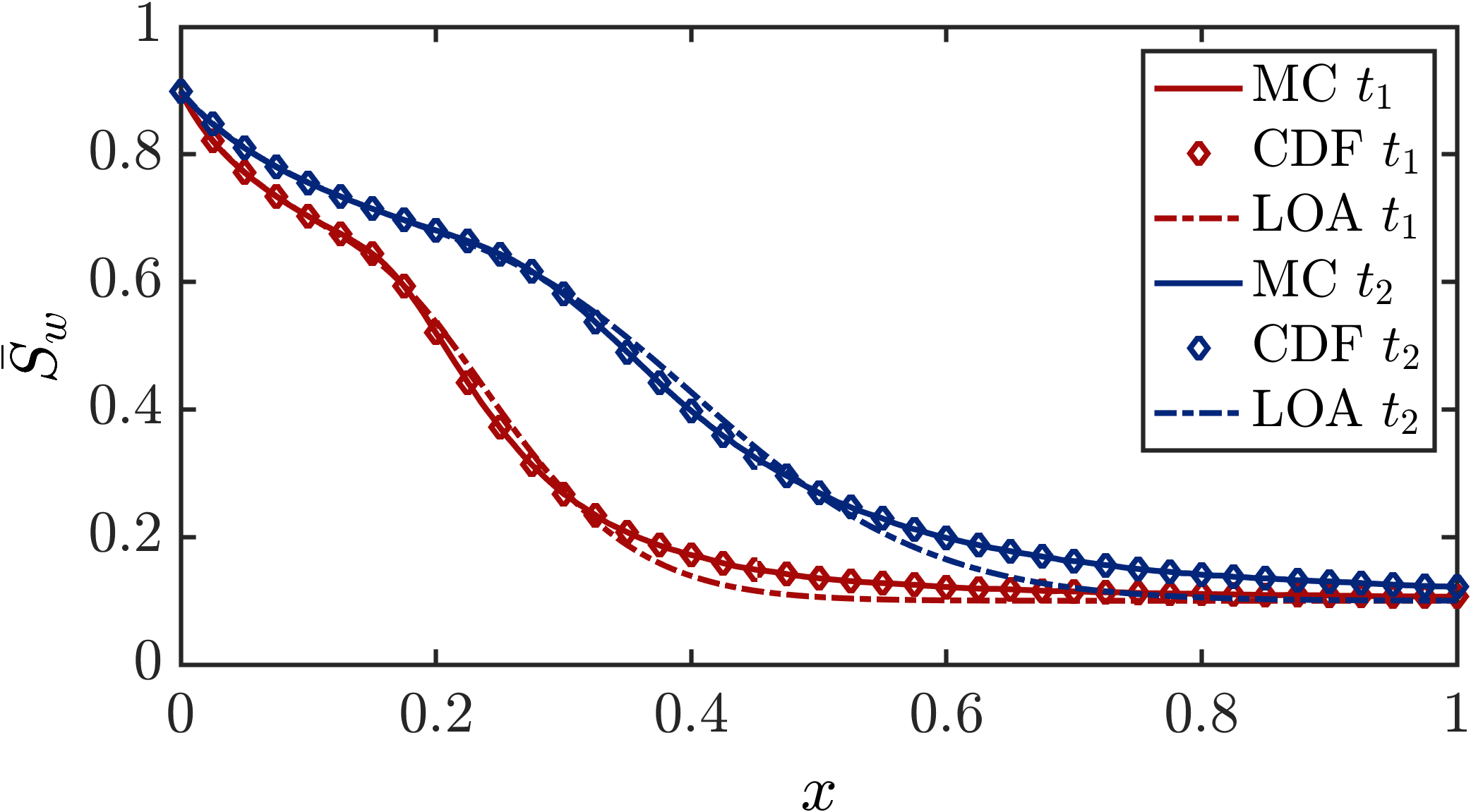}
\includegraphics[width=60mm]{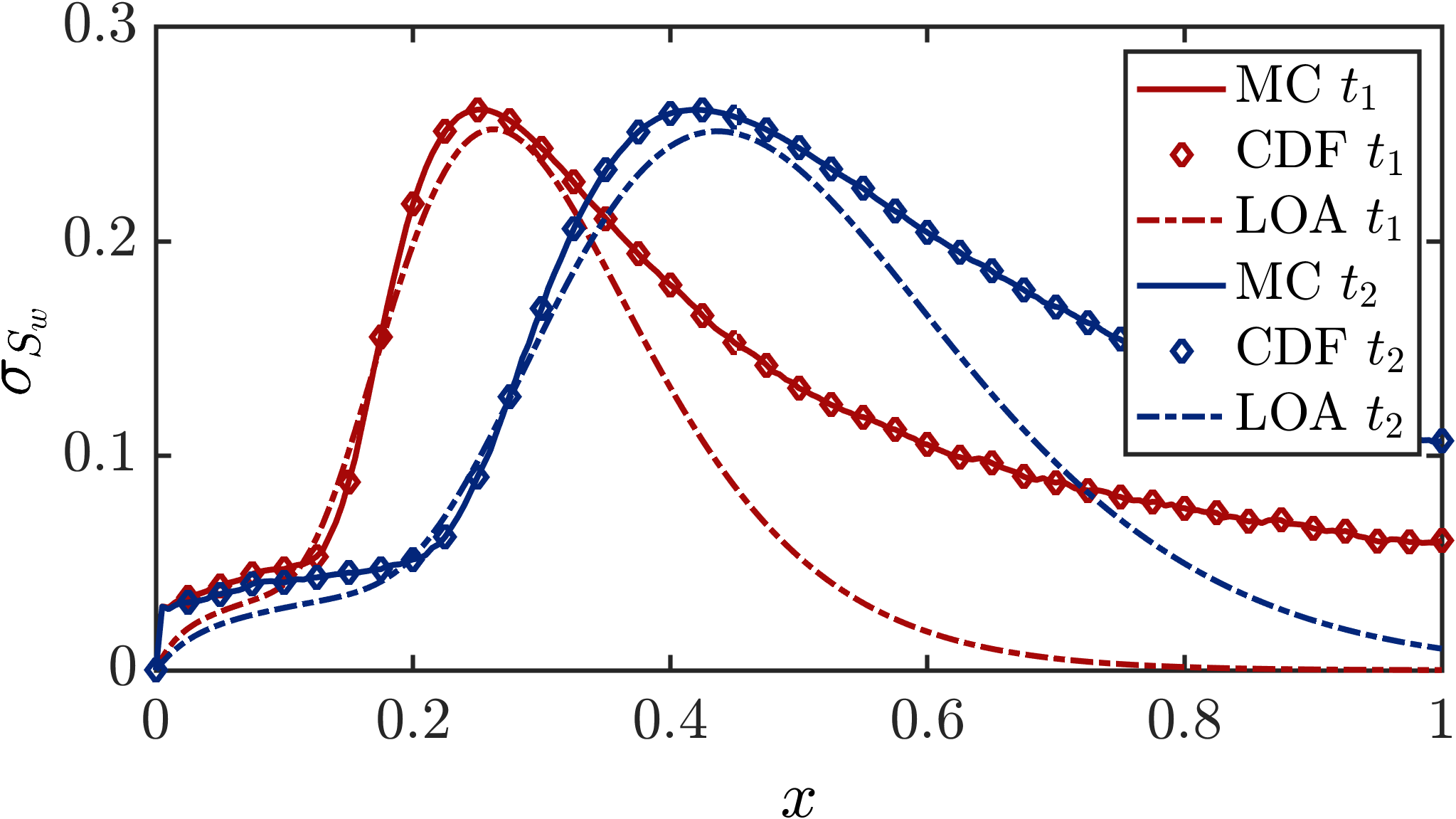}
	\caption{Approximations of the first two moments from LOA, MC and CDF method for the horizontal flooding with $\phi(x)$ random at dimensionless times $t_1=0.15, t_2=0.25$. For all cases, correlation length is $\lambda_{\phi} = 0.5L, L=1$ and variance is increasing as $\sigma_{\phi}^2=10^{-5}, 10^{-4}, 10^{-3}, 10^{-2}$, respectively. By increasing the variance, LOA approximations deviate from those of Monte Carlo, whereas approximations of the CDF method stay in agreement with those of Monte Carlo. }
	\label{fig:LOA horizontal_phi random}
\end{figure}
%%%%%%%%%------------------------  LOA horizontal------- PHI random--- correlation length
%%%%%%%%%-------------------------------  LOA downdip------- PHI random
%
\begin{figure}[htbp]
	\centering
\includegraphics[width=60mm]{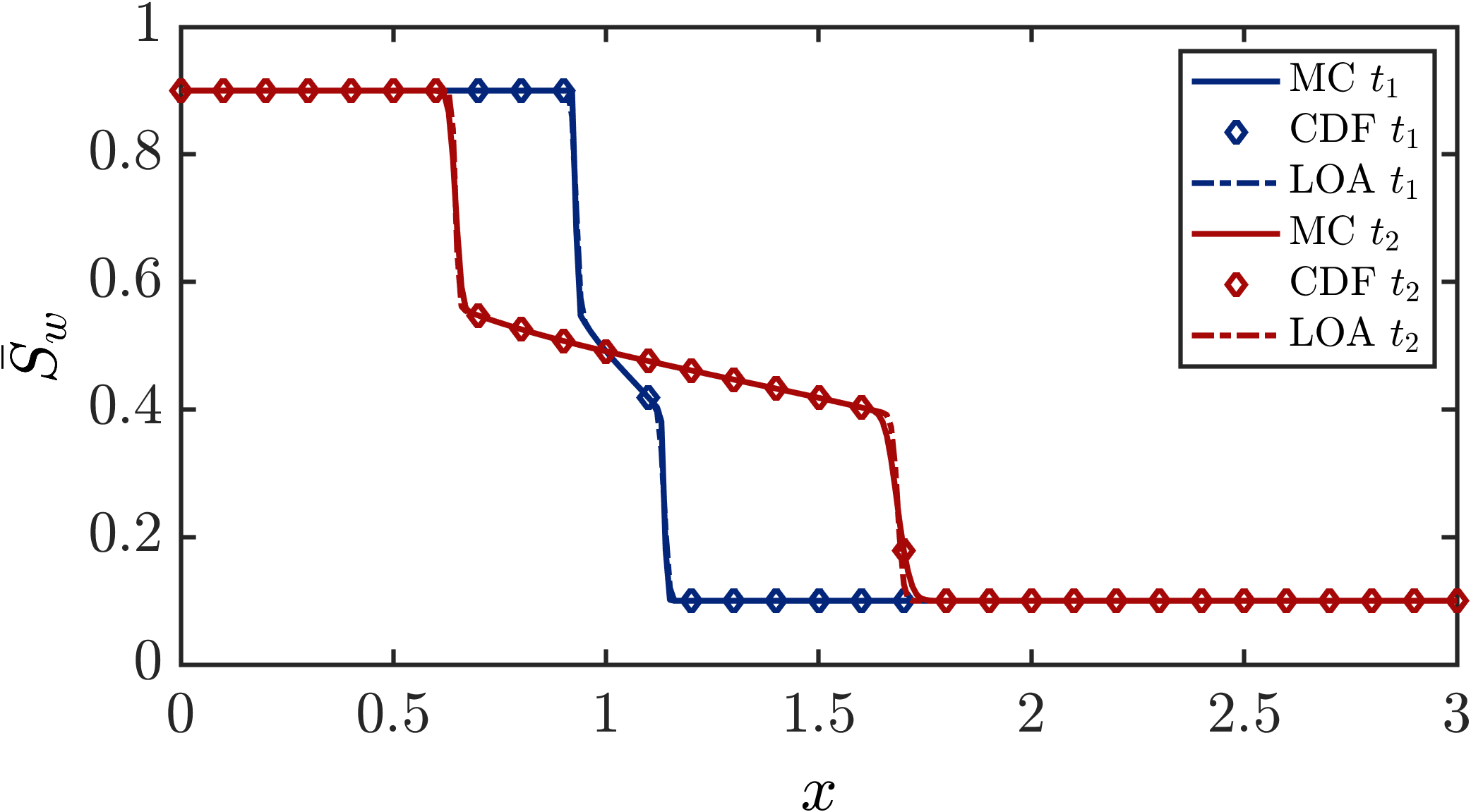}
\includegraphics[width=60mm]{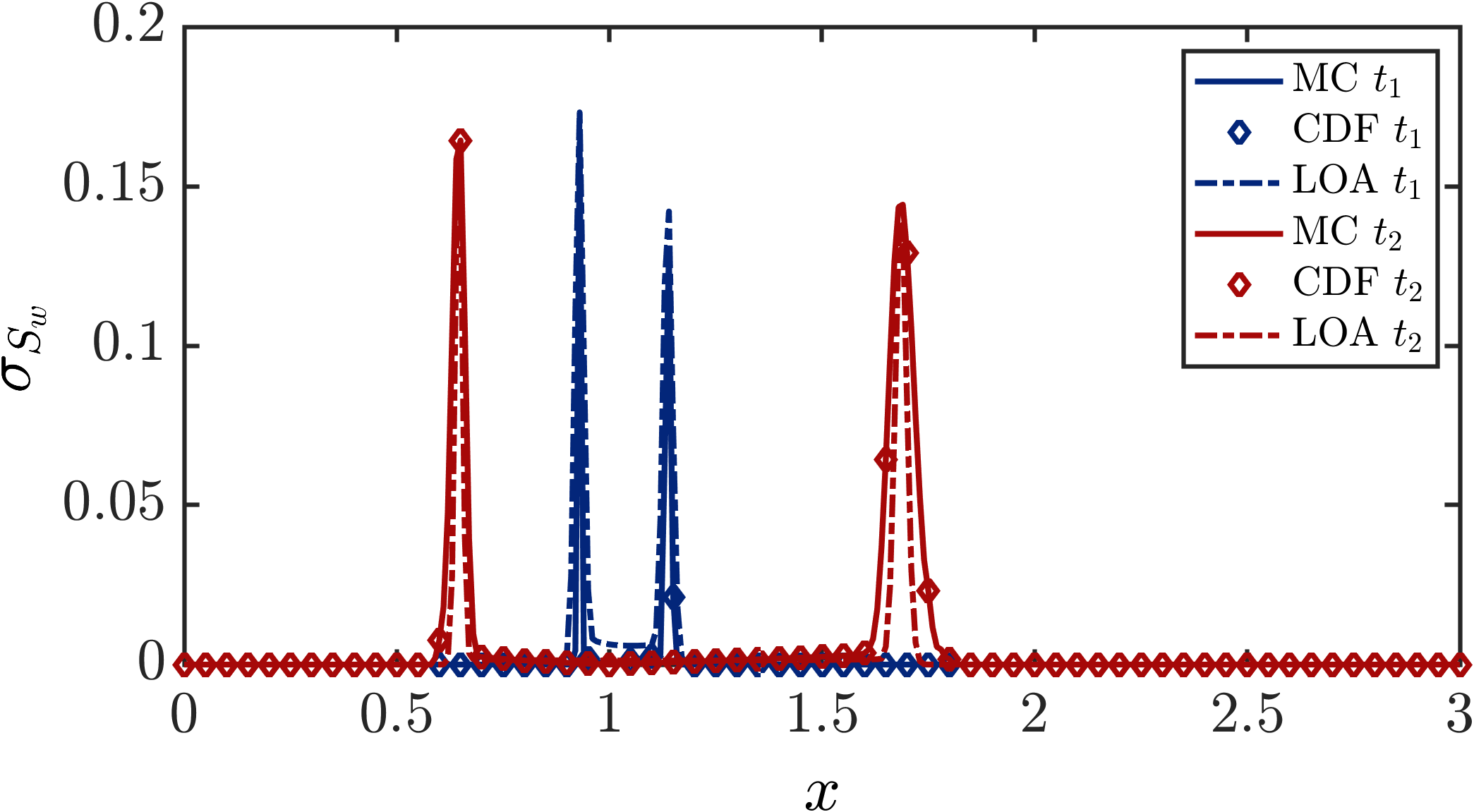}
%  	\caption{SME Downdip, $\phi(x)$ random, $\sigma^2_{\phi}=0.0001$, $\lambda_{\phi}=0.01 L$}
% \includegraphics[width=70mm]{figures/meant1t2_var1e_3.png}
% \includegraphics[width=70mm]{figures/stdt1t2_var1e_3.png}
\includegraphics[width=60mm]{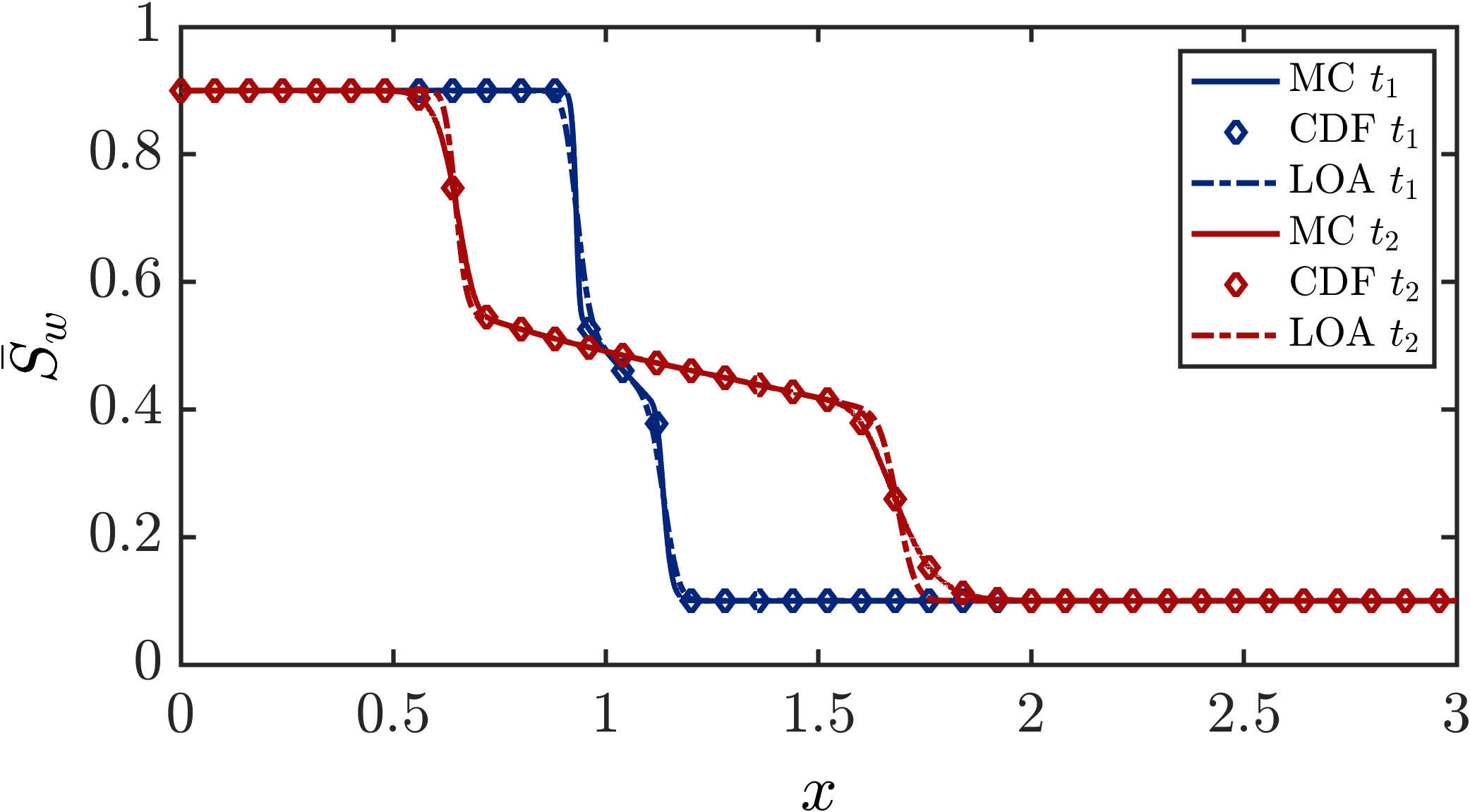}
\includegraphics[width=60mm]{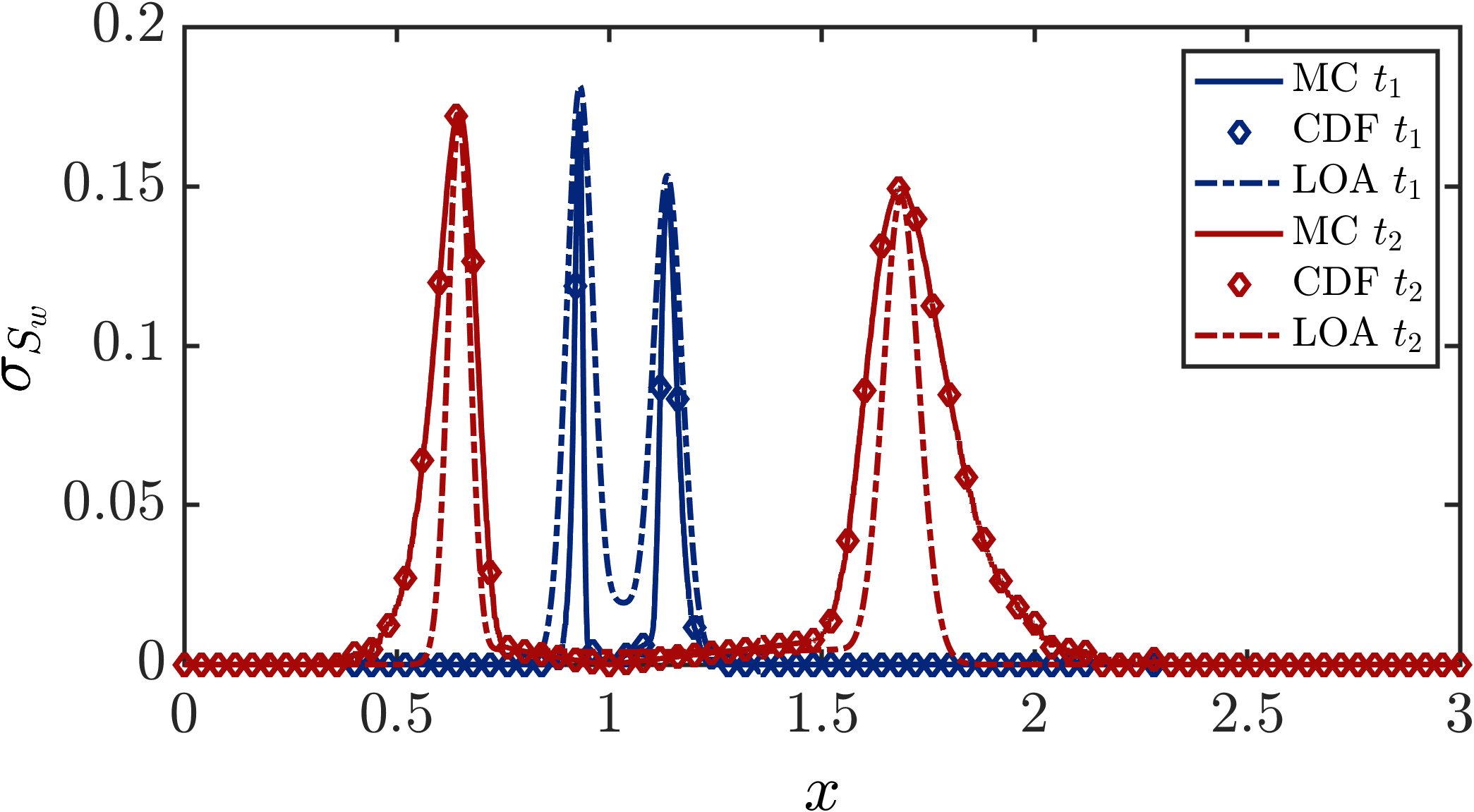}
%  	\caption{SME Downdip, $\phi(x)$ random, $\sigma^2_{\phi}=0.001$, $\lambda_{\phi}=0.01 L$}
 	%%%%%%%%%%%%%%%%%%%%%%%%%%%%%%%%%%%%%%%%%%%%%%%%%%%%%%%%%%%%%%%%
\includegraphics[width=58mm]{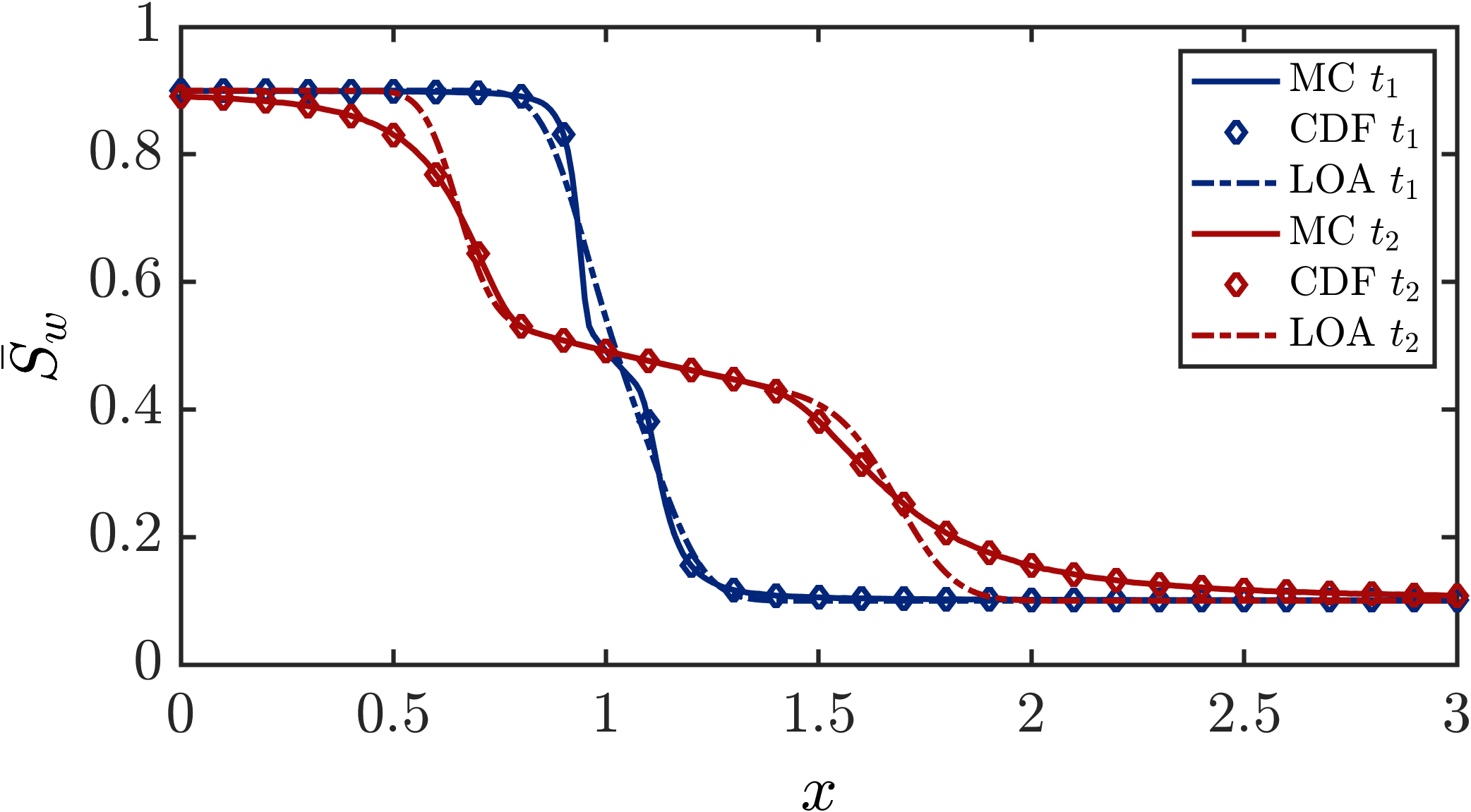}
\includegraphics[width=60mm]{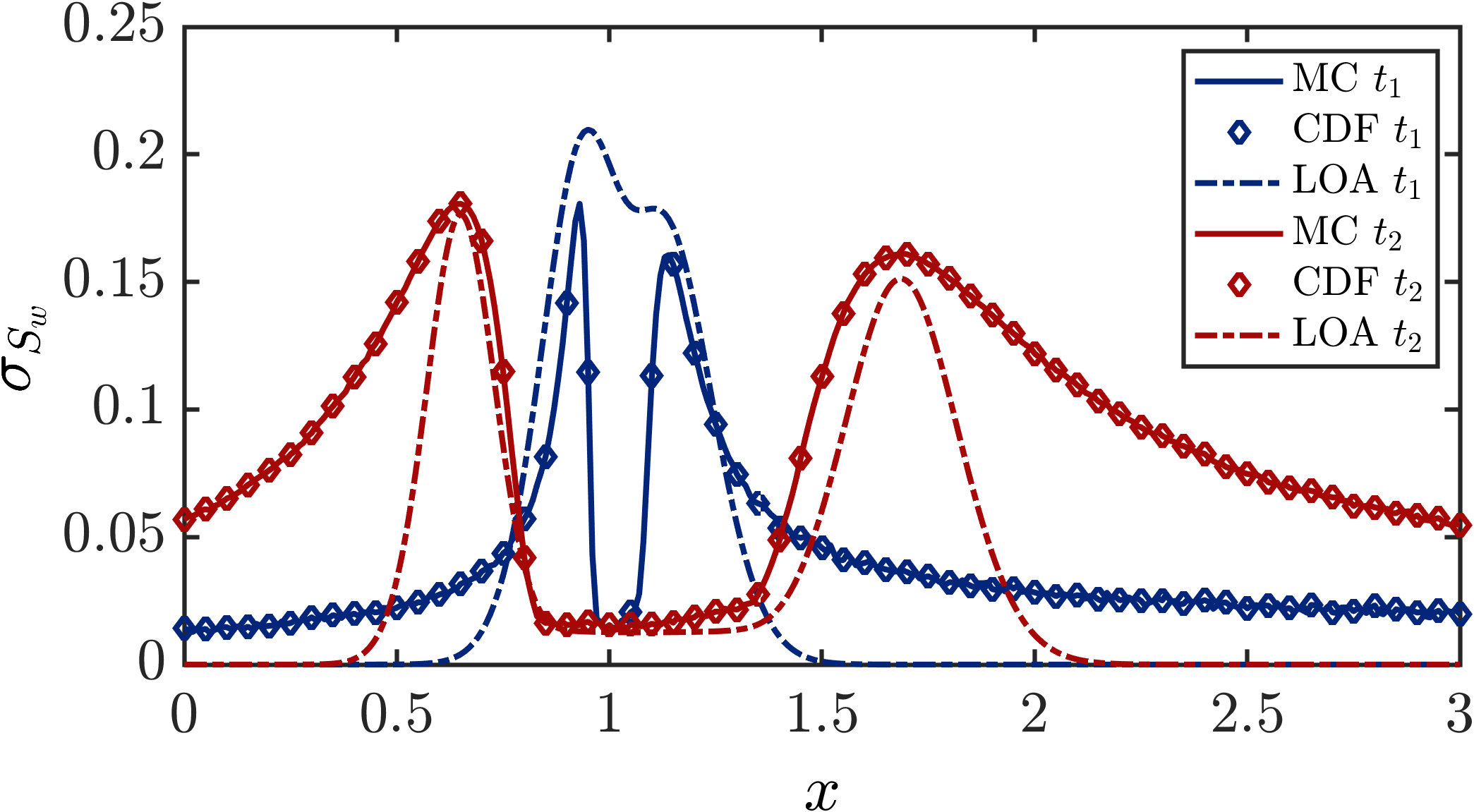}
%  	\caption{SME Downdip, $\phi(x)$ random, $\sigma^2_{\phi}=0.01$, $\lambda_{\phi}=0.01 L$}
	%%%%%%%%%%%%%%%%%%%%%%%%%%%%%%%%%%%%%%
	 	%%%%%%%%%%%%%%%%%%%%%%%%%%%%%%%%%%%%%%%%%%%%%%%%%%%%%%%%%%%%%%%%
\includegraphics[width=58mm]{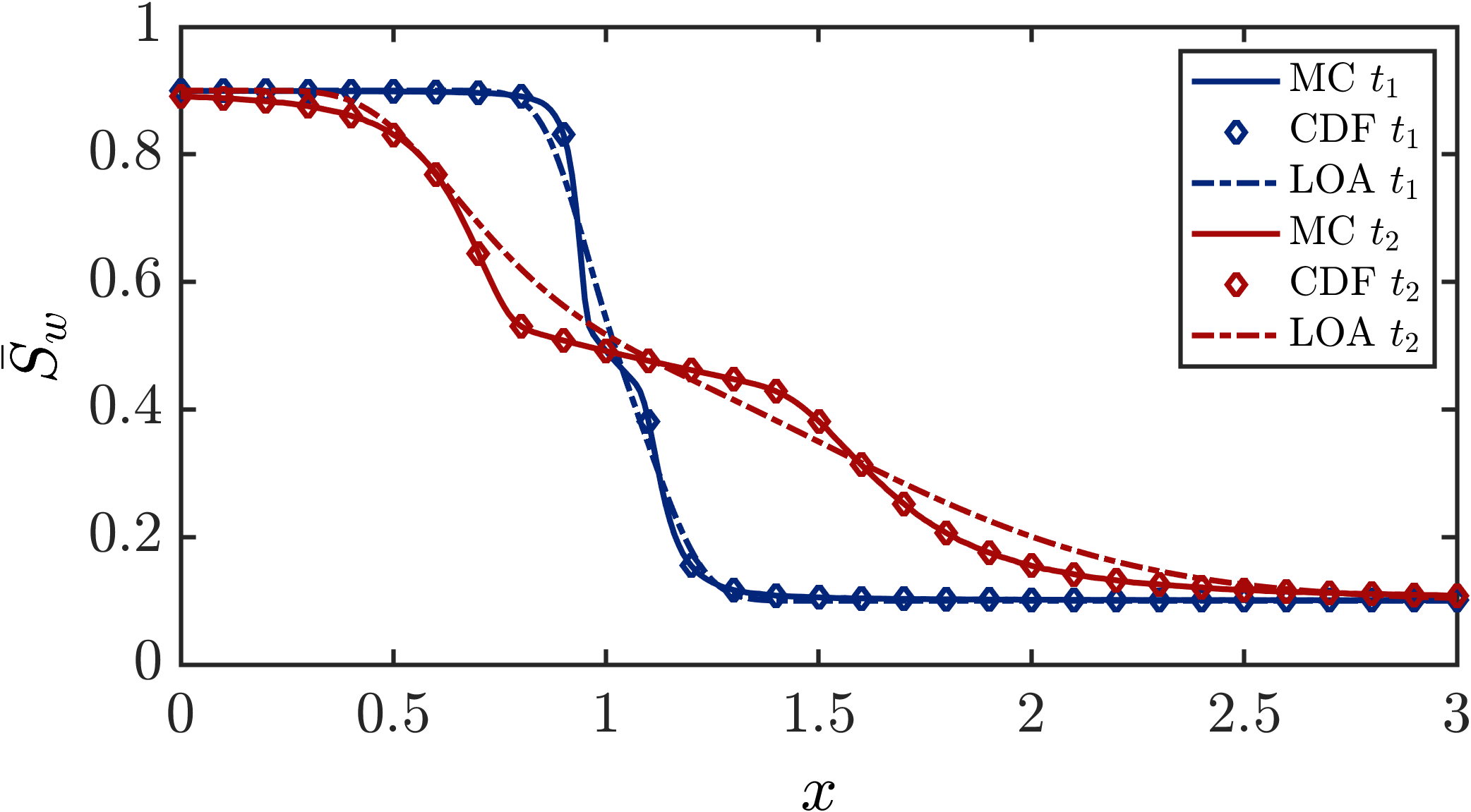}
\includegraphics[width=60mm]{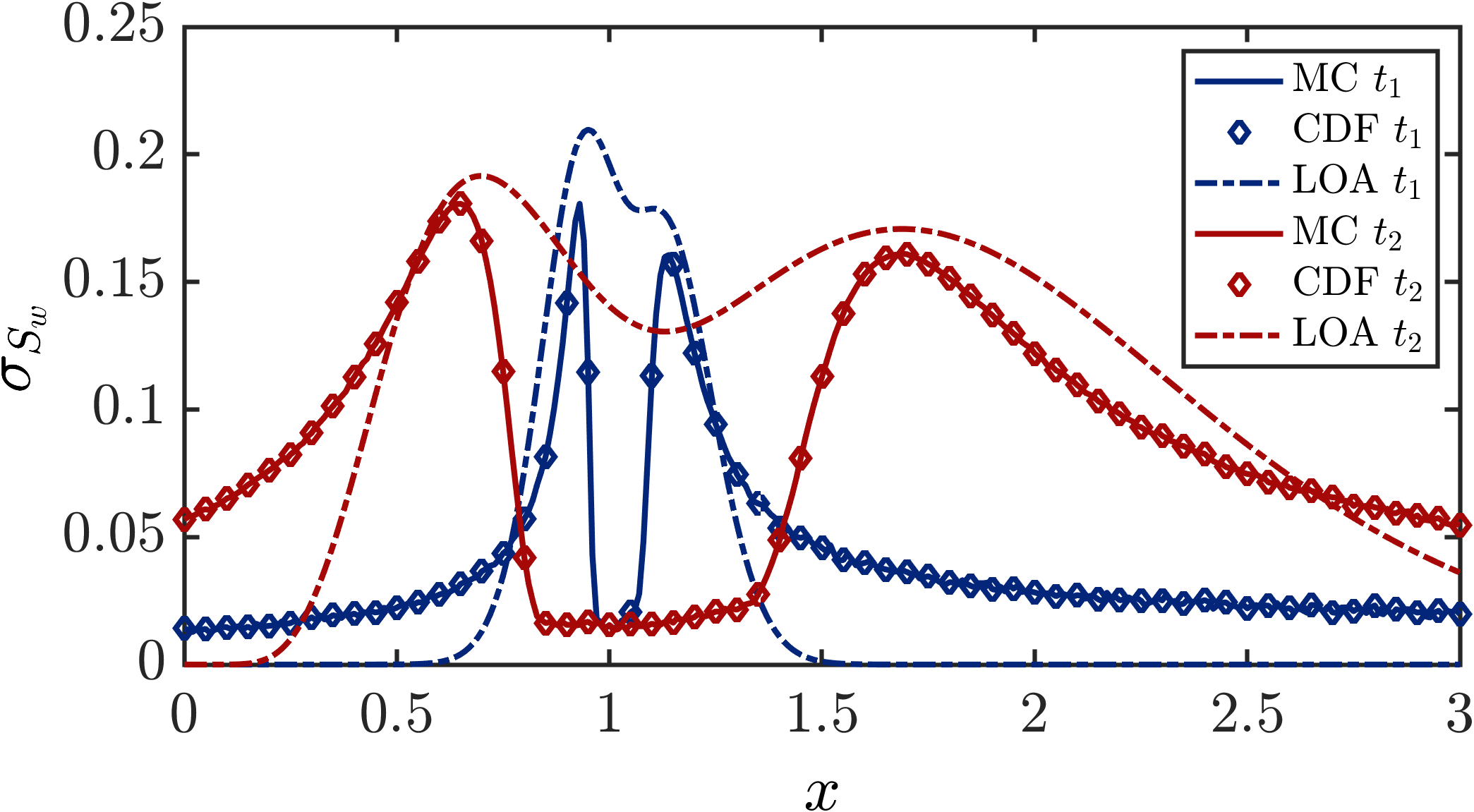}
%  	\caption{SME Downdip, $\phi(x)$ random, $\sigma^2_{\phi}=0.01$, $\lambda_{\phi}=0.1 L$}
	%%%%%%%%%%%%%%%%%%%%%%%%%%%%%%%%%%%%%%
% % \includegraphics[width=60mm,height=45mm]{figures/2_mean_LOA.png}
% % \includegraphics[width=60mm,height=45mm]{figures/2_std_LOA.png}\\
% \includegraphics[width=70mm]{figures/meannn.png}
% \includegraphics[width=70mm]{figures/stddd.png}
%  	\caption{SME Downdip, $\phi(x)$ random, $\sigma^2_{\phi}=0.01$, $\lambda_{\phi}=0.1 L$}
	%%%%%%%%%%%%%%%%%%%%%%%%%%%%%%%%%%%%%%%%%%%%%%%%%%%%%
% 	\includegraphics[width=70mm]{figures/mean_LOAt1t2var1e_2l5e_1.png}
% \includegraphics[width=70mm]{figures/std_LOAt1t2var1e_2l5e_1.png}
\includegraphics[width=60mm]{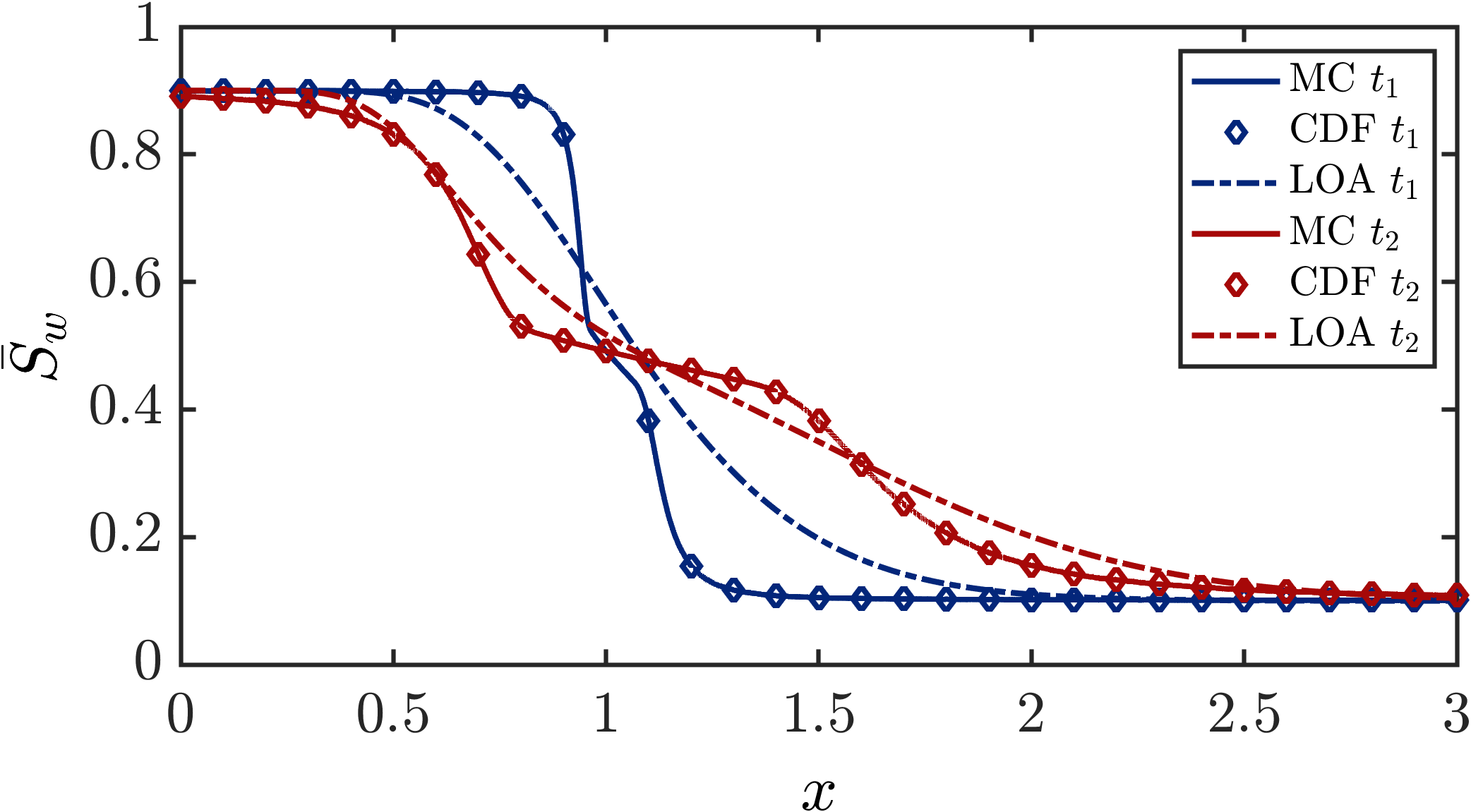}
\includegraphics[width=60mm]{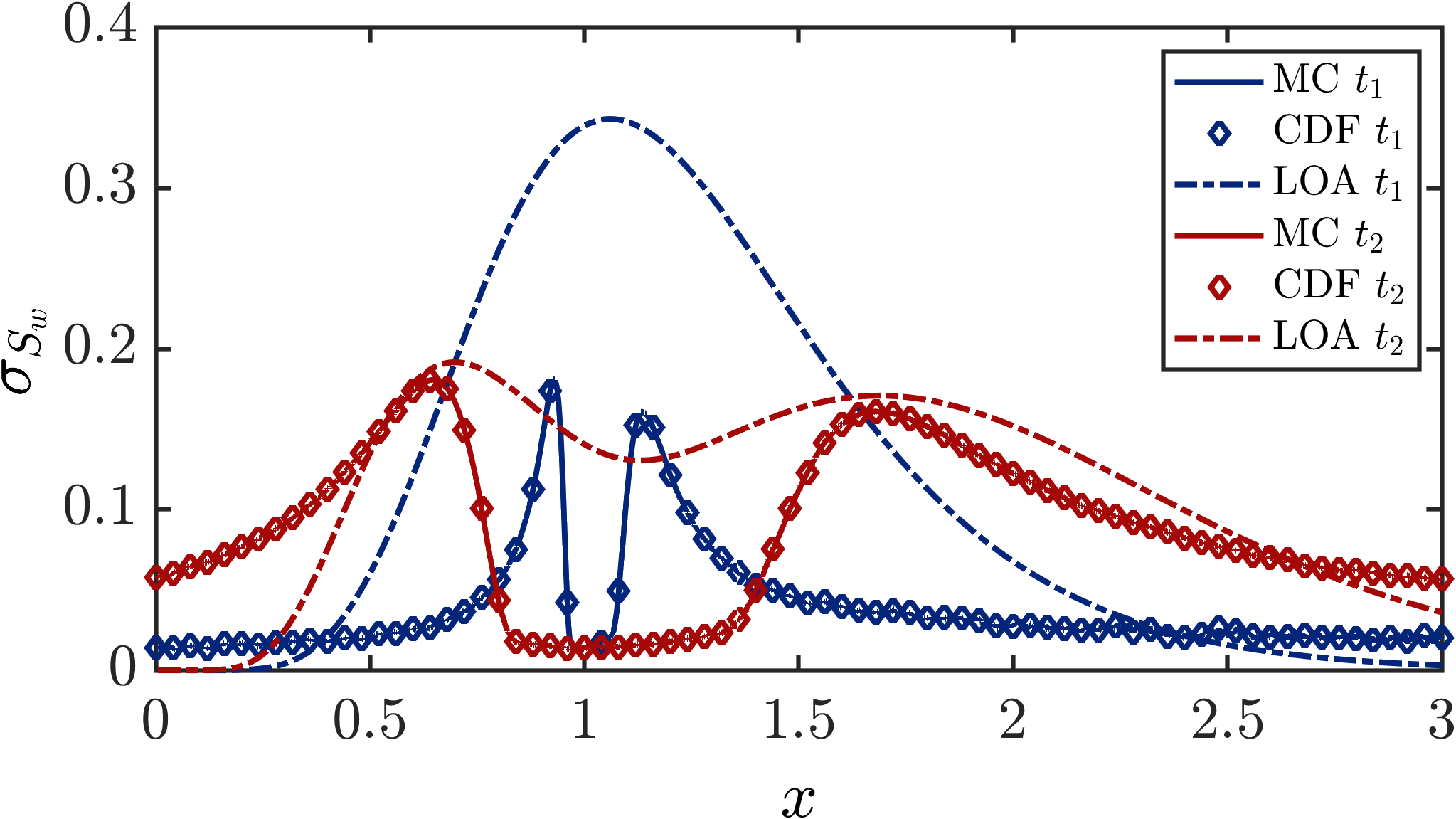}
\caption{Low-order approximation of downdip flooding, $\phi(x)$ random, (1) $\sigma^2_{\phi}=0.0001$, $\lambda_{\phi}=0.01 L$, (2) $\sigma^2_{\phi}=0.001$, $\lambda_{\phi}=0.01 L$,(3) $\sigma^2_{\phi}=0.01$, $\lambda_{\phi}=0.01 L$, (4) $\sigma^2_{\phi}=0.01$,$\lambda_{\phi}=0.1 L$, (5) $\sigma^2_{\phi}=0.01$, $\lambda_{\phi}=0.5 L$, $t_1=0.02, t_2=0.1$}
% , \Delta x=0.01, \Delta t=0.01$
	\label{fig:SME_downdip_phi_random}
\end{figure}
%%%%%%%%%%%%%%%%%%%%%%%%%%%%%%%%%%%%%%%%%%%%%%%%%%%%%%%%%%%%%%%%%%%%%%%%%%%%%%%%%%%%%
%%%                                 CONVERGENCE
%%%%%%%%%%%%%%%%%%%%%%%%%%%%%%%%%%%%%%%%%%%%%%%%%%%%%%%%%%%%%%%%%%%%%%%%%%%%%%%%%%%
\section{Convergence} \label{section: convergence}
\begin{figure}[htbp]
	\centering
\includegraphics[width=58mm]{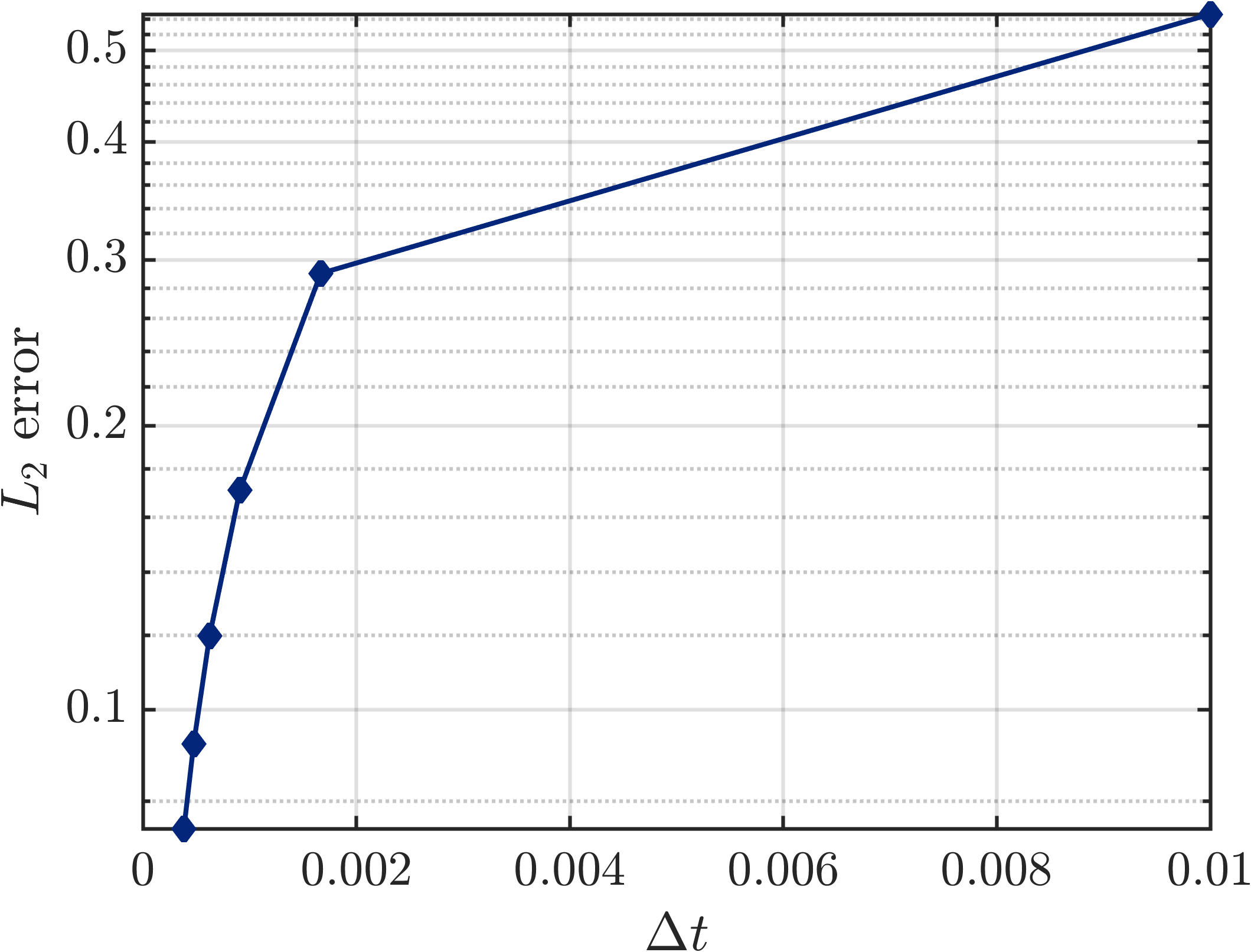}
\includegraphics[width=60mm]{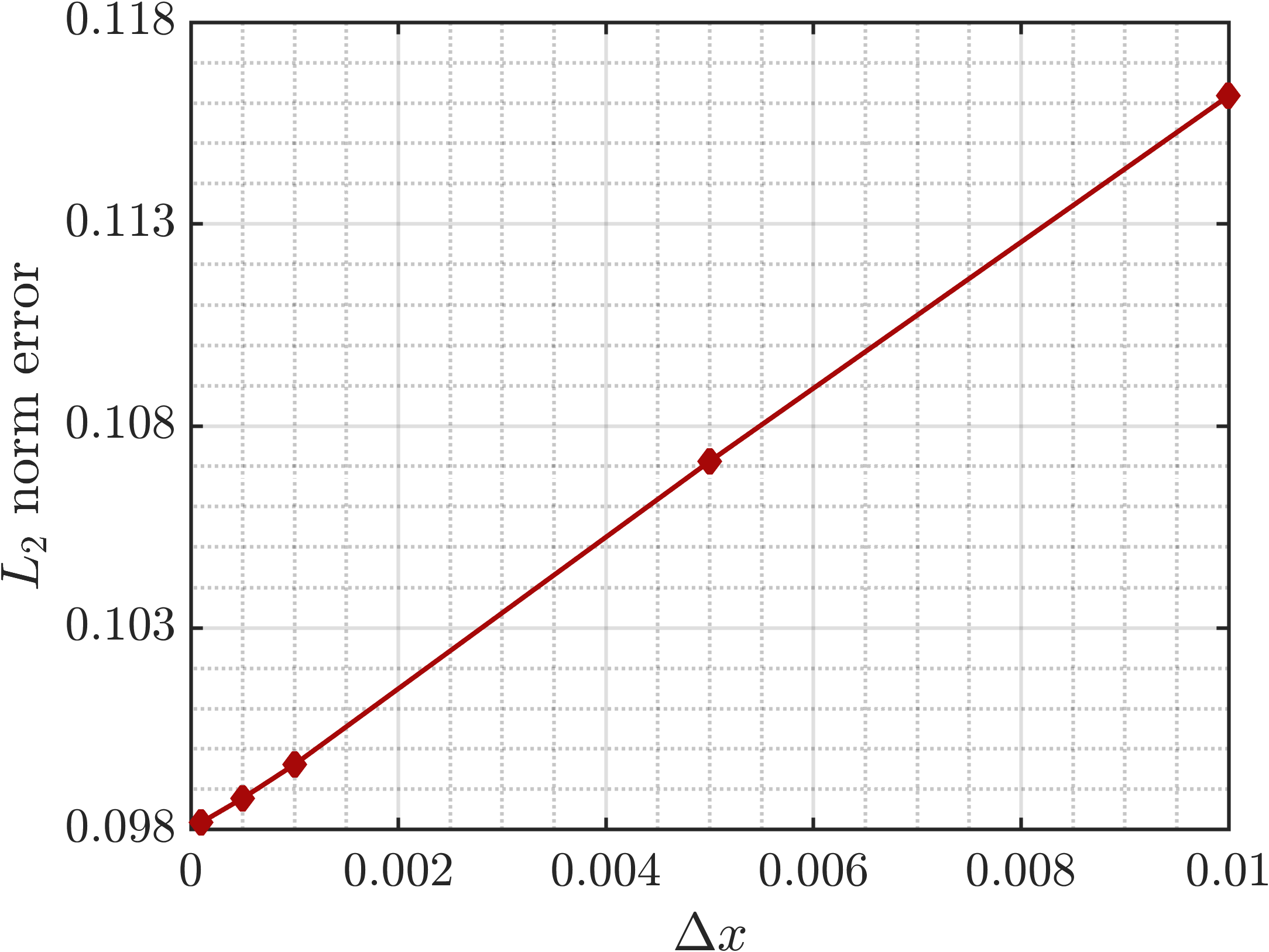}\\

 \includegraphics[width=60mm]{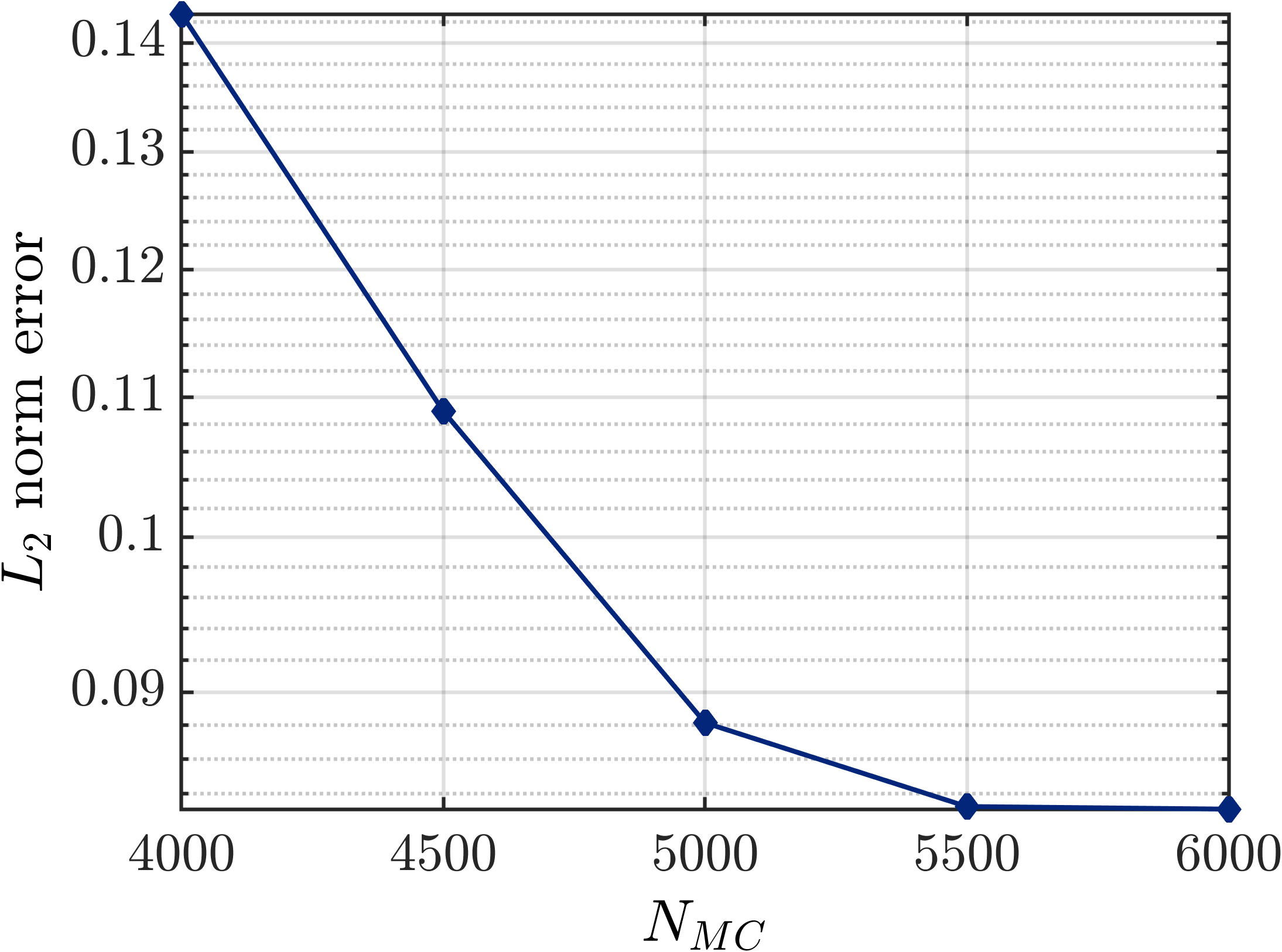}
	\caption{Grid convergence study for temporal grid size at constant $\Delta x= 0.001$ (top left), spatial grid size at constant $\Delta t=0.01$ (top right), and number of MC realizations required for the moments to converge, and for the spatial and temporal evolution of the CDF plots to be smooth.}
% 	\includegraphics[width=60mm]{figures/errorboth2.png}
% 	\includegraphics[width=60mm]{figures/std_downdip_qrandom.png}
% 	\caption{Grid convergence study for temporal grid size at constant $\Delta x= 0.001$ (top left), spatial grid size at constant $\Delta t=0.01$ (top right), and number of MC realizations required for the moments to converge, and for the spatial and temporal evolution of the CDF plots to be smooth.}
	\label{fig:convergence}
\end{figure}

For the MC simulations, $N_{MC} = 3000-5000$ were required for the mean and standard deviation profiles to converge, with a grid size of $\Delta x=0.0005, \Delta t=0.01$. Also, a large number of realizations were essential for the CDF plots to be smooth, and to be a perfectly horizontal line in the shock region. Consequently, both shock and rarefaction regions would be matching with those predicted by the method of characteristics. Regarding the KDE estimations of the CDF of random fields in the MD method, $N=1000-3000$ realization are critical for the resulting $F_s(s;x,t)$ to be matching with the corresponding of the MC benchmark. As Fig. \ref{fig:convergence} represents, by increasing the number of MC samplings, the error will keep declining up to $N_MC=5500$, after which increasing the number of trails was not observed to decrease the error. For the sensitivity of solutions to the spatial and temporal grid sizes, usually for $\Delta x<0.001$ and $\Delta t<0.001$ the lowest error was achieved for most cases. The number of realizations required for MC to converge increases with order of the moment. We experimented with skewness and kurtosis, and even though the results are not included here, they usually need up to $N_{MC}=10^5$ realizations in order to converge.

%
%%%%%%%%%%%%%%%%%%%%%%%%%%%%%%%%%%%%%%%%%%%%%%%%%%%%%%%%%%%%%%%%%%%%%%%%%%%%%%%%%%%
%%%                      ACCURACY AND EFFICIENCY OF THE CDF METHOD
%%%%%%%%%%%%%%%%%%%%%%%%%%%%%%%%%%%%%%%%%%%%%%%%%%%%%%%%%%%%%%%%%%%%%%%%%%%%%%%%%%%
\section{Accuracy and Efficiency of the Method of Distributions} \label{section:accuracy}
% The discontinuities existing in saturation profiles of homogeneous media disappear in the case of heterogeneous media. This is due to heterogeneity fingering, or the so-called "heterogeneity-induced dispersion". This finding is confirmed with Monte Carlo simulations. 
\begin{figure}[htbp]
	\centering
\includegraphics[width=55mm]{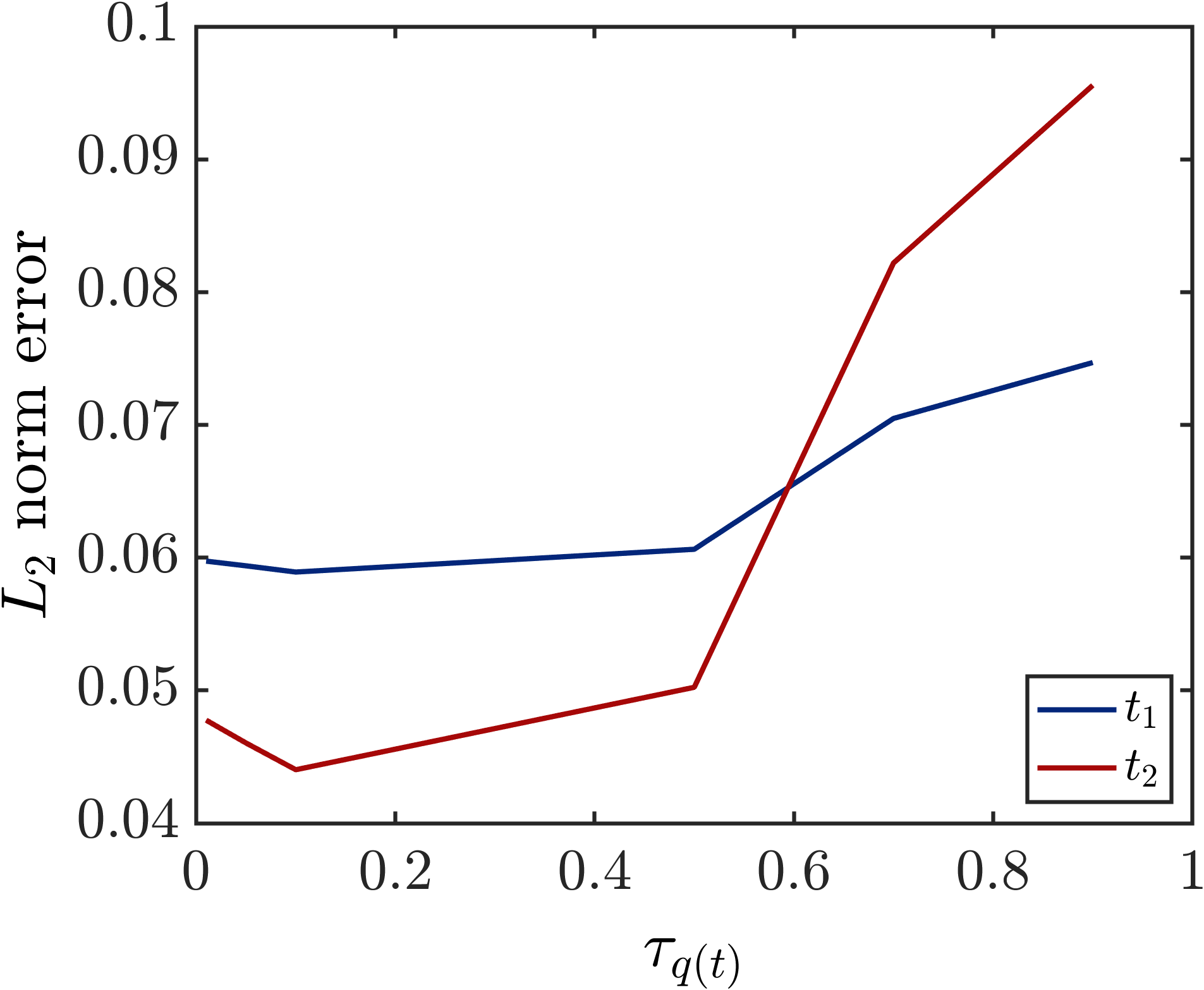}
\includegraphics[width=55mm]{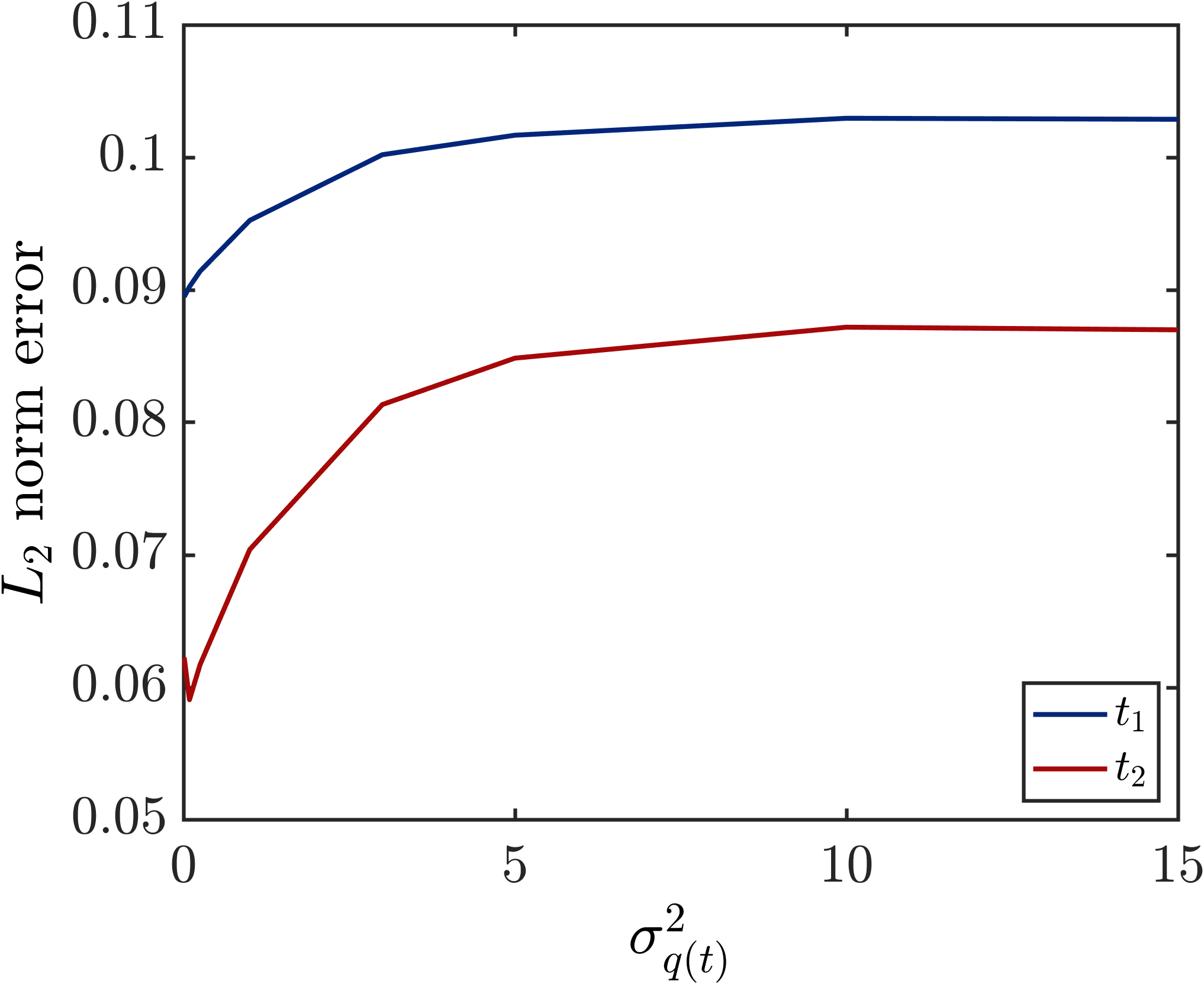}
	\caption{$L_2$ norm error of $F_s(S_w)$ computed at two specific dimensionless times, using Monte Carlo and CDF method for $q(t)$ as the sole source of uncertainty. Sensitivity to the correlation length and variance of the input random variable is represented. }
	\label{fig:caccuraccy q random}
\end{figure}
\begin{figure}[htbp]
	\centering
\includegraphics[width=60mm]{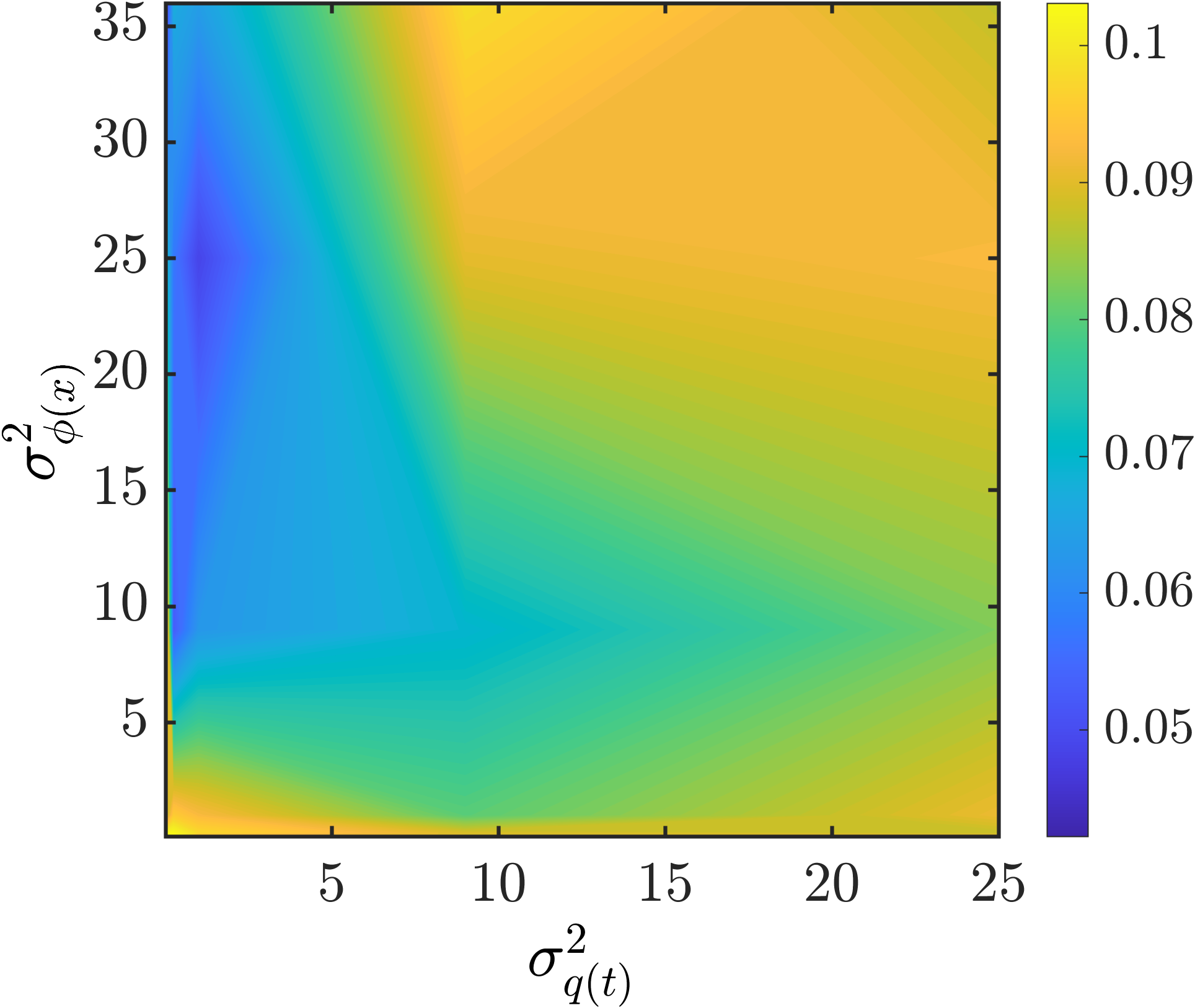}
\hspace{.5cm}
\includegraphics[width=58mm]{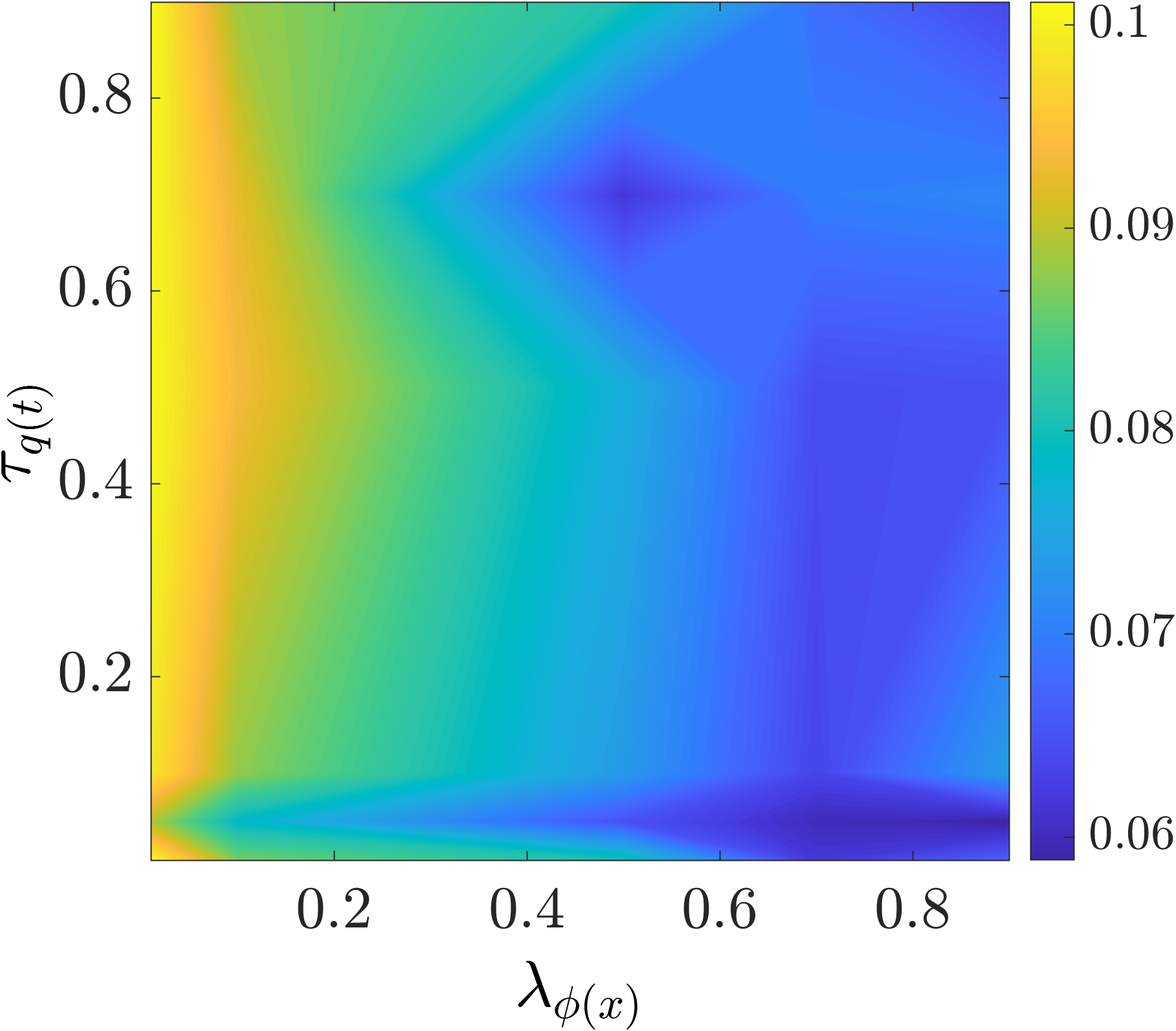}
	\caption{$L_2$ norm error of $F_s(S_w)$ computed at a specific time using Monte Carlo and CDF method for the test cases with both $\phi(x)$ and $q(t)$ random. Sensitivity to the correlation length/time and variances of both input random variables quantifies accuracy of our CDF method. }
	\label{fig:accuracy Q_phi random}
\end{figure}
The $L_2$ norm of error measures the discrepancy between average $F_s(S,x,t)$ along the domain, obtained from Monte Carlo and the corresponding results from the method of distributions. As shown in Fig. \ref{fig:caccuraccy q random} and Fig. \ref{fig:accuracy Q_phi random} . In Fig. \ref{fig:accuracy Q_phi random}, while experimenting with the sensitivity of the $F_s(s)$ to the correlation length/time of the input stochastic fields, for the $\lambda_{\phi(x)} > 0.5 L$ less error is observed. Also, initial values of correlation time, i.e. $\tau_q(t) < 0.5 T$ incur a high error compared to setting the correlation time higher than this threshold. Overall, the errors are usually bounded by $0.1$ for the grid sizes $\Delta x =0.001, \Delta t=0.01$ that we have used in the accuracy experiments. Less error could be obtained if the grid sizes, specially $\Delta x$ was selected to be smaller, however, the MC simulations would take a long time as well. Also, while experimenting  with the values of variances of both input random fields, as Fig. \ref{fig:accuracy Q_phi random} represents, the highest error is observed when both variances are high.
% \begin{align}
%   &\bar{\epsilon} = \dfrac{1}{L}\bigintsss_0^L \vert F_{{s}_{CDF}}(S;x,t) - F_{{s}_{MC}}(S;x,t) \vert dx
% \end{align}

In terms of the efficiency, our method is usually two orders of magnitude faster than the corresponding MC experiment. The only time consuming part in our CDF method, is generating the CDF of input random field, i.e. $F_{q(t)}(t) $, $F_{\phi(x)}(x)$, $F_{q(t)(t),\phi(x)}(x,t)$ using KDE. However, this process is usually two orders of magnitude faster than the Monte Carlo process of generating 5000 samples, solving the nonlinear saturation equation using Godunov method, and then KDE post-processing of the resulting saturation field to find $F_s(s)$. In our CDF method, once the CDF of underlying random field is generated, then a negligible computational cost is associated with the "semi-analytical" computation of the $F_s(s)$ using the formula we derived before. To this end, the CDF method is always considerably faster than corresponding Monte Carlo scheme.
\section{Conclusion} \label{section:conclusion}
We developed a semi-analytical formula for vertical heterogeneous reservoirs. The test cases included horizontal flooding, updip flooding, downdip flooding, and pure gravity segregation in an inverted gravity column, with sealed top and bottom boundaries. For the updip case, the results were very similar to those of the horizontal flooding, except that the rarefaction region is shorter (Fig.~(\ref{fig:frac_flow_curves_with_concave_convex_hulls}), and hence the results are not included here for the sake of brevity). We developed our CDF methodology in one dimension by extending a previous study (\cite{wang2013cdf}) for only horizontal reservoirs with a random injection flux as the sole source of uncertainty, where the analysis was done in the limit of $\sigma_q=0$, i.e. the deterministic case. Our method takes the gravitational forces into account, and hence deals with the more complicated spatio-temporal characteristics when two shocks are present in the problem. Also, we covered all cases of random correlated injection flux and/or random correlated porosity field, for a wide range of correlation lengths/times and variances, with a reasonable accuracy for the CDF $F_s(s)$. The above-mentioned uncertain fields lead to the uncertainty in the state variable $s(x,t)$ described by the nonlinear Buckley Leverett problem, with a random velocity field. The CDF method propagates uncertainty from input to the output state variable by converting the original stochastic nonlinear problem to a deterministic problem for the CDF of saturation, where the uncertainty is encapsulated in the coefficients and initial/boundary conditions in a linear fashion, making the problem more tractable. Having found the CDF of saturation, the full probabilistic description of saturation is available and hence we can easily find the higher order moments, i.e. skewness and kurtosis which provide critical information about tails of the distribution. To the best of our knowledge, this is the first work that converts the nonlinear stochastic multi-shock hyperbolic conservation laws to a linear deterministic equation for the CDF of state variable, and hence the methodology  is able to quantify the uncertainty for the more complicated physical scenarios than single-shock physics in horizontal reservoirs. Additionally, we experimented with both uniform and non-uniform initial conditions in our test cases, and the CDF method was shown to match with the Monte Carlo counterparts for the downdip case and gravity column with non-uniform boundary conditions. Also, no closure approximations were required as the CDF method has its boundary conditions straightforwardly defined. Moreover, we compared our results against the so-called low-order approximation (LOA) of SME method, and the CDF method was presented to match the results of the benchmark Monte Carlo, whereas the SME results were valid only for very small random parameters, i.e. variance and correlation length/time. Our LOA study extends the previous studies done for the horizontal cases to the more general scenarios of gravitational forces included.
We presented the results for the imbibition process, however, we can easily extend this methodology to the drainage process by leveraging the descriptions in the flux function and Fig.~\ref{fig:frac_flow_curves_with_concave_convex_hulls}. In that case, the updip flooding needs treatment of two shocks, while downdip flooding will have only one shock and one rarefaction zone, contrary to the current study. Extension of this work to higher dimensions can be done by manipulating the interaction of waves in multiple dimensions using a kinetic defect term, which satisfies the entropy condition in all dimensions. A comprehensive error analysis confirmed the robustness of the developed method in this work.
%  The degree of boundary flux uncertainty is encapsulated in the coefficient of variation $CV=\dfrac{\sigma_q}{\mu_q}$. For small CV, the saturation profile approaches step function, while for the larger CV, the saturation profile spreads showing higher predictive uncertainty 
%%%%%%%%%%%%%%%%%%%%%%%%%%%%%%%%%%%%%%%%%%%%%%%%%%%%%%%%%%%%%%%%%%%%%%%%%%%%%%%%%%%
%%%                      ACKNOWLEDGEMENTS
%%%%%%%%%%%%%%%%%%%%%%%%%%%%%%%%%%%%%%%%%%%%%%%%%%%%%%%%%%%%%%%%%%%%%%%%%%%%%%%%%%%
% \acknowledgements
\section*{Acknowledgments}
Financial support from Reservoir Simulation Industrial Affiliates Consortium at Stanford University (SUPRI-B) is gratefully acknowledged.

%% appendix sections are then done as normal sections and after Acknowledgements



%\begin{thebibliography}{-8}


%\small{
%\bibitem{key}

%...

%}

%\end{thebibliography}


\bibliographystyle{apalike}

\bibliography{main}

\begin{thebibliography}{}

\bibitem[Aziz, 1979]{aziz1979petroleum}
Aziz, K. (1979).
\newblock Petroleum reservoir simulation.
\newblock {\em Applied Science Publishers}, 476.

\bibitem[Botev et~al., 2010]{botev2010kernel}
Botev, Z.~I., Grotowski, J.~F., Kroese, D.~P., et~al. (2010).
\newblock Kernel density estimation via diffusion.
\newblock {\em The annals of Statistics}, 38(5):2916--2957.

\bibitem[Brenier and Jaffr{\'e}, 1991]{brenier1991upstream}
Brenier, Y. and Jaffr{\'e}, J. (1991).
\newblock Upstream differencing for multiphase flow in reservoir simulation.
\newblock {\em SIAM journal on numerical analysis}, 28(3):685--696.

\bibitem[Brooks and Corey, 1964]{brooks1964hydraulic}
Brooks, R.~H. and Corey, A.~T. (1964).
\newblock Hydraulic properties of porous media.
\newblock {\em Hydrology papers (Colorado State University); no. 3}.

\bibitem[Buckley et~al., 1942]{buckley1942mechanism}
Buckley, S.~E., Leverett, M., et~al. (1942).
\newblock Mechanism of fluid displacement in sands.
\newblock {\em Transactions of the AIME}, 146(01):107--116.

\bibitem[Cheng et~al., 2019]{cheng2019efficient}
Cheng, M., Narayan, A., Qin, Y., Wang, P., Zhong, X., and Zhu, X. (2019).
\newblock An efficient solver for cumulative density function-based solutions
  of uncertain kinematic wave models.
\newblock {\em Journal of Computational Physics}, 382:138--151.

\bibitem[Dongxiao et~al., 1999a]{dongxiao1999stochastic}
Dongxiao, Z., Tchelepi, H., et~al. (1999a).
\newblock Stochastic analysis of immiscible two-phase flow in heterogeneous
  media.
\newblock {\em SPE journal}, 4(04):380--388.

\bibitem[Dongxiao et~al., 1999b]{dongxiao1999moment}
Dongxiao, Z., Winter, C.~L., et~al. (1999b).
\newblock Moment-equation approach to single phase fluid flow in heterogeneous
  reservoirs.
\newblock {\em SPE Journal}, 4(02):118--127.

\bibitem[Fan et~al., 2012]{fan2012fully}
Fan, Y., Durlofsky, L.~J., and Tchelepi, H.~A. (2012).
\newblock A fully-coupled flow-reactive-transport formulation based on element
  conservation, with application to co2 storage simulations.
\newblock {\em Advances in Water Resources}, 42:47--61.

\bibitem[Fuks et~al., 2019]{fuks2019analysis}
Fuks, O., Ibrahima, F., Tomin, P., and Tchelepi, H.~A. (2019).
\newblock Analysis of travel time distributions for uncertainty propagation in
  channelized porous systems.
\newblock {\em Transport in Porous Media}, 126(1):115--137.

\bibitem[Gelhar and Gelhar, 1993]{gelhar1993stochastic}
Gelhar, L.~W. and Gelhar, L. (1993).
\newblock {\em Stochastic subsurface hydrology}, volume 390.
\newblock Prentice-Hall Englewood Cliffs, NJ.

\bibitem[Graham and McLaughlin, 1989]{graham1989stochastic}
Graham, W. and McLaughlin, D. (1989).
\newblock Stochastic analysis of nonstationary subsurface solute transport: 2.
  conditional moments.
\newblock {\em Water Resources Research}, 25(11):2331--2355.

\bibitem[Ibrahima et~al., 2015]{ibrahima2015distribution}
Ibrahima, F., Meyer, D.~W., and Tchelepi, H.~A. (2015).
\newblock Distribution functions of saturation for stochastic nonlinear
  two-phase flow.
\newblock {\em Transport in porous media}, 109(1):81--107.

\bibitem[Ibrahima and Tchelepi, 2017]{ibrahima2017multipoint}
Ibrahima, F. and Tchelepi, H.~A. (2017).
\newblock Multipoint distribution of saturation for stochastic nonlinear
  two-phase transport.
\newblock {\em SIAM/ASA Journal on Uncertainty Quantification}, 5(1):353--377.

\bibitem[Ibrahima et~al., 2018]{ibrahima2018efficient}
Ibrahima, F., Tchelepi, H.~A., and Meyer, D.~W. (2018).
\newblock An efficient distribution method for nonlinear two-phase flow in
  highly heterogeneous multidimensional stochastic porous media.
\newblock {\em Computational Geosciences}, 22(1):389--412.

\bibitem[Jarman and Russell, 2003]{jarman2003eulerian}
Jarman, K.~D. and Russell, T.~F. (2003).
\newblock Eulerian moment equations for 2-d stochastic immiscible flow.
\newblock {\em Multiscale modeling \& simulation}, 1(4):598--608.

\bibitem[Kitanidis, 1988]{kitanidis1988prediction}
Kitanidis, P.~K. (1988).
\newblock Prediction by the method of moments of transport in a heterogeneous
  formation.
\newblock {\em Journal of Hydrology}, 102(1-4):453--473.

\bibitem[Kwok and Tchelepi, 2008]{kwok2008convergence}
Kwok, F. and Tchelepi, H.~A. (2008).
\newblock Convergence of implicit monotone schemes with applications in
  multiphase flow in porous media.
\newblock {\em SIAM Journal on Numerical Analysis}, 46(5):2662--2687.

\bibitem[LeVeque et~al., 2002]{leveque2002finite}
LeVeque, R.~J. et~al. (2002).
\newblock {\em Finite volume methods for hyperbolic problems}, volume~31.
\newblock Cambridge university press.

\bibitem[Li and Tchelepi, 2014]{li2014unconditionally}
Li, B. and Tchelepi, H.~A. (2014).
\newblock Unconditionally convergent nonlinear solver for multiphase flow in
  porous media under viscous force, buoyancy, and capillarity.
\newblock {\em Energy Procedia}, 59:404--411.

\bibitem[Li and Tchelepi, 2015]{li2015nonlinear}
Li, B. and Tchelepi, H.~A. (2015).
\newblock Nonlinear analysis of multiphase transport in porous media in the
  presence of viscous, buoyancy, and capillary forces.
\newblock {\em Journal of Computational Physics}, 297:104--131.

\bibitem[Li et~al., 2013]{li2013influence}
Li, B., Tchelepi, H.~A., and Benson, S.~M. (2013).
\newblock Influence of capillary-pressure models on co2 solubility trapping.
\newblock {\em Advances in water resources}, 62:488--498.

\bibitem[Li et~al., 2009]{li2009efficient}
Li, H., Zhang, D., et~al. (2009).
\newblock Efficient and accurate quantification of uncertainty for multiphase
  flow with the probabilistic collocation method.
\newblock {\em Spe Journal}, 14(04):665--679.

\bibitem[Li et~al., 2005]{li2005conditional}
Li, L., Tchelepi, H.~A., et~al. (2005).
\newblock Conditional statistical moment equations for dynamic data integration
  in heterogeneous reservoirs.
\newblock In {\em SPE Reservoir Simulation Symposium}. Society of Petroleum
  Engineers.

\bibitem[Lie, 2019]{lie2019introduction}
Lie, K.-A. (2019).
\newblock {\em An introduction to reservoir simulation using MATLAB/GNU
  Octave}.
\newblock Cambridge University Press.

\bibitem[Likanapaisal, 2010]{likanapaisal2010statistical}
Likanapaisal, P. (2010).
\newblock {\em Statistical moment equations for forward and inverse modeling of
  multiphase flow in porous media}.
\newblock PhD thesis, Stanford University.

\bibitem[Meyer et~al., 2010]{meyer2010joint}
Meyer, D.~W., Jenny, P., and Tchelepi, H.~A. (2010).
\newblock A joint velocity-concentration pdf method for tracer flow in
  heterogeneous porous media.
\newblock {\em Water Resources Research}, 46(12).

\bibitem[Meyer and Tchelepi, 2010]{meyer2010particle}
Meyer, D.~W. and Tchelepi, H.~A. (2010).
\newblock Particle-based transport model with markovian velocity processes for
  tracer dispersion in highly heterogeneous porous media.
\newblock {\em Water Resources Research}, 46(11).

\bibitem[Meyer et~al., 2013]{meyer2013fast}
Meyer, D.~W., Tchelepi, H.~A., and Jenny, P. (2013).
\newblock A fast simulation method for uncertainty quantification of subsurface
  flow and transport.
\newblock {\em Water Resources Research}, 49(5):2359--2379.

\bibitem[M{\"u}ller et~al., 2013]{muller2013multilevel}
M{\"u}ller, F., Jenny, P., and Meyer, D.~W. (2013).
\newblock Multilevel monte carlo for two phase flow and buckley--leverett
  transport in random heterogeneous porous media.
\newblock {\em Journal of Computational Physics}, 250:685--702.

\bibitem[Orr et~al., 2007]{orr2007theory}
Orr, F.~M. et~al. (2007).
\newblock {\em Theory of gas injection processes}, volume~5.
\newblock Tie-Line Publications Copenhagen.

\bibitem[Pettersson and Tchelepi, 2014]{pettersson2014stochastic}
Pettersson, P. and Tchelepi, H. (2014).
\newblock Stochastic galerkin method for the buckley-leverett problem in
  heterogeneous formations.
\newblock In {\em ECMOR XIV-14th European Conference on the Mathematics of Oil
  Recovery}.

\bibitem[Pope, 1985]{pope1985pdf}
Pope, S.~B. (1985).
\newblock Pdf methods for turbulent reactive flows.
\newblock {\em Progress in energy and combustion science}, 11(2):119--192.

\bibitem[Rubin, 2003]{rubin2003applied}
Rubin, Y. (2003).
\newblock {\em Applied stochastic hydrogeology}.
\newblock Oxford University Press.

\bibitem[Tchelepi, 1995]{tchelepi1995viscous}
Tchelepi, H.~A. (1995).
\newblock Viscous fingering, gravity segregation and permeability heterogeneity
  in two-dimensional and three-dimensional flows.

\bibitem[Van~Genuchten, 1980]{van1980closed}
Van~Genuchten, M.~T. (1980).
\newblock A closed-form equation for predicting the hydraulic conductivity of
  unsaturated soils.
\newblock {\em Soil science society of America journal}, 44(5):892--898.

\bibitem[Wang et~al., 2013]{wang2013cdf}
Wang, P., Tartakovsky, D.~M., Jarman~Jr, K., and Tartakovsky, A.~M. (2013).
\newblock Cdf solutions of buckley--leverett equation with uncertain
  parameters.
\newblock {\em Multiscale Modeling \& Simulation}, 11(1):118--133.

\bibitem[Wang and Tchelepi, 2013]{wang2013trust}
Wang, X. and Tchelepi, H.~A. (2013).
\newblock Trust-region based solver for nonlinear transport in heterogeneous
  porous media.
\newblock {\em Journal of Computational Physics}, 253:114--137.

\bibitem[Winter et~al., 2003]{winter2003moment}
Winter, C.~L., Tartakovsky, D., and Guadagnini, A. (2003).
\newblock Moment differential equations for flow in highly heterogeneous porous
  media.
\newblock {\em Surveys in Geophysics}, 24(1):81--106.

\bibitem[Yang et~al., 2019]{yang2019probabilistic}
Yang, H.~J., Boso, F., Tchelepi, H.~A., and Tartakovsky, D.~M. (2019).
\newblock Probabilistic forecast of single-phase flow in porous media with
  uncertain properties.
\newblock {\em Water Resources Research}.

\bibitem[Yousefzadeh, 2020]{yousefzadeh2020numerical}
Yousefzadeh, M. (2020).
\newblock {\em Numerical Simulation of Fluid-Mineral Interaction and Reactive
  Transport in Porous and Fractured Media}.
\newblock PhD thesis, Stanford University.

\bibitem[Yousefzadeh and Battiato, 2017]{yousefzadeh2017physics}
Yousefzadeh, M. and Battiato, I. (2017).
\newblock Physics-based hybrid method for multiscale transport in porous media.
\newblock {\em Journal of Computational Physics}, 344:320--338.

\bibitem[Zaleski and Panfilov, 2017]{zaleski2017model}
Zaleski, S. and Panfilov, M. (2017).
\newblock Model of kinematic waves for gas--liquid segregation with phase
  transition in porous media.
\newblock {\em Journal of Fluid Mechanics}, 829:659--680.

\bibitem[Zhang et~al., 2000]{zhang2000stochastic}
Zhang, D., Li, L., Tchelepi, H., et~al. (2000).
\newblock Stochastic formulation for uncertainty analysis of two-phase flow in
  heterogeneous reservoirs.
\newblock {\em Spe Journal}, 5(01):60--70.

\end{thebibliography}

\end{document}